\documentclass[a4paper,11pt]{article}%
\usepackage{amssymb}
\usepackage[hiresbb,truetex]{graphicx}
\usepackage{amsbsy}
\usepackage{amsmath}
\usepackage{cite}
\usepackage{slashed}
\usepackage{amsfonts}
\usepackage{hyperref}
\usepackage{xtab}%
\providecommand{\U}[1]{\protect\rule{.1in}{.1in}}
\hypersetup{
colorlinks   = true,
linkcolor    = blue,
citecolor    = blue,
}
\newcommand*{\doi}[1]{\href{{http://dx.doi.org/#1}}{doi: #1}}

\ifx\pdfoutput\relax\let\pdfoutput=\undefined\fi
\newcount\msipdfoutput
\ifx\pdfoutput\undefined\else
\ifcase\pdfoutput\else
\ifx\paperwidth\undefined\else
\ifdim\paperheight=0pt\relax\else\pdfpageheight\paperheight\fi
\ifdim\paperwidth=0pt\relax\else\pdfpagewidth\paperwidth\fi
\fi\fi\fi

\begin{document}
\thispagestyle{empty} \setcounter{page}{0}

\begin{flushright}
June 2026\newline
\end{flushright}

\vskip3.5 true cm

\begin{center}
{\huge The Polymorphic Chiral Anomaly}\\[2.5cm]

{\large Christopher Smith}$^{1}$\vspace{0.5cm}\\[9pt]\smallskip
{\small \textsl{\textit{Laboratoire de Physique Subatomique et de Cosmologie,
}}}\linebreak{\small \textsl{\textit{Universit\'{e} Grenoble-Alpes,
CNRS/IN2P3, Grenoble INP, 38000 Grenoble, France}.}} \\[1.9cm]%
\textbf{Abstract}\smallskip
\end{center}

\begin{quote}
\noindent The chiral anomaly famously manifests in a rich variety of forms,
from abelian and singlet to consistent or covariant. In this paper, all these
realizations are described in detail, along with their properties and
phenomenological applications. Central to this presentation is a novel
expression for the fully generic chiral anomaly, derived with either massive
or massless fermions, that incorporates not only the standard triangle but
also the box and pentagon diagrams. From this master expression, the various
traditional forms of the anomaly are then transparently derived. This provides
a powerful tool, technically and conceptually, driving two further objectives.
First, the topological aspects of each form are dutifully described while
bypassing the differential language entirely, save for Stokes' theorem.
Second, to make sure anyone interested can truly reproduce all the results in
a reasonable amount of time, a FeynCalc implementation of the relevant
calculations is provided. Ultimately, this simplified and unified description
of all the forms of the chiral anomaly highlights the underlying conceptual
beauty, and offers a comprehensive grasp of the physics at play.

\let\thefootnote\relax\footnotetext{\newline$^{1}\;$chsmith@lpsc.in2p3.fr}
\end{quote}

\newpage

\setcounter{tocdepth}{2}

\tableofcontents

\newpage

\section{Introduction}

One of the most profound discoveries of the past century was Emy Noether's
theorem~\cite{Noether:1918zz} relating conservation laws to symmetry
principles. Doing so, they no longer rest on ad-hoc axioms but on the
geometric invariances of the theory, be that over space-time or for some
internal degrees of freedom, and bestows them with manifest robustness. These
laws have to hold if the geometry says so, no matter the physical objects
actually present. In particular, there is a priori no reason for them to be
different for elementary particles or for macroscopic systems.\ This view
dramatically shattered nearly sixty years ago with the discovery that
quantization itself can break symmetries, a phenomenon nowadays called an
anomaly. A posteriori, it should have been expected that the inherent
fuzziness, and at the same time step-wise, nature of the quantum world could
wreak havoc on the delicate geometric invariances. This said, fortunately, not
all classical invariances fall due to quantum effects, and crucial conserved
quantities like the energy or electric charge do survive, but others rather
unexpectedly do not with measurable consequences. This is what makes their
study particularly fascinating, and at times mysterious, requiring to dive
deep into the inner working of symmetries and quantization.

Since the discovery of the chiral anomaly in
1969~\cite{Adler:1969gk,Bell:1969ts}, many others have been identified, the
best known being the scale or trace anomaly~\cite{Coleman:1970je} and its
local version the Weyl anomaly~\cite{Capper:1974ic}, the Witten global $SU(2)$
anomaly~\cite{Witten:1982fp}, the gravitational anomalies arising for fermions
living in curved space-time~\cite{Alvarez-Gaume:1983ihn}, the parity
anomaly~\cite{Redlich:1983dv,Niemi:1983rq,Haldane:1988zza} showing up in low
dimensional system and topological insulators and other discrete symmetry
anomalies~\cite{Ibanez:1991hv,Preskill:1991kd,Banks:1991xj}, while recent
works concentrate on those of extended
symmetries~\cite{Gaiotto:2014kfa,Gaiotto:2017yup} and non-invertible
symmetries~\cite{Costa:2024wks}. They have also found many applications both
in particle phenomenology or in model-building, not least through the
requirement of anomaly cancellation or anomaly matching~\cite{tHooft:1979rat}.
In the present work, we will deal exclusively with the chiral anomaly arising
in the presence of fermions charged under some global and/or gauge symmetries,
in flat space-time. As said, historically, this was the first anomaly
discovered. It is also the one treated in most quantum field theory textbooks
via the famous triangle diagrams, and for which an extensive choice of
excellent reviews is already available (see e.g.
Refs.~\cite{Harvey:2005it,Bilal:2008qx}), so one may wonder what remains to be
said that warrants the present work. We feel this is called for because the
chiral anomaly comes in many guises, which are usually not clearly identified.
Even when they are, their definitions always rest on either rather advanced
mathematics, or results that have been obtained decades ago using somewhat
outdated calculational techniques. The net effect is that these various forms
are often confused, especially in practical applications, and the beauty of
their unified description obscured. So, our main goal here is to fill this gap.

To be specific, to make sense of our table of content, and because the
nomenclature has evolved with time and still varies greatly, let us list the
various forms of the chiral anomaly to be discussed here along with their main properties:

\begin{itemize}
\item \textbf{Abelian}: This is the well-known anomaly of the axial current in
QED, derived from the simple triangle diagram. It is also called the ABJ
anomaly in honor of its discoverers Adler, and Bell and Jackiw in
1969~\cite{Adler:1969gk,Bell:1969ts}. It explains why the $\pi^{0}%
\rightarrow\gamma\gamma$ decay occurs even if light quarks are taken as
massless in a first approximation.

\item \textbf{Singlet}: A simple generalization of the abelian anomaly to a
non-abelian gauge symmetry, with the axial current still being a singlet under
the gauge group. The main novelty is the presence of box and pentagon
diagrams. In the context of QCD, it is also called the axial anomaly, and is
at the root of the large mass of the singlet meson $\eta^{\prime}%
$~\cite{tHooft:1976rip}. Mathematically, it is closely associated with
so-called Chern-Simons forms~\cite{Chern:1974ft}, and thereby to the topology
of the gauge group, the existence of instanton
configurations~\cite{Belavin:1975fg}, the $\theta$
vacua~\cite{Callan:1976je,Jackiw:1976pf}, and the strong CP puzzle. It is also
related via the Atiyah-Singer theorem to the index of the Dirac
operator~\cite{Atiyah:1968mp}.

\item \textbf{Consistent}: This is the anomaly in a gauge current. It is
particularly dangerous since it breaks the gauge symmetry itself. In its
simplest form, it is induced by a single massless Weyl fermion. Being defined
from a gauge variation, it satisfies the Wess-Zumino
condition~\cite{Wess:1971yu}, which is essentially the gauge group algebra. In
the Standard Model or in its extensions, all these anomalies must be absent or
compensate each other since we need the gauge symmetry to ensure predictivity.
Mathematically, the consistent anomaly can be derived from the singlet anomaly
in six dimensions via the so-called Stora-Zumino chain of descent
equations~\cite{Stora:1983ct,Zumino:1983ew}.

\item \textbf{Non-abelian}: The consistent anomaly but in the context of a
$SU(N)_{L}\otimes SU(N)_{R}$ model, in which massive Dirac fermions can live.
Its defining characteristics is, when written in terms of axial and vector
gauge bosons $V_{\mu},A_{\mu}=A_{\mu}^{R}\pm A_{\mu}^{L}$ with $A_{\mu}^{L,R}$
associated to $SU(N)_{L,R}$, to be fully symmetric not only under the separate
permutations of $V_{\mu}$ and $A_{\mu}$, but also under $V_{\mu}%
\leftrightarrow A_{\mu}$ interchanges. Phenomenologically, this anomaly does
not have any other interest than being a necessary step towards the next one.

\item \textbf{Bardeen}: The non-abelian anomaly with the condition that the
vector symmetry is preserved replacing the invariance under $V_{\mu
}\leftrightarrow A_{\mu}$ interchanges~\cite{Bardeen:1969md}. It is still
consistent provided the so-called Bardeen counterterms are added to the
$SU(N)_{L}\otimes SU(N)_{R}$ Lagrangian. When the axial symmetry is
spontaneously broken, it takes the form of the Wess-Zumino-Witten
action~\cite{Wess:1971yu,Witten:1983tw}, and offers a new perspective on the
$\pi^{0}\rightarrow\gamma\gamma$ process. Also, when expressed for a singlet
vector current, it provides a topological interpretation to baryon
number~\cite{Skyrme:1962vh}, via the Goldstone-Wilczek
current~\cite{Goldstone:1981kk}.

\item \textbf{Covariant}: The consistent anomaly is not gauge invariant, and
is expressed in terms of the divergence of a gauge-variant constituent
current. This current can be corrected to make it covariant, thereby defining
the covariant anomaly. These corrections are uniquely defined from the
associated Bardeen-Zumino polynomial~\cite{Bardeen:1984pm}.
Phenomenologically, the electroweak anomaly of the baryon plus lepton number
current, which is essentially vectorial, is of this kind. When the symmetry
group is reducible down to an anomalous global axial symmetry and a conserved
vector gauge symmetry, the covariant anomaly becomes identical to the abelian
or singlet anomaly, thereby closing the cycle of chiral anomalies.
\end{itemize}

Our goal is to present all the topics listed above using the simplest language
possible. The strategy to achieve this rests on the combination of the
following three ingredients. First, it is often stated that the various
incarnations of the chiral anomaly can all be obtained from the calculation of
the same loop diagrams, upon imposing appropriate conditions. There is not
much doubt that this is true, but to our knowledge, it has never been proven
to work in practice. We will present for the first time the full calculation
of the most general form of the chiral anomaly, both for a massless or massive
fermions, and including the box and pentagon diagrams. The calculation is
actually not that complicated, though there are a number of surprises and
interesting issues lurking around. Having that result at hand, it is then easy
to derive all the possible explicit forms the chiral anomaly can take, or any
form in-between since it ends up being more of a continuum of anomalies than
separate stops. From a pedagogical stand-point, this gives a powerful
organizing tool, and provides clear insights into the nature and properties of
each form.

Once identified, it remains to connect these various forms to the usual
definitions. For that, we feel something is missing in the literature. Indeed,
there is always a gap between the diagram calculation, usually limited to that
of the triangle using undergraduate calculation tools, and the discussion of
concepts like the Wess-Zumino consistency condition, Bardeen counterterms,
Chern-Simons forms, or Bardeen-Zumino polynomials, for which a good command of
differential geometry and BRST symmetry appears necessary. Though the native
language of gauge theories is indeed differential, this extra layer of
complexity is not really required in a first approach. As will be seen, most
results can be derived rather simply using the usual tensorial formalism,
partial integration, and Stokes' theorem, which suffices to capture the
essence of the underlying topological structures~\cite{Tao2008}. In this
sense, the situation is not that different than for classical
electromagnetism, for which one also usually avoids differential forms
entirely. Here, sticking to the tensor formalism will bring the formal aspects
of anomalies at the same basic level as diagrams, allowing one to compare the
two and build a solid yet intuitive grasp of the physics at play. Having seen
how these formal concepts materialize in practice, one is then in a good
position to rediscover them using more advanced mathematical machinery.

For our rather pedestrian approach advertised in the previous two points to
make sense, calculations must be tractable and truly reproducible in a
reasonable amount of time. Yet, computing by hand the $d$-dimensional trace of
the ten Dirac matrices occurring for the pentagon diagrams or the gauge
variation of the seven-dimensional Chern-Simons form is certainly not advised.
That is why only the triangle diagram is usually treated, or why one reverts
to the differential formalism. But, modern computers open another route, and
this will be our third ingredient. Specifically, we design our presentation
with the understanding that anyone wishing to do so can (somewhat) easily
perform the calculations with the help of the Mathematica package
FeynCalc~\cite{Mertig:1990an,Shtabovenko:2016sxi,Shtabovenko:2020gxv,Shtabovenko:2023idz}%
, and this in its off-the-shelf version. Further, we do provide a companion
notebook where all the relevant calculations are already set up. In practice,
let us stress that we paid attention to keep the present text self-contained
and we will not refer to this notebook for details, but we do skip
intermediate steps when they amount to executing a trivial set of FeynCalc
commands. At the end, this sets a particular tempo to the presentation that we
think should satisfy both those looking for a qualitative overview of the
subject, those interested only in the original results, or those willing to
dive deep into the intricacies, even for the most advanced concepts.

This work is organized as a pedagogical review, but as said above, it does
contain many original results. The diagrammatic expression of the general
chiral anomaly has never been obtained before, nor its declination into
particular forms. Importantly, this way of deriving them is more than a
technical aspect. It truly manifests a certain point of view on anomalies,
being at their core ambiguities that need to be resolved through physical
conditions, in the same spirit as for the renormalization program. These
ambiguities will be shown to arise in five different ways, from the momentum
routing in loop diagrams, from the initial position of $\gamma_{5}$ in
dimensional regularization, from subtraction points in dispersion relations,
from counterterms or local interactions, or from adding total derivatives to
the Chern-Simons forms, all of which being strictly equivalent. Also, we will
explore in details the difference between the calculation of anomalies for
massless or massive fermions, which requires to go beyond the usual
calculation already for the triangle diagram, and refer to Sutherland-Veltman
theorem\textbf{~}\cite{Sutherland:1967vf,Veltman:1967ceb} to interpret the
various forms of the anomaly. On the mathematical side, the various topics
related to Chern-Simons forms are clearly not original, but their derivation
using a straightforward tensor formalism is absent in the literature. Yet,
these developments provide a coherent basis of reference, using modern
conventions, and do provide clear insights. And finally, last but not least, a
comprehensive number of applications are included, most of which seldom or
never discussed in introductory reviews.

As can be seen from the Table of Content, this paper is organized rather
sequentially, going through the various forms of the chiral anomalies listed
above. We start with the simple abelian and singlet anomalies, for which a
Pauli-Villars regulator is sufficient, and discuss the Sutherland-Veltman
theorem as well as the properties of the box and pentagon diagrams. Then, the
calculation of the most generic chiral anomaly is performed using dimensional
regularization, for a massless fermion. It is this result that is then
declined into the various forms in the following sections. In each case, we
start with the definition of the specific form of the anomaly, followed by its
diagrammatic derivation and main properties, and we end with phenomenological
applications. Those lie somewhat out of the main flow and can be skipped in a
first reading. Being slightly lengthy, the final form of the generic chiral
anomaly is written down explicitly in the Appendix.

\section{The abelian anomaly}

To set the stage, consider the fermionic kinetic term 
$\bar{\psi}(i \slashed \partial-m)\psi$. 
When $m=0$, it is invariant under the global $U(1)_{V}\otimes U(1)_{A}$ 
symmetry, corresponding to $\psi\rightarrow e^{i\alpha}\psi$
and $\psi\rightarrow e^{i\beta\gamma_{5}}\psi$. The associated Noether
currents are $V^{\mu}=\bar{\psi}\gamma^{\mu}\psi$ and $A^{\mu}=\bar{\psi
}\gamma^{\mu}\gamma_{5}\psi$, and both are classically conserved,
$\partial_{\mu}V^{\mu}=\partial_{\mu}A^{\mu}=0$. For a massive fermion, the
vector current remains conserved, but not the axial one since upon using the
Dirac equation, $\partial_{\mu}A^{\mu}=2imP$ with the pseudoscalar current
$P=\bar{\psi}\gamma_{5}\psi$. In quantum field theory, the classical
conservation equations $\partial_{\mu}V^{\mu}=0$ and $\partial_{\mu}A^{\mu
}=2imP$ are called \textbf{Ward identities}. They are to be interpreted within
matrix elements, and lead to infinitely many relationships or identities at
that level.

Let us be slightly more specific. In general, the $U(1)_{V}$ symmetry is made
local and identified with the electromagnetic interactions, leading to the
usual QED Lagrangian
\begin{equation}
\mathcal{L}_{QED}=-\frac{1}{4}F_{\mu\nu}F^{\mu\nu}+\bar{\psi}(i\slashed D-m)\psi\ ,
\end{equation}
where $D^{\mu}=\partial^{\mu}+iQ_{V}A^{\mu}$ and $Q_{V}=-e$ for the electron
(beware not to confuse the axial current and the photon field). As is
well-known, the Ward identity $\partial_{\mu}V^{\mu}=0$ guarantees the
masslessness of the photon, by forcing the vacuum polarization $i\Pi^{\mu\nu
}(q^{2})=\langle A^{\nu}(q)|A^{\mu}(q)\rangle$ to be transverse,
\begin{equation}
q_{\mu}\Pi^{\mu\nu}(q^{2})=0\Rightarrow\Pi^{\mu\nu}(q^{2})=(q^{\mu}q^{\nu
}-g^{\mu\nu}q^{2})\Pi(q^{2})\ .
\end{equation}
By contrast, the axial symmetry is not meant to be local since it is already
broken classically by the mass term. At best, it could be an approximate
global symmetry.

Historically, such a situation was encountered in early studies of the strong
force. To construct an effective theory involving nucleons and pions, those
were assigned into an $SU(2)$ doublet $\psi_{N}$ and triplet $\vec{\pi}$,
respectively. This isospin symmetry is global, and allows for either an axial
or pseudoscalar $\pi\bar{N}N$ coupling. Both are equivalent classically, as
can be seen integrating by part and using the Dirac equation 
$i \slashed \partial\psi_{N}=m\psi_{N}$:
\begin{equation}
\partial_{\mu}\vec{\pi}\cdot\bar{\psi}_{N}\gamma^{\mu}\gamma_{5}\vec{\sigma
}\psi_{N}\overset{\text{Classical}}{=}2im_{N}\vec{\pi}\cdot\bar{\psi}%
_{N}\gamma_{5}\vec{\sigma}\psi_{N}\ , \label{PionAxial}%
\end{equation}
where $\vec{\sigma}$ are the Pauli matrices. This is nothing but the classical
Ward identity $\partial_{\mu}A^{\mu}=2imP$. While it is indeed satisfied for
tree-level matrix elements, with e.g.
\begin{equation}
\langle\bar{N}N|\partial_{\mu}\vec{\pi}\cdot\bar{\psi}_{N}\gamma^{\mu}%
\gamma_{5}\vec{\sigma}\psi_{N}|\vec{\pi}\rangle=2im_{N}\langle\bar{N}%
N|\vec{\pi}\cdot\bar{\psi}_{N}\gamma_{5}\vec{\sigma}\psi_{N}|\vec{\pi}%
\rangle\ ,
\end{equation}
for on-shell nucleons, it fails when computing $\pi^{0}\rightarrow\gamma
\gamma$ via a nucleon loop,
\begin{equation}
\langle\gamma\gamma|\partial_{\mu}\vec{\pi}\cdot\bar{\psi}_{N}\gamma^{\mu
}\gamma_{5}\vec{\sigma}\psi_{N}|\pi^{0}\rangle\ll2im_{N}\langle\gamma
\gamma|\vec{\pi}\cdot\bar{\psi}_{N}\gamma_{5}\vec{\sigma}\psi_{N}|\pi
^{0}\rangle\ .
\end{equation}
Only the pseudoscalar coupling is able to account for the observed $\pi
^{0}\rightarrow\gamma\gamma$ rate. The situation is even more puzzling in the
quark picture, since in a first approximation, those can be taken as massless
but $\pi^{0}\rightarrow\gamma\gamma$ must nevertheless be non-zero since it is observed.

The only way out is to accept that at the loop level, where true quantum field
effects begin to be felt, the classical Ward identity associated to the local
$U(1)_{V}$ symmetry survives, but not that associated to the global $U(1)_{A}%
$. In other words, the latter symmetry must not survive quantization: it must
have what is called an \textbf{anomaly}. The purpose of this section is to
show that when carefully calculated, the loop amplitude indeed predicts an
extra term in the axial Ward identity. We will do this first for the simple
abelian $U(1)_{V}$ and $U(1)_{A}$ in the present section, and then generalize
to the case of a non-abelian, but still vectorial, gauge symmetry in the next section.

\subsection{ABJ triangles}

\begin{figure}[t]
\centering\includegraphics[width=0.60\textwidth]{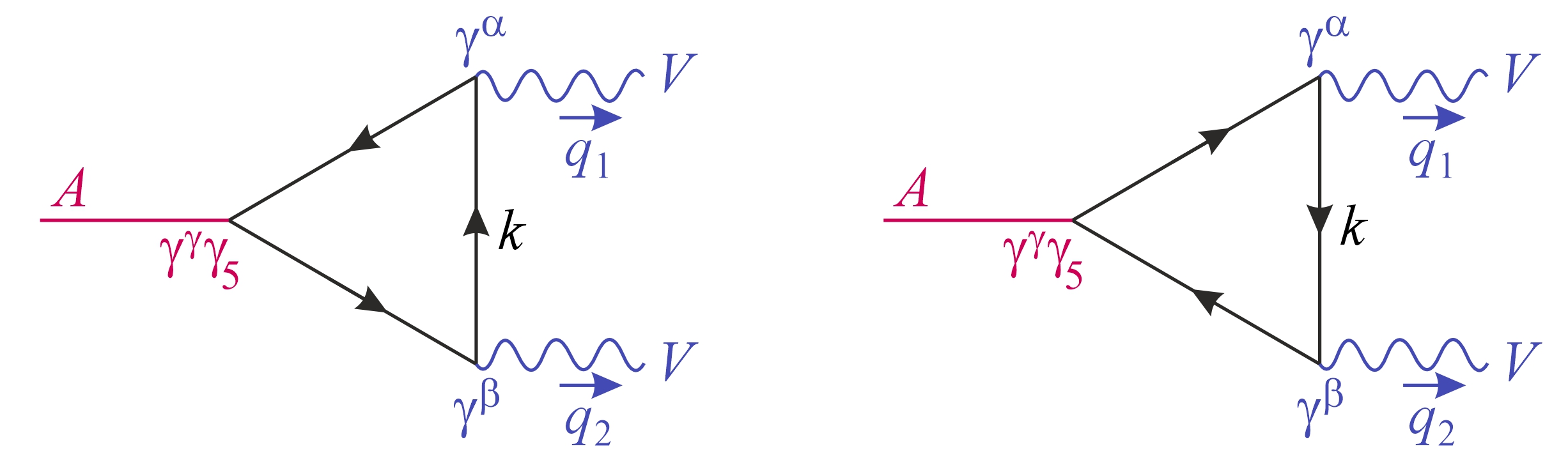}\caption{The
triangle diagrams corresponding to Eq.~(\ref{TAVV}).}%
\label{Fig1}%
\end{figure}

If $\partial_{\mu}A^{\mu}=2imP$ were true, it would remain so in any process,
and in particular for $A\rightarrow VV$ which arises via a fermion loop with
$V$ the vector current to which photons are coupled in QED. Specifically,
consider the diagrams in Fig.~\ref{Fig1}, with $q_{1}$ and $q_{2}$ the
outgoing momenta of the two photons, whose amplitude is (setting $e=1$)%
\begin{equation}
\mathcal{T}_{AVV}^{\gamma\alpha\beta}=\int\frac{d^{4}k}{(2\pi)^{4}%
}(-1)\operatorname*{Tr}\left[  \frac{i}{\slashed k-\slashed q_{1}-
\slashed q_{2}-m}(-i\gamma^{\beta})\frac{i}{\slashed k- \slashed
q_{1}-m}(-i\gamma^{\alpha})\frac{i}{\slashed k-m}\gamma^{\gamma}
\gamma_{5}\right]  +(1,\alpha\leftrightarrow2,\beta)\ \ .
\label{TAVV}
\end{equation}
This amplitude is not immediately UV finite and needs to be regulated. We
choose here the \textbf{Pauli-Villars regularization}~\cite{Pauli:1949zm}, so
we add the same two triangle loops with $m$ replaced by a large $M$, intended
to be sent to infinity at the end. Doing the Dirac algebra in $d=4$
dimensions, the calculation proceeds without difficulty, and we find a finite
result for the divergence%
\begin{equation}
i(q_{1}+q_{2})_{\gamma}\mathcal{T}_{AVV}^{\gamma\alpha\beta}=-\frac{m^{2}%
C_{0}(m^{2})-M^{2}C_{0}(M^{2})}{\pi^{2}}\varepsilon^{\alpha\beta\rho\sigma
}q_{1,\rho}q_{2,\sigma}\overset{M\rightarrow\infty}{=}-\frac{2m^{2}C_{0}%
(m^{2})+1}{2\pi^{2}}\varepsilon^{\alpha\beta\rho\sigma}q_{1,\rho}q_{2,\sigma
}\ ,
\end{equation}
where $C_{0}(m^{2})$ is the three-point scalar loop function obeying%
\begin{equation}
\underset{m\rightarrow\infty}{\lim}C_{0}(m^{2})=\frac{-1}{2m^{2}%
}\ ,\ \ \underset{m\rightarrow0}{\lim}m^{2}C_{0}(m^{2})=0\ .
\end{equation}
Sending the regulator mass to infinity has left a finite term, and the second
identity above shows that it survives in the $m\rightarrow0$ limit. This
contradicts the classical Ward identity $\partial_{\mu}A^{\mu}=0$. By
contrast, one can check that
\begin{equation}
-i(q_{1})_{\alpha}\mathcal{T}_{AVV}^{\gamma\alpha\beta}=-i(q_{2})_{\beta
}\mathcal{T}_{AVV}^{\gamma\alpha\beta}=0\ , \label{VTAVV}%
\end{equation}
showing that the vector Ward identity is preserved%
\begin{equation}
\partial_{\mu}V^{\mu}=0\ , \label{WardV}%
\end{equation}
and the QED gauge symmetry is maintained. In practice, if we couple the axial
current to some physical particle $\pi$ like in Eq.~(\ref{PionAxial}), the
spurious photon polarization states cancel out when computing the total
$\pi\rightarrow\gamma\gamma$ decay rate thanks to Eq.~(\ref{VTAVV}).

Let us now repeat the same calculation with $P=\bar{\psi}\gamma_{5}\psi$ in
place of $A^{\gamma}=\bar{\psi}\gamma^{\gamma}\gamma_{5}\psi$. Still using a
Pauli-Villars regulator, the amplitude is easily computed and we find:%
\begin{equation}
\mathcal{T}_{PVV}^{\alpha\beta}=i\frac{mC_{0}(m^{2})-MC_{0}(M^{2})}{2\pi^{2}%
}\varepsilon^{\alpha\beta\rho\sigma}q_{1,\rho}q_{2,\sigma}%
\overset{M\rightarrow\infty}{=}i\frac{mC_{0}(m^{2})}{2\pi^{2}}\varepsilon
^{\alpha\beta\rho\sigma}q_{1,\rho}q_{2,\sigma}\ . \label{TPVV}%
\end{equation}
This time, the regulator term disappears in the limit $M\rightarrow\infty$.
Putting things together, we can write%
\begin{equation}
i(q_{1}+q_{2})_{\gamma}\mathcal{T}_{AVV}^{\gamma\alpha\beta}=2im\mathcal{T}%
_{PVV}^{\alpha\beta}-\frac{1}{2\pi^{2}}\varepsilon^{\alpha\beta\rho\sigma
}q_{1,\rho}q_{2,\sigma}\ . \label{AnoAbelian}%
\end{equation}

At one loop, the classical Ward identity $\partial_{\mu}A^{\mu}=2imP$ is thus
not satisfied. There is an extra term, whose origin is in the need to regulate
the superficially divergent axial loop amplitude. Putting this result in
operator form and inserting back a $e$ coupling for each vector field, the
anomalous Ward identity satisfied by the axial current is%
\begin{equation}
\partial_{\mu}A^{\mu}=2imP-\frac{e^{2}}{8\pi^{2}}F_{\mu\nu}\tilde{F}^{\mu\nu
}\ , \label{AnoAbelian2}%
\end{equation}
where $\tilde{F}^{\mu\nu}=\varepsilon^{\mu\nu\rho\sigma}F_{\rho\sigma}/2$ and
we have used $\langle\gamma(q_{1},\alpha)\gamma(q_{2},\beta)|F_{\mu\nu}%
\tilde{F}^{\mu\nu}|0\rangle=4\varepsilon_{\rho\alpha\sigma\beta}(-iq_{1}%
^{\rho})(-iq_{2}^{\sigma})\varepsilon_{q_{1}}^{\ast,\alpha}\varepsilon_{q_{2}%
}^{\ast\beta}$. This is the famous result of Adler~\cite{Adler:1969gk}, Bell
and Jackiw~\cite{Bell:1969ts} obtained in 1969, known as the \textbf{ABJ
anomaly}, also called the \textbf{abelian anomaly}. It was proven soon after
that Eq.~(\ref{AnoAbelian2}) actually captures the whole
anomaly~\cite{Adler:1969er}: this equation remains identical at all orders
(though parameters get renormalized). In the following, we will explore this
result in more details, and rederive it using a range of different theoretical
tools. From that, it will become clear that the anomaly is not tied to a
specific calculation method, but truly represent a fundamental physical
effect: the incompatibility of a symmetry with quantization itself.

\subsection{UV divergences and surface terms}

The previous derivation using the Passarino-Veltman loop integral reduction is
so straightforward that it masks one important feature. The anomaly actually
comes from a surface term, and that term depends on the chosen regularization
scheme. Specifically, let us directly take the divergence of the loop
amplitude in Eq.~(\ref{TAVV}), and split
\begin{equation}
i(\slashed q_{1}+ \slashed q_{2})\gamma_{5}=i(\slashed k-m)\gamma_{5}+i\gamma_{5}
(\slashed k- \slashed q_{1}- \slashed q_{2}-m)+2im\gamma_{5}\ . \label{PropSplit}
\end{equation}
Doing a similar trick for the $(1,\alpha\leftrightarrow2,\beta)$ triangle, the
$2im\gamma_{5}$ terms alone reproduce $\mathcal{T}_{PVV}^{\alpha\beta}$ and
account for the $2imP$ in Eq.~(\ref{AnoAbelian2}). The other terms are now all
two-propagator loop amplitudes, which can be grouped as
\begin{equation}
i(q_{1}+q_{2})_{\gamma}\mathcal{T}_{AVV}^{\gamma\alpha\beta}=2im\mathcal{T}%
_{PVV}^{\alpha\beta}+\Delta_{12}^{\alpha\beta}+\Delta_{21}^{\beta\alpha}\ ,
\end{equation}
with%
\begin{equation}
\Delta_{ij}^{\mu\nu}=\int\frac{d^{4}k}{(2\pi)^{4}}\operatorname*{Tr}\left[
\frac{i}{\slashed k- \slashed q_{i}- \slashed q_{j}-m}\gamma^{\mu}
\frac{i}{\slashed k- \slashed q_{j}-m}\gamma^{\nu}\gamma_{5}-
\frac{i}{\slashed k- \slashed q_{i}-m}\gamma^{\mu}
\frac{i}{\slashed k-m}\gamma^{\nu}\gamma_{5}\right]  \ . \label{SurfT}%
\end{equation}
If we could shift $k\rightarrow k+q_{j}$ in the first term, it would cancel
the second, and $\Delta_{12}^{\alpha\beta}=\Delta_{21}^{\beta\alpha}=0$. The
problem though is that these integrals are divergent, and such shifts leave
behind a surface term. This is the origin of the anomalous term in the Ward
identity, as demonstrated in most textbooks (see e.g.
Ref.~\cite{Weinberg:1996kr}). A one-dimensional example shows this
intuitively: if one integrates two copies of some bell-shape function $f(x)$,
but displaces their center by some amount $a$, then one gets the same result
only once the integration range extends beyond the reach of the function. If
the integral of $f(x)$ diverges, no finite integration range suffices. In
practice, one needs to regularize the function, for example set a cut-off at
large $|x|$, and the integral over the interval $[-\Lambda,\Lambda]$ depends
on the position of the function within that interval.

The surface term depends on the momentum mismatch between the two triangles,
and this makes the situation way worse. Indeed, we can choose at will
the loop momentum routing of each triangle of Fig.~\ref{Fig1}, so the momentum
shift, and thus the anomalous term, is actually totally ambiguous! Later, this
ambiguity will be worked out in detail, and we will see that the generic triangle
amplitude depends on two free real parameters. Now, this does not mean all
predictivity is lost, but that anomalies have to be dealt with in a way very
similar to how UV divergences are treated. Those do not make QED unpredictive because
they are absorbed into the free parameters of the theory, as encoded by physical
renormalization conditions. Similarly, even if the anomalous diagrams are finite,
some physical conditions have to be prescribed on 
$\mathcal{T}_{AVV}^{\gamma\alpha\beta}$ or its derivatives to ensure predictivity. 
This is implicitly what was done to get the non-ambiguous result for the abelian
anomaly in Eq.~(\ref{AnoAbelian2}). By adopting a Pauli-Villars regulator, we
made sure the vector symmetry is preserved, see Eq.~(\ref{VTAVV}), and as will
be proven in due time, this is sufficient to fix all ambiguities. So, anomalies
represent a kind of middle ground, between finite loop processes like for
example that for the anomalous magnetic moment $g-2 = \alpha/\pi$, and infinite
loop diagrams like the vacuum polarization and the electron self-energy, whose
infinities are absorbed into the QED free parameters and field normalizations.

\subsection{Sutherland-Veltman theorem\label{SVT}}

In the $m\rightarrow\infty$ limit, the whole $AVV$ amplitude vanishes
identically since the $m$ and $M$ contributions cancel each other,
\begin{equation}
i(q_{1}+q_{2})_{\gamma}\mathcal{T}_{AVV}^{\gamma\alpha\beta}%
\overset{m\rightarrow\infty}{=}0\ .
\end{equation}
This is a manifestation of the \textbf{Sutherland-Veltman theorem~}%
\cite{Sutherland:1967vf,Veltman:1967ceb}. Indeed, consider the most general
form for the matrix element $\langle\gamma\gamma|A\rangle$. Because of parity,
it must involve the $\varepsilon$ tensor, so Lorentz invariance predicts the
structure
\begin{align}
\langle\gamma(q_{1},\alpha)\gamma(q_{2},\beta)|A^{\gamma}(q_{3})\rangle &
=\varepsilon_{\alpha,1}^{\ast}\varepsilon_{\beta,2}^{\ast}\times\left(
\frac{{}}{{}}f_{1}\varepsilon^{\alpha\beta\rho\sigma}q_{1,\rho}q_{2,\sigma
}q_{1}^{\gamma}+f_{2}\varepsilon^{\alpha\gamma\rho\sigma}q_{1,\rho}%
q_{2,\sigma}q_{1}^{\beta}+f_{3}\varepsilon^{\alpha\beta\gamma\sigma
}q_{1,\sigma}\right.  \nonumber\\
& \ \ \ \ \ \ \ \ \ \ \ \ \ \ \left.  +f_{4}\varepsilon^{\alpha\beta\rho
\sigma}q_{1,\rho}q_{2,\sigma}q_{2}^{\gamma}+f_{5}\varepsilon^{\beta\gamma
\rho\sigma}q_{1,\rho}q_{2,\sigma}q_{2}^{\alpha}+f_{6}\varepsilon^{\alpha
\beta\gamma\sigma}q_{2,\sigma}\frac{{}}{{}}\right)  \ ,
\end{align}
where we have discarded $q_{i}^{\mu}\varepsilon_{\mu,i}^{\ast}=0$. If we set
$q_{1}^{2}=q_{2}^{2}=0$ but leave $q_{3}^{2}=(q_{1}+q_{2})^{2}=2q_{1}\cdot
q_{2}$ free, the form-factors are unknown functions $f_{i}\equiv f_{i}%
(q_{3}^{2})$. This general form is then reduced by imposing vector gauge
invariance (i.e., the term in parathesis must vanish when contracted with
either $q_{1,\alpha}$ or $q_{2,\beta}$) and Bose symmetry under the
interchange of the photons $(q_{1},\alpha)\leftrightarrow(q_{2},\beta)$:
\begin{align}
\langle\gamma(q_{1},\alpha)\gamma(q_{2},\beta)|A^{\gamma}(q_{3})\rangle &
=f_{1}(q_{1}+q_{2})^{\gamma}\varepsilon^{\alpha\beta\rho\sigma}q_{1,\rho
}q_{2,\sigma}\varepsilon_{\alpha,1}^{\ast}\varepsilon_{\beta,2}^{\ast
}\nonumber\\
&  +f_{2}((\varepsilon^{\gamma\beta\rho\sigma}q_{2}^{\alpha}-\varepsilon
^{\gamma\alpha\rho\sigma}q_{1}^{\beta})q_{1,\rho}q_{2,\sigma}-q_{1}\cdot
q_{2}\varepsilon^{\alpha\beta\gamma\sigma}(q_{1}-q_{2})_{\sigma}%
)\varepsilon_{\alpha,1}^{\ast}\varepsilon_{\beta,2}^{\ast}\ .\ \label{SVFF}%
\end{align}
Now, contracting with $q_{3,\gamma}$ to take the divergence $\partial_{\gamma
}A^{\gamma}$, we find that
\begin{equation}
\langle\gamma(q_{1},\alpha)\gamma(q_{2},\beta)|\partial_{\gamma}A^{\gamma
}(q_{3})\rangle=q_{3}^{2}(f_{1}+f_{2})\varepsilon^{\alpha\beta\rho\sigma
}q_{1,\rho}q_{2,\sigma}\varepsilon_{\alpha,1}^{\ast}\varepsilon_{\beta
,2}^{\ast}\ .\label{SVFF2}%
\end{equation}
This matrix element thus vanishes linearly if $q_{3}^{2}\rightarrow0$, or
equivalently if $m^{2}\rightarrow\infty$ since what really matters is the
dimensionless ratio $q_{3}^{2}/m^{2}$. The only caveat is for the form-factor
to be regular functions of $q_{3}^{2}$, i.e., they should not have IR
singularities like poles in $1/q_{3}^{2}$. Though we will see in
Sec.~\ref{IRUV} that something deep is hidden here, for now, this appears as a
perfectly reasonable assumption which obviously holds for the $AVV$ amplitude
of Eq.~(\ref{TAVV}).

This theorem leads to a possible source of confusion. Since $\langle
\gamma\gamma|\partial_{\mu}A^{\mu}|0\rangle\rightarrow0$, we also have that
$2im\langle\gamma\gamma|P|0\rangle\rightarrow(\alpha/4\pi)\langle\gamma
\gamma|F\tilde{F}|0\rangle$ when $m\rightarrow\infty$. Parametrically, the
pseudoscalar triangle equals the anomaly in that limit, even though it is not
itself anomalous! Pushing further, one could be tempted to state that the
Sutherland-Veltman theorem is proof that something is missing in the classical
Ward identity, arguing that without the anomaly, $\langle\gamma\gamma
|P|0\rangle$ would be directly related to $\langle\gamma\gamma|\partial_{\mu
}A^{\mu}|0\rangle$ which goes to zero too fast to explain the observed
$\pi^{0}\rightarrow\gamma\gamma$ rate.

This argument is actually wrong, both physically and historically. This
theorem did provide the $\pi^{0}\rightarrow\gamma\gamma$ puzzle from which the
anomaly emerged, but on a different basis. As said at the beginning of this
section, historically, it was realized very early that $\pi^{0}\bar{\psi
}\gamma_{5}\psi$ and $\partial_{\mu}\pi^{0}\bar{\psi}\gamma^{\mu}\gamma
_{5}\psi$ predict very different scaling for $\pi^{0}\rightarrow\gamma\gamma$,
with only the former being able to account (surprisingly well) for the
observed rate if protons run in the loop~\cite{Steinberger:1949wx}. Though
puzzling, this was not viewed as too serious, especially as Schwinger showed
in the early fifties~\cite{Schwinger:1951nm} how to recover the pseudoscalar
result starting from the axial coupling (this will be discussed in
Sec.~\ref{SecPointSplit}). The situation only became critical with further
developments in the description of the strong interactions. By the end of the
sixties, the pion was no longer viewed as a normal pseudoscalar field, but
rather as the Goldstone boson associated to the spontaneous breaking of the
axial part of a chiral symmetry. Not only is the pion massless in a first
approximation, its field is also directly coupled to the axial current via the
Goldstone theorem, $\langle0|A^{\mu}|\pi\left(  q\right)  \rangle=iF_{\pi
}q^{\mu}$. This is known as the PCAC relation, with $F_{\pi}$ related to the
chiral symmetry breaking scale. What Sutherland and Veltman showed is that
PCAC, together with the decomposition in Eq.~(\ref{SVFF}), forbid $\pi
^{0}\rightarrow\gamma\gamma$ to happen at the observed rate. The solution is
to amend the PCAC relation, which is essentially the massless classical Ward
identity, because the axial symmetry does not survive quantization. For more
information, we refer to the excellent historical account in
Ref.~\cite{Adler:2004qt}.

\section{The singlet (axial) anomaly\label{Singlet}}

The previous calculation can be immediately generalized to the case of a
vectorial non-abelian gauge symmetry provided the axial current is a gauge
singlet. In this statement, vectorial means that the usual QED vertices
$-ie\gamma_{\mu}$ are replaced by $-ig\gamma_{\mu}T^{a}$, with $T^{a}$ the
gauge generators in the representation carried by the fermions\footnote{Here,
we switch convention with respect to QED where $D_{\mu}=\partial_{\mu
}+iQA_{\mu}$ with $Q=-e$ the electron electric charge. For non-abelian groups,
we define the covariant derivative as $D_{\mu}=\partial_{\mu}-igA_{\mu}%
^{a}T^{a}$, such that $[D_{\mu},D_{\nu}]=-ig\mathbf{F}_{\mu\nu}$ with
$\mathbf{F}_{\mu\nu}=\partial_{\mu}\mathbf{A}_{\nu}-\partial_{\nu}%
\mathbf{A}_{\mu}-ig[\mathbf{A}_{\mu},\mathbf{A}_{\nu}]$.}. The calculation
proceeds without change but for the occurrence of the quadratic Casimir
invariant $\langle T^{a}T^{b}\rangle=\mathcal{I}_{2}\delta^{ab}$ after
summation over the fermion $SU(N)$ indices, and we find:%
\begin{equation}
i(q_{1}+q_{2})_{\gamma}\mathcal{T}_{AVV}^{\gamma\alpha\beta,ab}=2im\mathcal{T}%
_{PVV}^{\alpha\beta,ab}-\frac{g^{2}}{2\pi^{2}}\mathcal{I}_{2}\delta
^{ab}\varepsilon^{\alpha\beta\rho\sigma}q_{1,\rho}q_{2,\sigma}\ ,
\end{equation}
where the amplitudes are related to those in Eqs.~(\ref{TAVV}) and~(\ref{TPVV}%
) as $\mathcal{T}_{AVV}^{\gamma\alpha\beta,ab}=\mathcal{I}_{2}\delta
^{ab}\mathcal{T}_{AVV}^{\gamma\alpha\beta}$ and $\mathcal{T}_{PVV}%
^{\gamma\alpha\beta,ab}=\mathcal{I}_{2}\delta^{ab}\mathcal{T}_{PVV}%
^{\gamma\alpha\beta}$. The extra term is now called the \textbf{singlet
anomaly}, or in the particular case of QCD, the \textbf{axial anomaly}.

\begin{figure}[t]
\centering\includegraphics[width=0.90\textwidth]{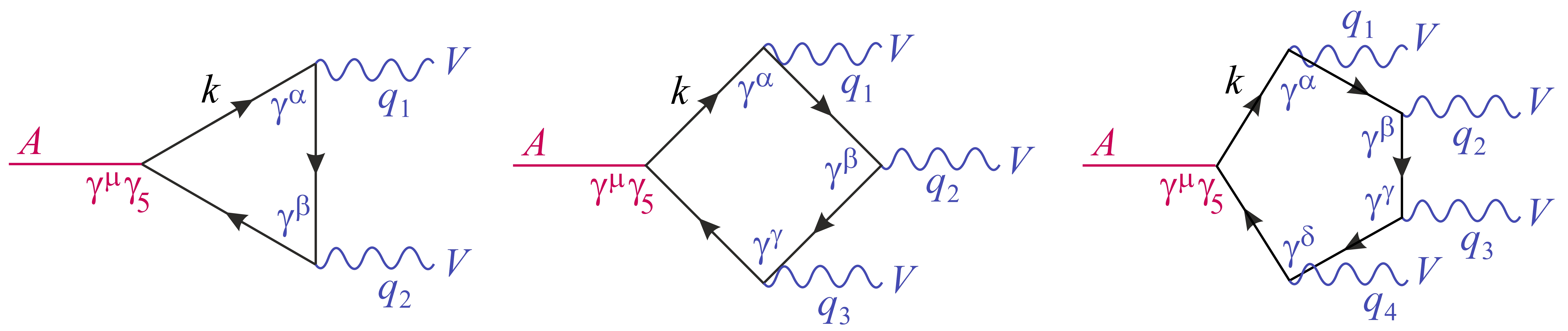}\caption{Main
topologies to be considered for the singlet anomaly. The depicted triangle,
box, and pentagon diagrams are understood to represent the 2, 6, and 24
possible permutations of the vector fields.}%
\label{Fig2}%
\end{figure}

There is, however, a subtlety to construct the operator form of this anomaly.
Since $A^{\mu}=\bar{\psi}^{i}\gamma^{\mu}\gamma_{5}\psi_{i}$ and $P=\bar{\psi
}^{i}\gamma_{5}\psi_{i}$ are gauge invariant (summation over $i=1,...n$ is
understood, with $n$ the dimension of the fermion representation, so there are
actually $2n$ triangle diagrams), we would like to write%
\begin{equation}
\partial_{\mu}A^{\mu}=2imP-\frac{g^{2}}{8\pi^{2}}\langle\mathbf{F}_{\mu\nu
}\mathbf{\tilde{F}}^{\mu\nu}\rangle\ , \label{AnoSinglet1}%
\end{equation}
where $\mathbf{A}_{\mu}=A_{\mu}^{a}T^{a}$ and $\mathbf{F}_{\mu\nu}=F_{\mu\nu
}^{a}T^{a}$. But when the gauge field is non abelian, the field strength is
not linear in the gauge field, $F_{\mu\nu}^{a}=\partial_{\mu}A_{\nu}%
^{a}-\partial_{\nu}A_{\mu}^{a}+gf^{abc}A_{\mu}^{b}A_{\nu}^{c}$, where the
structure constant is defined as usual from $[T^{a},T^{b}]=if^{abc}T^{c}$. The
quartic term vanishes since by cyclicity of the trace, $\langle\mathbf{A}%
_{\mu}\mathbf{A}_{\nu}\mathbf{A}_{\rho}\mathbf{A}_{\sigma}\rangle
=\langle\mathbf{A}_{\sigma}\mathbf{A}_{\mu}\mathbf{A}_{\nu}\mathbf{A}_{\rho
}\rangle$, so that $\varepsilon^{\mu\nu\rho\sigma}\langle\mathbf{A}_{\sigma
}\mathbf{A}_{\mu}\mathbf{A}_{\nu}\mathbf{A}_{\rho}\rangle=-\varepsilon^{\mu
\nu\rho\sigma}\langle\mathbf{A}_{\mu}\mathbf{A}_{\nu}\mathbf{A}_{\rho
}\mathbf{A}_{\sigma}\rangle$, hence $\varepsilon^{\mu\nu\rho\sigma}%
\langle\mathbf{A}_{\mu}\mathbf{A}_{\nu}\mathbf{A}_{\rho}\mathbf{A}_{\sigma
}\rangle=0$. So, Eq.~(\ref{AnoSinglet1}) supposes a coherent presence of the same anomaly in box
diagrams\footnote{We would have preferred to called these \textit{square
diagrams}, but this could have caused confusion when referred to as
\textit{square(d) amplitude}.}, and its absence in the pentagon diagrams, see
Fig.~\ref{Fig2}. As this is rarely presented explicitly, let us now check
that this is indeed the case.

\subsection{Boxes and pentagons}

The calculation of the cubic and quartic terms would be quite cumbersome but
for a few tricks. First, one should understand how the Pauli-Villars
regularization works. Consider first the unregulated loop amplitudes. Though
the exact expressions for the square and pentagon loop amplitudes are very
complicated, their leading term in a $1/m$ expansions are actually simple and
straightforward to compute:
\begin{subequations}
\label{AtoV}%
\begin{align}
\mathcal{T}(A^{\mu}V_{q_{1}}^{\alpha,a}V_{q_{2}}^{\beta,b})  &  =g^{2}%
\frac{i\varepsilon^{\alpha\beta\mu\nu}(q_{1\nu}-2q_{2\nu})}{8\pi^{2}}\langle
T^{b}T^{a}\rangle+\mathcal{O}(m^{-2})\ ,\\
\mathcal{T}(A^{\mu}V_{q_{1}}^{\alpha,a}V_{q_{2}}^{\beta,b}V_{q_{3}}^{\gamma
,c})  &  =g^{3}\frac{i\varepsilon^{\alpha\beta\gamma\mu}}{8\pi^{2}}\langle
T^{c}T^{b}T^{a}\rangle+(5~perm.)+\mathcal{O}(m^{-2})\ ,\\
\mathcal{T}(A^{\mu}V_{q_{1}}^{\alpha,a}V_{q_{2}}^{\beta,b}V_{q_{3}}^{\gamma
,c}V_{q_{4}}^{\delta,d})  &  =\mathcal{O}(m^{-2})\ ,
\end{align}
where $q_{i}$ are the outgoing momenta of the vector fields. The pentagon
amplitude is necessarily of higher order because it needs five indices, and
that requires the presence of momenta in the numerator hence additional powers
of $m$ in the denominator. Notice that to get these amplitudes, we use naive
dimensional regularization, which means performing the algebra in four
dimensions but the loop integration in $d$ dimension (as we did in the
previous section). In particular, we systematically enforce the Schouten
identity,%
\end{subequations}
\begin{equation}
\varepsilon_{\mu\nu\rho\sigma}(k_{1}^{\alpha}k_{2}^{\mu}k_{3}^{\nu}k_{4}%
^{\rho}k_{5}^{\sigma}-k_{2}^{\alpha}k_{1}^{\mu}k_{3}^{\nu}k_{4}^{\rho}%
k_{5}^{\sigma}-k_{3}^{\alpha}k_{2}^{\mu}k_{1}^{\nu}k_{4}^{\rho}k_{5}^{\sigma
}-k_{4}^{\alpha}k_{2}^{\mu}k_{3}^{\nu}k_{1}^{\rho}k_{5}^{\sigma}-k_{5}%
^{\alpha}k_{2}^{\mu}k_{3}^{\nu}k_{4}^{\rho}k_{1}^{\sigma})=0\ ,
\label{Schouten}%
\end{equation}
expressing the fact that five vectors in four dimensions must be linearly
dependent (there are many such identities as some momenta can be replaced by
metric tensors).

The leading $\mathcal{O}(m^{0})$ terms in Eq.~(\ref{AtoV}) are finite but do
depend on our rather peculiar calculation scheme. Changing scheme is in
general a delicate procedure, but not here: to switch to the Pauli-Villars
scheme, we simply need to subtract the same loops with $m\rightarrow M$, and
take the $M\rightarrow\infty$ limit. Importantly, we have to do this even
though the amplitudes happen to be finite. The net effect of this subtraction
is obviously to entirely kill the $\mathcal{O}(m^{0})$ terms in
Eq.~(\ref{AtoV}). With the Pauli-Villars scheme, the regulated amplitudes
automatically start at $\mathcal{O}(1/m^{2})$ and thus necessarily satisfy
Sutherland-Veltman theorem at leading order since the divergence of a null
amplitude obviously gives zero.

This is the key since from Eq.~(\ref{AnoSinglet1}), it is now clear that there
can be no anomaly in the vector currents, while that in the axial current can
be extracted entirely from the $PVV$, $PVVV$, and $PVVVV$ amplitudes by
multiplying their leading $\mathcal{O}(m^{-1})$ term by $2im$. Using the same
regularization technique, the mass expansions are easy to compute
\begin{subequations}
\label{PtoV}%
\begin{align}
\mathcal{T}(PV_{q_{1}}^{\alpha,a}V_{q_{2}}^{\beta,b})  &  =-g^{2}%
\frac{i\varepsilon^{\alpha\beta\gamma\delta}q_{1\gamma}q_{2\delta}}{8\pi^{2}%
m}\langle T^{b}T^{a}\rangle+(1~perm.)+\mathcal{O}(m^{-3})\ ,\\
\mathcal{T}(PV_{q_{1}}^{\alpha,a}V_{q_{2}}^{\beta,b}V_{q_{3}}^{\gamma,c})  &
=g^{3}\frac{i\varepsilon^{\alpha\beta\gamma\delta}(q_{1\delta}+q_{3\delta}%
)}{8\pi^{2}m}\langle T^{c}T^{b}T^{a}\rangle+(5~perm.)+\mathcal{O}(m^{-3})\ ,\\
\mathcal{T}(PV_{q_{1}}^{\alpha,a}V_{q_{2}}^{\beta,b}V_{q_{3}}^{\gamma
,c}V_{q_{4}}^{\delta,d})  &  =g^{4}\frac{i\varepsilon^{\alpha\beta\gamma
\delta}}{8\pi^{2}m}\langle T^{d}T^{c}T^{b}T^{a}\rangle+(23~perm.)+\mathcal{O}%
(m^{-3})\ .
\end{align}
Remains to use $SU(N)$ identities to sum over the permutations. For the
pseudoscalar box, plugging in $\langle T^{a}T^{b}T^{c}\rangle=i\mathcal{I}%
_{2}f^{abc}/2+\mathcal{I}_{3}d^{abc}/4$, only the $\mathcal{I}_{2}$ term
survives thanks to the antisymmetric property of the leading term, and we find%
\end{subequations}
\begin{equation}
\mathcal{T}(PV_{q_{1}}^{\alpha,a}V_{q_{2}}^{\beta,b}V_{q_{3}}^{\gamma
,c})=g^{3}\frac{\mathcal{I}_{2}f^{abc}}{4\pi^{2}m}\varepsilon^{\alpha
\beta\gamma\delta}(q_{1,\delta}+q_{2,\delta}+q_{3,\delta})+\mathcal{O}%
(m^{-3})\ ,
\end{equation}
This $\mathcal{O}(m^{-1})$ term gives back the cubic term of the singlet
anomaly, Eq.~(\ref{AnoSinglet1}), once multiplied by $-2im$. Being
proportional to $f^{abc}$, it would vanish for photons, which is a particular
case of Furry's theorem discussed later on. For the pentagon, the summation
vanishes because the leading term is totally antisymmetric (one can show this
by hand or using quartic $SU(N)$ identities, see the appendix of
Ref.~\cite{Quevillon:2018mfl}) and the $PVVVV$ amplitude starts at
$\mathcal{O}(1/m^{3})$. It does not contribute to the anomaly, in agreement
with Eq.~(\ref{AnoSinglet1}). We leave it to the reader to check that
Eq.~(\ref{AnoSinglet1}) also works beyond the leading order in the $1/m$
expansion, when there is no contribution from the anomaly.

\subsection{Vector Ward identities}\label{VectorWard}

A final aspect of these amplitudes will prove useful later on. The vector Ward
identity is immediately satisfied by the $PVV$ amplitude and, thanks to the
Pauli-Villars regularization, by the $AVV$ amplitude. In other words,
\begin{equation}
q_{1\alpha}\mathcal{T}(PV_{q_{1}}^{\alpha,a}V_{q_{2}}^{\beta,b})=q_{2\beta
}\mathcal{T}_{1PI}(PV_{q_{1}}^{\alpha,a}V_{q_{2}}^{\beta,b})=0\ ,
\label{VWardT}%
\end{equation}
and similarly for $\mathcal{T}(A^{\mu}V_{q_{1}}^{\alpha,a}V_{q_{2}}^{\beta
,b})$. Yet, this property is lost for the box and pentagon diagrams, already
at the leading $\mathcal{O}(m^{-1})$ for the $PVVV$ amplitude. At first sight,
this is worrisome, the vector Ward identity must be satisfied since this
ensures unphysical polarization state cancel out when computing physical decay rates.

\begin{figure}[t]
\centering\includegraphics[width=0.90\textwidth]{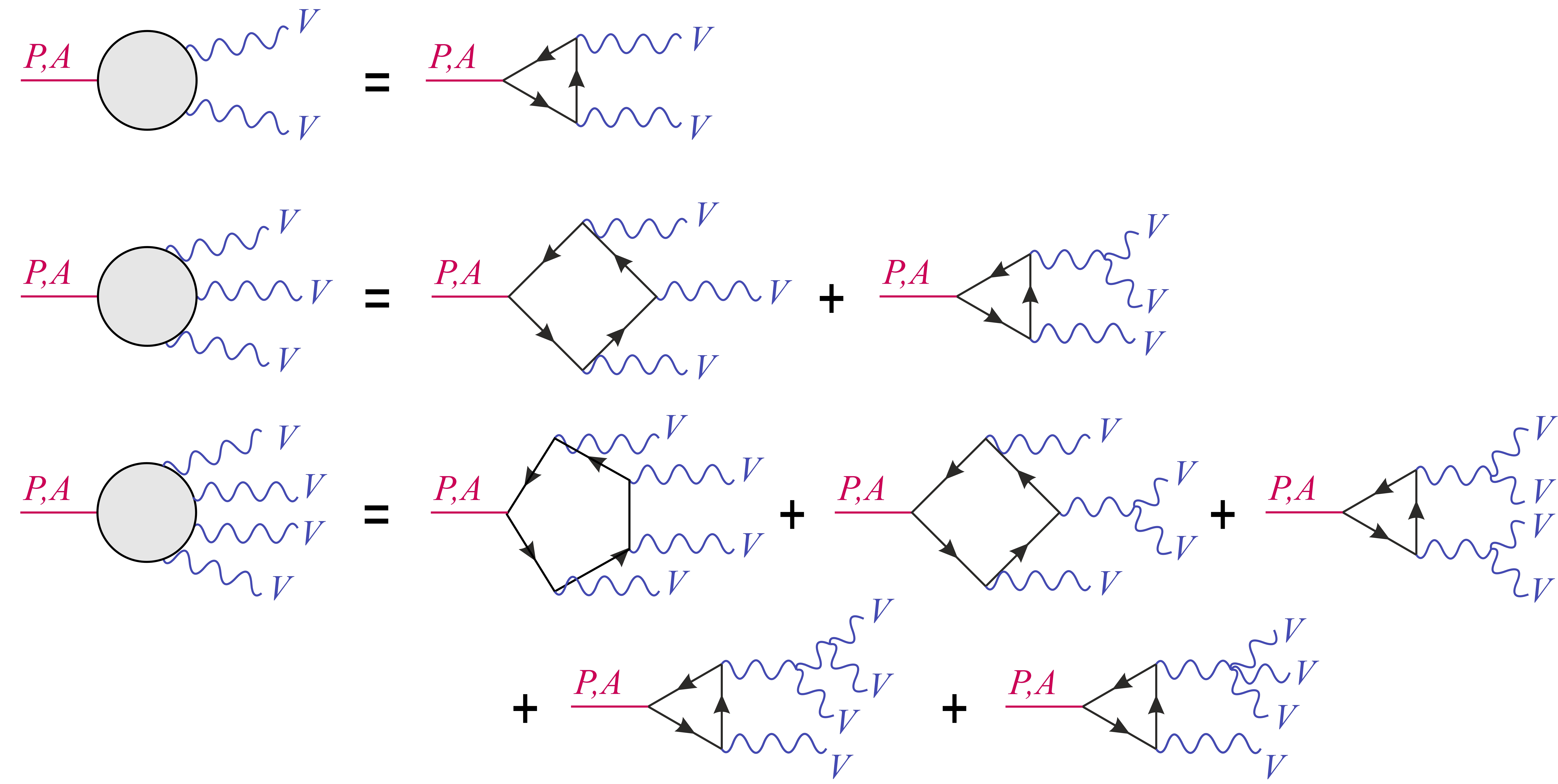}\caption{The various
1PI and non-1PI contributions to the full $P,A\rightarrow VV$, $P,A\rightarrow
VVV$, and $P,A\rightarrow VVVV$ amplitudes, where $P=\bar{\psi}\gamma_{5}\psi
$, $A^{\mu}=\bar{\psi}\gamma^{\mu}\gamma_{5}\psi$, and the $V$'s are
non-abelian gauge vector fields. Notice that all vacuum polarization graphs
have been discarded since they involve $\langle T^{a}\rangle=0$.}%
\label{Fig3}%
\end{figure}

Actually, one should remember that for $V$ a non-abelian gauge field, the Ward
identity is satisfied only at the level of the total or physical amplitudes,
which include all the non-1PI diagrams involving the cubic and quartic gauge
couplings among $V$'s, see Fig.~\ref{Fig3}. For the $P\rightarrow VVV$ case,
adding these contributions and contracting with $q_{1\alpha}$ (the situation
is similar for $q_{2\beta}$ and $q_{3\gamma}$), it is easy to check explicitly
that the contraction reduces to%
\begin{align}
q_{1\alpha}\mathcal{T}_{Full}(PV_{q_{1}}^{\alpha,a}V_{q_{2}}^{\beta,b}%
V_{q_{3}}^{\gamma,c})  &  =q_{1\alpha}\mathcal{T}_{1PI}(PV_{q_{1}}^{\alpha
,a}V_{q_{2}}^{\beta,b}V_{q_{3}}^{\gamma,c})-igf^{acd}\mathcal{T}%
_{1PI}(PV_{q_{2}}^{\beta,b}V_{q_{1}+q_{3}}^{\gamma,d})\nonumber\\
&  \ \ \ \ \ \ \ \ \ \ \ \ \ \ \ \ \ \ \ \ \ \ \ \ \ \ \ \ \ \ -igf^{abd}%
\mathcal{T}_{1PI}(PV_{q_{3}}^{\gamma,c}V_{q_{1}+q_{2}}^{\beta,d}%
)\overset{}{=}0\ . \label{VWardB}%
\end{align}
The first equality requires only the triangle Ward identity of
Eq.~(\ref{VWardT}), while the last is obtained plugging in the pseudoscalar
amplitude of Eq.~(\ref{PtoV}). For the $P\rightarrow VVVV$ process, the
algebra is quite cumbersome because the full amplitude involves five
topologies with a total of 26 diagrams, see Fig.~\ref{Fig3}. To help those
willing to go through the calculation, it can be expressed as
\begin{align}
\mathcal{T}_{Full}(PV_{q_{1}}^{\alpha,a}V_{q_{2}}^{\beta,b}V_{q_{3}}%
^{\gamma,c}V_{q_{4}}^{\delta,d})  &  =\mathcal{T}_{1PI}(PV_{q_{1}}^{\alpha
,a}V_{q_{2}}^{\beta,b}V_{q_{3}}^{\gamma,c}V_{q_{4}}^{\delta,d})\nonumber\\
+\sum_{6~perm.}  &  \mathcal{T}_{1PI}(PV_{q_{1}}^{\alpha,a}V_{q_{2}}^{\beta
,b}V_{q_{3}+q_{4}}^{\mu,e})\frac{\mathcal{T}_{g}(V_{-q_{3}-q_{4}}^{\mu
,e}V_{q_{3}}^{\gamma,c}V_{q_{4}}^{\delta,d})}{(q_{3}+q_{4})^{2}}\nonumber\\
+\sum_{4~perm.}  &  \mathcal{T}_{1PI}(PV_{q_{1}}^{\alpha,a}V_{q_{2}%
+q_{3}+q_{4}}^{\mu,e})\frac{\mathcal{T}_{g}(V_{-q_{2}-q_{3}-q_{4}}^{\mu
,e}V_{q_{2}}^{\beta,b}V_{q_{3}}^{\gamma,c}V_{q_{4}}^{\delta,d})}{(q_{2}%
+q_{3}+q_{4})^{2}}\nonumber\\
+\sum_{3~perm.}  &  \mathcal{T}_{1PI}(PV_{q_{1}+q_{2}}^{\mu,e}V_{q_{3}+q_{4}%
}^{\nu,f})\frac{\mathcal{T}_{g}(V_{-q_{1}-q_{2}}^{\mu,e}V_{q_{1}}^{\alpha
,a}V_{q_{2}}^{\beta,b})}{(q_{1}+q_{2})^{2}}\frac{\mathcal{T}_{g}%
(V_{-q_{3}-q_{4}}^{\nu,f}V_{q_{3}}^{\gamma,c}V_{q_{4}}^{\delta,d})}%
{(q_{3}+q_{4})^{2}}\nonumber\\
+\sum_{4~perm.}  &  \mathcal{T}_{1PI}(PV_{q_{1}}^{\alpha,a}V_{q_{2}%
+q_{3}+q_{4}}^{\mu,e})\nonumber\\
&  \times\left[  \sum_{3~perm.}\frac{\mathcal{T}_{g}(V_{-q_{2}-q_{3}-q_{4}%
}^{\mu,e}V_{q_{2}}^{\beta,b}V_{q_{3}+q_{4}}^{\nu,f})}{(q_{2}+q_{3}+q_{4})^{2}%
}\frac{\mathcal{T}_{g}(V_{-q_{3}-q_{4}}^{\nu,f}V_{q_{3}}^{\gamma,c}V_{q_{4}%
}^{\delta,d})}{(q_{3}+q_{4})^{2}}\right]  \ ,
\end{align}
where $\mathcal{T}_{g}$ are the cubic and quartic gauge couplings derived from
the usual Yang-Mills Lagrangian $\langle\mathbf{F}_{\mu\nu}\mathbf{F}^{\mu\nu
}\rangle$, summations are carried over inequivalent permutations of the four
momenta $q_{1,2,3,4}$, and summation over repeated $SU(N)$ indices is again
understood. Contracting that expression by $q_{1\alpha}$ and using the
triangle and box Ward identities, Eqs.~(\ref{VWardT}) and~(\ref{VWardB}),
along with the Jacobi identity, there are many cancellations and the final
expression collapses to%
\begin{align}
q_{1\alpha}\mathcal{T}_{Full}(PV_{q_{1}}^{\alpha,a}V_{q_{2}}^{\beta,b}%
V_{q_{3}}^{\gamma,c}V_{q_{4}}^{\delta,d})  &  =q_{1\alpha}\mathcal{T}%
_{1PI}(PV_{q_{1}}^{\alpha,a}V_{q_{2}}^{\beta,b}V_{q_{3}}^{\gamma,c}V_{q_{4}%
}^{\delta,d})-igf^{eab}\mathcal{T}_{1PI}(PV_{q_{3}}^{\gamma,c}V_{q_{4}%
}^{\delta,d}V_{q_{1}+q_{2}}^{\beta,e})\nonumber\\
-igf^{eac}\mathcal{T}_{1PI}  &  (PV_{q_{2}}^{\beta,b}V_{q_{4}}^{\delta
,d}V_{q_{1}+q_{3}}^{\gamma,e})-igf^{ead}\mathcal{T}_{1PI}(PV_{q_{2}}^{\beta
,b}V_{q_{3}}^{\gamma,c}V_{q_{1}+q_{4}}^{\delta,e})\overset{}{=}0\ ,
\label{VWardP}%
\end{align}
where the final equality holds for the specific amplitudes of Eq.~(\ref{PtoV}).

These identities are very general since they essentially rely on the
properties of the three- and four-gauge-boson vertices. As such, they
translate the non-abelian gauge symmetry, which is preserved by our
regularization scheme. Being valid for any physical process, they hold at all
orders in $1/m$, for both the pseudoscalar and axial matrix elements.
Crucially, one can now identify in Eqs.~(\ref{VWardT}),\ (\ref{VWardB}),
and~(\ref{VWardP}) the matrix elements of $D_{\mu}V^{\mu,a}\equiv\partial
_{\mu}V^{\mu,a}+igf^{abc}V_{\mu}^{b}V^{\mu,c}$ with $V^{a,\mu}=\bar{\psi
}\gamma^{\mu}T^{a}\psi$. Without surprise, for a non-abelian gauge symmetry,
the Ward identity in Eq.~(\ref{WardV}) must take the covariant form
\begin{equation}
D_{\mu}V^{\mu,a}=0\ .
\end{equation}
Still, to build one's intuition, it is worth to keep in mind how the covariant
derivative sums up all the diagrams of Fig.~\ref{Fig3}.

\subsection{Chern-Simons form}\label{SecChernSimons}

The singlet anomaly shares a very important property with the abelian anomaly:
both are total derivatives. For the singlet case, setting $g=1$ from now on,
this can be written out as:%
\begin{equation}
\partial_{\mu}A^{\mu}=2imP-\frac{1}{16\pi^{2}}\partial_{\mu}G^{\mu
}\ ,\ \ G^{\mu}=4\varepsilon^{\mu\nu\rho\sigma}\left\langle \mathbf{A}_{\nu
}\partial_{\rho}\mathbf{A}_{\sigma}-\frac{i}{3}\mathbf{A}_{\nu}\left[
\mathbf{A}_{\rho},\mathbf{A}_{\sigma}\right]  \right\rangle \ .
\label{AnoSinglet2}%
\end{equation}
Mathematically, the anomalous term is proportional to $\partial_{\mu}G^{\mu
}=\left\langle \varepsilon^{\mu\nu\rho\sigma}\mathbf{F}_{\mu\nu}%
\mathbf{F}_{\rho\sigma}\right\rangle $, called the \textbf{Chern-Simons
term}~\cite{Chern:1974ft}, with $G^{\mu}$ being the Chern-Simons current. Note
that if we shift the current to $A^{\mu}\rightarrow\mathcal{A}^{\mu}=A^{\mu
}+1/(16\pi^{2})G^{\mu}$, it is possible to recover a conserved current in the
$m\rightarrow0$ limit. Yet, the current $\mathcal{A}^{\mu}$ is then not a
gauge singlet since $G^{\mu}$ is not gauge-invariant, and it cannot be used to
construct physical couplings. We will comment more on this fact when
discussing the covariant anomaly in Sec.~\ref{SecCovAno}.

The quantity $\left\langle \varepsilon^{\mu\nu\rho\sigma}\mathbf{F}_{\mu\nu
}\mathbf{F}_{\rho\sigma}\right\rangle $ is very special. To see this, consider
first the following series of mathematical objects in even space-time
dimensions $d=2n$:
\begin{subequations}
\label{ChernPoly}%
\begin{align}
n  &  =1:\left\langle \varepsilon^{\mu\nu}\mathbf{F}_{\mu\nu}\right\rangle
=\partial_{\mu}G_{1}^{\mu}\\
n  &  =2:\left\langle \varepsilon^{\mu\nu\rho\sigma}\mathbf{F}_{\mu\nu
}\mathbf{F}_{\rho\sigma}\right\rangle =\partial_{\mu}G_{3}^{\mu}\\
n  &  =3:\left\langle \varepsilon^{\mu\nu\rho\sigma\alpha\beta}\mathbf{F}%
_{\mu\nu}\mathbf{F}_{\rho\sigma}\mathbf{F}_{\alpha\beta}\right\rangle
=\partial_{\mu}G_{5}^{\mu}\ ,\\
n  &  =4:\left\langle \varepsilon^{\mu\nu\rho\sigma\alpha\beta\gamma\delta
}\mathbf{F}_{\mu\nu}\mathbf{F}_{\rho\sigma}\mathbf{F}_{\alpha\beta}%
\mathbf{F}_{\gamma\delta}\right\rangle =\partial_{\mu}G_{7}^{\mu}\ ,\nonumber
\end{align}
These particular polynomial in the field strength are obviously gauge
invariant, but they are also the only way to contract field strengths without
raising any of their indices with the metric tensor. This means they must have
a global topological character, which translates into them being expressible
as total divergences so that their integral over some $2n$-dimensional volume
$V$ depends only on the value of $G_{2n-1}^{\mu}$ at the boundary $\partial
V$. This is confirmed by an explicit calculation plugging in $\mathbf{F}%
_{\mu\nu}=\partial_{\mu}\mathbf{A}_{\nu}-\partial_{\nu}\mathbf{A}_{\mu
}-i[\mathbf{A}_{\mu},\mathbf{A}_{\nu}]$, which gives
\end{subequations}
\begin{subequations}
\label{CSforms}%
\begin{align}
G_{1}^{\mu}  &  =2\varepsilon^{\mu\nu}\left\langle \mathbf{A}_{\nu
}\right\rangle \ ,\\
G_{3}^{\mu}  &  =4\varepsilon^{\mu\nu\rho\sigma}\left\langle \mathbf{A}_{\nu
}\partial_{\rho}\mathbf{A}_{\sigma}-\frac{2}{3}i\mathbf{A}_{\nu}%
\mathbf{A}_{\rho}\mathbf{A}_{\sigma}\right\rangle \ ,\\
G_{5}^{\mu}  &  =8\varepsilon^{\mu\nu\rho\sigma\alpha\beta}\left\langle
\mathbf{A}_{\nu}\partial_{\rho}\mathbf{A}_{\sigma}\partial_{\alpha}%
\mathbf{A}_{\beta}+\frac{3}{2}i\mathbf{A}_{\alpha}\mathbf{A}_{\beta}%
\mathbf{A}_{\rho}\partial_{\nu}\mathbf{A}_{\sigma}-\frac{3}{5}\mathbf{A}_{\nu
}\mathbf{A}_{\rho}\mathbf{A}_{\sigma}\mathbf{A}_{\alpha}\mathbf{A}_{\beta
}\right\rangle \ ,\\
G_{7}^{\mu}  &  =16\varepsilon^{\mu\nu\rho\sigma\alpha\beta\gamma\delta
}\left\langle \mathbf{A}_{\nu}\partial_{\rho}\mathbf{A}_{\sigma}%
\partial_{\alpha}\mathbf{A}_{\beta}\partial_{\gamma}\mathbf{A}_{\delta}%
+\frac{8}{5}i\mathbf{A}_{\beta}\mathbf{A}_{\gamma}\mathbf{A}_{\delta}%
\partial_{\nu}\mathbf{A}_{\sigma}\partial_{\rho}\mathbf{A}_{\alpha}-\frac
{4}{5}i\mathbf{A}_{\sigma}\partial_{\nu}\mathbf{A}_{\alpha}\mathbf{A}_{\beta
}\partial_{\rho}\mathbf{A}_{\gamma}\mathbf{A}_{\delta}\right. \nonumber\\
&
\ \ \ \ \ \ \ \ \ \ \ \ \ \ \ \ \ \ \ \ \ \ \ \ \ \ \ \ \ \ \ \ \ \ \ \ \ \ \ \ \ \left.
+2\mathbf{A}_{\alpha}\mathbf{A}_{\beta}\mathbf{A}_{\gamma}\mathbf{A}_{\delta
}\mathbf{A}_{\rho}\partial_{\nu}\mathbf{A}_{\sigma}+\frac{4}{7}i\mathbf{A}%
_{\nu}\mathbf{A}_{\rho}\mathbf{A}_{\sigma}\mathbf{A}_{\alpha}\mathbf{A}%
_{\beta}\mathbf{A}_{\gamma}\mathbf{A}_{\delta}\right\rangle \ .
\end{align}
Notice though that the algebra quickly becomes cumbersome and various
identities have to be called in, like for example $\varepsilon^{\mu\nu
\rho\sigma\alpha\beta}\left\langle \mathbf{A}_{\alpha}\partial_{\mu}%
\mathbf{A}_{\beta}\mathbf{A}_{\rho}\partial_{\nu}\mathbf{A}_{\sigma
}\right\rangle =0$ by antisymmetry, as well as $SU(N)$ identities to cancel
the totally antisymmetric sum of the trace of $2n$ generators. It is also
useful to have the equivalent expressions in which $\partial_{\mu}%
\mathbf{A}_{\nu}$ is traded for $\mathbf{F}_{\mu\nu}$, which are
\end{subequations}
\begin{subequations}
\label{CSactions}%
\begin{align}
G_{1}^{\mu}  &  =2\varepsilon^{\mu\nu}\left\langle \mathbf{A}_{\nu
}\right\rangle \ ,\\
G_{3}^{\mu}  &  =4\varepsilon^{\mu\nu\rho\sigma}\left\langle \frac{1}%
{2}\mathbf{A}_{\nu}\mathbf{F}_{\rho\sigma}+\frac{1}{3}i\mathbf{A}_{\nu
}\mathbf{A}_{\rho}\mathbf{A}_{\sigma}\right\rangle \ ,\\
G_{5}^{\mu}  &  =8\varepsilon^{\mu\nu\rho\sigma\alpha\beta}\left\langle
\frac{1}{4}\mathbf{A}_{\nu}\mathbf{F}_{\rho\sigma}\mathbf{F}_{\alpha\beta
}+\frac{1}{4}i\mathbf{A}_{\nu}\mathbf{A}_{\rho}\mathbf{A}_{\sigma}%
\mathbf{F}_{\alpha\beta}-\frac{1}{10}\mathbf{A}_{\nu}\mathbf{A}_{\rho
}\mathbf{A}_{\sigma}\mathbf{A}_{\alpha}\mathbf{A}_{\beta}\right\rangle \ ,\\
G_{7}^{\mu}  &  =16\varepsilon^{\mu\nu\rho\sigma\alpha\beta\gamma\delta
}\left\langle \frac{1}{8}\mathbf{A}_{\nu}\mathbf{F}_{\rho\sigma}%
\mathbf{F}_{\alpha\beta}\mathbf{F}_{\gamma\delta}+\frac{1}{10}i\mathbf{A}%
_{\nu}\mathbf{A}_{\rho}\mathbf{A}_{\sigma}\mathbf{F}_{\alpha\beta}%
\mathbf{F}_{\gamma\delta}+\frac{1}{20}i\mathbf{A}_{\nu}\mathbf{F}_{\rho\sigma
}\mathbf{A}_{\alpha}\mathbf{A}_{\beta}\mathbf{F}_{\gamma\delta}\right.
\nonumber\\
&
\ \ \ \ \ \ \ \ \ \ \ \ \ \ \ \ \ \ \ \ \ \ \ \ \ \ \ \ \ \ \ \ \ \ \ \ \ \ \ \ \ \left.
-\frac{1}{10}\mathbf{A}_{\nu}\mathbf{A}_{\rho}\mathbf{A}_{\sigma}%
\mathbf{A}_{\alpha}\mathbf{A}_{\beta}\mathbf{F}_{\gamma\delta}-\frac{1}%
{35}i\mathbf{A}_{\nu}\mathbf{A}_{\rho}\mathbf{A}_{\sigma}\mathbf{A}_{\alpha
}\mathbf{A}_{\beta}\mathbf{A}_{\gamma}\mathbf{A}_{\delta}\right\rangle \ .
\end{align}
With other techniques, one can show that the coefficient of the non-derivative
term of $G_{2n-1}^{\mu}$ can be obtained from~(see e.g.
Ref.~\cite{Stone:2012ud})%
\end{subequations}
\begin{equation}
c_{n}=2^{n}i^{n-1}\frac{\left(  n-1\right)  !n!}{(2n-1)!}\ . \label{cn}%
\end{equation}

The topological aspect of anomalies is a vast topic which quickly becomes
rather technical. In the present review, we want to avoid the differential
language altogether, though we will not refrain from "rediscovering" some
interesting results using the more pedestrian but familiar tensor notation.
One piece of information we shall take for granted though is \textbf{Chern's
theorem}. It states that the integrals of the polynomials in the curvature of
Eq.~(\ref{ChernPoly}) are quantized and measures the winding of the gauge
configurations over the boundary of the integration volume. This theorem is
itself part of a series of theorems, now all interpreted on the basis of the
capability of some differential forms at detecting singularities or holes in
the underlying space. Its oldest incarnation is the familiar Gauss-Bonnet
theorem, dating back to the XIX century, which relates the integral of the
Gaussian curvature of a surface to its Euler characteristic.

Let us illustrate more precisely what are those winding numbers, from a
physics point of view. The statement of Chern's theorem is that, once in the
Euclidian and properly normalized, the integral of $\partial_{\mu}%
G_{2n-1}^{\mu}$ is quantized. The situation is particularly simple in two
dimensions since the Chern-Simons form collapses to the gauge potential, which
needs to be that of a $U(1)$ theory since otherwise $\left\langle
\mathbf{A}_{\nu}\right\rangle =0$. The boundary is then a loop, i.e., a
one-dimensional circle $S^{1}$, and the integral of the Chern-Simons form is
just a Wilson loop. It is quantized because of the need to have a consistent
gauge potential after going round the loop, with the winding number denoting
quite appropriately the number of times it winds around $S^{1}$. Intuitively,
the picture is very similar to that for Dirac monopole charge quantization.
Though that system is four-dimensional, the monopole acts as a space-time
singularity around which the gauge potential can also wind around.

Similarly, in higher dimensions, we have for example,
\begin{align}
\frac{1}{32\pi^{2}}\int_{V}d^{4}x\ \left\langle \varepsilon^{\mu\nu\rho\sigma
}\mathbf{F}_{\mu\nu}\mathbf{F}_{\rho\sigma}\right\rangle  &  =\frac{1}%
{32\pi^{2}}\int_{\partial V=S^{3}}d\sigma_{\mu}\ G_{3}^{\mu}\nonumber\\
&  =\frac{i}{24\pi^{2}}\int_{\partial V=S^{3}}d\sigma_{\mu}\ \varepsilon
^{\mu\nu\rho\sigma}\left\langle \mathbf{A}_{\nu}\mathbf{A}_{\rho}%
\mathbf{A}_{\sigma}\right\rangle =\nu\in \mathbb{Z} \ . 
\label{CSInteg1}
\end{align}
In the second line, we assumed $V$ is our Wick-rotated four-dimensional
space-time, so its boundary can be identified as a three-sphere $S^{3}$ of
infinite radius. The integral thus depends only on the values of the fields at
infinity. Though the field strength vanishes there, the gauge field may not.
All we need is for the gauge field to collapse onto a pure gauge configuration
$\mathbf{A}_{\mu}\rightarrow i\boldsymbol{\Omega}^{\dagger}\partial_{\mu
}\boldsymbol{\Omega}$. Thus, accounting for the fact that $\mathbf{F}_{\mu\nu}=0$
on $\partial V$, only the last term of the Chern-Simons form contributes. This
final integral is often written in the $A_{0}=0$ gauge, in which case
$d\sigma_{\mu}\ \varepsilon^{\mu\nu\rho\sigma}\rightarrow d^{3}x\ \varepsilon
^{ijk}$ and it becomes purely three-dimensional. A similar treatment can be
repeated for the other cases, with the result
\begin{equation}
\frac{1}{(4\pi)^{n}n!}\int_{S^{2n-1}}d\sigma_{\mu}\ G_{2n-1}^{\mu}=\frac
{c_{n}}{(4\pi)^{n}n!}\int_{S^{2n-1}}d\sigma_{\mu}\ \varepsilon^{\mu\nu
_{1}...\nu_{2n-1}}\langle(i\boldsymbol{\Omega}^{\dagger}\partial_{\nu_{1}%
}\boldsymbol{\Omega})...(i\boldsymbol{\Omega}^{\dagger}\partial_{\nu_{2n-1}%
}\boldsymbol{\Omega})\rangle=\nu\in\mathbb{Z}\ . \label{CSInteg2}%
\end{equation}
We shall not demonstrate why this integral produces an integer for that
specific coefficient. The power of $\pi$ certainly makes sense given the
surface of a $S^{2n-1}$ sphere. Notice also that the coefficient of the
integral over $\langle(\boldsymbol{\Omega}^{\dagger}\partial\boldsymbol{\Omega
})^{2n-1}\mathbb{\rangle}$ collapses to $i/(2\pi)$ for $n=1$, as expected from
the Aharonov--Bohm effect, to $1/(24\pi^{2})$ for $n=2$, as will be confirmed
below, and to $-i/(480\pi^{3})$ for $n=3$, as relevant for the WZW action to
be discussed in Sec.~\ref{SecWZW}. All this shows that pure gauge configurations
fall into equivalence classes. If two different pure gauge configurations
$\boldsymbol{\Omega}(x)$ and $\boldsymbol{\Omega}^{\prime}(x)$ can be continuously
transformed into each other (i.e., are homotopic), we say that these
configurations are related by small gauge transformations, and one can show
(see e.g. Ref.~\cite{Baez:1995sj}) that the winding number does not change. By
contrast, for a non-abelian group, it is possible to construct gauge
configurations that are only connected by large gauge transformations, of
different winding numbers.

What remains to be done is to explicitly construct gauge configurations with
non-zero winding number. It is sufficient to do that in $SU(2)$, since
embedding those inside that of a larger $SU(N)$ representation would lead to
the same winding. Then, the strategy can be illustrated using $U(1)$ as an
example. First, its elements can be viewed as $S^{1}\rightarrow S^{1}$ maps.
Indeed, on one hand, we can construct any of its elements as $\omega=a+ib$
with $a^{2}+b^{2}=1$, thus mapping $S^{1}\rightarrow U(1)$. At the same time,
any $\omega\in U(1)$ can be written as $\omega=\exp i\phi$, thereby defining a
$U(1)\rightarrow S^{1}$ map. Altogether, $U(1)$ thus maps $S^{1}\rightarrow
S^{1}$. In such a case, there are equivalence classes of maps, each made of
gauge configurations wrapping the circle $\nu$ times. In practice, to get such
a non-trivial winding configuration, we can construct%
\begin{equation}
\omega=\frac{x^{2}-d^{2}}{x^{2}+d^{2}}+i\frac{2dx}{x^{2}+d^{2}}\ ,
\end{equation}
for some parameter $d>0$. As $x$ goes from $-\infty$ to $+\infty$, $\arg
\omega$ goes from $0$ to $2\pi$, and $\arg\omega^{\nu}$ from $0$ to $2\pi\nu$.
As a result, $i\omega^{\nu\dagger}\partial\omega^{\nu}/\partial x$ integrates
to $2\pi\nu$, in agreement with Eq.~(\ref{CSInteg1}). 

\begin{figure}[t]
\centering\includegraphics[width=0.90\textwidth]{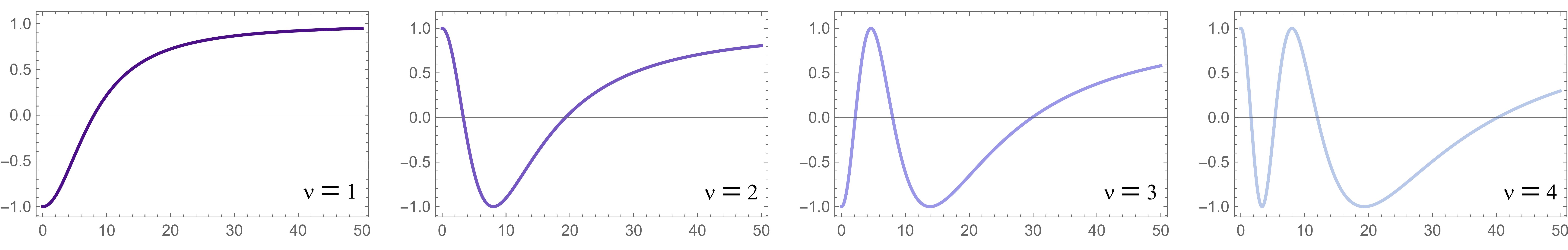}\caption{The
evolution of $\left\langle \boldsymbol{\Omega}^{\nu}\right\rangle /2$ as a
function of $r$, for a fixed value $d=8$, and for the first few values of
$\nu$. The corresponding winding number $\nu$ is apparent in the number of
times this function switches between $\pm1$. For $\nu=1$, $\left\langle
\boldsymbol{\Omega}\right\rangle /2=a$ as given in Eq.~(\ref{JackReb}). At points
where $\left\langle \boldsymbol{\Omega}^{\nu}\right\rangle $ reaches $\pm1$,
$r/x_{i}\times\left\langle \boldsymbol{\Omega}^{\nu}\sigma^{i}\right\rangle $
vanishes.}%
\label{FigSU2}%
\end{figure}

The case of $\boldsymbol{\Omega}\in SU(2)$ is totally analogous. We can write on
one hand $\boldsymbol{\Omega}=a+i\boldsymbol{\sigma}\cdot\mathbf{b}$ with
$a^{2}+|\mathbf{b}|^{2}=1$, thereby defining a $S^{3}\rightarrow SU(2)$ map,
and on the other, the Lie group $SU(2)$ is compact and made of $\exp
(i\boldsymbol{\sigma}\cdot\boldsymbol{\phi})$ matrices, so its elements themselves can be
viewed as $SU(2)\rightarrow S^{3}$ maps. Altogether, these induced
$S^{3}\rightarrow S^{3}$ maps fall within equivalent classes made of
configurations wrapping the $S^{3}$ sphere $\nu$ times. By analogy with the
$U(1)$ case above, the standard $SU(2)$ example of a $\nu=1$ configuration
starts with~\cite{Jackiw:1976pf}
\begin{equation}
\boldsymbol{\Omega}=a+i\boldsymbol{\sigma}\cdot\mathbf{b\ ,\ \ }a=\frac{r^{2}-d^{2}%
}{r^{2}+d^{2}}\ ,\ \ \mathbf{b}=\frac{2d\mathbf{x}}{r^{2}+d^{2}}\ ,
\label{JackReb}%
\end{equation}
with $d>0$ some free parameter and $r^{2}=x_{1}^{2}+x_{2}^{2}+x_{3}^{2}$. In
this approach, the $S^{3}$ sphere is viewed as the union of $S_{r}^{2}$
spheres with radius $r=[0,+\infty\lbrack$. Indeed, the Cartesian coordinates
of the points on the sphere $S^{3}$ of radius $R$ embedded in $\mathbb{R}^{4}$
satisfy $x_{1}^{2}+x_{2}^{2}+x_{3}^{2}=r^{2}=R^{2}-x_{4}^{2}$. When
$R\rightarrow\infty$, we have to integrate $x_{1,2,3}$ over the $S_{r}^{2}$
sphere of radius $r$, and then integrate $r$ from zero to infinity. We can now
see what makes this gauge configuration special. It goes from $\boldsymbol{\Omega
}\rightarrow-\mathbf{1}$ as $r\rightarrow0$ to $\boldsymbol{\Omega}\rightarrow
+\mathbf{1}$ as $r\rightarrow\infty$, two configurations that certainly cannot
be joined infinitesimally. To check this, it suffices to directly compute
$\partial\boldsymbol{\Omega}/\partial x_{i}$ and perform the antisymmetric
summation:%
\begin{equation}
\frac{1}{24\pi^{2}}\int_{0}^{\infty}dr\int_{S_{r}^{2}}d^{3}x\ \varepsilon
^{ijk}\left\langle \boldsymbol{\Omega}^{\dagger}\partial_{i}\boldsymbol{\Omega\Omega
}^{\dagger}\partial_{j}\boldsymbol{\Omega\Omega}^{\dagger}\partial_{k}%
\boldsymbol{\Omega}\right\rangle =-\frac{1}{2\pi^{2}}\int_{0}^{\infty}%
dr\int_{S_{r}^{2}}d^{3}x\ \left(  \frac{2d}{d^{2}+r^{2}}\right)  ^{3}=-1\ ,
\label{JackReb2}%
\end{equation}
since the $d^{3}x$ integral simply gives the $S_{r}^{2}$ surface $4\pi r^{2}$.
From this, any other winding number $\nu\in\mathbb{Z}$ can be obtained 
using $\boldsymbol{\Omega}_{\nu}(x)=(\boldsymbol{\Omega}(x))^{\nu}$,
which switches $\nu$ times between $-1$ and $1$ as $r$ varies between $0$ and
$\infty$, see Fig.~\ref{FigSU2}.

Another famous parametrization of $S^{3}$ is more intuitive, and closely
related to the \textbf{BPST instanton} configuration~\cite{Belavin:1975fg}. In
this case, one directly uses the $SU(2)$ parameters as Cartesian coordinates
for $S^{3}\subset\mathbb{R}^{4}$:%
\begin{equation}
\boldsymbol{\Omega}=\frac{x_{0}+i\boldsymbol{\sigma}\cdot\mathbf{x}}{x}=\frac{x_{\mu
}\bar{\sigma}^{\mu}}{x}\rightarrow\boldsymbol{\Omega}^{\dagger}=\frac{x_{\mu
}\sigma^{\mu}}{x}\rightarrow\boldsymbol{\Omega}^{\dagger}\partial_{\mu
}\boldsymbol{\Omega}=\frac{x^{\nu}\sigma_{\mu\nu}}{x^{2}}\ , \label{BPST1}%
\end{equation}
where $x^{2}=x_{\mu}x_{\nu}\delta^{\mu\nu}$ is the fixed squared radius of
$S^{3}$, $\sigma^{\mu}\equiv\left(  1,-i\boldsymbol{\sigma}\right)  $ and
$\bar{\sigma}^{\mu}\equiv\left(  1,i\boldsymbol{\sigma}\right)  $ such that
$\sigma^{\mu}\bar{\sigma}^{\nu}+\sigma^{\nu}\bar{\sigma}^{\mu}=2\delta^{\mu
\nu}$, and $\sigma^{\mu}\bar{\sigma}^{\nu}-\sigma^{\nu}\bar{\sigma}^{\mu
}\equiv2\sigma^{\mu\nu}$, $\bar{\sigma}^{\mu}\sigma^{\nu}-\bar{\sigma}^{\nu
}\sigma^{\mu}\equiv2\bar{\sigma}^{\mu\nu}$. Those are the Euclidian versions
of the usual definitions for the Lorentz group, with $\Sigma_{\mu\nu}%
=\sigma_{\mu\nu}/2$ and $\bar{\sigma}_{\mu\nu}/2$ for the fundamental
representations satisfying the $SO(4)$ algebra
\begin{equation}
\left[  \Sigma_{\mu\nu},\Sigma_{\rho\sigma}\right]  =\delta_{\nu\rho}%
\Sigma_{\mu\sigma}-\delta_{\nu\sigma}\Sigma_{\mu\rho}-\delta_{\mu\rho}%
\Sigma_{\nu\sigma}+\delta_{\mu\sigma}\Sigma_{\nu\rho}\ . \label{SO4Alg}%
\end{equation}
This configuration indeed winds over the $S^{3}$ sphere:%
\begin{equation}
\frac{1}{24\pi^{2}}\int_{S^{3}}d\sigma_{\mu}\ \varepsilon^{\mu\nu\rho\sigma
}\left\langle \boldsymbol{\Omega}^{\dagger}\partial_{\nu}\boldsymbol{\Omega\Omega
}^{\dagger}\partial_{\rho}\boldsymbol{\Omega\Omega}^{\dagger}\partial_{\sigma
}\boldsymbol{\Omega}\right\rangle =-\frac{1}{2\pi^{2}}\int_{S^{3}}d\sigma_{\mu
}\frac{x^{\mu}}{x^{4}}=-1\ .
\end{equation}
This integral is independent of the radius $x^{2}$, which can now be sent to
infinity. The conjugate configuration $\boldsymbol{\Omega}\partial_{\mu
}\boldsymbol{\Omega}^{\dagger}=x^{\nu}\bar{\sigma}_{\mu\nu}/x^{2}$ gives $\nu=+1$.
Intuitively, this covering is a trivial identity map wrapping the sphere just
once. To see this, notice that in spherical coordinates, $x^{\mu}%
/x=(\cos\theta_{1},\sin\theta_{1}\cos\theta_{2},\sin\theta_{1}\sin\theta
_{2}\cos\theta_{3},\sin\theta_{1}\sin\theta_{2}\sin\theta_{3})$ is a radial
unit vector. It coincides with the coordinates on $S^{3}$, so the integral
gives its surface $2\pi^{2}$. Analytically, this is confirmed writing the
differential surface element $d\sigma_{\mu}=\varepsilon_{\mu\nu\rho\sigma
}\partial x^{\nu}/\partial\theta_{1}\partial x^{\rho}/\partial\theta
_{2}\partial x^{\sigma}/\partial\theta_{3}$.

\subsection{$\theta$ vacua and instantons}\label{Sethetavac}

\begin{figure}[t]
\centering\includegraphics[width=0.38\textwidth]{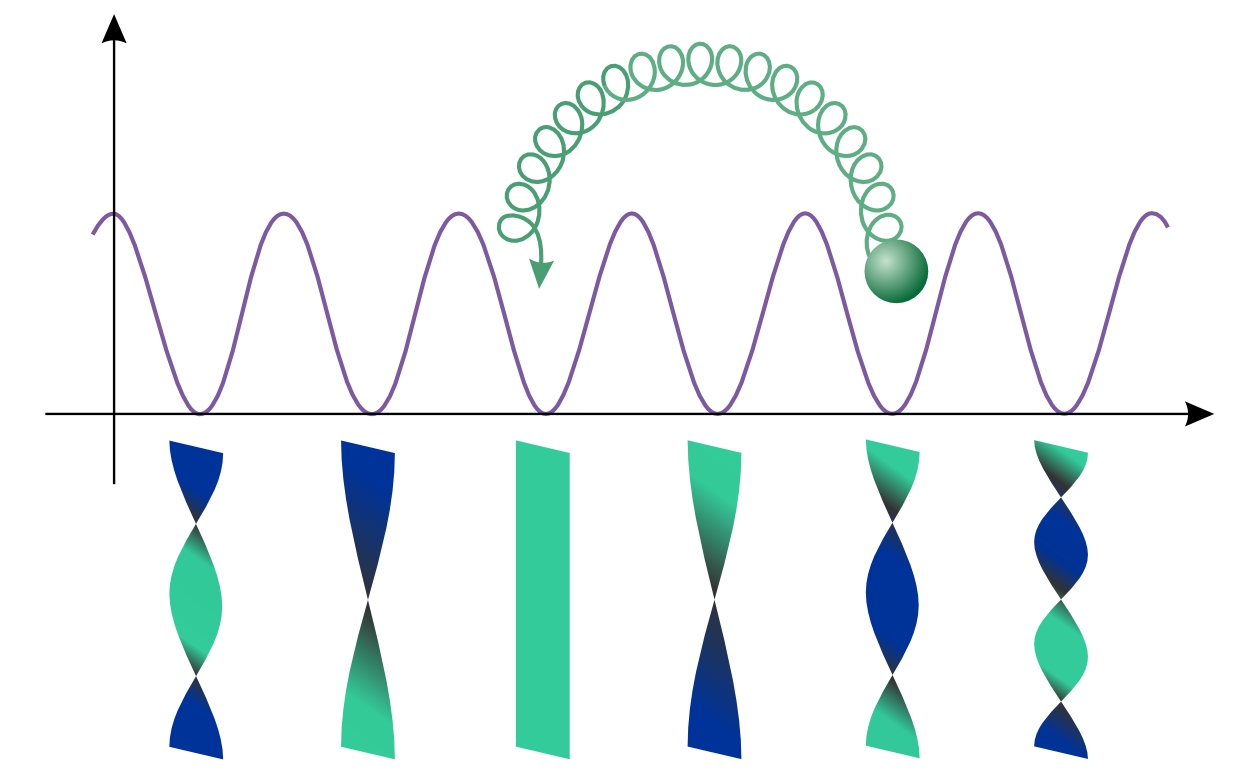}\caption{Naive
representation of the violation of time reversal induced by the presence of
$\theta$ vacua.}%
\label{FigTheta}%
\end{figure}

From the previous considerations, we see that the vacuum of a non-abelian
gauge theory is quite complicated. There are infinitely many topologically
distinct pure gauge configurations $|\nu\rangle$ with $\nu\in\mathbb{Z}$. 
At the same time, those vacua are not physically distinguishable since they
are related by large gauge transformations. In fact, a large gauge
transformation $\boldsymbol{\Omega}(x)$ able to move from $|\nu\rangle$ to
$|\nu+1\rangle$ can, once repeated, get to $|\nu+k\rangle$ for any $k$. This
sort of periodicity permits the use of \textbf{Bloch's theorem}: in a periodic
potential, the wavefunction can be expressed in terms of plane wave modulated
by periodic functions. In our case, periodicity is in gauge space but the idea
remains the same. We can take $|\nu\rangle$ as basis functions, and express
the true vacuum under the form%
\begin{equation}
|\theta\rangle=\sum_{\nu}e^{i\theta\nu}|\nu\rangle\ ,
\end{equation}
where $\theta$ is a free parameter denoting a particular $\boldsymbol{\theta}%
$\textbf{ vacuum}. The reason for the plane-wave factors $e^{i\theta\nu}$ is
to ensure periodicity, i.e., gauge invariance, up to a phase. As seen
previously, a large gauge transformation shifts $\boldsymbol{\Omega}_{n}%
|\nu\rangle\rightarrow|\nu+n\rangle$. But then, acting on $|\theta\rangle$ and
counter-shifting the summation index $\nu\rightarrow\nu-n$ gives
$\boldsymbol{\Omega}_{n}|\theta\rangle=\exp(in\theta)|\theta\rangle$. This is an
important piece of information because starting from a $|\theta\rangle$ vacuum
for a given $\theta$, no gauge-invariant operator $\mathcal{O}$ can move us to
one with a different $\theta$. Indeed, if $\mathcal{O}=\boldsymbol{\Omega}%
_{n}\mathcal{O}\boldsymbol{\Omega}_{n}^{\dagger}$, then we have a
\textbf{super-selection rule} since $\langle\theta^{\prime}|\mathcal{O}%
|\theta\rangle=e^{i(\theta-\theta^{\prime})n}\langle\theta^{\prime
}|\mathcal{O}|\theta\rangle$ imposes $\theta=\theta^{\prime}$. Each value of
$\theta$ defines a different theory, totally secluded from all the other
theories with different values of $\theta$. Further, there is one value of
$\theta$ that corresponds to Nature, and it has to be specified along with the
other free parameters of the Standard Model. In practice, this dependence on
$\theta$ can be moved into the Lagrangian as the so-called $\boldsymbol{\theta
}$ \textbf{term} since, from Eq.~(\ref{CSInteg1}),
\begin{equation}
S_{\theta}=\int_{V}d^{4}x\frac{\theta}{32\pi^{2}}\left\langle \varepsilon
^{\mu\nu\rho\sigma}\mathbf{F}_{\mu\nu}\mathbf{F}_{\rho\sigma}\right\rangle
\rightarrow e^{iS_{\theta}}|\nu\rangle=e^{i\theta\nu}|\nu\rangle\ .
\label{Stheta}%
\end{equation}
In this picture, it is clear that $\theta$ is periodic since $|\theta\rangle$
and $|\theta+2k\pi\rangle$ specify the same vacuum. Very naively, we also
understand why the $\theta$ term violates time-reversal $T$, hence $CP$ since
$CPT$ holds. If one imagines a process where some gauge fields are created
from a vacuum $|\nu\rangle$ and absorbed back into the vacuum $|\nu^{\prime
}\rangle$, as depicted in Fig.~\ref{FigTheta}, the time-reversed process would
see the gauge bosons emerging from $|\nu^{\prime}\rangle$, but there is no
need for it to end up in $|\nu\rangle$, any other vacuum is fine. The two
processes differ if $\theta\neq2k\pi$, and $T$ is broken.

The need to construct the physical vacuum $|\theta\rangle$ as a linear
combination of all the topologically-distinct vacua $|\nu\rangle$ can be
understood from the existence of \textbf{tunnelling effects}, induced by
instantons. In other words, there exists so-called \textbf{instanton
configurations} which are finite-energy Euclidian solutions interpolating
between topologically distinct vacua. To see this, let us start with the
Yang-Mills Lagrangian. After a Wick rotation, we can write (assuming $\nu>0$,
otherwise switch signs such that $S$ stays positive)%
\begin{align}
S  &  =\frac{1}{2g^{2}}\int_{V}d^{4}x\left\langle \mathbf{F}_{\mu\nu
}\mathbf{F}^{\mu\nu}\right\rangle \nonumber\\
&  =\frac{1}{4g^{2}}\int d^{4}x\left\langle (\mathbf{F}_{\mu\nu}%
-\mathbf{\tilde{F}}_{\mu\nu})(\mathbf{F}^{\mu\nu}-\mathbf{\tilde{F}}^{\mu\nu
})\right\rangle +\frac{1}{4g^{2}}\int d^{4}x\left\langle \varepsilon^{\mu
\nu\rho\sigma}\mathbf{F}_{\mu\nu}\mathbf{F}_{\rho\sigma}\right\rangle
\geq\frac{8\pi^{2}\nu}{g^{2}}\ ,
\end{align}
since the first term is necessarily positive (using
inequalities involving a topological quantity to derive bounds is called the
\textbf{Bogomol'nyi trick}). Field configurations $\mathbf{A}_{\mu}$ such that
$\mathbf{F}_{\mu\nu}=\pm\mathbf{\tilde{F}}_{\mu\nu}$ saturate this bound, and
coincidentally in four dimensions, also automatically solve the equation of
motion thanks to the Bianchi identity, $D_{\mu}\mathbf{\tilde{F}}^{\mu\nu
}=D_{\mu}\mathbf{F}^{\mu\nu}=0$. These are the instanton configurations that
can transition between the $\nu$ vacua (their explicit form is derived at the
end of this section). For instance, for $\nu=+1$, and including the $\theta$
term, we can use an instanton to connect $|\nu\rangle$ and $|\nu+1\rangle$:%
\begin{equation}
e^{-S_{inst}}=e^{-8\pi^{2}/g^{2}}e^{i\theta}=\langle\nu+1|\nu\rangle_{\theta
}\ . \label{OneInst}%
\end{equation}
Beware not to confuse the $\nu$ of the instanton configuration, and the $\nu$
labelling the vacua. Notice also that in the Heisenberg picture, the matrix
element should read $\langle\nu+1|\exp(-H\Delta T)|\nu\rangle_{\theta}$ with
$\Delta T=T_{out}-T_{in}\rightarrow\infty$ since the instanton is supposed to
asymptote to pure gauge configurations at early and late times. For a more
careful derivation, see e.g. Ref.~\cite{Coleman:1978ae}. In the Euclidian, the
$\theta$ term has the same $i$ factor as in Minkowski (where it comes from
$e^{iS}$) thanks to the Levi-Civita tensor. However, the instanton transition
classically interpreted in Minkowski space does not have positive energy at
all time. Instead, as is well-known from quantum mechanics, the above
expression should be understood as the one-instanton probability to tunnel
between topologically distinct states, $|\nu\rangle\rightarrow|\nu+1\rangle$.
This is why instantons are sometimes called pseudoparticles. In any case,
because of the exponential factor, this action is very suppressed when $g$ is
small, like for $SU(2)_{L}$, but could become important for QCD at low energy
where $g$ becomes large.

More generally, the vacuum-to-vacuum transition amplitude for a given $\theta
$, saturated by one instanton configuration (with $\nu=1$) takes the form%
\begin{equation}
\langle\theta|\theta\rangle_{one-inst.}\sim\int DA_{\mu}\times e^{-S[A_{\mu}%
]}\times e^{iS_{\theta}}\overset{A_{\mu}\rightarrow A_{\mu}^{inst}}{\sim
}e^{-8\pi^{2}/g^{2}}e^{i\theta}\ . \label{OneInstVac}%
\end{equation}
If we sum this for $n$ instantons with $\nu=1$ and $\bar{n}$ anti-instanton
with $\nu=-1$, assuming they are so separated that the above approximation is
valid individually for each of them (the so-called \textbf{dilute instanton
gas} approximation), we get~(see e.g.
Refs.~\cite{Coleman:1978ae,Khoze:2025auv})%
\begin{equation}
\langle\theta|\theta\rangle_{many-inst.}\sim\sum_{n,\bar{n}=0}^{\infty}%
\frac{1}{n!}\frac{1}{\bar{n}!}(e^{-8\pi^{2}/g^{2}})^{n+\bar{n}}e^{i\theta
(n-\bar{n})}=\exp(2e^{-8\pi^{2}/g^{2}}\cos\theta)\ . \label{InstVac}%
\end{equation}
There are many complicated normalization factors here but we shall not enter
into the details. All we need is that once properly normalized and in the
Euclidian, this transition is the vacuum effective action $\exp(-S_{eff})$,
and $S_{eff}$ is just the energy$\ $times the four-volume (remember that for a
classical point particle, $t\rightarrow it$ changes the sign of the kinetic
energy). In other words,
\begin{equation}
E(\theta)\sim\log\langle\theta|\theta\rangle_{many-inst.}\sim e^{-8\pi
^{2}/g^{2}}\cos\theta\ . \label{Etheta}%
\end{equation}
Notice that the periodic nature of $E(\theta)$ is not directly related to the
presence of periodic vacua. In particular, we have seen previously that
$\theta$ is a free parameter and worlds with different $\theta$ values cannot
communicate. We could thus be living in a world whose vacuum energy is not
minimal in absolute term, but this is totally inconsequential.

To close this section, let us detail the \textbf{BPST instanton} configuration
(after Belayev, Polyakov, Schwartz, Tyupkin, Ref.~\cite{Belavin:1975fg}) for
the $SU(2)$ gauge group. This is where the parametrization of Eq.~(\ref{BPST1}%
) is interesting because the $\sigma^{\mu\nu}$ and $\bar{\sigma}^{\mu\nu}$
matrices are anti-self-dual and self-dual, $\sigma^{\mu\nu}\varepsilon_{\mu
\nu\rho\sigma}=-2\bar{\sigma}_{\rho\sigma}$ and $\bar{\sigma}^{\mu\nu
}\varepsilon_{\mu\nu\rho\sigma}=+2\sigma_{\rho\sigma}$. It thus suffices to
modulate the pure gauge configurations of Eq.~(\ref{BPST1}) away from the
$x\rightarrow\infty$ boundary while maintaining the self-duality properties to
get a true instanton field configuration. For instance, we can take%
\begin{equation}
\mathbf{A}_{\mu}^{inst}=\frac{x^{2}}{x^{2}+\rho^{2}}\boldsymbol{\Omega}%
\partial_{\mu}\boldsymbol{\Omega}^{\dagger}=\frac{\bar{\sigma}_{\mu\nu}x^{\nu}%
}{x^{2}+\rho^{2}}\ ,\ \mathbf{F}_{\mu\nu}^{inst}=\partial_{\mu}\mathbf{A}%
_{\nu}^{inst}-\partial_{\nu}\mathbf{A}_{\mu}^{inst}-[\mathbf{A}_{\mu}%
^{inst},\mathbf{A}_{\nu}^{inst}]=\frac{-2\rho^{2}\bar{\sigma}_{\mu\nu}}%
{(x^{2}+\rho^{2})^{2}}\ ,\label{SU2inst}%
\end{equation}
for a $\nu=1$ instanton, where $\rho>0$ is a free parameter, the instanton
size. One can check that such a modulation does not upset the equation of
motion,%
\begin{equation}
D^{\mu}\mathbf{F}_{\mu\nu}^{inst}=\partial^{\mu}\mathbf{F}_{\mu\nu}%
^{inst}-[\mathbf{A}^{inst,\mu},\mathbf{F}_{\mu\nu}^{inst}]=\frac{8x^{\mu}%
\rho^{2}\bar{\sigma}_{\mu\nu}}{(x^{2}+\rho^{2})^{3}}-[\bar{\sigma}^{\mu\rho
},\bar{\sigma}_{\mu\nu}]\frac{x_{\rho}}{x^{2}+\rho^{2}}\frac{-2\rho^{2}%
}{(x^{2}+\rho^{2})^{2}}=0\ ,
\end{equation}
where we used the $SO(4)$ algebra Eq.~(\ref{SO4Alg}). These vector potentials
can be put in the usual form $\mathbf{A}_{\mu}=A_{\mu}^{i}\sigma^{i}/2$ by
introducing \textbf{'t Hooft matrices} $\eta_{\mu\nu}^{i}$ via $\bar{\sigma
}_{\mu\nu}=\eta_{\mu\nu}^{i}\sigma_{i}~$\cite{tHooft:1976snw}. As a check, we
can this time compute the winding number directly from the volume integral of
$\varepsilon^{\mu\nu\rho\sigma}\mathbf{F}_{\mu\nu}^{inst}\mathbf{F}%
_{\rho\sigma}^{inst}$. Using $\bar{\sigma}^{\mu\nu}\bar{\sigma}_{\mu\nu
}=-12\times\mathbf{1}$, we indeed get back the same winding number as for the
pure gauge case:%
\begin{equation}
\frac{1}{32\pi^{2}}\int_{V}d^{4}x\ \left\langle \varepsilon^{\mu\nu\rho\sigma
}\mathbf{F}_{\mu\nu}^{inst}\mathbf{F}_{\rho\sigma}^{inst}\right\rangle =3\int
r^{3}dr\ \left(  \frac{-2\rho^{2}}{(r^{2}+\rho^{2})^{2}}\right)  ^{2}=1\ ,
\end{equation}
independently of $\rho$.

To get to the integral in Eq.~(\ref{InstVac}), one must sum over
configurations with many (separated) instantons and anti-instantons, but also
on the free parameters describing different instanton "shapes" called
\textbf{collective coordinates}: the center of $\mathbf{A}_{\mu}^{inst}$ is at
$x=0$, but it could be somewhere else, the parameter $\rho$ for the instanton
size, and its orientation in group space (rigid $SU(2)$ rotations), for a
total of eight collective coordinates over which one has to integrate (for a
detailed introduction to instantons, see e.g. Ref.~\cite{Vandoren:2008xg}).
The integral over $\rho$ is particularly delicate, since it naively diverges
as $\rho\rightarrow\infty$ (when the instanton becomes big). This is a
complicated problem because $\rho$ is related to the scale invariance of the
Yang-Mills action ($\mathbf{A}_{\mu}^{inst}$ of all sizes are necessarily
solutions), but this does not survive renormalization, with an explicit scale
appearing in the infrared. Ultimately, a connection with the confinement
mechanism is suspected, though not proven.

\subsubsection{Fermionic instanton interactions}

In the presence of massless fermions, the $\theta$ term can trivially be
rotated away via a $U(1)_{A}$ rotation. Indeed, under the chiral rotation
$\psi\rightarrow e^{i\beta\gamma_{5}}\psi$, the Lagrangian varies by
$\beta\partial_{\mu}A^{\mu}$, with $\partial_{\mu}A^{\mu}$ satisfying the
anomalous Ward identity Eq.~(\ref{AnoSinglet2}) with $m=0$. What is less clear
a priori is how the presence of massless fermions affects the previous
discussion. Instanton tunnelling certainly still exist, but as we now discuss,
they cease to occur between vacua but bridges states of different $U(1)_{A}$ charges.

Let us phrase the possibility of rotating away the $\theta$ term via a chiral
transformation in slightly different terms (see
Ref.~\cite{Callan:1976je,Jackiw:1976pf}). Instead of true axial
transformations, consider the conserved but not gauge invariant current
$\mathcal{A}_{\mu}=A_{\mu}+g^{2}/(16\pi^{2})G_{\mu}$ introduced after
Eq.~(\ref{AnoSinglet2}). The charge $Q_{5}$ obtained by a spatial integration
over $\mathcal{A}_{\mu}$ varies under a large gauge transformation
$\boldsymbol{\Omega}$ of winding number $\nu$, and Eq.~(\ref{CSInteg1}) tells us
that $\boldsymbol{\Omega}Q_{5}\boldsymbol{\Omega}^{\dagger}=Q_{5}-2\nu$. If we choose
a reference state $|0\rangle$ such that $Q_{5}|0\rangle=0$, then $Q_{5}%
|\nu\rangle=2\nu|\nu\rangle$. Thus, performing a $U(1)_{\mathcal{A}}$ rotation
of angle $\alpha$ changes $\theta$ into
\begin{equation}
\exp(i\alpha Q_{5})|\theta\rangle=\sum_{\nu}e^{i(\theta+2\alpha)\nu}%
|\nu\rangle=|\theta+2\alpha\rangle\ ,
\end{equation}
so $Q_{5}$ relates theories with different $\theta$ values. Those are no
longer distinguishable and the $\theta$ vacua collapse to a single theory.

Whatever the point of view, the crucial consequence is that $E(\theta)$ loses
its dependence on $\theta$, and this removes all possible dependence on
$\theta$ in observables. Indeed, $Q_{5}$ is conserved since $\partial^{\mu
}\mathcal{A}_{\mu}=0$, so it must commute with the Hamiltonian, $0=\langle
\nu^{\prime}|[Q_{5},H]|\nu\rangle=(\nu^{\prime}-\nu)\langle\nu^{\prime}%
|H|\nu\rangle$. This means that $\langle\nu^{\prime}|H|\nu\rangle=\delta
_{\nu\nu^{\prime}}$, instanton tunnelling transitions are no longer possible
between vacua of different winding numbers, and we get $\langle\theta
|\theta\rangle_{one-inst.}=0$. Now, in the path integral formalism, matrix
elements are represented as weighted averages (a brief introduction is in
Sec.~\ref{SecFujikawa}). For example,
\begin{equation}
\langle\theta|\langle\mathbf{F}_{\mu\nu}\mathbf{\tilde{F}}^{\mu\nu}%
\rangle|\theta\rangle=\int DA_{\mu}\times\langle\mathbf{F}_{\mu\nu
}\mathbf{\tilde{F}}^{\mu\nu}\rangle\times e^{-S[A_{\mu}]}\times e^{iS_{\theta
}}=16\pi^{2}\frac{\partial}{\partial\theta}\langle\theta|\theta\rangle\ .
\label{dEtheta}%
\end{equation}
While without massless quarks, $\langle\theta|\langle\mathbf{F}_{\mu\nu
}\mathbf{\tilde{F}}^{\mu\nu}\rangle|\theta\rangle_{many-inst.}\sim\sin\theta$
is obtained as the derivative of the energy $E(\theta)$, we now have
$\langle\theta|\langle\mathbf{F}_{\mu\nu}\mathbf{\tilde{F}}^{\mu\nu}%
\rangle|\theta\rangle_{many-inst.}=0$, and similarly for any matrix element of
$\langle\mathbf{F}_{\mu\nu}\mathbf{\tilde{F}}^{\mu\nu}\rangle$ with additional
operators. The $\theta$ term cannot lead to any observable effects anymore.

On the other hand, tunnelling between quark states of different axial charges
are now permitted. Naively, in the presence of fermions, $\nu$ takes the new
meaning of being the axial number since $Q_{5}|\nu\rangle=2\nu|\nu\rangle$.
So, when instantons drive the tunnelling to a state of different winding
number, some fermions must be emitted to compensate for the change in axial
charge. This costs no energy since they are massless. Actually, along with
$\langle\nu^{\prime}|\nu\rangle\sim\delta_{\nu\nu^{\prime}}$, we obviously
have $\langle\nu^{\prime}+a|\mathcal{O}_{A}|\nu\rangle\sim\delta_{\nu
\nu^{\prime}}$ for $\mathcal{O}_{A}$ an operator of axial charge $a$.

Phenomenologically, there are two important applications of this effect.
First, it makes the $\eta^{\prime}$ meson much heavier than the other
pseudoscalar mesons like the $\pi$ or $K$. As will be detailed in Sec.~\ref{SecWZW}, the
QCD Lagrangian restricted to the $u,d,s$ flavors and in the absence of quark
mass terms has the chiral symmetry $U(3)_{L}\otimes U(3)_{R}$ since strong
interactions do not care about flavors. This symmetry is spontaneously broken
down to $U(3)_{V}$, so one would expect a nonet of pseudoscalar Goldstone
bosons associated to the $U(3)_{A}$ broken generators. The axial $U(1)_{A}$
symmetry is anomalous though, so the singlet state is presumably heavier. This
is confirmed in the instanton language by the existence of an axial operator
$\mathcal{O}_{A}$ that creates self-interactions between $U(1)_{A}$ singlet
currents. Specifically, $\mathcal{O}_{A}$ must be a $SU(3)_{L}\otimes
SU(3)_{R}$ singlet, so it must have the form~\cite{Shifman:1979uw}
\begin{equation}
\mathcal{O}_{A}=\varepsilon^{ijk}\bar{q}_{L}^{i}\bar{q}_{L}^{j}\bar{q}_{L}%
^{k}\times\varepsilon^{lmn}q_{R}^{l}q_{R}^{m}q_{R}^{n}\ \ \text{with\ }%
q=u,d,s\ ,
\end{equation}
which directly becomes a mass term for the $\eta^{\prime}$ for two flavors.
Another application is electroweak baryogenesis. In that case, the axial
operator has to be invariant under the SM flavor symmetry $SU(3)^{5}$
exhibited in the absence of Yukawa couplings, see Sec.~\ref{SecSMFlavorSym}. 
It is made of twelve left-handed fermion weak doublets in the flavor-singlet
combination (see e.g. Ref.~\cite{Smith:2011rp})
\begin{equation}
\mathcal{O}_{A}=\varepsilon^{ijk}\ell^{i}\ell^{j}\ell^{k}\times\varepsilon
^{abc}q^{a}q^{b}q^{c}\times\varepsilon^{def}q^{d}q^{e}q^{f}\times
\varepsilon^{lmn}q^{l}q^{m}q^{n}\ \text{with\ }\ell=(\nu_{L},e_{L}),q=(u_{L},d_{L})\ ,
\end{equation}
where it is understood that color and flavor antisymmetric contractions are
sufficiently entwined. This operator thus breaks $\mathcal{B}+\mathcal{L}$ by
three units. Though at colliders, the instanton tunnelling is suppressed by
$\exp(-2\pi/(\alpha\sin\theta_{W}))$ and there is no hope to access it, at
higher temperature, similar gauge configurations called \textbf{sphalerons}
are much more effective and make this effect potentially relevant for
electroweak baryogenesis~\cite{Rubakov:1996vz}.

The capability of instantons to induce fermionic transitions was first
demonstrated by 't Hooft in 1976~\cite{tHooft:1976snw,tHooft:1976rip}. His
calculation is somewhat technical, but the main idea can be stated rather
simply. Let us consider the $SU(2)$ instanton of Eq.~(\ref{SU2inst}). In that
background, we can construct a special left-handed Weyl spinor called a
\textbf{zero mode}:%
\begin{equation}
(\xi_{\alpha})_{i}=\frac{\varepsilon_{\alpha i}}{(x^{2}+\rho^{2})^{3/2}}%
\zeta\ ,
\end{equation}
where $\zeta$ is a free constant. Its crucial feature is the $\varepsilon$
tensor which entangles the gauge $SU(2)$ index $i$ with the Weyl spinor index
$\alpha$, which is also an $SU(2)$ index since we are in the Euclidian and
$SO(4)\sim SU(2)\times SU(2)$. Thanks to this feature, $\xi_{\alpha}$ is a
non-trivial spinor that obeys the Weyl equation (see e.g.
Ref.~\cite{Vandoren:2008xg}):%
\begin{equation}
\bar{\sigma}^{\mu}D_{\mu}\xi=(\bar{\sigma}^{\mu})^{\dot{\alpha}\alpha
}(\mathbf{1}\partial_{\mu}-\mathbf{A}_{\mu}^{inst})_{i}^{~j}(\xi_{\alpha}%
)_{j}=(\bar{\sigma}^{\mu})^{\dot{\alpha}\alpha}\frac{-3x_{\mu}\varepsilon
_{\alpha i}-(\bar{\sigma}_{\mu\nu})_{\alpha i}x^{\nu}}{(x^{2}+\rho^{2})^{5/2}%
}\zeta=0\ ,
\end{equation}
where we used where the fact that $(\bar{\sigma}_{\mu\nu})_{ij}=(\bar{\sigma
}_{\mu\nu})_{ji}$ when both its $SU(2)$ indices are down, and $(\bar{\sigma
}^{\mu})^{\dot{\alpha}\alpha}(\bar{\sigma}_{\mu\nu})_{\alpha i}=-3(\bar
{\sigma}^{\mu})^{\dot{\alpha}\alpha}\varepsilon_{\alpha i}$. From this, a
similar right-handed spinor can be shown to be a zero mode in a single
anti-instanton background. Now, in the presence of massless fermions, the
Lagrangian should include a $\psi i \slashed D\psi$ term, and the path 
integral for the vacuum-to-vacuum transition of Eq.~(\ref{OneInstVac}) 
should also be carried over fermions (see Sec.~\ref{SecFujikawa}).
Performing that integration first brings a factor $\det(i \slashed D)$, 
which vanishes since we have just seen that in the instanton background,
we can construct a fermionic zero mode. The idea behind the effective fermion
interactions is then to insert in Eq.~(\ref{OneInstVac}) an operator
$\mathcal{O}_{A}$ involving enough fermion fields to annihilate all the zero
modes in the path integral, so as to avoid the $\det(i \slashed D)=0$ factor. 
The instanton transition forces this operator to acquire a
vacuum expectation value $\langle\theta|\mathcal{O}_{A}|\theta\rangle$ and the
axial symmetry is broken. There is no Goldstone boson though because the phase
of this matrix element is not dynamical: it is simply $\theta$, which we are
still forced to keep in the theory to define a gauge-invariant vacuum (and
ensure cluster decomposition, see Ref.~\cite{Callan:1976je}).

\subsubsection{Topological susceptibility and singlet mass}\label{SecSuscTopo}

From a differential point of view, gauge fields are best understood as
one-forms, and the Chern-Simons current as the dual to a three-form. In the
usual tensor formalism, this means that the rank-three antisymmetric field%
\begin{equation}
C_{\nu\rho\sigma}=\frac{1}{3}\left\langle \mathbf{A}_{\nu}%
\overleftrightarrow{\partial}_{\rho}\mathbf{A}_{\sigma}-\mathbf{A}_{\rho
}\overleftrightarrow{\partial}_{\nu}\mathbf{A}_{\sigma}-\mathbf{A}_{\nu
}\overleftrightarrow{\partial}_{\sigma}\mathbf{A}_{\rho}-2i\mathbf{A}_{\nu
}\left[  \mathbf{A}_{\rho},\mathbf{A}_{\sigma}\right]  \right\rangle \ ,
\label{DualCS}%
\end{equation}
is dual to the three-dimensional Chern-Simons form (we shall drop the
subscript 3 here), $G_{\mu}=(1/3!)\varepsilon_{\mu\nu\rho\sigma}C^{\nu
\rho\sigma}$, in the same sense that the electromagnetic field strengths
$F_{\mu\nu}$ and $\tilde{F}_{\mu\nu}$ are dual and the free Maxwell equations
invariant under $(\vec{E},\vec{B})\rightarrow(\vec{B},-\vec{E})$. The interest
of constructing $C_{\mu\nu\rho}$ is to evidence a hidden long-range force of
topological origin, induced by instantons. Let us see how this comes about.

A peculiar property of dualization is to transform the divergences
$\partial_{\mu}G^{\mu}$, analog to the Lorenz condition for a vector field,
into the field strength of the dual field:
\begin{equation}
\partial^{\mu}G_{\mu}=\frac{1}{4!}\varepsilon_{\mu\nu\rho\sigma}F^{C,\mu
\nu\rho\sigma}\ ,\ \ F_{\mu\nu\rho\sigma}^{C}=\partial_{\mu}C_{\nu\rho\sigma
}+\partial_{\nu}C_{\rho\mu\sigma}+\partial_{\rho}C_{\mu\nu\sigma}%
+\partial_{\sigma}C_{\nu\mu\rho}\ .
\end{equation}
In the present case, the three-form field does not have much dynamics because
$F_{\mu\nu\rho\sigma}^{C}$ has only one non-zero component $F_{\mu\nu
\rho\sigma}^{C}=\varepsilon_{\mu\nu\rho\sigma}F_{0123}^{C}$. We call this
component \textit{electric} by analogy with the $F_{0i}$ entries of the
Maxwell strength tensor, and we will see in Sec.~\ref{ElectricTheta} that $F_{0123}^{C}$ is a
constant background electric field. Of course, neither $G_{\mu}$ nor
$C_{\mu\nu\rho}$ have independent dynamics since ultimately it is the gauge
fields that drive them. Actually, it is once instantons enter the game that
this formalism becomes particularly appealing. We have seen that they give a
$\theta$ dependence to the vacuum energy, Eq.~(\ref{Etheta}), and that this
drives matrix elements of $dE(\theta)/d\theta\sim\left\langle \varepsilon
^{\mu\nu\rho\sigma}\mathbf{F}_{\mu\nu}\mathbf{F}_{\rho\sigma}\right\rangle $,
Eq.~(\ref{dEtheta}). This also means that%
\begin{equation}
\left.  \frac{d^{2}E(\theta)}{d\theta^{2}}\right\vert _{\theta=0}=\left(
\frac{g^{2}}{16\pi^{2}}\right)  ^{2}\int d^{4}x\ \langle\theta|T\{\partial
^{\mu}G_{\mu}(x)\partial^{\nu}G_{\nu}(0)\}|\theta\rangle_{\theta=0}\neq0\ .
\label{ddEtheta}%
\end{equation}
In Minkowski space and after a Fourier transform, this matrix element is
related to a quantity called the \textbf{topological susceptibility}%
\begin{equation}
\chi(p^{2})=-i\left(  \frac{g^{2}}{16\pi^{2}}\right)  ^{2}\int d^{4}%
x~e^{ip\cdot x}~\langle\theta|T\{\partial^{\mu}G_{\mu}(x)\partial^{\nu}G_{\nu
}(0)\}|\theta\rangle_{\theta=0}\ . \label{TopoSusc}%
\end{equation}

Effectively, the non-zero value $\chi(0)$ generated by instantons can be
represented by a kinetic term for the three-form field%
\begin{equation}
\frac{1}{2\chi(0)}(\partial^{\mu}G_{\mu})^{2}=-\frac{1}{2\chi(0)}\frac{1}%
{4!}F_{\mu\nu\rho\sigma}^{C}F^{C,\mu\nu\rho\sigma}\ , \label{KinC}%
\end{equation}
where we have rescaled $\partial^{\mu}G_{\mu}$ and $F_{\mu\nu\rho\sigma}^{C}$
by $g^{2}/(16\pi^{2})$. In other words, there appear a new long-range force of
topological origin~\cite{Luscher:1978rn}. It is long range because of the
$U(1)$ gauge symmetry of the $C$ field kinetic term in Eq.~(\ref{KinC}): it is
invariant under
\begin{equation}
C_{\mu\nu\rho}\rightarrow C_{\mu\nu\rho}+\partial_{\mu}\Lambda_{\nu\rho
}+\partial_{\nu}\Lambda_{\rho\mu}+\partial_{\rho}\Lambda_{\mu\nu}\ ,
\end{equation}
with $\Lambda_{\mu\nu}=-\Lambda_{\mu\nu}$ a space-time dependent tensor of
gauge parameters. Alternatively, this can be seen by computing the propagator
of $C_{\mu\nu\rho}$, after including a gauge-fixing term $(\partial^{\mu
}C_{\mu\nu\rho})^{2}/(2\xi)$ to invert the kinetic term (see
Ref.~\cite{Plantier:2025hcm}):%
\begin{align}
\mathcal{P}_{\mu\nu\rho,\alpha\beta\gamma}(k)  &  =\int d^{4}x~e^{ik\cdot
x}\langle\theta|T\{C_{\mu\nu\rho}(x)C_{\alpha\beta\gamma}(0)\}|\theta
\rangle_{\theta=0}\nonumber\\
&  =-i\varepsilon_{\mu\nu\rho\sigma}\varepsilon_{\alpha\beta\gamma\delta
}\left(  \frac{\xi}{k^{2}}g^{\sigma\delta}+\frac{1-\xi}{k^{2}}\frac{k^{\sigma
}k^{\delta}}{k^{2}}\right)  \ .
\end{align}
In the Landau gauge $\xi=0$, where only transverse degrees of freedom
propagate, only the $k_{\sigma}k_{\delta}/k^{4}$ piece survives. Killing off
the $\varepsilon$ tensors, this term then corresponds to the $\langle
\theta|T\{G_{\sigma}(x)G_{\delta}(0)\}|\theta\rangle_{\theta=0}$ propagator,
$\mathcal{P}_{\sigma,\delta}(k)\sim k_{\sigma}k_{\delta}/k^{4}$. Contracted
twice\footnote{Be careful though that moving the derivatives outside the
matrix element is not necessarily permitted but can generate Schwinger terms,
see Eq.~(\ref{SchwingerTerms}).} to get to Eq.(\ref{TopoSusc}), it collapses
to a contact term since $k^{\sigma}k^{\delta}\mathcal{P}_{\sigma,\delta
}(k)\sim1$. Once accounting for $\chi(0)$ in the kinetic term, this is in
agreement with Eq.~(\ref{ddEtheta}). All this may be a bit puzzling at first
sight. One may wonder why all the $C$ field is capable of doing is to generate
a contact term if it is long range. Later on, in Sec.~\ref{ElectricTheta}, we
will see that the $C$ field actually does not propagate true degrees of
freedom, but can be viewed as a kind of constant electric field background,
whence its long-range nature.

\begin{figure}[t]
\centering\includegraphics[width=0.34\textwidth]{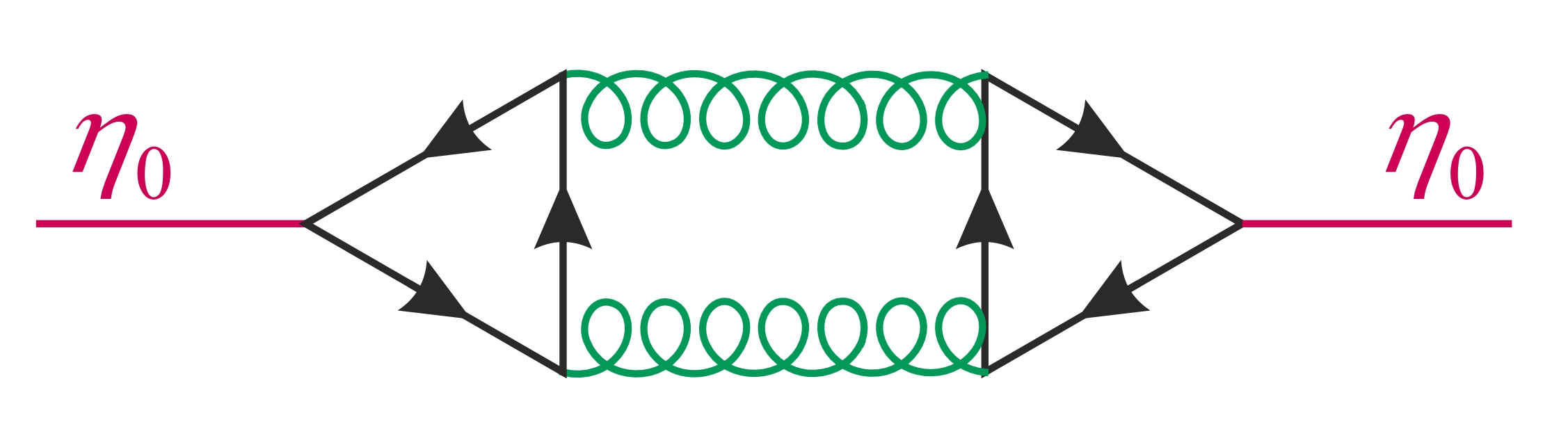}\caption{Naive
representation of the origin of the singlet $\eta_{0}$ mass. One should
understand that what is meant by the two gluon states is actually the
non-perturbative effect of gluonic instantons, generating an effective
long-range force.}%
\label{FigGG}%
\end{figure}

Indirectly, the topological susceptibility drives the mass of the singlet
pseudoscalar meson\footnote{We cannot denote this state as $\eta^{\prime}$
because the physical $\eta^{\prime}$ state is a mixture of the flavor singlet
$\eta_{0}$ and flavor octet $\eta_{8}$, see Eq.~(\ref{hhmixing}).} $\eta_{0}$.
If this state is the Goldstone boson associated to the axial current,
Goldstone theorem predicts a non-zero matrix element $\langle0|\partial^{\mu
}A_{\mu}|\eta_{0}\rangle\sim F_{\eta}$, with $F_{\eta}$ the symmetry breaking
scale (and decay constant of the $\eta_{0}$ meson). But then, the singlet
anomaly Eq.~(\ref{AnoSinglet1}) states that $\left\langle \varepsilon^{\mu
\nu\rho\sigma}\mathbf{F}_{\mu\nu}\mathbf{F}_{\rho\sigma}\right\rangle $ can
also create singlet mesons. To account for this, we can construct the
effective Lagrangian (for the sign and normalization of the three-form kinetic
term, we follow Ref.~\cite{Plantier:2025hcm})%
\begin{equation}
\mathcal{L}_{eff}=-\frac{F_{\eta}^{4}}{2\chi(0)}\frac{1}{4!}F_{\mu\nu
\rho\sigma}^{C}F^{C,\mu\nu\rho\sigma}+\frac{1}{4!}F_{\eta}\eta_{0}%
\varepsilon_{\mu\nu\rho\sigma}F^{C,\mu\nu\rho\sigma}+\frac{1}{2}\partial_{\mu
}\eta_{0}\partial^{\mu}\eta_{0}\ , \label{EffCS1}%
\end{equation}
where we have used Eq.~(\ref{KinC}) and given that $\chi(0)$ has
mass-dimension four, introduced the only (besides $\chi(0)$ itself) available
dimensional scale $F_{\eta}^{4}$ to have a proper kinetic term of
mass-dimension four. Notice that this is not so innocent since in terms of the
underlying gluon fields, $F^{C,\mu\nu\rho\sigma}$ is actually of
mass-dimension four. Yet, we insist on treating it as a new field. Since it is
in any case not really propagating, it can be integrated out using its
classical EoM, and this produces a mass term for the singlet,%
\begin{equation}
F_{\eta}^{2}m_{\eta}^{2}=\chi(0)\ . \label{VeneWitt1}%
\end{equation}
This is the famous \textbf{Witten-Veneziano relation}%
~\cite{Veneziano:1979ec,Witten:1979vv}. It shows that what drives the
$\eta_{0}$ mass is not really its coupling to the anomaly, but rather the
presence of instantons. In other words, a Feynman diagram representation like
in Fig.~\ref{FigGG} is not a realistic representation because non-perturbative gluonic
instanton configurations must be involved in between the two triangles.

From this derivation, two questions immediately come to
mind~\cite{Witten:1979vv}. First, if the axial symmetry is anomalous, there is
no reason to expect $\langle0|\partial^{\mu}A_{\mu}|\eta_{0}\rangle$ to be
non-zero. Certainly, $\eta_{0}$ cannot be the Goldstone boson of an inexistent
symmetry. The solution to that puzzle relies on large $N_{C}$ arguments, with
$N_{C}$ the number of QCD colors. When it goes to infinity, the anomaly
disappears and $\eta_{0}$ is indeed a true Goldstone boson. So, it makes sense
for it to remain coupled to the axial current for finite $N_{C}$. Second, we
have seen in the previous section that in the presence of massless fermions,
which are required for the axial symmetry to exist, the topological
susceptibility $\chi(p^{2})$ vanishes at $p^{2}=0$. One may thus wonder how
$m_{\eta}$ could be related to the topological susceptibility of a quark-less
version of QCD. To answer this, it is necessary to go beyond the naive
integration of the $C$ field in Eq.~(\ref{EffCS1}), and consider the coupled
system made of the Chern-Simons form and the singlet $\eta_{0}$. It is
actually easier to switch back to the dual picture, in which the effective
Lagrangian is~\cite{DiVecchia:1980yfw,Rosenzweig:1979ay}
\begin{equation}
\mathcal{L}_{eff}=\frac{F_{\eta}^{4}}{2\chi(0)}(\partial^{\mu}G_{\mu}%
)^{2}+F_{\eta}\eta_{0}\partial^{\mu}G_{\mu}+\frac{1}{2}\partial_{\mu}\eta
_{0}\partial^{\mu}\eta_{0}-\frac{\bar{m}_{\eta_{0}}^{2}}{2}\eta_{0}^{2}\ .
\label{LEffTopo}%
\end{equation}
The instantonic glueball $\partial^{\mu}G_{\mu}$ looks like a non-propagating
scalar field. Its mass is of the wrong sign though, $\bar{m}_{CS}=-F_{\eta
}^{4}/\chi(0)$, because it truly is a three-form
field~\cite{DiVecchia:1980yfw}. The $\bar{m}_{\eta_{0}}^{2}$ mass term is that
purely induced by quark masses. If we remember that the singlet has the quark
content $\eta_{0}=(\bar{u}u+\bar{d}d+\bar{s}s)/\sqrt{3}$, its mass can be
extracted from%
\begin{equation}
\frac{i}{q^{2}-\bar{m}_{\eta_{0}}^{2}}=\frac{1}{\sqrt{3}}\left(
\sum_{q=u,d,s}\frac{i}{q^{2}-2Bm_{q}}\right)  \frac{1}{\sqrt{3}}%
\Rightarrow\bar{m}_{\eta_{0}}^{2}=\frac{6B}{\omega_{m}}\ ,
\label{Eq6Bw}
\end{equation}
where $B$ is a hadronic parameter relating quark mass to meson masses squared
(see Sec.~\ref{SecMesons}), and%
\begin{equation}
\omega_{m}=\frac{1}{m_{u}}+\frac{1}{m_{d}}+\frac{1}{m_{s}}\ .
\end{equation}
Now, if we truly interpret $\partial^{\mu}G_{\mu}$ as a scalar field, it mixes
with the singlet $\eta_{0}$ so a Dyson resummation is needed to get the
physical singlet mass and the topological susceptibility of the full theory.
The geometric series then resolve
into~\cite{DiVecchia:1980yfw,DiVecchia:2017xpu}
\begin{subequations}
\label{DysonCS}%
\begin{align}
\frac{i}{q^{2}-m_{\eta_{0},\rm{phys}}^{2}}  &  =\frac{i}{q^{2}-\bar{m}_{\eta_{0}}^{2}%
}\left(  1-\frac{F_{\eta}}{q^{2}-\bar{m}_{CS}^{2}}\frac{F_{\eta}}{q^{2}%
-\bar{m}_{\eta_{0}}^{2}}\right)  ^{-1}\rightarrow m_{\eta_{0},\rm{phys}}^{2}%
=\frac{6B_{0}}{\omega_{m}}+\frac{\chi(0)}{F_{\eta}^{2}}\ ,\ \\
\chi_{\rm{full}}(q)  &  =\frac{F_{\eta}^{4}}{q^{2}-\bar{m}_{CS}^{2}}\left(
1-\frac{F_{\eta}}{q^{2}-\bar{m}_{\eta_{0}}^{2}}\frac{F_{\eta}}{q^{2}-\bar
{m}_{CS}^{2}}\right)  ^{-1}\rightarrow\frac{1}{\chi_{\rm{full}}(0)}=\frac{1}%
{\chi(0)}+\frac{\omega_{m}}{6B_{0}F_{\eta}^{2}}\ .
\end{align}
These equations exhibit all the right features: the physical singlet mass is dominated
by the gluonic topological susceptibility if any of the quark mass vanishes,
while $\chi_{full}(0)$ does indeed go to zero in that limit, as well as if
instantons are turned off and $\chi(0)\rightarrow0$.

\section{The chiral anomaly}

The abelian and singlet anomaly calculations treat the external currents in a
very asymmetric way: one represents a global symmetry current, while the
others stand for genuine gauge interactions. This is clearly inadequate to
deal with the case in which anomalies would jeopardy a given gauge theory,
that is, where all the vertices arise from the same gauge current. In
practice, this situation is significantly more complicated to address for
several reasons. This will be detailed in the next section to get a clear
picture of what needs to be tackled. In the following section, the actual
calculation of the triangle, box, and pentagon amplitudes is presented.

\subsection{Gauge anomalies}

Whenever a gauge current is anomalous, the gauge symmetry itself does not
survive to quantization. This is very serious because, as said in the
introduction, Ward identities are essential to ensure predictivity. Obviously,
the main danger comes from UV divergent diagrams. We need the Ward identities
to ensure all infinities disappear once only a restricted number of
renormalization conditions are set. In the presence of gauge anomalies, the
Ward identity receives extra terms, the renormalization procedure fails, and
the theory should be discarded. Yet, this situation is interesting for several
reasons. First, the way in which anomalies break the gauge symmetry is very
peculiar, and this will teach us a lot about the intimate structure of gauge
theories. Second, the SM is consistent only thanks to the accidental
cancellations of the gauge anomalies showing up for each of its fermions, so
it is worth to know about them. Third, gauge anomalies for models with more
than one gauge field, or more precisely such that the gauge group can be split
into subgroups, describe many situations of relevance phenomenologically.
Whenever it is possible to "move" all the anomaly on the currents of one
subgroup, leaving the other conserved, we get a viable gauge theory simply by
not coupling physical gauge fields to the anomalous currents. In this way, we
will be able to describe many more anomalous situations than that of the
global axial current coupled to gauge vector fields of the previous two sections.

Before embarking into the calculations of diagrams, it is worth to think about
what to expect. First, we know that anomalous diagrams suffer from finite UV
ambiguities. In the previous section, we had to use Ward identities to get a
definite result (actually, we used a Pauli-Villars regularization, which
automatically imposes the vector Ward identities). Obviously, this path is
closed if all legs of our diagrams are meant to stand for the same gauge
field. Fortunately, we will see that the quite natural symmetry requirements
under the interchange of the fields are in general sufficient to fix all the
ambiguities. For this to work though, it will prove crucial to fully encode
the whole ambiguity, i.e., to introduce enough free parameters to cover all
possible outcomes of the diagram calculation whatever the choice of
regularization scheme, momentum routing,... Also, we will have to live with
the fact that the anomalous divergences are not gauge invariant, and cannot be
expressed in operator form using only the field strengths.

Another complexity comes from the very nature of Ward identities as
differential relationships between correlators or Green functions. Previously,
to go from $\partial_{\mu}\langle\gamma\gamma|A^{\mu}|0\rangle$ to
$\langle\gamma\gamma|\partial_{\mu}A^{\mu}|0\rangle$, we simply moved the
derivative within the matrix element. Yet, there is implicitly a time-ordered
product, which do depend on time. So, we should actually write%
\end{subequations}
\begin{equation}
\partial^{\mu}\langle0|T\{A_{\mu}V_{\nu}V_{\rho}\}|0\rangle=\langle
0|T\{\partial^{\mu}A_{\mu}V_{\nu}V_{\rho}\}|0\rangle+\langle0|(\partial
_{0}T)\{A_{0}V_{\nu}V_{\rho}\}|0\rangle\ . \label{SchwingerTerms}%
\end{equation}
If we express the time ordering in terms of step functions, its derivative
brings delta functions times $\langle0|T\{[A_{0},V_{\nu}]V_{\rho}\}|0\rangle$
and $\langle0|T\{[A_{0},V_{\rho}]V_{\nu}\}|0\rangle$. For the abelian and
singlet anomalies of the previous section, the axial current is a gauge
singlet so these commutators trivially vanish, but they do not in general and
are called \textbf{Schwinger terms}. A well-known example comes from the QED
one-particle irreducible vertex function $\Gamma^{\mu}\varepsilon_{\mu
}^{\lambda}(k)=\langle e^{+}(p)e^{-}(p^{\prime})|\gamma(k)\rangle$, which in
momentum space satisfies
\begin{equation}
k_{\mu}\Gamma^{\mu}\left[  k;p,-\left(  p+k\right)  \right]  =S_{F}%
^{-1}\left(  p+k\right)  -S_{F}^{-1}\left(  p\right)  ^{-1}\ ,
\end{equation}
with $S_{F}$ the full electron propagator $\langle0|T\{\bar{\psi}_{e}\psi
_{e}\}|0\rangle$, including the electron self energy\footnote{By the way, this
Ward identity illustrates very well the point made at the beginning of this
section. By relating the UV infinities in the photon-fermion vertex correction
and the fermion wavefunction correction, it ensures the electromagnetic
coupling stays the same for all charged fermions.}. Similarly, if we start
from $\langle0|T\{\mathbf{A}_{\mu}\mathbf{A}_{\nu}\mathbf{A}_{\rho}%
\}|0\rangle$, taking the divergence will generates Schwinger terms
proportional to the structure constant via $[T^{a},T^{b}]=if^{abc}T^{c}$
together with the two-point function $\langle0|T\{\mathbf{A}_{\nu}%
\mathbf{A}_{\rho}\}|0\rangle$, i.e., the vacuum polarization. All this also
applies to four and five-point correlators, for which Schwinger terms produce the
configuration-space equivalent of the momentum-space covariant derivatives
already encountered in Sec.~\ref{VectorWard}.

Yet another complication comes from the nature of the gauge theory itself. If
it is meant to be anomalous, it needs to be a \textbf{chiral theory} because a
vector theory like QED cannot have gauge anomalies. To see this, consider a
fermion loop to which $n$ photons are attached:%
\begin{equation}
\mathcal{M}_{1\rightarrow n}=\int\frac{d^{4}k}{(2\pi)^{4}}\operatorname*{Tr}%
\left[  S_{F}(k)\gamma^{\alpha_{1}}S_{F}(k-q_{1})\gamma^{\alpha_{2}}%
S_{F}(k-q_{1}-q_{2})...\gamma^{\alpha_{n}}\right]  \ .
\end{equation}
There are $(n-1)!$ such diagrams, among which there will be that corresponding
to the loop momentum running in the opposite direction%
\begin{equation}
\mathcal{M}_{n\rightarrow1}=\int\frac{d^{4}k}{(2\pi)^{4}}\operatorname*{Tr}%
\left[  \gamma^{\alpha_{n}}...S_{F}(k+q_{1}+q_{2})\gamma^{\alpha_{2}}%
S_{F}(k+q_{1})\gamma^{\alpha_{1}}S_{F}(k)\right]  \ .
\end{equation}
Inserting $\mathbf{1}=\mathrm{C}^{-1}\mathrm{C}$ in between all factors in the
Dirac trace for $\mathcal{M}_{n\rightarrow1}$, with the charge conjugation
matrices $\mathrm{C}=-i\gamma^{2}\gamma^{0}$ satisfying $\mathrm{C}\gamma
_{\mu}\mathrm{C}^{-1}=-\gamma_{\mu}^{T}$ and thus $\mathrm{C}S_{F}%
(p)\mathrm{C}^{-1}=S_{F}(-p)^{T}$, and using the invariance of the Dirac trace
under matrix transposition, one finds $\mathcal{M}_{n\rightarrow1}%
=(-1)^{n}\mathcal{M}_{1\rightarrow n}$ after switching the integration
variable $k\rightarrow-k$. This proves that once summed over permutations of
the photon fields, the one-loop amplitude with an odd number of photons
vanishes, a result known as \textbf{Furry's theorem}~\cite{Furry:1937zz}.

All this to say that to get a non-trivial triangle amplitude involving the
same gauge field on all its legs, we need to consider chiral gauge theories,
for instance those whose gauge couplings involve the projector $P_{L,R}%
=(1\mp\gamma_{5})/2$. But then, we cannot have a mass term for the fermion
since $\psi_{L}$ and $\psi_{R}$ do not have the same gauge charges, and
$m\bar{\psi}_{L}\psi_{R}$ would already break the gauge symmetry at the
classical level. Not only do we need to set the fermion mass to zero, but we
also have to give up the simple Pauli-Villars regularization procedure. Though
this brings some serious technical complications, it will prove extremely
rewarding. First, we will obtain a general expression for the gauge anomalies,
from which their various incarnations (consistent, covariant...) can be
deduced, and their properties studied. Then, it will allow us to go beyond the
Sutherland-Veltman theorem, that is, beyond vector current conservation. This
is very relevant phenomenologically already in the SM, because the baryon
plus lepton number current is vectorial yet it has an electroweak anomaly.

\subsection{Chiral triangles, boxes, and pentagons}\label{SecChiralAno}

\begin{figure}[t]
\centering\includegraphics[width=0.80\textwidth]{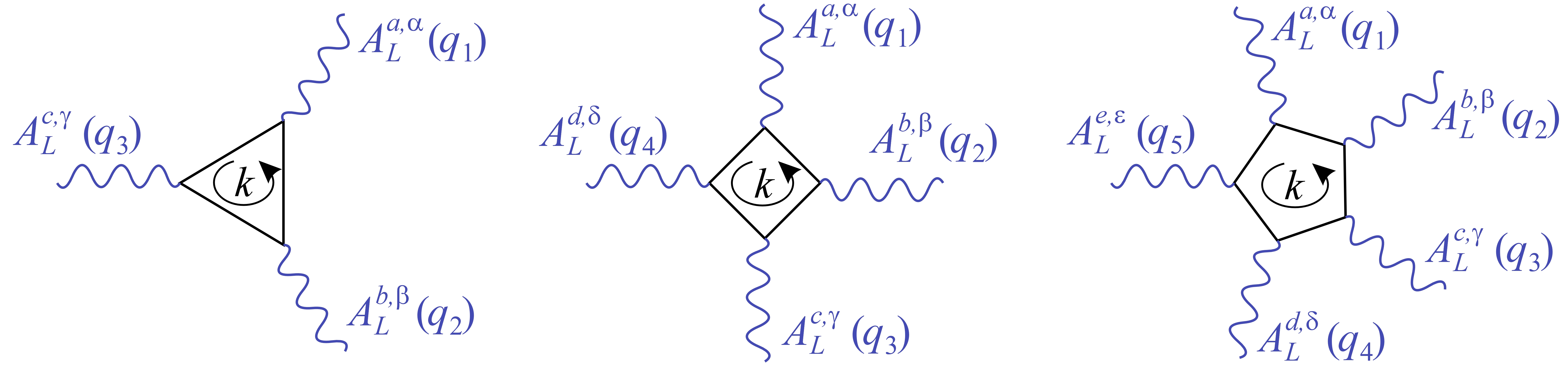}\caption{Triangle,
box, and pentagon amplitudes for the chiral anomaly. All external momenta are
outgoing. Summation is understood under the permutations of the gauge
vertices, as well as over the fermionic states making up a representation of
$SU(N)$.}%
\label{Fig4}%
\end{figure}

In the simplest setting, we consider a purely chiral $SU(N)$ gauge theory,
coupled to a set of left-handed fermions. With our convention $D_{\mu}%
\psi=\partial_{\mu}\psi-i\mathbf{A}_{\mu}^{L}P_{L}\psi$, so the gauge coupling
is $\bar{\psi}\gamma^{\mu}P_{L}\mathbf{A}_{\mu}^{L}\psi$ with unit coupling
constant, $\mathbf{A}_{\mu}^{L}=A_{\mu}^{L,a}T^{a}$ with $T^{a}$ the group
generators in the representation carried by the massless fermions, and the
corresponding Feynman rule is $-i\gamma^{\mu}P_{L}$. Since the gauge symmetry
is non abelian, we must consider the three, four, and five-point diagrams
shown in Fig.~\ref{Fig4}.

Let us start with the triangle amplitude:%
\begin{align}
\mathcal{T}_{LLL}^{\alpha\beta\gamma,abc}  &  =(-1)\int\frac{d^{4}k}%
{(2\pi)^{4}}\operatorname*{Tr}\left[  \frac{i}{
\slashed k- \slashed q_{1}- \slashed q_{2}}(-i\gamma^{\beta}P_{L})\frac{i}{
\slashed k- \slashed q_{1}}(-i\gamma^{\alpha}P_{L})\frac{i}{\slashed k}
(-i\gamma^{\gamma}P_{L})\right]  \langle T^{c}T^{b}T^{a}\rangle\nonumber\\
&  +(1,\alpha,a\overset{}{\leftrightarrow}2,\beta,b)\ . \label{LoopTri}%
\end{align}
We use $4\operatorname*{Tr}(T^{a}T^{b}T^{c})=2i\mathcal{I}_{2}f^{abc}%
+\mathcal{I}_{3}d^{abc}$ to immediately discard the part proportional to
$f^{abc}$, corresponding to the Schwinger terms. The cubic Casimir invariant
$\mathcal{I}_{3}$ depends on the representation, and is often called the
\textbf{anomaly coefficient} of the representation in which the fermions are
embedded (further discussed in Sec.~\ref{ConsistSM}).

To calculate this loop, one way to proceed is to introduce some parameters to
keep track of the choice of momentum routing, split the divergences
$i(q_{1}+q_{2})_{\alpha}\mathcal{T}_{LLL}^{\alpha\beta\gamma,abc}$ into pairs
of terms that superficially cancel out under momentum shifts, and calculate
the corresponding surface term by cutting off the momentum integral. This is
the technique used e.g. in Weinberg's book~\cite{Weinberg:1996kr}, which we
shall not repeat here. Also, that method does not give us the original
$\mathcal{T}_{LLL}^{\alpha\beta\gamma,abc}$ amplitude, which we will need later
on. So, we switch to dimensional regularization. As noted already by 't Hooft
and Veltman~\cite{tHooft:1972tcz}, making sense of $\gamma_{5}$ in $d\neq4$
dimensions in general breaks its anticommuting properties (which is something
used to get to Eq.~(\ref{PropSplit})). The momentum routing ambiguity is then
replaced by that in the position of $\gamma_{5}$ prior to moving to $d$ dimensions.

In practice, the first step, still in $d=4$, is to move all the $P_{L}$ to the
left, and discard the purely vectorial contribution thanks to Furry's theorem.
There is then a unique $\gamma_{5}$ occurring in the Dirac trace. To keep
track of the ambiguity in going to dimension $d$, we follow
Ref.~\cite{Chanowitz:1979zu,Elias:1982ea}\ (see also
Ref.~\cite{Novotny:1994yx}) and split this trace as%
\begin{align}
(\slashed k- \slashed q_{1}- \slashed q_{2})\gamma^{\beta}
(\slashed k- \slashed q_{1})\gamma^{\alpha}\slashed k\gamma^{\gamma}\gamma_{5}  &  
\rightarrow(1-a_{1}-a_{2})(\slashed k- \slashed q_{1}- \slashed q_{2})\gamma^{\beta}(
\slashed k- \slashed q_{1})\gamma^{\alpha} \slashed k\gamma^{\gamma}\gamma_{5}\nonumber\\
&  \ \ \ \ +a_{1}(\slashed k- \slashed q_{1}- \slashed q_{2})\gamma^{\beta}(
\slashed k- \slashed q_{1})\gamma^{\alpha}\gamma_{5} \slashed k\gamma^{\gamma}\nonumber\\
&  \ \ \ \ +a_{2}(\slashed k- \slashed q_{1}- \slashed q_{2})\gamma^{\beta}\gamma_{5}(
\slashed k- \slashed q_{1})\gamma^{\alpha} \slashed k\gamma^{\gamma}\ , \label{DRano}%
\end{align}
and similarly for the second triangle with independent $a_{3,4}$ coefficients.
These traces are computed using the so-called Breit-Maison-'t Hooft-Veltman
scheme~\cite{Breitenlohner:1977hr,tHooft:1972tcz} as implemented in
FeynCalc~\cite{Mertig:1990an,Shtabovenko:2016sxi,Shtabovenko:2020gxv,Shtabovenko:2023idz}%
, and the loop integral is carried in dimensional regularization. Setting
$q_{1}^{2}=q_{2}^{2}=0$ only at the very end of the calculation (and keeping
$q_{3}^{2}=2q_{1}\cdot q_{2}\neq0$) gives%
\begin{equation}
\mathcal{T}_{LLL}^{\alpha\beta\gamma,abc}\cong\frac{\mathcal{I}_{3}d^{abc}%
}{32\pi^{2}}\left(  (1-2a_{2})\varepsilon^{\alpha\beta\gamma\mu}q_{1\mu
}-(1-2a_{1})\varepsilon^{\alpha\beta\gamma\mu}q_{2\mu}-\frac{q_{1}^{\beta
}\varepsilon^{\alpha\gamma\mu\nu}q_{1\mu}q_{2\nu}}{q_{1}\cdot q_{2}}%
+\frac{q_{2}^{\alpha}\varepsilon^{\beta\gamma\mu\nu}q_{1\mu}q_{2\nu}}%
{q_{1}\cdot q_{2}}\right)  \ . \label{AmpTri}%
\end{equation}
The symbol $\cong$ reflects the neglect of some terms that do not contribute
to divergences, like imaginary parts. This will be discussed in
Sec.~\ref{IRUV}. In the course of the computation, the $a_{3,4}$ coefficients
are fixed with respect to the $a_{1,2}$ to ensure Furry's theorem, i.e., that
$\mathcal{T}_{LLL}^{\alpha\beta\gamma,abc}$ has no $f^{abc}$ term. For
triangle diagrams, the distinction between normal and covariant derivatives is
not important because the difference amounts to non-anomalous vacuum
polarization graphs (see Fig.~\ref{Fig3}), that is, to Schwinger terms. Taking
the divergences in momentum space amounts to contracting with the
corresponding momentum, and we write somewhat symbolically:
\begin{subequations}
\label{AnoLLL}%
\begin{align}
D_{\alpha}^{L}\mathcal{T}_{LLL}^{\alpha\beta\gamma,abc}  &  =-i\frac
{\mathcal{I}_{3}d^{abc}}{16\pi^{2}}a_{1}\varepsilon^{\beta\gamma\mu\nu}%
q_{1\mu}q_{2\nu}\ ,\\
D_{\beta}^{L}\mathcal{T}_{LLL}^{\alpha\beta\gamma,abc}  &  =-i\frac
{\mathcal{I}_{3}d^{abc}}{16\pi^{2}}a_{2}\varepsilon^{\gamma\alpha\mu\nu
}q_{1\mu}q_{2\nu}\ ,\\
D_{\gamma}^{L}\mathcal{T}_{LLL}^{\alpha\beta\gamma,abc}  &  =-i\frac
{\mathcal{I}_{3}d^{abc}}{16\pi^{2}}\left(  1-a_{1}-a_{2}\right)
\varepsilon^{\alpha\beta\mu\nu}q_{1\mu}q_{2\nu}\ .
\end{align}
This is an important result. The full amplitude includes IR singular terms
since $q_{3}^{2}=(q_{1}+q_{2})^{2}=2q_{1}\cdot q_{2}$, so all three legs of
the triangle cannot be put on-shell simultaneously. Though its divergences are
totally safe, these IR singular terms leave their marks, and make it
impossible to choose $a_{1,2}$ so as to cancel the three divergences simultaneously.

Let us now turn our attention to the box diagrams shown in Fig.~\ref{Fig4},
whose amplitude is the sum of six terms:%
\end{subequations}
\begin{align}
\mathcal{T}_{LLLL}^{\alpha\beta\gamma\delta,abcd}  &  =\nonumber\\
(-1)  &  \int\frac{d^{4}k}{(2\pi)^{4}}\operatorname*{Tr}\left[  \frac{1}{
\slashed k- \slashed q_{1}- \slashed q_{2}- \slashed q_{3}}\gamma^{\gamma}P_{L}
\frac{1}{\slashed k- \slashed q_{1}- \slashed q_{2}}\gamma^{\beta}P_{L}
\frac{1}{\slashed k- \slashed q_{1}}\gamma^{\alpha}P_{L}
\frac{1}{\slashed k}\gamma^{\delta}P_{L}\right]  \langle T^{d}T^{c}T^{b}T^{a}\rangle\nonumber\\
&  +(1,\alpha,a\overset{}{\leftrightarrow}2,\beta,b\overset{}{\leftrightarrow
}3,\gamma,c)\ . \label{LoopBox}%
\end{align}
Moving all the $P_{L}$ to the right, there is a $VVVV$ and $VVVA$
contribution. The former is not anomalous and generates interactions that do not
interest us here (they already exist for an abelian
theory like QED, and describe light-by-light scattering
effects~\cite{Quevillon:2018mfl}). The anomaly is in the $VVVA$ piece, for
which the prescription of Eq.~(\ref{DRano}) can readily be adapted. Following
exactly the same steps as for the triangle diagram, we obtain the rather
compact result%
\begin{align}
\mathcal{T}_{LLLL}^{\alpha\beta\gamma\delta,abcd}  &  \cong-i\frac
{\mathcal{I}_{3}d^{abe}f^{cde}}{32\pi^{2}}\left(  (2b_{2}-1)\varepsilon
^{\alpha\beta\gamma\delta}+\left(  \frac{q_{3}^{\beta}}{q_{23}}-\frac
{q_{1}^{\beta}}{q_{12}}\right)  \frac{q_{2}^{\gamma}\varepsilon^{\alpha\delta
q_{1}q_{2}}-q_{2}^{\alpha}\varepsilon^{\gamma\delta q_{2}q_{3}}}{q_{13}%
}\right. \nonumber\\
&
\ \ \ \ \ \ \ \ \ \ \ \ \ \ \ \ \ \ \ \ \ \ \ \ \ \ \ \ \ \ \ \ \ \ \ \ \ \ \ \ \ \ \ \left.
-\frac{q_{1}^{\beta}+q_{3}^{\beta}}{q_{13}}\varepsilon^{\alpha\gamma\delta
q_{2}}-\frac{q_{2}^{\gamma}q_{2}^{\delta}}{q_{13}}\left(  \frac{\varepsilon
^{\alpha\beta q_{1}q_{2}}}{q_{12}}+\frac{\varepsilon^{\alpha\beta q_{2}q_{3}}%
}{q_{23}}\right)  \right) \nonumber\\
&  +(q_{2},\beta,b\overset{}{\leftrightarrow}q_{3},\gamma,c)~\&~(b_{2}%
\overset{}{\rightarrow}b_{3})\nonumber\\
&  +(q_{2},\beta,b\overset{}{\leftrightarrow}q_{4},\delta,d)~\&~(b_{2}%
\overset{}{\rightarrow}b_{4})\ , \label{AmpBox}%
\end{align}
where $q_{ij}=q_{i}\cdot q_{j}$ and $q_{4}=-q_{1}-q_{2}-q_{3}$. The three
arbitrary constants $b_{2,3,4}$ are the only surviving combinations of the
initial $18$ free parameters, coming from the six box diagrams each with three
parameters after half of them are fixed to enforce Furry's theorem (i.e., such
that terms not involving $d\times f$ invariants cancel out). The strange
numbering $b_{i}$, $i=2,3,4$ will later allow us to use $b_{i}$ to tune the
anomaly on the $i^{th}$ leg. Beware that though straightforward, the
calculation is not easy because the loop integration has to be done keeping
the external legs off-shell. Further, once going on-shell, many tricky
simplifications follow from systematically enforcing the Schouten identity
Eq.~(\ref{Schouten}). Finally, the Jacobi identity%
\begin{equation}
f^{cde}d^{abe}+f^{ade}d^{bce}+f^{bde}d^{ace}=0\ ,
\end{equation}
is called to reduce the amplitude to a combination of only three linearly
independent $SU(N)$ tensors.

The last step to get the box contribution to the anomaly is to take the
covariant derivatives\footnote{Compared to Eq.~(\ref{VWardB}) and
Fig.~\ref{Fig3}, let us stress that here, contracting the full amplitude by
$q_{1\alpha}$ would not coincide with the covariant derivative since gauge
invariance is broken already in the triangle diagrams.}, as we did in
Eq.~(\ref{VWardB}). For example, that with respect to the $q_{1}$ current
involves three triangle amplitudes (the other cases are in the
Appendix~\ref{AppChiral}):%
\begin{align}
D_{\alpha}^{L}\mathcal{T}_{LLLL}^{\alpha\beta\gamma\delta,abcd}  &
=-iq_{1\alpha}\mathcal{T}_{LLLL}^{\alpha\beta\gamma\delta,abcd}-f^{ade}%
\mathcal{T}(L_{q_{1}+q_{4}}^{\delta,e}L_{q_{2}}^{\beta,b}L_{q_{3}}^{\gamma
,c})_{a^{23}}\nonumber\\
&  \ \ \ -f^{ace}\mathcal{T}(L_{q_{1}+q_{3}}^{\gamma,e}L_{q_{2}}^{\beta
,b}L_{q_{4}}^{\delta,d})_{a^{24}}-f^{abe}\mathcal{T}(L_{q_{1}+q_{2}}^{\beta
,e}L_{q_{3}}^{\gamma,c}L_{q_{4}}^{\delta,d})_{a^{34}}\ ,
\end{align}
where $q_{4}=-q_{1}-q_{2}-q_{3}$, $L^{\alpha,a}=A_{\alpha}^{L,a}T^{a}$, and in
subscript are indicated the corresponding triangle arbitrary parameters,
numbered according to the outgoing momenta. Using Eqs.~(\ref{AmpTri})
and~(\ref{AmpBox}), all the IR singularities nicely cancel out and we find%
\begin{equation}
D_{\alpha}\mathcal{T}_{LLLL}^{\alpha\beta\gamma\delta,abcd}=-\frac
{\mathcal{I}_{3}\varepsilon^{\beta\gamma\delta\mu}}{16\pi^{2}}(%
\begin{array}
[c]{ccc}%
q_{1} & q_{2} & q_{3}%
\end{array}
)_{\mu}\cdot V_{1}\ , \label{AnoLLLL}%
\end{equation}
with%
\begin{equation}
V_{1}=\left(
\begin{array}
[c]{ccc}%
1-b_{4}-a_{1}^{24}-a_{1}^{34} & b_{3}-a_{1}^{34} & a_{1}^{24}-b_{2}\\
1-a_{1}^{24}-a_{2}^{24}-a_{1}^{34} & a_{2}^{23}-a_{1}^{34} & a_{2}^{23}%
+a_{1}^{24}+a_{2}^{24}-1\\
1-a_{1}^{24}-a_{1}^{34}-a_{2}^{34} & 1-a_{1}^{23}-a_{1}^{34}-a_{2}^{34} &
a_{1}^{24}-a_{1}^{23}%
\end{array}
\right)  \cdot\left(
\begin{array}
[c]{c}%
d^{ade}f^{bce}\\
d^{ace}f^{bde}\\
d^{abe}f^{cde}%
\end{array}
\right)  \ .
\end{equation}
This is the most general form of the divergence, to be interpreted as the
cubic contribution to $(D_{\mu}^{L}J_{L}^{\mu})^{a}$. Of course, the twelve
$a_{1,2}^{ij}$ parameters have to be set in a way consistent with the choices
made for the triangle amplitudes in Eq.~(\ref{AnoLLL}). Notice the peculiar
feature that the Lorentz structures do not involve only the momentum
corresponding to the divergence, with e.g. $q_{2}$ and $q_{3}$ occurring in
Eq.~(\ref{AnoLLLL}). We will see later that it is absolutely required.

Using the present technique is not very practical for the pentagon diagrams of
Fig.~\ref{Fig4}. Calculating the full off-shell five-point amplitude is a
quite formidable task. Notice though that it is immediately finite, so it is
not anomalous and does not involve any arbitrary parameters. Yet, it is not
vanishing because its covariant derivative also involves four box amplitudes.
For example,
\begin{align}
D_{\alpha}^{L}\mathcal{T}_{LLLLL}^{\alpha\beta\gamma\delta\varepsilon,abcde}
&  =-iq_{1\alpha}\mathcal{T}_{LLLLL}^{\alpha\beta\gamma\delta\varepsilon
,abcde}-f^{aeg}\mathcal{T}(L_{q_{1}+q_{5}}^{\varepsilon,g}L_{q_{2}}^{\beta
,b}L_{q_{3}}^{\gamma,c}L_{q_{4}}^{\delta,d})_{b^{234}}\nonumber\\
&  \ \ \ -f^{adg}\mathcal{T}(L_{q_{1}+q_{4}}^{\delta,g}L_{q_{2}}^{\beta
,b}L_{q_{3}}^{\gamma,c}L_{q5}^{\varepsilon,e})_{b^{235}}-f^{acg}%
\mathcal{T}(L_{q_{1}+q_{3}}^{\gamma,g}L_{q_{2}}^{\beta,b}L_{q_{4}}^{\delta
,d}L_{q5}^{\varepsilon,e})_{b^{245}}\nonumber\\
&  \ \ \ -f^{abg}\mathcal{T}(L_{q_{1}+q_{2}}^{\beta,g}L_{q_{3}}^{\gamma
,c}L_{q_{4}}^{\delta,d}L_{q5}^{\varepsilon,e})_{b^{345}}\ , \label{CovPen}%
\end{align}
so the pentagon diagrams must have IR singular terms to cancel those of the
box amplitudes, in the same way the box amplitudes have just the right terms
to cancel those of the triangle amplitudes. We will see later how to obtain
the final expression for $D_{\alpha}^{L}\mathcal{T}_{LLLLL}^{\alpha\beta
\gamma\delta\varepsilon,abcde}$ and the other divergences rather trivially,
and simply quote the result here:
\begin{equation}
D_{\alpha}^{L}\mathcal{T}_{LLLLL}^{\alpha\beta\gamma\delta\varepsilon
,abcde}=\frac{i\mathcal{I}_{3}\varepsilon^{\beta\gamma\delta\varepsilon}%
}{16\pi^{2}}\left(
\begin{array}
[c]{l}%
(b_{2}^{234}+b_{3}^{234}+b_{4}^{234}-b_{2}^{245}-b_{3}^{245}-b_{4}%
^{245})d^{bfg}f^{aef}f^{cdg}\\
+(b_{2}^{245}+b_{3}^{245}+b_{4}^{245}-b_{2}^{235}-b_{3}^{235}-b_{4}%
^{235})d^{bfg}f^{adf}f^{ceg}\\
+(b_{2}^{234}+b_{2}^{345}+b_{3}^{345}+b_{4}^{345}-1)d^{cfg}f^{aef}f^{bdg}\\
+(1-b_{2}^{235}-b_{2}^{345}-b_{3}^{345}-b_{4}^{345})d^{cfg}f^{adf}f^{beg}\\
+(b_{2}^{345}-b_{2}^{234})d^{dfg}f^{aef}f^{bcg}+(b_{2}^{245}-b_{2}%
^{345})d^{dfg}f^{acf}f^{beg}\\
+(1-b_{2}^{245}-b_{3}^{245}-b_{4}^{245}-b_{3}^{345}-b_{4}^{345}+b_{3}%
^{235})d^{efg}f^{adf}f^{bcg}\\
+(b_{2}^{245}+b_{4}^{245}+b_{3}^{345}+b_{4}^{345}-1)d^{efg}f^{acf}f^{bdg}\\
+(1-b_{2}^{245}-b_{3}^{245}-b_{4}^{245}-b_{4}^{345})d^{efg}f^{abf}f^{cdg}%
\end{array}
\right)  \ . \label{M0Pen}%
\end{equation}
Again, the thirty $b_{2,3,4}^{ijk}$ constants are not additional parameters,
but have to be set in accordance with the three $b_{i}$ constants of the cubic
term, exactly like the $a_{1,2}^{ij}$ are all fixed in terms of the $a_{1,2}$
of the quadratic term. This means that the quartic contribution to $(D_{\mu
}^{L}J_{L}^{\mu})^{a}$ is entirely fixed by the cubic one. The final result is
quite complicated though, but this is because with five $SU(N)$ indices, there
are many combinations of Casimir invariants. For example, $SU(3)$ has a
total of 32 independent combinations~\cite{Dittner:1971fy} of $d^{abc}$,
$f^{abc}$, and $\delta^{ab}$, of which we need to keep $11$ independent
$d\times f\times f$ tensors.

Altogether, Eqs.~(\ref{AnoLLL}),~(\ref{AnoLLLL}), and~(\ref{M0Pen}) represent
the most general form for the chiral anomaly. In the following sections, we will
explore how further constraints translate into specific values for all the
coefficients, thereby giving specific anomalies like the consistent, Bardeen,
or covariant anomalies, or any form interpolating between them.

\subsection{IR versus UV singularities}\label{IRUV}

A crucial feature of the calculation of the loop amplitudes for massless
fermions is the presence of IR singularities. If we were to simply remove them
from Eqs.~(\ref{AmpTri}) and (\ref{AmpBox}), it would be possible to choose
the arbitrary parameters to ensure the naive Ward identity holds, gauge
invariance would be preserved, and there would be no anomaly.

One may wonder where those IR singularities originate from. After all,
anomalies are usually thought of as a UV\ phenomenon, brought in by surface
terms. In our computation of Eq.~(\ref{AmpTri}), we used a very specific
property of dimensional regularization. All loop integrals are first reduced
to scalar ones using the Passarino-Veltman procedure~\cite{Passarino:1978jh},
and then set to zero on-shell because%
\begin{equation}
\int\frac{d^{d}k}{(2\pi)^{d}}\frac{1}{k^{n}}=0\ . \label{DR0}%
\end{equation}
Indeed, this integral converge in the UV provided $d<n$, and in the IR
provided $d>n$. By analytical continuation to complex $d$, it thus vanishes
everywhere. In this sense, the IR pole cancels against the UV one. This is
sufficient to kill most, but not all IR singularities.

With this in mind, one may think these singularities are an artifact of our
regularization, sticking to the view that anomalies must have UV origins. This
would be a mistake, anomalies do not care about UV or IR. They represent an
inconsistency between a symmetry and quantization, and as such, they will show
up wherever they can find some space to seep in given the specific techniques
adopted to compute the matrix element of three or four currents. With our
dimensional regularization scheme, they show up partly in the UV, via the
arbitrary coefficients, and partly in the IR, via the singularities. Another
way to give physical content to these singularities is in the context of the
Sutherland-Veltman theorem. Obviously, the limit $m\rightarrow\infty$ is no
longer appropriate since $m=0$, but the decomposition in Eq.~(\ref{SVFF}) and
the limit $q_{3}^{2}\rightarrow0$ of Eq.~(\ref{SVFF2}) still make sense.
However, we can no longer conclude that the divergences of $\mathcal{T}%
_{LLL}^{\alpha\beta\gamma,abc}$ are all vanishing in that limit because the
form-factors in Eq.~(\ref{SVFF}) now have singularities in $1/q_{3}^{2}$.

Actually, this opens the way to a quite different treatment of anomalies based
on dispersion
relations~\cite{Dolgov:1971ri,Nishijima:1974qe,Frishman:1980dq,Horejsi:1985qu}%
. The $f_{1,2}$ form-factors in Eq.~(\ref{SVFF}) are analytic functions of
$q_{1,2,3}^{2}\in\mathbb{C}$, so we can call in Cauchy theorem to express them as integral over their
absorptive part (for a review, see e.g. Ref.~\cite{Kubis:2025zji}). For
example, setting $q_{1}^{2}=q_{2}^{2}=0$, we can write%
\begin{equation}
f_{i}(q_{3}^{2})=\frac{1}{\pi}\int_{0}^{\infty}ds\frac{\operatorname{Im}%
f_{i}(s)}{s-q_{3}^{2}}\ . \label{DispRel}%
\end{equation}
Here, our $\mathcal{T}_{LLL}^{\alpha\beta\gamma,abc}$ amplitude in
Eq.~(\ref{AmpTri}) is real because we have already discarded all the scalar
loop integrals whose real part vanish at $q_{3}^{2}=0$ via Eq.~(\ref{DR0}%
)\footnote{Imaginary parts of loop amplitudes cannot contribute to the
divergences in Eq.~(\ref{AnoLLL}) because, using Cutkoski rules, they arise
from on-shell fermions in the loop, for which naive Ward identities are
fulfilled.}. Had we kept them, the $1/q_{3}^{2}$ poles we found would be
accompanied by corresponding delta function singularities $\delta(q_{3}^{2})$
in the imaginary parts of the form-factors, in agreement with
Eq.~(\ref{DispRel}) and the usual formula $(x-i\varepsilon)^{-1}%
=\mathcal{P}(1/x)+i\pi\delta(x)$.

In addition, there is an inherent ambiguity in constructing the dispersion
integrals, depending on whether we consider $f_{i}(q_{3}^{2})$ or
$f_{i}^{\prime}(q_{3}^{2})=(f_{i}(q_{3}^{2})-f_{i}(q_{0}^{2}))/(q_{3}%
^{2}-q_{0}^{2})$ for some fixed subtraction point $q_{0}^{2}$. Both are
analytic functions, so both can be represented via dispersion integrals.
Plugging $f_{i}^{\prime}$ in Eq.~(\ref{DispRel}) and assuming
$\operatorname{Im}f^{\prime}(q_{0}^{2})=0$ then gives
\begin{equation}
f_{i}(q_{3}^{2})=f_{i}(q_{0}^{2})+\frac{q_{3}^{2}-q_{0}^{2}}{\pi}\int%
_{0}^{\infty}ds\frac{\operatorname{Im}f_{i}(s)}{(s-q_{3}^{2})(s-q_{0}^{2})}\ .
\end{equation}
The ambiguity in $q_{0}^{2}$ plays the same role as that in the position of
$\gamma_{5}$ in dimensional regularization, or in the momentum routing using a
cut-off procedure. It also shares the same UV origin, since subtracting the
dispersion relation makes it converge faster. In practice, in all cases, the
free parameters allow to move the anomaly around the legs of the triangle diagram.

The full analysis of the triangle amplitude using the dispersive approach can
be found in many places, see e.g. Refs.~\cite{Bertlmann:1996xk,Ioffe:2006ww},
so it will not be repeated here. Though this method is of limited use, as its
extension to the box or pentagon diagrams would be a horrendous task, it
provides a very enlightening alternative view on the anomalies, and clearly
illustrates their propensity at creeping in wherever they can to ultimately
break the classical Ward identities.

\section{The consistent anomaly}

The \textbf{consistent anomaly} is that which spreads out the anomaly equally
onto all the legs of the triangle, box, and pentagon diagrams, in agreement
with the expected Bose symmetry of a diagram involving the same gauge
currents~\cite{Bardeen:1969md}.

For the quadratic terms, this fixes $a_{1}=a_{2}=1/3$, so that all three
divergences involve the same factor
\begin{subequations}
\label{TriLLL}%
\begin{align}
D_{\alpha}^{L}\mathcal{T}_{LLL}^{\alpha\beta\gamma,abc}  &  =-i\frac
{\mathcal{I}_{3}d^{abc}}{48\pi^{2}}\varepsilon^{\beta\gamma\mu\nu}q_{1\mu
}q_{2\nu}\ ,\ \\
D_{\beta}^{L}\mathcal{T}_{LLL}^{\alpha\beta\gamma,abc}  &  =-i\frac
{\mathcal{I}_{3}d^{abc}}{48\pi^{2}}\varepsilon^{\gamma\alpha\mu\nu}q_{1\mu
}q_{2\nu}\ ,\ \\
D_{\gamma}^{L}\mathcal{T}_{LLL}^{\alpha\beta\gamma,abc}  &  =-i\frac
{\mathcal{I}_{3}d^{abc}}{48\pi^{2}}\varepsilon^{\alpha\beta\mu\nu}q_{1\mu
}q_{2\nu}\ .
\end{align}
This is quite striking compared to the coefficient for the abelian or singlet
anomaly, which are always multiples of one-half. The calculation for the $RRR$
configuration gives the same result, but for the opposite sign from moving all
the $P_{R}$ to the left. At the operator level, the triangle graph
contribution to the consistent anomaly is thus\footnote{Diagrams were
calculated with the gauge vertices $-i\gamma^{\mu}P_{L}$, and need to be
multiplied by $i$ to convert one vertex into a $\bar{\psi}\gamma^{\mu}%
P_{L}\psi$ current insertion.}%
\end{subequations}
\begin{equation}
(D_{\mu}^{R,L}J_{L,R}^{\mu})_{Triangle}^{a}=\pm\frac{\varepsilon^{\mu\nu
\rho\sigma}}{24\pi^{2}}\left\langle T^{a}\partial_{\mu}\left(  \mathbf{A}%
_{\nu}^{L,R}\partial_{\rho}\mathbf{A}_{\sigma}^{L,R}\right)  \right\rangle \ .
\label{TriC}%
\end{equation}

For the box, we first have to set $a_{k}^{ij}=1/3$ by consistency with the
choice made for the triangle diagrams. Bose symmetry is then sufficient to
uniquely fix $b_{i}=1/4$, so that
\begin{subequations}
\label{AnoLLLL2}%
\begin{align}
D_{\alpha}^{L}\mathcal{T}_{LLLL}^{\alpha\beta\gamma\delta}  &  =\frac
{\mathcal{I}_{3}}{192\pi^{2}}(d^{ade}f^{bce}-d^{ace}f^{bde}+d^{abe}%
f^{cde})\varepsilon^{\alpha\beta\gamma\delta}q_{1\alpha}\ ,\\
D_{\beta}^{L}\mathcal{T}_{LLLL}^{\alpha\beta\gamma\delta}  &  =\frac
{\mathcal{I}_{3}}{192\pi^{2}}(d^{ade}f^{bce}-d^{ace}f^{bde}-3d^{abe}%
f^{cde})\varepsilon^{\alpha\beta\gamma\delta}q_{2\beta}\ ,\\
D_{\gamma}^{L}\mathcal{T}_{LLLL}^{\alpha\beta\gamma\delta}  &  =\frac
{\mathcal{I}_{3}}{192\pi^{2}}(d^{ade}f^{bce}+3d^{ace}f^{bde}+d^{abe}%
f^{cde})\varepsilon^{\alpha\beta\gamma\delta}q_{3\gamma}\ ,\\
D_{\delta}^{L}\mathcal{T}_{LLLL}^{\alpha\beta\gamma\delta}  &  =\frac
{\mathcal{I}_{3}}{192\pi^{2}}(-3d^{ade}f^{bce}-d^{ace}f^{bde}+d^{abe}%
f^{cde})\varepsilon^{\alpha\beta\gamma\delta}q_{4\delta}\ .
\end{align}
These divergences are identical under permutations of the currents. It is only
because we made a choice of three independent $d\times f$ tensors that this is
not manifest. Also, thanks to the Bose symmetry, only the momentum
corresponding to the divergence occurs. These two features allow
Eq.~(\ref{AnoLLLL2}) to be reproduced from the operator%
\end{subequations}
\begin{equation}
(D_{\mu}^{R,L}J_{L,R}^{\mu})_{box}^{a}=\mp i\frac{\varepsilon^{\mu\nu
\rho\sigma}}{96\pi^{2}}\left\langle T^{a}\partial_{\mu}\left(  \mathbf{A}%
_{\nu}^{L,R}[\mathbf{A}_{\rho}^{L,R},\mathbf{A}_{\sigma}^{L,R}]\right)
\right\rangle \ , \label{BoxC}%
\end{equation}
where $(D_{\mu}^{L,R}J_{L,R}^{\mu})^{a}=\partial_{\mu}J_{L,R}^{a,\mu}%
-f^{abc}A_{\mu}^{L,R,b}J_{L,R}^{c,\mu}$. Notice that setting $a_{k}^{ij}=1/3$
corresponds to symmetrizing the prescription of Eq.~(\ref{DRano}), and
similarly, setting $b_{i}=1/4$ symmetrizes that for the box amplitude which
involves four terms.

Finally, plugging in the Bose symmetric result $b_{l}^{ijk}=1/4$ in the
quartic covariant divergences, they all vanish identically:%
\begin{equation}
(D_{\mu}^{R,L}J_{R,L}^{\mu})_{pentagon}^{a}=0\ . \label{PenC}%
\end{equation}
This is actually the only value for which they all do so, and as already emphasized 
for the cancellation of IR singularities, this requires a quite delicate interplay 
between the five-point diagram and various permutations of the four-point diagrams, 
see Eq.~(\ref{CovPen}).

As a final comment, we should stress that the consistent anomaly does not do
justice to the coefficients we fought so hard to include, setting them all to
common values. In the following, we will encounter more interesting
situations, in which the full complexity of Eq.~(\ref{AnoLLLL}) or
(\ref{M0Pen}) is put to good use.

\subsection{The consistency condition}

The Bose symmetry suffices to resolve all the ambiguities of the chiral
anomaly. Yet, the final result has a more profound consistency property, at
the origin of its name. A full description quickly becomes quite technical,
but for completeness, let us give the main idea.

Imagine that the fermion is integrated out to form the quantum effective
action $\Gamma\lbrack A]$, where $A$ can stand for $\mathbf{A}_{\mu}^{L}$ or
$\mathbf{A}_{\mu}^{R}$. This effective action can be thought of as a
Lagrangian but with all the vertices among gauge bosons induced by fermion
loops included as local effective couplings. Under a gauge transformation with
gauge parameter $\Lambda(x)$, the field transforms as $A_{\mu}^{a}\rightarrow
A_{\mu}^{a}+(D_{\mu}\Lambda)^{a}$, and the variation of our new Lagrangian is
generically:%
\begin{equation}
\Gamma\lbrack A]\rightarrow\Gamma\lbrack A]+\int dx\frac{\delta\Gamma\lbrack
A]}{\delta A^{a,\mu}(x)}(D^{\mu}\Lambda(x))^{a}=\Gamma\lbrack A]+\int
dx\Lambda^{a}(x)\left(  -D^{\mu}\frac{\delta\Gamma\lbrack A]}{\delta A^{a,\mu
}(x)}\right)  \ . \label{EffAg}%
\end{equation}
When $\Gamma\lbrack A]$ is the classical action $S[A]$, the variation vanishes
and the classical current $J_{\mu}^{a}=\delta S[A]/\delta A^{a,\mu}(x)$ is
conserved. This is Noether's theorem. At the loop level, the full current is
still defined from the variation of the action, but its divergence no longer
vanishes because of the anomaly, which we precisely define as%
\begin{equation}
J_{\mu}^{a}(x)\equiv\frac{\delta\Gamma\lbrack A]}{\delta A^{a,\mu}%
(x)}\rightarrow D^{\mu}J_{\mu}^{a}(x)\equiv\mathcal{A}^{a}(x)\neq0\ .
\label{ConsCurr}%
\end{equation}
The quantity $\mathcal{A}^{a}$ is called the \textbf{consistent anomaly}.
Because it arises from the gauge variation of the action, it automatically
exhibits the Bose symmetry since all the currents are treated equally. For an
analogy, think for example about the Feynman rule for a $(\partial_{\mu}%
\phi\partial^{\mu}\phi)^{2}$ vertex: it is necessarily symmetric under the
permutations of the four $\phi$.

The important point is that once the anomaly is defined directly from the
gauge variation of the action, it must satisfy the gauge group algebra. If we
define the functional differential operator of Eq.~(\ref{EffAg}) as%
\begin{equation}
\mathcal{G}^{a}(x)=-\left(  D_{\mu}\frac{\delta}{\delta A_{\mu}(x)}\right)
^{a}=-\frac{\partial}{\partial x^{\mu}}\frac{\delta}{\delta A_{\mu}^{a}%
(x)}-f^{abc}A_{\mu}^{b}(x)\frac{\delta}{\delta A_{\mu}^{c}(x)}\ ,
\end{equation}
then it obeys $[i\mathcal{G}^{a}(x),i\mathcal{G}^{b}(y)]=if^{abc}\delta
^{4}(x-y)\mathcal{G}^{c}(x)$. Applying this identity to $\Gamma\lbrack A]$,
the anomaly is found to obey
\begin{equation}
\mathcal{G}^{a}(x)\mathcal{A}^{b}(y)-\mathcal{G}^{b}(y)\mathcal{A}%
^{a}(x)=f^{abc}\delta^{4}(x-y)\mathcal{A}^{c}(x)\ , \label{WZConsCond}%
\end{equation}
which is called \textbf{Wess-Zumino consistency condition}~\cite{Wess:1971yu}.
It requires the consistent anomaly to be gauge-dependent, but in a very
specific way, consistent with the algebra of the gauge group.

In practice, thanks to its non-linearity in the gauge fields, the consistency
condition is sufficiently constraining to fix the cubic and quartic terms in
$D_{\mu}^{R,L}J_{R,L}^{\mu}$ once the triangle graph is fixed by Bose
symmetry. We will not present the method to solve the consistency condition
here, because an explicit calculation using the current formalism would be
terribly cumbersome. It becomes simpler adopting the BRST differential
language, but this would bring us too far afield (see e.g.
Ref.~\cite{Weinberg:1996kr}). So, simply quoting the result, the consistent
anomaly is%
\begin{equation}
(D_{\mu}^{L,R}J_{L,R}^{\mu})^{a}=\pm\frac{\varepsilon^{\mu\nu\rho\sigma}%
}{24\pi^{2}}\left\langle T^{a}\partial_{\mu}\left(  \mathbf{A}_{\nu}%
^{L,R}\partial_{\rho}\mathbf{A}_{\sigma}^{L,R}-\frac{i}{2}\mathbf{A}_{\nu
}^{L,R}\mathbf{A}_{\rho}^{L,R}\mathbf{A}_{\sigma}^{L,R}\right)  \right\rangle
\ , \label{AnoCons}%
\end{equation}
in agreement with Eqs.~(\ref{TriC}), (\ref{BoxC}) and~(\ref{PenC}). The Bose
symmetry is thus sufficient to restrict the most general chiral anomaly down
to precisely $\mathcal{A}^{a}$ of Eq.~(\ref{ConsCurr}).

\subsection{Chain of descent equations}\label{ChainofDescent}

There is another particularly elegant method to derive the cubic and quartic
terms of the consistent anomaly once the quadratic one is known, thereby
proving that Eqs.~(\ref{TriC}), (\ref{BoxC}) and~(\ref{PenC}) indeed satisfy
the consistency condition. It relies on a differential relationship between
the consistent anomaly and the Chern-Simons form. We have seen in Sec.~\ref{SecChernSimons} that
the singlet anomaly is a total derivative, $\partial_{\mu}G^{\mu}%
=\langle\mathbf{F}_{\mu\nu}\mathbf{\tilde{F}}^{\mu\nu}\rangle$. Now, let us
take $\langle\mathbf{F}_{\mu\nu}\mathbf{\tilde{F}}^{\mu\nu}\rangle$ or one of
its $2n$-dimensional equivalent as a Lagrangian. Using Stokes theorem, its
integral over some $2n$-dimensional volume $V$ gives that of $G_{2n-1}^{\mu}$
over the boundary $\partial V$:%
\begin{equation}
S_{CS}^{2n-1}[\mathbf{A}]=\frac{1}{(4\pi)^{n}n!}\int_{\partial V}d\sigma_{\mu
}\ G_{2n-1}^{\mu}[\mathbf{A}]\ .
\end{equation}
This is called a $2n-1$ dimensional \textbf{Chern-Simons theory}, here defined
on the $2n-1$ dimensional boundary $\partial V$.

This is a very special theory. Its quite unusual equations of motion will be
derived later (in the next section for $n=3$, and in Sec.~\ref{SecCovAno} more generally), and
we concentrate here on its gauge properties. Under a generic non-infinitesimal
gauge transformation%
\begin{equation}
\mathbf{A\rightarrow A}^{\Omega}=\boldsymbol{\Omega}^{\dagger}\mathbf{A\Omega
}+i\boldsymbol{\Omega}^{\dagger}\partial_{\mu}\boldsymbol{\Omega\ ,}%
\end{equation}
the Chern-Simons Lagrangian varies as%
\begin{align}
\delta G_{2n-1}^{\mu}[\boldsymbol{\Omega},\mathbf{A}]  &  =G_{2n-1}^{\mu
}[\mathbf{A}^{\Omega}]-G_{2n-1}^{\mu}[\mathbf{A}]\nonumber\\
&  =\partial_{\nu}G_{2n-2}^{\mu\nu}[\boldsymbol{\Omega},\mathbf{A}]+c_{n}%
\varepsilon^{\mu\nu_{1}...\nu_{2n-1}}\langle(i\boldsymbol{\Omega}^{\dagger
}\partial_{\nu_{1}}\boldsymbol{\Omega})...(i\boldsymbol{\Omega}^{\dagger}\partial
_{\nu_{2n-1}}\boldsymbol{\Omega})\rangle\ . \label{VarCS}%
\end{align}
where $c_{n}$ is given in Eq.~(\ref{cn}). This apparently complicated result
can be understood simply. First, note that it is consistent with the
expression in Eq.~(\ref{CSInteg2}) for the winding number of a pure gauge
configuration. As explained there, the pure gauge term vanishes under small
gauge transformations, i.e., those connected to the identity, because they
have a trivial winding number. The Chern-Simons action must then be gauge
invariant since it is obtained from a combination of field strengths. This
forces the rest of the variation of $G_{2n-1}^{\mu}$ to be at most a total
derivative, $\partial_{\nu}G_{2n-2}^{\mu\nu}$ for some $G_{2n-2}^{\mu\nu}$.
Indeed, the integral of $\partial_{\nu}G_{2n-2}^{\mu\nu}$ over $\partial V$
is, via Stokes theorem, equal to that of $G_{2n-2}^{\mu\nu}$ over
$\partial\partial V$, so it necessarily vanishes because $\partial\partial
V=\varnothing$, i.e., a boundary has no boundary\footnote{If one thinks in
three dimensions, the surface hermetically enclosing some volume cannot have
any hole, hence can have no boundary.}. For large gauge transformations, the
last term does not vanish but gives an integer $\Delta\nu$ once divided by
$(4\pi)^{n}n!$. Under such transformations, the gauge field is moved to a
different equivalence class, characterized by a $\Delta\nu$ change in its
winding number. This is precisely what we expect from Eq.~(\ref{CSInteg2}):
though $\left\langle \varepsilon^{\mu\nu\rho\sigma}\mathbf{F}_{\mu\nu
}\mathbf{F}_{\rho\sigma}\right\rangle $ appears invariant under
$\mathbf{F\rightarrow F}^{\Omega}=\boldsymbol{\Omega}^{\dagger}\mathbf{F\Omega}$,
its integral is shifted non-continuously, by a finite amount proportional to
$\Delta\nu$.

Let us now concentrate on small gauge transformations, and check explicitly
that the gauge variation is a total derivative. For that, it suffices to
perform the transformation $\boldsymbol{\Omega}=\exp(-i\boldsymbol{\Lambda})$ and keep
only terms up to first order in $\boldsymbol{\Lambda}$, which amounts to
$\mathbf{A}_{\mu}\rightarrow\mathbf{A}_{\mu}+\partial_{\mu}\boldsymbol{\Lambda
}-i[\mathbf{A}_{\mu},\boldsymbol{\Lambda}]$, and we find
\begin{subequations}
\label{VarCS2}%
\begin{align}
\delta G_{3}^{\mu}[\boldsymbol{\Lambda},\mathbf{A}]  &  =4\varepsilon^{\mu\nu
\rho\sigma}\left\langle \partial_{\nu}\boldsymbol{\Lambda}\partial_{\rho
}\mathbf{A}_{\sigma}\right\rangle \ ,\\
\delta G_{5}^{\mu}[\boldsymbol{\Lambda},\mathbf{A}]  &  =8\varepsilon^{\mu\nu
\rho\sigma\alpha\beta}\left\langle \partial_{\nu}\boldsymbol{\Lambda}%
\partial_{\rho}\left(  \mathbf{A}_{\sigma}\partial_{\alpha}\mathbf{A}_{\beta
}-\frac{i}{2}\mathbf{A}_{\sigma}\mathbf{A}_{\alpha}\mathbf{A}_{\beta}\right)
\right\rangle \ ,\\
\delta G_{7}^{\mu}[\boldsymbol{\Lambda},\mathbf{A}]  &  =16\varepsilon^{\mu\nu
\rho\sigma\alpha\beta\gamma\delta}\left\langle \partial_{\nu}\boldsymbol{\Lambda
}\partial_{\rho}\left(  \mathbf{A}_{\sigma}\partial_{\alpha}\mathbf{A}_{\beta
}\partial_{\gamma}\mathbf{A}_{\delta}-\frac{2}{5}\mathbf{A}_{\sigma}%
\mathbf{A}_{\alpha}\mathbf{A}_{\beta}\mathbf{A}_{\gamma}\mathbf{A}_{\delta
}\right.  \right. \nonumber\\
&  \ \ \ \ \ \ \ \ \ \ \ \ \ \ \ \ \left.  \left.  +\frac{4}{5}i\mathbf{A}%
_{\sigma}\mathbf{A}_{\alpha}\mathbf{A}_{\beta}\partial_{\delta}\mathbf{A}%
_{\gamma}+\frac{1}{5}i\mathbf{A}_{\sigma}\mathbf{A}_{\alpha}\partial_{\delta
}\mathbf{A}_{\beta}\mathbf{A}_{\gamma}+\frac{3}{5}i\mathbf{A}_{\sigma}%
\partial_{\delta}\mathbf{A}_{\alpha}\mathbf{A}_{\beta}\mathbf{A}_{\gamma
}\right)  \right\rangle \ .
\end{align}
Because $\partial_{\nu}\boldsymbol{\Lambda}$ is always accompanied by a total
derivative in the trace, integrating by part these variations gives zero after
the antisymmetric Lorentz contraction, while the surface term cancels out
because $\partial\partial V=\varnothing$. However, something interesting
happens if one allows $S_{CS}^{2n-1}[\mathbf{A}]$ to be defined on some
generic $2n-1$ dimensional surface $S$ whose boundary is not trivial. The
surface term no longer vanishes, and it is precisely proportional to the
consistent anomaly in $2n-2$ dimensions! For example, the five-dimensional
Chern-Simons action varies by%
\end{subequations}
\begin{equation}
\int_{S}\delta G_{5}^{\mu}[\boldsymbol{\Lambda},\mathbf{A}]d\sigma_{\mu}%
=\int_{\partial S}d\sigma_{\mu\nu}\ 8\varepsilon^{\mu\nu\rho\sigma\alpha\beta
}\left\langle \boldsymbol{\Lambda}\partial_{\rho}\left(  \mathbf{A}_{\sigma
}\partial_{\alpha}\mathbf{A}_{\beta}-\frac{i}{2}\mathbf{A}_{\sigma}%
\mathbf{A}_{\alpha}\mathbf{A}_{\beta}\right)  \right\rangle \ . \label{CSvar3}%
\end{equation}
And indeed, if we identify $\partial S$ with our four-dimensional space-time
and take the variation $\delta/\delta\Lambda^{a}$, we recover
Eq.~(\ref{AnoCons}), up to the normalization. The reason for that unexpected
result is simple: this surface term represents the gauge variation of the
Chern-Simons action. Albeit a bit weird, this is a true action and a valid
gauge transformation. So, it must satisfy the algebra of the gauge group, that
is, the consistency condition. Though the overall normalization of the anomaly
cannot be fixed relative to that of the surface term\footnote{Actually, it was
demonstrated in Ref.~\cite{Alvarez-Gaume:1983ict} that there is a way to fix
the normalization of the Chern-Simons action so that it gives back the
correctly normalized consistent anomaly. The proof is well beyond our scope
though. It relies on constructing first a specific $2n$-dimensional Dirac
operator, on which to apply the Atiyah-Singer theorem to relate its zero modes
to the $2n$ singlet anomaly.}, what we can now read off Eq.~(\ref{CSvar3}) is
the relative size of the quadratic, cubic and quartic terms, as unambiguously
fixed by the consistency condition. This proves the equivalence of that
condition with the Bose symmetry leading to Eqs.~(\ref{TriC}), (\ref{BoxC})
and~(\ref{PenC}).

The only price to pay to apply this method to the four-dimensional case is
that we had to start in five dimensions, or even six if one considers the
original singlet anomaly expressed in terms of field strengths. This was
necessary to end up with a quadratic term like $\mathbf{A}_{\sigma}%
\partial_{\alpha}\mathbf{A}_{\beta}$, matching that of the triangle
calculation. The consistent anomaly in two or six dimensions can be similarly
read off Eq.~(\ref{VarCS2}). Historically, this connection between the
consistent anomaly in $2n-2$ dimension and the singlet anomaly in $2n$
dimensions was discovered by Zumino in the early eighties~\cite{Zumino:1983rz}%
. Soon after, a whole chain of differential relationships was formulated, the
\textbf{Stora-Zumino chain of descent equations~}%
\cite{Zumino:1983ew,Stora:1983ct,Manes:1985df}. It starts by relating the
singlet anomaly in $2n$ dimension to the Chern-Simons form in $2n-1$
dimension. Then, at each step, one takes the gauge variation and expresses it
as a total derivative to extract a surface term. This first gives the
consistent anomaly in $2n-2$ dimensions. At the next level, one gets the
consistency condition, valid locally up to a total derivative, and so on until
only gauge parameters are left after $2n$ steps. In practice, looking back at
the consistency condition in Eq.~(\ref{WZConsCond}), it is clear that the
tensorial formalism is not well-suited to tackle compositions of gauge
variations, and more advanced mathematical tools are necessary, see e.g.
Ref.~\cite{Bertlmann:1996xk}.

The need to start from a higher dimensional setting can also be understood
differently. Consider a genuine four-dimensional Lagrangian $\mathcal{L}%
_{local}[\mathbf{A}_{\mu}]$ that is not gauge invariant. Computing the gauge
variation of the corresponding action as in Eq.~(\ref{EffAg}), we know from
Noether's theorem that the current $J_{\mu}^{a}=\delta S[A]/\delta A^{a,\mu}$
is not conserved. Obviously, $D^{\mu}J_{\mu}^{a}$ would satisfy the
consistency condition, but this would not signal a true anomaly because it
suffices to add to $\mathcal{L}_{local}[\mathbf{A}_{\mu}]$ some counterterms
to kill off all the non-gauge-invariant couplings, after which the divergence
of the current vanishes again. Why renormalization conditions would impose
those counterterms is another matter, the point is that it is possible at
least in principle to get rid of all such anomalies in this way. We call them
\textbf{trivial anomalies}, because they correspond to trivial solutions of
the consistency condition.

By contrast, the one-loop triangle, box, and pentagon amplitudes generate
non-local contributions to the effective action. In some sense, the loop never
really shrinks to a single point but retains some spread over space-time. If
anything, that is precisely what their IR singularities are meant to tell us.
As a result, these contributions cannot be compensated by adding some
non-gauge invariant counterterms in the Lagrangian. We will see later that at
most, counterterms may reshuffle the anomalies around, in a way totally
equivalent as choosing different values for the $a_{i}$, $b_{i}$ parameters,
but it is not possible to get rid of the anomaly simultaneously in all the
gauge currents. Technically, non-local couplings, involving fields at
different space-time points, could do the job, but such counterterms are not
permitted. It is in this sense that the consistent anomaly represents a
non-trivial solution to the consistency condition. Now, the Chern-Simons trick
to get back this non-trivial solution directly (at tree level) from a local
Lagrangian is to start with more than four dimensions. Indeed, though local in
five dimensions, the gauge variation of this action is a whole
four-dimensional surface term. As such, it is clearly not local. This
procedure thus must give a non-trivial solution to the consistency condition,
and since there is only one, must give the consistent anomaly in four dimensions.

\subsection{$\theta$ vacua and background electric fields\label{ElectricTheta}%
}

We have seen in Sec.~\ref{SecSuscTopo} that the $n=2$ Chern-Simons form has an equivalent
description in terms of a three-form field $C_{\mu\nu\rho}$. Now that we
better understand its gauge properties from the previous section, we can move
on to explore in more details its dynamics. In doing so, we will gain a dual
view on the $\theta$ term in QCD. Specifically, our starting point will be the
two free Lagrangians in four and two spacetime dimensions:%
\begin{equation}
\mathcal{L}_{4}=-\frac{1}{2}\frac{1}{4!}F_{\mu\nu\rho\sigma}^{C}F^{C,\mu
\nu\rho\sigma}\ \ \leftrightarrow\ \ \mathcal{L}_{2}=-\frac{1}{2}\frac{1}%
{2!}F_{\mu\nu}^{A}F^{A,\mu\nu}\ , \label{Class1}%
\end{equation}
from which the EoM are the usual $\partial^{\mu}F_{\mu\nu\rho\sigma}^{C}=0$
and $\partial^{\mu}F_{\mu\nu}^{A}=0$. The model in two dimensions, when
supplemented with fermions, is known as the \textbf{Schwinger model}%
~\cite{Schwinger:1962tn, Schwinger:1962tp}. It does not resemble QED much, but
it is solvable and was shown to exhibit many of the non-perturbative features
of gauge theories (see e.g. Refs.~\cite{Manton:1985jm,Hetrick:1988yg}, as well
as Ref.~\cite{DavidTong}). Without fermions, this $d=2$ model is still
interesting for us because its gauge dynamics is essentially identical to that
of the three-form gauge field in four dimensions~\cite{Aurilia:1980xj}.
Indeed, what makes these models special is the fact that their field strengths
$F_{\mu\nu}^{A}$ and $F_{\mu\nu\rho\sigma}^{C}$ are necessarily proportional
to the Levi-Civita tensors $\varepsilon_{\mu\nu}$ and $\varepsilon_{\mu\nu
\rho\sigma}$, respectively. In some sense, the dynamical content of both
theories is so limited that it can only be identical. To gain intuition, it is
then instructive to present both in parallel.

The Lagrangians in Eq.~(\ref{Class1}) are invariant under $U(1)$ gauge
symmetries, which take the infinitesimal forms%
\begin{equation}
C_{\mu\nu\rho}\rightarrow C_{\mu\nu\rho}+\partial_{\mu}\lambda_{\nu\rho
}+\partial_{\nu}\lambda_{\rho\mu}+\partial_{\rho}\lambda_{\mu\nu
}\ \ \leftrightarrow\ \ A_{\mu}\rightarrow A_{\mu}+\partial_{\mu}\lambda\ ,
\label{Class5}%
\end{equation}
where $\lambda$ and $\lambda_{\mu\nu}$ are functions of space-time, and
$\lambda_{\mu\nu}=-\lambda_{\nu\mu}$ (for more details on the Maxwell
three-form fields, see e.g. Ref.~~\cite{Plantier:2025hcm}). To identify the
physical degrees of freedom, it is necessary to take care of this invariance,
so adopting the temporal gauge and then imposing the Lorenz condition on the
remaining spatial indices:
\begin{subequations}
\label{Class2}%
\begin{align}
\text{Temporal gauge\ }  &  \text{:\ \ }C_{0ij}=0\leftrightarrow A_{0}=0\ ,\\
\text{Residual Lorenz condition\ }  &  \text{:\ \ }\partial^{i}C_{ijk}%
=0\leftrightarrow\partial^{1}A_{1}=0\ ,\\
\text{Maxwell's equation }  &  \text{:\ \ }\partial_{0}\partial_{0}%
C_{ijk}=0\leftrightarrow\partial_{0}\partial_{0}A_{1}=0\ ,
\end{align}
which sum up in both cases to the only solution being a constant electric
field $E=\dot{C}_{123}=F_{0123}^{C}$ or $E=\dot{A}_{1}=F_{01}$. There is not
enough room for an orthogonal magnetic field, and no wave can propagate.

In a vibrating string analogy, this constant would represent the zeroth
excitation mode~\cite{Jackiw:1977yn}, with the string sitting flat at some
position. The constant electric field is thus not truly a dynamical degree of
freedom, but rather specifies the underlying rest state or background. Yet, it
has to take some value, so there must remain some sort of dynamics in the
gauge fields. To extract it, consider the Wilson loop (actually, volume for
$C_{\mu\nu\rho}$),%
\end{subequations}
\begin{equation}
\phi(t)=\frac{1}{2\pi^{2}}\frac{1}{3!}\int_{S^{3}}d\sigma^{\mu\nu\rho}%
C_{\mu\nu\rho}\leftrightarrow\phi(t)=\frac{1}{2\pi}\int_{S^{1}}d\sigma^{\mu
}A_{\mu}\ , \label{Class3}%
\end{equation}
where $C_{\mu\nu\rho}(x_{1},x_{2},x_{3},t)$ and $A_{\mu}(x,t)$. The domain of
integration is on the boundary of (compactified) space-time, $S^{d-1}=\partial
V^{d}$ with $d$ the spacetime dimension. In the temporal gauge, both
expressions integrate over the spatial indices only,
\begin{equation}
\phi(t)=\frac{1}{2\pi^{2}}\frac{1}{3!}\int_{S^{3}}d^{3}x\varepsilon
^{ijk}C_{ijk}\leftrightarrow\phi(t)=\frac{1}{2\pi}\int_{S^{1}}dxA_{1}\ .
\label{Class4}%
\end{equation}
If we assume that $V^{d}$ is compact, homotopic to a $d$-dimensional sphere of
radius one, then $\dot{\phi}=E$. Being expressed in terms of the gauge field,
it is thus $\phi$ that correctly represents the only dynamical degree of freedom.

By construction, $\phi$ is invariant under the small gauge transformations of
Eq.~(\ref{Class5}), as can be seen invoking Stokes' theorem:%
\begin{equation}
\phi(t)=\frac{1}{2\pi^{2}}\frac{1}{3!}\int_{V^{4}}d\sigma^{\mu\nu\rho\sigma
}F_{\mu\nu\rho\sigma}^{C}=\frac{2}{\pi^{2}}\int_{V^{4}}d^{4}x\ F_{0123}%
^{C}\leftrightarrow\phi(t)=\frac{1}{2\pi}\int_{V^{2}}d\sigma^{\mu\nu}F_{\mu
\nu}^{A}=\frac{1}{\pi}\int_{V^{2}}d^{2}x\ F_{01}^{A}\ , \label{Class6}%
\end{equation}
and $d\sigma^{\mu\nu\rho\sigma}=d^{4}x\ \varepsilon^{\mu\nu\rho\sigma}$,
$d\sigma^{\mu\nu}=\varepsilon^{\mu\nu}d^{2}x$ (remember the 3-dimensional
sphere of unit radius has surface $2\pi^{2}$ and volume $\pi^{2}/2$). Acting
with $\partial_{0}$ on these integrals, the integrands are fixed since the
field strengths are constant in time, but not the boundaries, so $\dot{\phi}$
is driven entirely by the electric field on $\partial V^{d}$, in agreement
with Eq.~(\ref{Class3}). In its volume integral form, this also shows that the
electric field $E$ becomes infinitesimal as the volume grows\footnote{If
$S^{1}$ is given a radius $R$, one may wish to normalize Eq.~(\ref{Class4}) by
$(2\pi R)^{-1}$, so that $\dot{\phi}=E$ stays fixed.}. It should actually be
interpreted as an electric density, generating an energy density
$\mathcal{E}=E^{2}/2$. Classically, any energy $\mathcal{E}$ or electric field
$E$ is fine.

Something special happens under large gauge transformations. Let us start with
the $d=2$ case which is easier to visualize. A generic $U(1)$ transformation
is $\Omega=\exp(i\lambda)$. Since the boundary $S^{1}$ is a compact space
spanned from $x=0$ to $x=2\pi$ and $\Omega(x)$ must be univalued, it must
satisfy $\Omega(0)=\Omega(2\pi)$ which allows for $\lambda(2\pi)=\lambda
(0)+2\pi\nu$. These are gauge transformation winding $\nu$ times around
$S^{1}$ as $x$ goes from zero to $2\pi$. Technically, this is a manifestation
of the homotopy group of the $U(1)\simeq S^{1}\rightarrow S^{1}$ maps, the
same as for magnetic monopoles. Under the large gauge transformations, those
with $\nu\neq0$, the gauge field simply undergoes a constant shift. Indeed, if
$\lambda(2\pi)=\lambda(0)+2\pi\nu$, then $\partial_{1}\lambda\neq0$. For
example, one can take $\lambda(x)=\nu x$, so that $A_{1}$ and $A_{1}+\nu$ are
equivalent gauge field configurations, linked only by large gauge
transformations. This means the set of gauge inequivalent field configurations
can be taken within $0\leqslant A_{1}\leqslant1$, and also that there is a
countable infinity of vacua $|\nu\rangle$, exactly like in QCD.

Going back to $\phi(t)$, if we want this quantity to be invariant under large
transformations, it needs to be periodically identified as $\phi
(t)=\phi(t)+\nu$ because under $\Omega=\exp(i\lambda)$, the loop integral in
Eq.~(\ref{Class3}) changes by%
\begin{equation}
\frac{1}{2\pi}\int_{S^{1}}d\sigma^{\mu}i\Omega^{\dagger}\partial_{\mu}%
\Omega=\frac{1}{2\pi}\int_{S^{1}}dx\lambda(x)=\nu\ . \label{Class7}%
\end{equation}
In other words, it must behave as an angle. With this in mind, and given that
$\dot{\phi}=E$, the energy $\mathcal{E}=\dot{\phi}^{2}/2$ looks terribly like
that of a particle of unit mass moving on a circle of unit radius, whose
Lagrangian is%
\begin{equation}
\mathcal{L}_{\phi}=\frac{\dot{\phi}^{2}}{2}\ .
\end{equation}
Evidently, the classical equation of motion $\ddot{\phi}=0$ sets $\dot{\phi}$
to a constant, which is what we found earlier since $\dot{\phi}=E$. Do not
confuse though this purely abstract $S^{1}$ spanned by $\phi$ to that
corresponding to the boundary of spacetime. Rather, the former owes its
existence to the latter, since it is the topology of the gauge fields over
$S^{1}$ which gives $\phi$ its periodicity. By the way, if space is not
compact, then $\phi$ is rather void since we have to identify all of its
values to make it gauge invariant.

For $d=4$, we will not try to directly construct an analog geometric
representation. Instead, let us show that $\phi(t)$ in $d=4$ has the same
properties as in $d=2$ when actual Chern-Simons forms drive the $U(1)$ gauge
fields, $C_{\mu\nu\rho}=\varepsilon_{\mu\nu\rho\sigma}G_{3}^{\sigma}$ and
$A_{\mu}=\varepsilon_{\mu\nu}G_{1}^{\nu}$ with $G_{2n-1}^{\mu}$ given in
Eq.~(\ref{CSforms}). To avoid confusion, keep in mind that there are thus two
kinds of gauge fields in the game: the true gauge fields driving the
Chern-Simons forms, themselves driving the $U(1)$ gauge fields. For $d=2$,
both kinds of gauge fields are actually identical because $G_{1}^{\mu
}=2\varepsilon^{\mu\nu}\left\langle \mathbf{A}_{\nu}\right\rangle $, so the
true gauge fields need to be abelian and are immediately dual to the
Chern-Simons current. From a more mathematical perspective, one could even say
that it is thanks to this coincidence that $\phi$ has its integer periods in
$d=2$, since we can recognize in Eq.~(\ref{Class7}) the winding number of
Eq.~(\ref{CSInteg2}) for $n=1$. The same should be true also in $d=4$, which
means that we can use the gauge variations of the true gauge fields to
generate non-trivial winding configurations for the $U(1)$ three-form field
over $S^{3}$. Specifically, from Eqs.~(\ref{VarCS}), when the true gauge
fields undergo $\mathbf{A}_{\mu}\rightarrow\boldsymbol{\Omega}^{\dagger}%
\mathbf{A}_{\mu}\boldsymbol{\Omega}+i\boldsymbol{\Omega}^{\dagger}\partial_{\mu
}\boldsymbol{\Omega}^{\dagger}$, the Chern-Simons form varies as%
\begin{equation}
\delta G_{3}^{\mu}=-2\varepsilon^{\mu\nu\rho\sigma}\left\langle i\partial
_{\nu}(\mathbf{A}_{\rho}\partial_{\sigma}\boldsymbol{\Omega\Omega}^{\dagger
})+\frac{1}{3}\boldsymbol{\Omega}^{\dagger}\partial_{\nu}\boldsymbol{\Omega\Omega
}^{\dagger}\partial_{\rho}\boldsymbol{\Omega\Omega}^{\dagger}\partial_{\sigma
}\boldsymbol{\Omega}\right\rangle \ .
\end{equation}
The first term is a total derivative that do not affect $\phi(t)$ since
$S^{3}$ is already a boundary, $\partial S^{3}=\varnothing$. If we take
$\boldsymbol{\Omega}=\mathbf{1}+i\boldsymbol{\omega}$, this invariance match precisely
onto Eq.~(\ref{Class5}) with $\lambda_{\mu\nu}=\operatorname*{Tr}\left[
\mathbf{A}_{\mu}\partial_{\nu}\boldsymbol{\omega}-\mathbf{A}_{\nu}\partial_{\mu
}\boldsymbol{\omega}\right]  $ to first order in $\boldsymbol{\omega}$. By
construction, those leave $\phi(t)$ invariant. The second term, on the other
hand, cannot be put in the form Eq.~(\ref{Class5}). It truly shifts $C_{\mu
\nu\rho}$ by a non-trivial quantity, in which we recognize the winding number
of the non-abelian gauge field configuration, see Eq.~(\ref{CSInteg2}). These
shifts are quantized so $\phi(t)$ again needs to be periodically identified.
Upon adopting an adequate normalization for $G_{3}^{\mu}$, the $d=2$ and $d=4$
Chern-Simons models do indeed match precisely.

Having used field strengths with exactly as many indices as there are
dimensions, we should actually allow for another term at the renormalizable
level:%
\begin{equation}
\mathcal{L}_{4}=-\frac{1}{2}\frac{1}{4!}F_{\mu\nu\rho\sigma}^{C}F^{C,\mu
\nu\rho\sigma}-\frac{1}{4!}\theta\varepsilon_{\mu\nu\rho\sigma}F^{C,\mu\nu
\rho\sigma}\leftrightarrow\mathcal{L}_{2}=-\frac{1}{2}\frac{1}{2!}F_{\mu\nu
}^{A}F^{A,\mu\nu}-\frac{1}{2!}\theta\varepsilon_{\mu\nu}F^{A,\mu\nu}\ .
\end{equation}
Given that $F_{\mu\nu\rho\sigma}^{C}=E\varepsilon_{\mu\nu\rho\sigma}$ and
$F_{\mu\nu}^{A}=E\varepsilon_{\mu\nu}$, the Lagrangian for $\phi$ is modified
to%
\begin{equation}
\mathcal{L}_{4}=\mathcal{L}_{2}=\frac{E^{2}}{2}+\theta E\rightarrow
\mathcal{L}_{\phi}=\frac{\dot{\phi}^{2}}{2}+\theta\dot{\phi}\ .
\label{LwithTh}%
\end{equation}
Classically, this new term does not affect the equation of motion $\ddot{\phi
}=0$, and $\dot{\phi}$ can still take any value. This makes sense since these
terms are total derivatives, $\varepsilon_{\mu\nu\rho\sigma}F^{C,\mu\nu
\rho\sigma}=4\varepsilon_{\mu\nu\rho\sigma}\partial^{\mu}C^{\nu\rho\sigma}$
and $\varepsilon_{\mu\nu}F^{A,\mu\nu}=2\varepsilon_{\mu\nu}\partial^{\mu
}A^{\nu}$.

The situation drastically changes as soon as $\phi$ is interpreted as a
quantum variable. As usual in quantum mechanics, when a dynamical variable is
constrained, there will be quantized energy levels, so indirectly, the
periodicity $\phi(t)=\phi(t)+\nu$ forces the quantization of the background
electric field. To see this, the standard procedure is to compute the
canonical momentum from Eq.~(\ref{LwithTh}), construct the Hamiltonian
$\mathcal{H}$, and impose the canonical commutation relation $[\phi,\pi]=i$ to
express $\mathcal{H}$ as a differential operator:%
\begin{equation}
\pi\equiv\frac{\partial\mathcal{L}}{\partial\dot{\phi}}=\dot{\phi}%
+\theta\ \rightarrow\mathcal{H}_{\phi}=\dot{\phi}\pi-\mathcal{L}_{\phi}%
=\frac{\dot{\phi}^{2}}{2}=-\frac{1}{2}\left(  i\frac{\partial}{\partial\phi
}+\theta\right)  \ .
\end{equation}
Solutions to the Schrodinger equation are then%
\begin{equation}
\mathcal{H}_{\phi}\left\vert \psi(\phi)\right\rangle =\mathcal{E}\left\vert
\psi(\phi)\right\rangle \rightarrow\left\vert \psi(\phi)\right\rangle
=\exp(i\kappa\phi)\ ,\ \ \mathcal{E}=\frac{1}{4}\left(  \kappa-\theta\right)
^{2}\ ,
\end{equation}
with $2\pi\kappa\in\mathbb{Z}$ since $\left\vert \psi(\phi)\right\rangle =\left\vert \psi(\phi
+\nu)\right\rangle $. The background electric field is indeed quantized, but
its levels are shifted by $\theta$. In this sense, the $\theta$ term, now
emerging as an angular variable, is also acting as a background electric
field. This is a perfectly equivalent dual view on the $\theta$ term of QCD.
Keep in mind though that it strongly relies on the existence of the
topological susceptibility, which provides the kinetic term for the $C_{\mu
\nu\rho}$ field. In this sense, instantons are still playing the crucial role
of opening the door to the sensitivity to $\theta$.

\subsection{Gauge anomalies in the Standard Model}\label{ConsistSM}

At this stage, it is a good time to look a bit more into the anomaly
coefficients. For a given algebra and representation $\mathbf{R}$, the
quadratic and cubic Casimir invariants are defined in terms of the fully
symmetrized trace over two and three generators%
\begin{align}
\frac{1}{2!}\left\langle T_{\mathbf{R}}^{a}T_{\mathbf{R}}^{b}\right\rangle
_{\text{\textrm{sym}}}  &  =\left\langle T_{\mathbf{R}}^{a}T_{\mathbf{R}}%
^{b}\right\rangle \equiv I_{2}(\mathbf{R})\delta^{ab}\ ,\\
\frac{1}{3!}\left\langle T_{\mathbf{R}}^{a}T_{\mathbf{R}}^{b}T_{\mathbf{R}%
}^{c}\right\rangle _{\text{\textrm{sym}}}  &  =\frac{1}{2}\left\langle
T_{\mathbf{R}}^{a}\{T_{\mathbf{R}}^{b},T_{\mathbf{R}}^{c}\}\right\rangle
\equiv\frac{1}{4}\mathcal{I}_{3}(\mathbf{R})d^{abc}\ . \label{CubicC}%
\end{align}
The $\mathcal{I}_{3}(\mathbf{R})$ are the \textbf{anomaly coefficients},
normalized to one for the fundamental representation $\mathbf{F}$. One also
sometimes defines $d_{\mathbf{R}}^{abc}=\mathcal{I}_{3}(\mathbf{R})d^{abc}$
with $d^{abc}\equiv d_{\mathbf{F}}^{abc}$. Higher order invariants can be
constructed in a similar way, see e.g. Ref.~\cite{Quevillon:2018mfl}, where
one can also find a tabulated list for most Lie algebras and representations
of interest.

In practice, to compute a given $\mathcal{I}_{3}(\mathbf{R})$, it is usually
not necessary to explicitly construct the generators. One can rely instead on
the generic properties of Casimir invariants:%
\begin{align}
\mathcal{I}_{n}(\mathbf{R})  &  =(-1)^{n}\mathcal{I}_{n}(\mathbf{R}^{\dagger
})\ ,\label{Casi1}\\
\mathcal{I}_{n}(\mathbf{R}_{1}\oplus\mathbf{R}_{2})  &  =\mathcal{I}%
_{n}(\mathbf{R}_{1})+\mathcal{I}_{n}(\mathbf{R}_{2})\ ,\label{Casi2}\\
\mathcal{I}_{n}(\mathbf{R}_{1}\otimes\mathbf{R}_{2})  &  =\mathcal{I}%
_{n}(\mathbf{R}_{1})N(\mathbf{R}_{2})+\mathcal{I}_{n}(\mathbf{R}%
_{2})N(\mathbf{R}_{1})=\sum\mathcal{I}_{n}(\mathbf{R}_{i}^{\prime})\;,
\label{Casi3}%
\end{align}
where $\mathbf{R}_{1}\otimes\mathbf{R}_{2}=\sum_{i}\mathbf{R}_{i}^{\prime}$
and $N(\mathbf{R})$ is the dimension of the representation $\mathbf{R}$. For
example, if one is after $\mathcal{I}_{3}(\mathbf{6})$ in $SU(3)$, knowing
that $\mathcal{I}_{3}(\mathbf{3})=1$ by definition, it suffices to use
Eq.~(\ref{Casi3}) on $\mathbf{3}\otimes\mathbf{3}=\mathbf{6}\oplus
\boldsymbol{\bar{3}}$ to find $\mathcal{I}_{3}(\mathbf{6})=7$ since $\mathcal{I}%
_{3}(\boldsymbol{\bar{3}})=-1$ from Eq.~(\ref{Casi1}). This can be repeated
systematically to get the anomaly coefficient of any representation.

Notice also that Eq.~(\ref{Casi1}) immediately implies that $\mathcal{I}%
_{3}(\mathbf{R})=0$ for real representations. This is actually true for any
self-dual representation. Indeed, those are such that $T^{a}$ and $-T^{a\ast}$
are related by a similarity transformation, $T^{a}=S^{-1}(-T^{a\ast})S$, and
since the trace is invariant under $S$,%
\begin{equation}
\left\langle T_{\mathbf{R}}^{a}\{T_{\mathbf{R}}^{b},T_{\mathbf{R}}%
^{c}\}\right\rangle =\left\langle ST_{\mathbf{R}}^{a}S^{-1}\{ST_{\mathbf{R}%
}^{b}S^{-1},ST_{\mathbf{R}}^{c}S^{-1}\}\right\rangle =-\left\langle
T_{\mathbf{R}}^{a}\{T_{\mathbf{R}}^{b},T_{\mathbf{R}}^{c}\}\right\rangle \;,
\end{equation}
This means in particular that any fermion in the adjoint representation does
not contribute to the triangle anomaly. Also, $SU(2)$ as a whole has no
anomaly coefficient since all its representations are either real or
self-dual. For example, its generators in the fundamental representation
satisfy $\vec{\sigma}=(i\sigma_{2})(-\vec{\sigma}^{T})(i\sigma_{2})^{\dagger}$.

It is worth to know also that the anomaly coefficients vanish for all
orthogonal groups except $SO(6)$. To prove this, let us repeat the simple but
elegant argument of Ref.~\cite{Georgi:1972bb} (see also
Ref.~\cite{Georgi:1982jb}). Remember that for the $SO(n)$ Lie algebra, it is
always possible to represent the generators in terms of antisymmetric
hermitian matrices. Actually, instead of denoting the $n(n-1)/2$ generators as
$T_{\mathbf{R}}^{a}$, we can use $T_{\mathbf{R}}^{ij}$ with $i>j$,
$i,j=1,...,n$, and $T_{\mathbf{R}}^{ij}=-T_{\mathbf{R}}^{ji}$. This means that
the invariant tensor $d_{\mathbf{R}}^{abc}\rightarrow d_{\mathbf{R}%
}^{ij,kl,mn}$ must be antisymmetric under $i\leftrightarrow j$,
$k\leftrightarrow l$, or $m\leftrightarrow n$, but symmetric under
$ij\leftrightarrow kl\leftrightarrow mn$. The most general structure having
the right antisymmetric property is%
\begin{align}
d_{\mathbf{R}}^{ij,kl,mn}  &  \sim\delta^{ik}\delta^{jm}\delta^{ln}%
-\delta^{jk}\delta^{im}\delta^{ln}-\delta^{il}\delta^{jm}\delta^{kn}%
+\delta^{jl}\delta^{im}\delta^{kn}\nonumber\\
&  \ \ -\delta^{ik}\delta^{jn}\delta^{lm}+\delta^{jk}\delta^{in}\delta
^{lm}+\delta^{il}\delta^{jn}\delta^{km}-\delta^{jl}\delta^{in}\delta^{km}\;,
\end{align}
since it cannot contain $\delta^{ij}$, $\delta^{kl}$, or $\delta^{mn}$. It is
thus automatically antisymmetric under $ij\leftrightarrow kl\leftrightarrow
mn$, and $d_{\mathbf{R}}^{ij,kl,mn}=0$. The only exception is $SO(6)$ because
it is possible to use its six-dimensional antisymmetric tensor, $d_{\mathbf{R}%
}^{ij,kl,mn}\sim\varepsilon^{ijklmn}$, which has the desired mixed properties
under permutation. This is no surprise since $SO(6)\sim SU(4)$, which does
have a non-trivial $d$ tensor.

\subsubsection{QED and QCD}

Theories in which the gauge boson couples exclusively to the vector current
can be made anomaly free simply by moving the anomaly onto the axial current,
to which no gauge interactions are attached. QED and QCD fall into that class
of so-called \textbf{vector theories}.

From the point of view of the consistent anomaly, however, these theories do
have anomalies, but they always cancel out in observables. Indeed, the
vectorial coupling of a Dirac fermion can be split as%
\begin{equation}
\bar{\psi}\gamma_{\mu}T^{a}\psi=\bar{\psi}_{L}\gamma_{\mu}T^{a}\psi_{L}%
+\bar{\psi}_{R}\gamma_{\mu}T^{a}\psi_{R}=\bar{\psi}_{L}\gamma_{\mu}T^{a}%
\psi_{L}+\bar{\psi}_{R}^{\mathrm{C}}\gamma_{\mu}(-T^{aT})\psi_{R}^{\mathrm{C}%
}\;.
\end{equation}
For the last equality, $\psi_{R}$ is replaced by its left-handed
charge-conjugate field $\psi_{R}^{\mathrm{C}}=\mathrm{C}\bar{\psi}_{R}^{T}$,
using that $\mathrm{C}^{-1}\gamma_{\mu}\mathrm{C}=-\gamma_{\mu}^{T}$. In a
massless Dirac theory, $\psi_{L}$ and $\psi_{R}$ are independent degrees of
freedom and could a priori transform according to any representation of the
gauge groups. For vector theories, by definition, $\psi_{R}^{\mathrm{C}}$ and
$\psi_{L}$ transform according to dual representations, $\psi_{L}%
\sim\mathbf{R}$ and $\psi_{R}^{\mathrm{C}}\sim\mathbf{R}^{\dagger}$ say, which
then allows for a mass term $\bar{\psi}\psi=(\psi_{R}^{\mathrm{C}}%
)^{T}\mathrm{C}^{\dagger}\psi_{L}$. Now, the triangle diagram for a Dirac
fermion is the sum of the triangle diagram with $\psi_{L}$ and with $\psi
_{R}^{\mathrm{C}}$. But by Eq.~(\ref{Casi1}), the anomaly coefficients of
$\psi_{L}$ and $\psi_{R}^{\mathrm{C}}$ are opposite, so the two cancel out.
Said differently, a Dirac fermion is necessarily in a self dual representation
$\psi_{L}\oplus\psi_{R}^{C}\sim\mathbf{R\oplus\bar{R}}$, hence has a vanishing
anomaly coefficient. Fundamentally, this is nothing else than Furry's theorem,
adapted to non-abelian gauge interactions.

\subsubsection{Standard Model}

The SM is not a vector theory, but a \textbf{chiral theory} since left- and
right-handed fields have different quantum numbers. Yet, it is one of the
mysteries of the SM that though individual fermion species do generate
anomalies, those cancel out when summed over all quarks and leptons,
separately for each generation. This is described in details in most
textbooks, so we shall be very brief.

To prove anomaly cancellation in the SM, consider the triangle diagrams with
three $G_{SM}=SU(3)_{C}\otimes SU(2)_{L}\otimes U(1)_{Y}$ gauge bosons,
$V_{1}^{a}V_{2}^{b}V_{3}^{c}$, where $a$, $b$, $c$ denote $G_{SM}$ group
indices. The gauge quantum numbers of the SM fermions, all taken as
left-handed, are
\begin{subequations}
\label{SMfermions}%
\begin{align}
q_{L}  &  =\left(
\begin{array}
[c]{c}%
u_{L}\\
d_{L}%
\end{array}
\right)  \sim\left(  \mathbf{3\otimes2}\right)  _{+1/3}\ ,\ u_{R}^{\mathrm{C}%
}\sim\left(  \boldsymbol{\bar{3}\otimes1}\right)  _{-4/3}\ ,\ d_{R}^{\mathrm{C}%
}\sim\left(  \boldsymbol{\bar{3}\otimes1}\right)  _{+2/3}\ ,\\
\ell_{L}  &  =\left(
\begin{array}
[c]{c}%
\nu_{L}\\
e_{L}%
\end{array}
\right)  \sim\left(  \mathbf{1\otimes2}\right)  _{-1}\ ,\ e_{R}^{\mathrm{C}%
}\sim\left(  \mathbf{1\otimes1}\right)  _{+2}\ .
\end{align}
The gauge anomalies are tuned by $\langle T_{f}^{a}\{T_{f}^{b},T_{f}%
^{c}\}\rangle$, summed over all fermions $f$. Clearly, we can discard
triangles with one gluon or one weak boson since Pauli and Gell-Mann matrices
are traceless. We can also discard diagrams with three gluons since QCD is a
vector theory, and those with three weak bosons since the fundamental
representation of $SU(2)_{L}$ is self-dual. For each fermion family, the
remaining triangles are then tuned by
\end{subequations}
\begin{subequations}
\label{SMano}%
\begin{align}
\left\langle Y^{3}\right\rangle  &  =\sum_{f=~\text{all}}Y_{f}^{3}%
=N_{C}\left(  2\left(  \frac{1}{3}\right)  ^{3}+\left(  -\frac{4}{3}\right)
^{3}+\left(  \frac{2}{3}\right)  ^{3}\right)  +\left(  2\left(  -1\right)
^{3}+\left(  2\right)  ^{3}\right)  =0\;,\\
\left\langle Y\{T_{L}^{a},T_{L}^{b}\}\right\rangle  &  \sim\delta^{ab}%
\sum_{f=q_{L},\ell_{L}}Y_{f}\sim N_{C}\left(  \frac{1}{3}\right)  +\left(
-1\right)  =0\;,\\
\left\langle Y\{T_{C}^{a},T_{C}^{b}\}\right\rangle  &  \sim\delta^{ab}%
\sum_{f=q_{L},u_{R}^{\mathrm{C}},d_{R}^{\mathrm{C}}}Y_{f}\sim2\left(  \frac
{1}{3}\right)  +\left(  -\frac{4}{3}\right)  +\left(  \frac{2}{3}\right)
=0\;,
\end{align}
where the generators are $T_{L}^{a}=\sigma^{a}/2$ for the left-handed fermion
doublets, and $T_{C}^{a}=\lambda^{a}/2$ for quarks, $-\lambda^{aT}/2$ for
antiquarks, and $\left\langle \{T^{a},T^{b}\}\right\rangle =2\mathcal{I}%
_{2}\delta^{ab}$, and $N_{C}=3$ is the number of colors.

Indirectly, these anomaly cancellations impose particular relationships
between the gauge quantum numbers of quarks and leptons. In this respect, note
that $N_{C}$ is not truly a free parameter in Eq.~(\ref{SMano}). To actually
leave it free requires to allow also for the quark hypercharges to be
different. If we set them to $Y_{u,d,q}$ and impose $Y_{q}=-1-Y_{u}=1-Y_{d}$,
which enforce the $U(1)_{Y}$ symmetry on the SM Yukawa couplings, then the
cancellation of anomalies implies~\cite{Abbas:1990kd}%
\end{subequations}
\begin{equation}
Y_{q}=\frac{1}{N_{C}}\ ,\ Y_{u}=-1-\frac{1}{N_{C}}\ ,\ Y_{d}=1-\frac{1}{N_{C}%
}\ \rightarrow Q_{u}=\frac{1}{2}\left(  \frac{1}{N_{C}}+1\right)
\ ,\ Q_{d}=\frac{1}{2}\left(  \frac{1}{N_{C}}-1\right)  \ . \label{NcQuQd}%
\end{equation}
It is in this sense that the cancellation of anomalies forces the electric
charge of quarks to occur in multiples of $1/3$ when $N_{C}=3$.

\subsubsection{Grand Unified Theories}

On closer inspection, one can notice that the contributions of $q_{L}$,
$\bar{u}_{R}$, $\bar{e}_{R}$ compensate those of $\ell_{L}$, $\bar{d}_{R}$ in
the tree summations in Eqs.~(\ref{SMano}). With in addition the requirement
that quark hypercharges are in multiples of $1/3$, this can be understood by
embedding SM fermions into $SU(5)$ representations:
\begin{subequations}
\label{FermionSU5}%
\begin{align}
\boldsymbol{\bar{5}}  &  =(\boldsymbol{\bar{3}}\otimes\mathbf{1})_{2/3}\oplus
(\mathbf{1}\otimes\mathbf{2})_{-1}=d_{R}^{\mathrm{C}}\oplus\ell_{L}%
\;,\label{Fermion5}\\
\mathbf{10}  &  =(\boldsymbol{\bar{3}}\otimes\mathbf{1})_{-4/3}\oplus
(\mathbf{3}\otimes\mathbf{2})_{1/3}\oplus(\mathbf{1}\otimes\mathbf{1}%
)_{2}=u_{R}^{\mathrm{C}}\oplus q_{L}\oplus e_{R}^{\mathrm{C}}\ .
\end{align}
where color indices are implicit, as well as some $SU(2)$ conjugations and
conventional minus signs. These are called branching rules for the
representations of $SU(5)$ into that of its $SU(3)\otimes SU(2)\otimes U(1)$
subgroup (see e.g. Refs.~\cite{Slansky:1981yr,Georgi:1982jb}). In this
picture, the SM would emerge from a $SU(5)$ gauge theory, spontaneously broken
to $G_{SM}$ at the grand unification scale, somewhere around $10^{16}$~GeV.
That theory is anomaly free since the SM is, but this now takes the simpler
form
\end{subequations}
\begin{equation}
\mathcal{I}_{3}(\boldsymbol{\bar{5}})+\mathcal{I}_{3}(\mathbf{10})=0\ ,
\label{I10I5}%
\end{equation}
since the gauge group is no longer a factor group. Its generators are not
tensor products, and only one type of triangle diagrams exists.

To prove Eq.~(\ref{I10I5}) without explicitly constructing the generators is a
bit tricky though. First, we have to set $\mathcal{I}_{3}(\boldsymbol{\bar{5}})=1$
(or $-1$) by definition of the $d$ tensor of $SU(5)$. This $d$ tensor is then
normalized differently than that of QCD, and $\mathcal{I}_{3}(\boldsymbol{\bar{5}%
})$ is not related in a simple way to that of $\ell_{L}$ and $\bar{d}_{R}$ on
the basis of Eq.~(\ref{FermionSU5}). The question is then to get
$\mathcal{I}_{3}(\mathbf{10})$ given the convention $\mathcal{I}%
_{3}(\boldsymbol{\bar{5}})=1$. Tensor products are of no help here because
$\mathbf{5}\otimes\mathbf{5}=\mathbf{10}\oplus\mathbf{15}$, but we do not know
$\mathcal{I}_{3}(\mathbf{15})$ either. Instead, the trick is to compare the
value of the $d_{\mathbf{5}}$ and $d_{\mathbf{10}}$ tensors for a very
specific but simple generator: the electric charge. It can be read off the
particle contents in Eq.~(\ref{FermionSU5}) as%
\begin{align}
Q_{\boldsymbol{\bar{5}}}  &  =\operatorname*{diag}%
(1/3,1/3,1/3,-1,0)\ ,\label{Q10expl}\\
Q_{\mathbf{10}}  &  =\operatorname*{diag}%
(-2/3,-2/3,-2/3,2/3,-1/3,2/3,-1/3,2/3,-1/3,1)\;.
\end{align}
This proves Eq.~(\ref{I10I5}) since
\begin{equation}
\mathcal{I}_{3}(\mathbf{10})=\mathcal{I}_{3}(\mathbf{5})\times\frac
{\operatorname*{Tr}(Q_{\mathbf{10}}\{Q_{\mathbf{10}},Q_{\mathbf{10}}%
\})}{\operatorname*{Tr}(Q_{\mathbf{5}}\{Q_{\mathbf{5}},Q_{\mathbf{5}}%
\})}=\mathcal{I}_{3}(\mathbf{5})\;.
\end{equation}

Going one step higher in the unification chain provides another perspective on
the anomaly cancellation in $SU(5)$. We know that orthogonal groups have no
anomaly, so let us consider $SO(10)\rightarrow SU(5)\otimes U(1)$. This time,
all the fermions are embedded in a single sixteen-dimensional representation,
$\mathbf{16}=\overline{\mathbf{10}}\oplus\mathbf{5}\oplus\mathbf{1}$. A
right-handed neutrino perfectly fits in the extra singlet since it ends up
neutral under all the SM gauge group. So, provided we accept the existence of
that particle, $\mathcal{I}_{3}(\boldsymbol{\bar{5}})+\mathcal{I}_{3}%
(\mathbf{10})=0$ becomes an algebraic consequence of $\mathcal{I}%
_{3}(\mathbf{16})=0$, and so is the cancellation of all gauge anomalies in the
SM. Notice that in practice, knowing the branching rules of orthogonal groups
down to unitary groups provide a powerful tool to quickly compute anomaly coefficients.

\section{The non-abelian anomaly}

Imagine a massless Dirac fermion has its left and right components coupled to
two gauge fields $\mathbf{A}_{\mu}^{L,R}=A_{\mu}^{a}T^{a}$ as
\begin{equation}
\mathcal{L}_{\mathrm{fermion}}=\bar{\psi}_{L}iD^{L}\psi_{L}+\bar{\psi}%
_{R}iD^{R}\psi_{R}\ ,\ D_{\mu}^{L,R}=\partial_{\mu}-i\mathbf{A}_{\mu}^{L,R}\ .
\label{NAmodel1}%
\end{equation}
Classically, $\mathcal{L}_{\mathrm{fermion}}$ has the $SU(N)_{L}\otimes
SU(N)_{R}$ gauge symmetry thanks to the chiral currents being conserved%
\begin{equation}
J_{L,R}^{a,\mu}=\bar{\psi}_{L,R}\gamma^{\mu}T^{a}\psi_{L,R}\ \ ,\ \ \ D_{\mu
}^{L,R}J_{L,R}^{\mu}\overset{\mathrm{Classical}}{=}0\ .
\end{equation}
At the quantum level, the currents $J_{L}^{\mu}$ and $J_{R}^{\mu}$ are both
anomalous, with their consistent anomaly given in Eq.~(\ref{AnoCons}). In the
next section, we trade these chiral currents and gauge fields for the usual
axial and vector currents and gauge fields, and express the consistent anomaly
in terms of these quantities, thereby deriving what we shall call the
\textbf{non-abelian anomaly}. Both anomalies are identical at the fundamental
level, but the latter deserves a distinct name because it is specific to the
$SU(N)_{L}\otimes SU(N)_{R}$ model of Eq.~(\ref{NAmodel1}), while only a
single Weyl fermion suffices for the former.

Having a Dirac fermion at our disposal, it is time to turn on a Dirac mass term. 
Naively, this should not change much the
result. A mass term explicitly breaks the axial gauge symmetry, but that
should be easily taken care of by adding the appropriate term from the
classical Ward identity $\partial_{\mu}A^{\mu}=2imP$. For the anomalous part,
the mass term should not have any impact, in the spirit of anomalies being
"mostly UV". What is a bit puzzling though is that we saw in Sec.~\ref{IRUV} the central
role played by IR singularities. It is their presence that prevents us from
turning off the anomaly in all the currents simultaneously. Yet, a fermion
mass term automatically regulates all these IR divergences, so another
mechanism must take over. Schematically, in the massless case of the previous
section, we had%
\begin{gather}
\left(  \text{massless axial loop}\right)  \supset\left(  \text{UV
ambiguities}\right)  +\left(  \text{IR singularities}\right) \nonumber\\
\rightarrow D\left(  \text{massless axial loop}\right)  =\text{Anomaly\ .}
\label{Interp1}%
\end{gather}

In the massive case, the loops can only give the UV part, but we now have to
take care of the $2imP$ piece of the Ward identity. At the diagram level, it
translates into new loops involving a pseudoscalar current, and schematically,%
\begin{gather}
\left(  \text{massive axial loop}\right)  \supset\left(  \text{UV
ambiguities}\right) \nonumber\\
\rightarrow D\left(  \text{massive axial loop}\right)  -2im\left(
\text{pseudoscalar loop}\right)  =\text{Anomaly\ .} \label{Interp2}%
\end{gather}
The final anomalous terms will turn out to be identical, as they should. The
massive pseudoscalar loop contributions and the IR singularities of the
massless axial loops play exactly the same role, and this even if they enter
at different levels, before or after taking the divergence. Strictly speaking,
neither are anomalous in themselves, but in their presence, it is impossible
to enforce all the classical Ward identities. In the dispersive approach
mentioned in Sec.~\ref{IRUV}, one can even show that these pseudoscalar
contributions do generate delta functions if $m\rightarrow0$, generating the
IR poles in the massless limit. Sending $m\rightarrow0$ is thus not a trivial
matter, and it is all but smooth, explaining why the calculations end up being
so different with $m=0$ or $m\neq0$.

In practice, being able to recover the anomaly with massive fermions will
prove extremely valuable. All the loop calculations will become (almost)
trivial once free of those IR singularities. Actually, it will be easy to
reconstruct not only the consistent anomaly, but also the fully general
expression of the chiral anomaly for the triangle, box, and pentagon, keeping
track of all the free parameters.

\subsection{Axial and vector currents}\label{SecVAdefs}

Putting the chiral components of the fermions back together permits to define
vector and axial currents, $J_{V,A}^{a,\mu}=J_{R}^{a,\mu}\pm J_{L}^{a,\mu}$,
as well as vector and axial gauge fields $\mathbf{A}_{\mu}^{R,L}%
=\mathbf{V}_{\mu}\pm\mathbf{A}_{\mu}$ such that the covariant derivative is%
\begin{equation}
D_{\mu}=\partial_{\mu}-i\mathbf{A}_{\mu}^{R}P_{R}-i\mathbf{A}_{\mu}^{L}%
P_{L}=\partial_{\mu}-i\mathbf{V}_{\mu}-i\mathbf{A}_{\mu}\gamma_{5}\ .
\end{equation}
From this, we also define the so-called \textbf{Bardeen curvatures} as
$[D_{\mu},D_{\nu}]=-i\mathbf{F}_{\mu\nu}^{V}-i\gamma_{5}\mathbf{F}_{\mu\nu
}^{A}$ with
\begin{subequations}
\label{BardeenCurv}%
\begin{align}
\mathbf{F}_{\mu\nu}^{V}  &  =\partial_{\mu}\mathbf{V}_{\nu}-\partial_{\nu
}\mathbf{V}_{\mu}-i[\mathbf{V}_{\mu},\mathbf{V}_{\nu}]-i[\mathbf{A}_{\mu
},\mathbf{A}_{\nu}]\ ,\\
\mathbf{F}_{\mu\nu}^{A}  &  =\partial_{\mu}\mathbf{A}_{\nu}-\partial_{\nu
}\mathbf{A}_{\mu}-i[\mathbf{V}_{\mu},\mathbf{A}_{\nu}]-i[\mathbf{A}_{\mu
},\mathbf{V}_{\nu}]\ ,
\end{align}
such that $\mathbf{F}_{\mu\nu}^{R,L}=\mathbf{F}_{\mu\nu}^{V}\pm\mathbf{F}%
_{\mu\nu}^{A}=\partial_{\mu}\mathbf{A}_{\nu}^{R,L}-\partial_{\nu}%
\mathbf{A}_{\mu}^{R,L}-i[\mathbf{A}_{\mu}^{R,L},\mathbf{A}_{\nu}^{R,L}]$.

Despite the notation, it is important to realize that the active symmetry is
still $SU(N)_{L}\otimes SU(N)_{R}$, whose Lie algebra cannot be factorized
into $SU(N)_{V}\otimes SU(N)_{A}$. In particular, starting from $D_{\mu}%
^{L,R}J_{L,R}^{\mu}=\partial_{\mu}-i[\mathbf{A}_{\mu}^{L,R},J_{L,R}^{\mu}]$,
the covariant derivatives of the corresponding vector and axial currents are
mixed,%
\end{subequations}
\begin{equation}
D_{\mu}J_{V,A}^{\mu}=\partial_{\mu}J_{V,A}^{\mu}-i[\mathbf{V}_{\mu}%
,J_{V,A}^{\mu}]-i[\mathbf{A}_{\mu},J_{A,V}^{\mu}]\ .
\end{equation}
Similarly, $\delta\mathbf{A}_{\mu}^{L,R}=D_{\mu}^{L,R}\boldsymbol{\Lambda}_{L,R}$
under $SU(N)_{L}\otimes SU(N)_{R}$ translates into the complicated
transformations%
\begin{subequations}
\begin{align}
\delta\mathbf{V}_{\mu},\delta\mathbf{A}_{\mu}  &  =\partial_{\mu
}\boldsymbol{\Lambda}_{V,A}-i[\mathbf{V}_{\mu},\boldsymbol{\Lambda}_{V,A}%
]-i[\mathbf{A}_{\mu},\boldsymbol{\Lambda}_{A,V}]\ ,\label{GaugeAV}\\
\delta\mathbf{F}_{\mu\nu}^{V,A}  &  =-i[\mathbf{F}_{\mu\nu}^{V,A}%
,\boldsymbol{\Lambda}_{V,A}]-i[\mathbf{F}_{\mu\nu}^{A,V},\boldsymbol{\Lambda}%
_{A,V}]\ ,
\end{align}
with $\boldsymbol{\Lambda}_{R,L}=\boldsymbol{\Lambda}_{V}\pm\boldsymbol{\Lambda}_{A}$.

\begin{figure}[t]
\centering\includegraphics[width=0.60\textwidth]{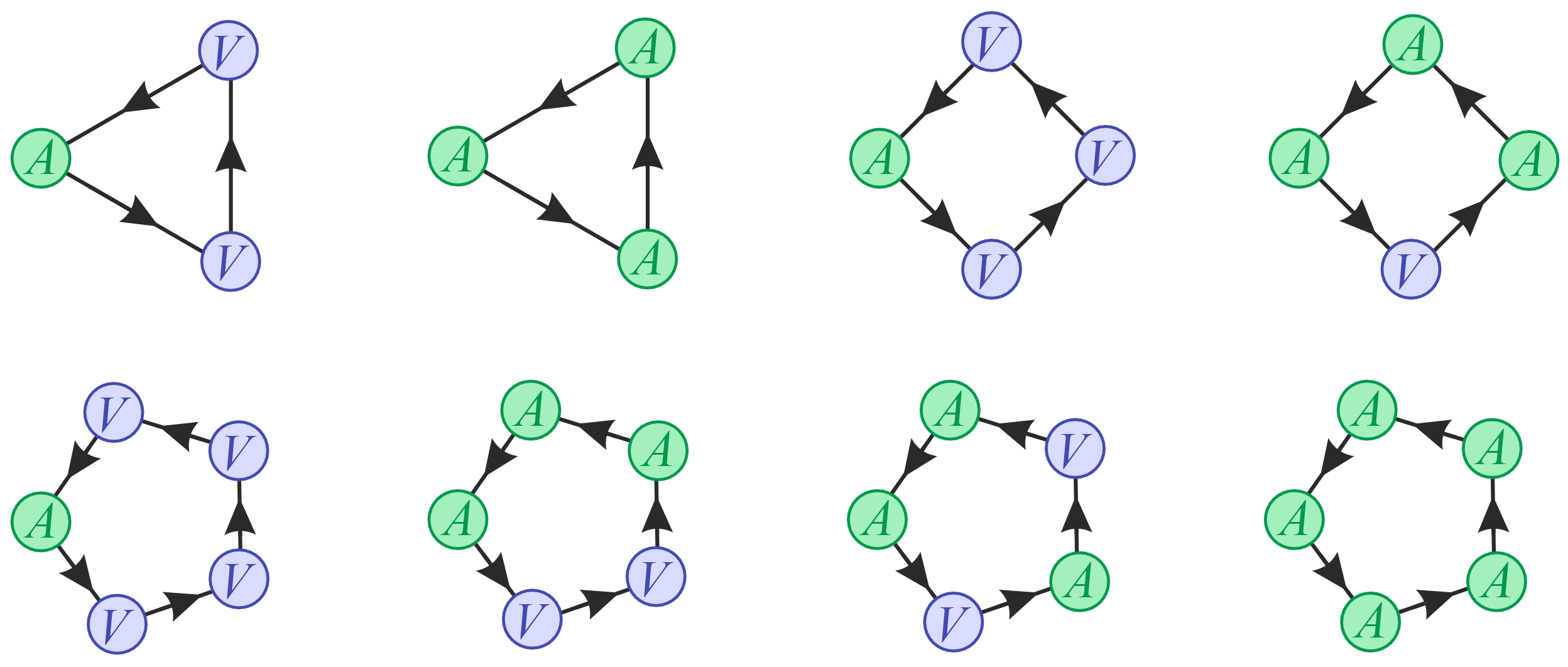}\caption{For the
fully Bose-symmetric consistent anomaly of Eq.~(\ref{AVC}), only the triangle
and box configurations in the top row exist, with identical covariant
divergences for any of the axial or vector currents. By contrast, for the
Bardeen anomaly of Eq.~(\ref{AVC2}), all the pentagon configurations also
exist, but the covariant divergences are identical separately for either the
vector or axial currents, with the former being further tuned to zero.}
\label{DiagsSUNxSUN}%
\end{figure}

With all these provisions in mind, we can work out the anomaly in $D_{\mu
}J_{V,A}^{\mu}=D_{\mu}^{R}J_{R}^{\mu}\pm D_{\mu}^{L}J_{L}^{\mu}$ simply by
plugging in the result in Eq.~(\ref{AnoCons}) expressed in terms of
$\mathbf{V}_{\mu}$ and $\mathbf{A}_{\mu}$. This gives the \textbf{non-abelian
anomaly}
\end{subequations}
\begin{equation}
(D_{\mu}J_{V,A}^{\mu})^{a}=-\frac{\varepsilon^{\mu\nu\rho\sigma}}{48\pi^{2}%
}\left\langle T^{a}\mathcal{A}_{V,A}\right\rangle \ ,
\end{equation}
with
\begin{subequations}
\label{AVC}%
\begin{align}
\mathcal{A}_{V}  &  =\mathbf{F}_{\mu\nu}^{A}\mathbf{F}_{\rho\sigma}%
^{V}+\mathbf{F}_{\mu\nu}^{V}\mathbf{F}_{\rho\sigma}^{A}-2\{\mathbf{A}_{\mu
},\mathbf{A}_{\nu},\mathbf{A}_{\rho},\mathbf{V}_{\sigma}\}-2\{\mathbf{V}_{\mu
},\mathbf{V}_{\nu},\mathbf{V}_{\rho},\mathbf{A}_{\sigma}\}\nonumber\\
&  \ \ \ \ +i\{\mathbf{F}_{\mu\nu}^{A},\mathbf{A}_{\rho},\mathbf{A}_{\sigma
}\}+i\{\mathbf{F}_{\mu\nu}^{A},\mathbf{V}_{\rho},\mathbf{V}_{\sigma
}\}+i\{\mathbf{F}_{\mu\nu}^{V},\mathbf{A}_{\rho},\mathbf{V}_{\sigma
}\}+i\{\mathbf{F}_{\mu\nu}^{V},\mathbf{V}_{\rho},\mathbf{A}_{\sigma}\}\ ,\\
\mathcal{A}_{A}  &  =\mathbf{F}_{\mu\nu}^{A}\mathbf{F}_{\rho\sigma}%
^{A}+\mathbf{F}_{\mu\nu}^{V}\mathbf{F}_{\rho\sigma}^{V}-2\mathbf{V}_{\mu
}\mathbf{V}_{\nu}\mathbf{V}_{\rho}\mathbf{V}_{\sigma}-2\{\mathbf{A}_{\mu
},\mathbf{A}_{\nu},\mathbf{V}_{\rho},\mathbf{V}_{\sigma}\}-2\mathbf{A}_{\mu
}\mathbf{A}_{\nu}\mathbf{A}_{\rho}\mathbf{A}_{\sigma}\nonumber\\
&  \ \ \ \ +i\{\mathbf{F}_{\mu\nu}^{A},\mathbf{V}_{\rho},\mathbf{A}_{\sigma
}\}+i\{\mathbf{F}_{\mu\nu}^{V},\mathbf{A}_{\rho},\mathbf{A}_{\sigma
}\}+i\{\mathbf{F}_{\mu\nu}^{A},\mathbf{A}_{\rho},\mathbf{V}_{\sigma
}\}+i\{\mathbf{F}_{\mu\nu}^{V},\mathbf{V}_{\rho},\mathbf{V}_{\sigma}\}\ ,
\end{align}
where $\{...\}$ denotes cyclic permutations, for example $\{X_{\mu\nu}%
,Y_{\rho},Z_{\sigma}\}=X_{\mu\nu}Y_{\rho}Z_{\sigma}+Z_{\mu}X_{\nu\rho
}Y_{\sigma}+Y_{\mu}Z_{\nu}X_{\rho\sigma}$. Notice that since there is no
quartic term in Eq.~(\ref{AnoCons}), their appearance here is an
artifact:\ they all disappear if the field strengths are expanded back using
their definitions in Eq.~(\ref{BardeenCurv}). In practice, the calculation of
these covariant divergences proceeds exactly as for the chiral anomaly in
Sec.~\ref{SecChiralAno}. Actually, in that calculation, the $P_{L}$
projectors were already expressed in terms of vector and axial couplings, see
Eq.~(\ref{DRano}). So, the quadratic and cubic terms in Eq.~(\ref{AVC}) have
the same expressions as in Eqs.~(\ref{AnoLLL}) and~(\ref{AnoLLLL}), setting
again the arbitrary parameters to their Bose-symmetric values $a_{k}^{ij}=1/3$
and $b_{i}=1/4$, but for a factor of two and a minus sign from $P_{L}%
=(1-\gamma_{5})/2$. As a result, the Bose symmetry is still present in
Eq.~(\ref{AVC}), and is enforced irrespective of the axial or vector character
of the currents. For example, the $(DJ_{V})VVA$, $(DJ_{V})AAA$, $(DJ_{A})VVV$,
and $(DJ_{A})AAV$ vertices all have exactly the same expression, which is just
twice that in Eq.~(\ref{AnoLLLL2}), see Fig.~\ref{DiagsSUNxSUN}.

\subsection{Massive loops and anomalies}

Turning on the fermion mass brings two important simplifications, that will
finally allow us to extend the calculation to pentagon diagrams. First, all IR
singularities disappear, and the calculation can be done on-shell throughout.
Second, there is no need to fully compute the loop integrals since we know
from the calculation of the singlet anomaly in Sec.~\ref{Singlet} that only
the leading term in a $1/m$ expansion can be anomalous. These two facts render
the computation trivial\footnote{Especially if FeynCalc loads, via FeynHelpers
\cite{Shtabovenko:2016whf}, the Package-X~functions~\cite{Patel:2016fam}.
Otherwise, it is first necessary to first calculate the leading terms of the
mass expansion of the various scalar loop functions.}, and still adopting the
procedure of Eq.~(\ref{DRano}) along with the BMHV scheme, we find:
\end{subequations}
\begin{subequations}
\label{MassiveAno}%
\begin{align}
\mathcal{\tilde{T}}_{AVV}^{\alpha\beta\gamma,abc}  &  =\mathcal{\tilde{T}%
}_{AAA}^{\alpha\beta\gamma,abc}=\frac{\mathcal{I}_{3}}{8\pi^{2}}d^{abc}\left(
\tilde{a}_{2}\varepsilon^{\alpha\beta\gamma\mu}q_{1\mu}-\tilde{a}%
_{1}\varepsilon^{\alpha\beta\gamma\mu}q_{2\mu}\right)  +\mathcal{O}%
(m^{-2})\ ,\\
\mathcal{\tilde{T}}_{AVVV}^{\alpha\beta\gamma\delta,abcd}  &  =\mathcal{\tilde
{T}}_{AAAV}^{\alpha\beta\gamma\delta,abcd}=i\frac{\mathcal{I}_{3}}{8\pi^{2}%
}(\tilde{b}_{2}d^{abe}f^{cde}-\tilde{b}_{3}d^{ace}f^{bde}+\tilde{b}_{4}%
d^{ade}f^{bce})\varepsilon^{\alpha\beta\gamma\delta}+\mathcal{O}(m^{-2})\ ,\\
\mathcal{\tilde{T}}_{AVVVV}^{\alpha\beta\gamma\delta,abcd}  &
=\mathcal{\tilde{T}}_{AAAVV}^{\alpha\beta\gamma\delta,abcd}=\mathcal{T}%
_{AAAAA}^{\alpha\beta\gamma\delta,abcd}=\mathcal{O}(m^{-2})\ ,
\end{align}
where coefficients are a priori different for each configuration of $A$ and
$V$ currents. As in the massless case, the arbitrary parameters $\tilde{a}%
_{i}$ and $\tilde{b}_{i}$ are combinations of the initial parameters occurring
in the Dirac traces (see Eq.~(\ref{DRano})) of the two triangle and six box
diagrams, after enforcing Furry's theorem. We denote them as well as the
amplitudes with a tilde to distinguish them from their massless counterparts.
These massive amplitudes are slightly more general than those in
Ref.~\cite{Chanowitz:1979zu}, keeping track of all possible ambiguities, and
we will see throughout this section that this is actually crucial. In this
respect, notice that if everything but the coefficients is discarded from
Eq.~(\ref{AmpTri}) and~(\ref{AmpBox}), we immediately get
Eq.~(\ref{MassiveAno}) once multiplied by $-2$ (to compensate for
$P_{L}=(1-\gamma_{5})/2$), as expected since the inherent UV ambiguity in
these amplitudes is the same whether the fermion is massive or massless. In
practice, one should understand that at this stage, this correspondence is
somewhat ambiguous since the coefficients are free, and rescaling them would
change the prefactors. Yet, it will prove very valuable to adopt the same
normalization for the massive and massless cases.

As said, it should not come as a surprise that these amplitudes closely
resemble the massless results. Yet, the disappearance of all IR singular terms
has profound consequences. In the absence of IR poles, Sutherland-Veltman
theorem can hold. We can tune all these $\mathcal{O}(m^{0})$ amplitudes to
zero by setting $\tilde{a}_{i}=0$, $\tilde{b}_{i}=0$. This does not mean the
anomaly disappears though, because once the fermion is massive, we have to
include pseudoscalar loop contributions whenever the covariant derivative with
respect to an axial current is taken, in agreement with the massive classical
Ward identity $\partial_{\mu}A^{\mu}=2imP$. There are seven such pseudoscalar
amplitudes to consider, see Fig.~\ref{FigPseudo}, and a straightforward
calculation gives for their $\mathcal{O}(m^{-1})$ terms:
\end{subequations}
\begin{subequations}
\label{PnV}%
\begin{align}
\mathcal{\tilde{T}}(P_{q_{3}}^{c}V_{q_{1}}^{\alpha,a}V_{q_{2}}^{\beta,b})  &
=-\frac{\mathcal{I}_{3}\varepsilon^{\alpha\beta\gamma\delta}}{16\pi^{2}%
m}d^{abc}q_{1\gamma}q_{2\delta}\ ,\\
\mathcal{\tilde{T}}(P_{q_{3}}^{c}A_{q_{1}}^{\alpha,a}A_{q_{2}}^{\beta,b})  &
=\frac{1}{3}\mathcal{\tilde{T}}(P_{q_{3}}^{c}V_{q_{1}}^{\alpha,a}V_{q_{2}%
}^{\beta,b})\ ,\\
\mathcal{\tilde{T}}(P_{q_{4}}^{d}V_{q_{1}}^{\alpha,a}V_{q_{2}}^{\beta
,b}V_{q_{3}}^{\gamma,c})  &  =i\frac{\mathcal{I}_{3}\varepsilon^{\alpha
\beta\gamma\delta}}{16\pi^{2}m}(d^{abe}f^{cde}q_{2\delta}-d^{ace}%
f^{bde}q_{3\delta}+d^{ade}f^{bce}q_{4\delta})\ ,\\
\mathcal{\tilde{T}}(P_{q_{4}}^{d}V_{q_{1}}^{\alpha,a}A_{q_{2}}^{\beta
,b}A_{q_{3}}^{\gamma,c})  &  =\frac{i\mathcal{I}_{3}\varepsilon^{\alpha
\beta\gamma\delta}}{24\pi^{2}m}(d^{abe}f^{cde}-d^{ace}f^{bde})q_{1\delta
}+\frac{1}{3}\mathcal{\tilde{T}}(P_{q_{4}}^{d}V_{q_{1}}^{\alpha,a}V_{q_{2}%
}^{\beta,b}V_{q_{3}}^{\gamma,c})\ ,\\
\mathcal{\tilde{T}}(P_{q_{5}}^{e}V_{q_{1}}^{\alpha,a}V_{q_{2}}^{\beta
,b}V_{q_{3}}^{\gamma,c}V_{q_{4}}^{\delta,d})  &  =-\frac{\mathcal{I}%
_{3}\varepsilon^{\alpha\beta\gamma\delta}}{16\pi^{2}m}d^{efg}(f^{adf}%
f^{bcg}-f^{acf}f^{bdg}+f^{abf}f^{cdg})\ ,\\
\mathcal{\tilde{T}}(P_{q_{5}}^{e}A_{q_{1}}^{\alpha,a}A_{q_{2}}^{\beta
,b}A_{q_{3}}^{\gamma,c}A_{q_{4}}^{\delta,d})  &  =-\frac{1}{3}\mathcal{\tilde
{T}}(P_{q_{5}}^{e}V_{q_{1}}^{\alpha,a}V_{q_{2}}^{\beta,b}V_{q_{3}}^{\gamma
,c}V_{q_{4}}^{\delta,d})\ ,\\
\mathcal{\tilde{T}}(P_{q_{5}}^{e}A_{q_{1}}^{\alpha,a}A_{q_{2}}^{\beta
,b}V_{q_{3}}^{\gamma,c}V_{q_{4}}^{\delta,d})  &  =\frac{\mathcal{I}%
_{3}\varepsilon^{\alpha\beta\gamma\delta}}{24\pi^{2}m}f^{cdg}(d^{afg}%
f^{bef}-d^{bfg}f^{aef})-\mathcal{\tilde{T}}(P_{q_{5}}^{e}A_{q_{1}}^{\alpha
,a}A_{q_{2}}^{\beta,b}A_{q_{3}}^{\gamma,c}A_{q_{4}}^{\delta,d})\ .
\end{align}
For these amplitudes, external states are on-shell, all momenta are outgoing,
and summation over repeated color indices is understood. Also, since we use
the Feynman rule $-i\gamma^{\mu}\gamma^{5}$ for the axial vertex $\bar{\psi
}\gamma^{\mu}\gamma^{5}\mathbf{A}^{\mu}\psi$, that for the pseudoscalar vertex
$\bar{\psi}\gamma^{5}\mathbf{P}\psi$ should be $-i\gamma_{5}$ to preserve
$\partial^{\mu}A_{\mu}^{a}=2imP^{a}$. Beware that though the loop calculation
is rather trivial, the simplification of the $SU(N)$ traces over five
generators is still tricky analytically. The above results were
obtained by numerically projecting the amplitudes on an explicit basis of
five-index tensors made of the $d^{abc}$, $f^{abc}$, and $\delta^{ab}$
invariants~\cite{Dittner:1971fy}.

\begin{figure}[t]
\centering\includegraphics[width=0.80\textwidth]{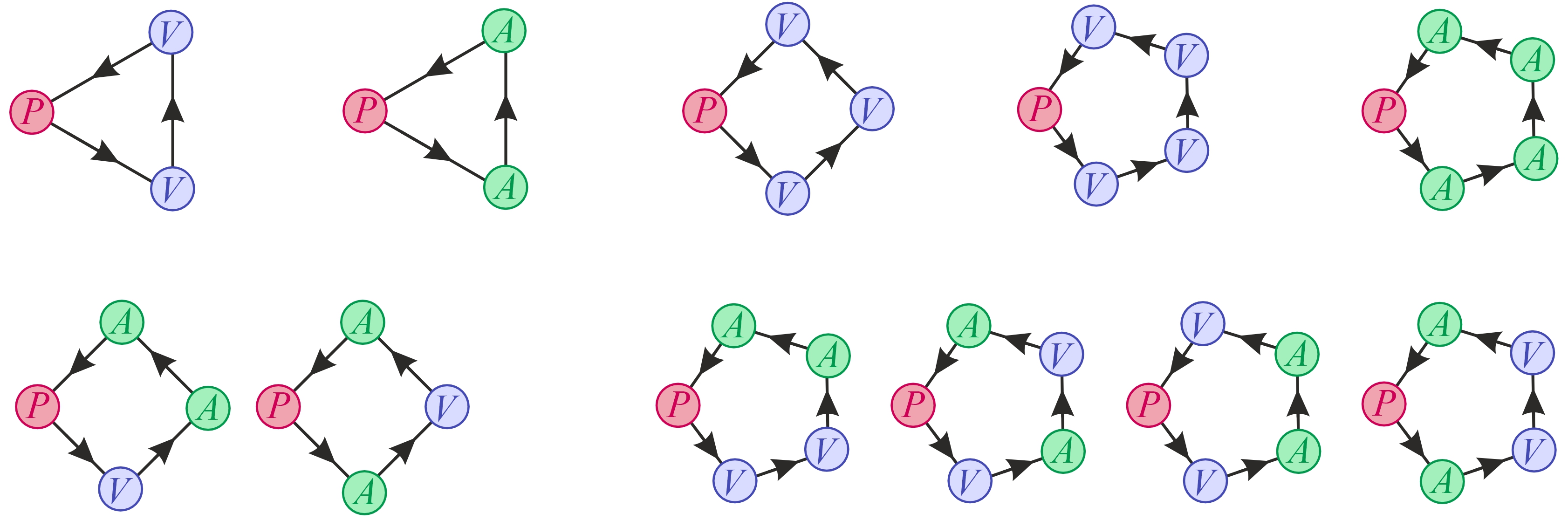}\caption{The
configurations of axial and vector currents entering in the seven possible
pseudoscalar loop amplitudes of Eq.~(\ref{PnV}).}
\label{FigPseudo}
\end{figure}

Let us now see in details how to reconstruct the result for the general
chiral anomaly derived for massless fermions in Sec.~\ref{SecChiralAno}
from the massive loop amplitudes. There are three
steps. One first plugs the expressions in Eq.~(\ref{MassiveAno}) in all the
explicit expressions of the covariant derivatives given in
Appendix~\ref{AppChiral}. This produces expressions similar to
Eq.~(\ref{AnoLLL}),~(\ref{AnoLLLL}), and~(\ref{M0Pen}), but with $\tilde
{a}_{i},\tilde{b}_{i}$ instead of $a_{i},b_{i}$, and all the $+1$ and $-1$
terms removed since these covariant derivatives vanish if $\tilde{a}%
_{i},\tilde{b}_{i}=0$. The next step is then to subtract $2im$ times the
appropriate pseudoscalar amplitudes of Eq.~(\ref{PnV}). This adds back some
multiples of $1$ and $1/3$ in various places. For example, for the triangles,
we arrive after these two steps at
\end{subequations}
\begin{subequations}
\label{AnoVVA}%
\begin{align}
D_{\alpha}\mathcal{\tilde{T}}_{VVA}^{\alpha\beta\gamma,abc}  &  =\frac
{i\mathcal{I}_{3}d^{abc}}{8\pi^{2}}\tilde{a}_{1}\varepsilon^{\beta\gamma\mu
\nu}q_{1\mu}q_{2\nu}\ ,\\
D_{\beta}\mathcal{\tilde{T}}_{VVA}^{\alpha\beta\gamma,abc}  &  =\frac
{i\mathcal{I}_{3}d^{abc}}{8\pi^{2}}\tilde{a}_{2}\varepsilon^{\gamma\alpha
\mu\nu}q_{1\mu}q_{2\nu}\ ,\\
D_{\gamma}\mathcal{\tilde{T}}_{VVA}^{\alpha\beta\gamma,abc}-2im\mathcal{\tilde
{T}}_{VVP}^{\alpha\beta}  &  =\frac{i\mathcal{I}_{3}d^{abc}}{8\pi^{2}}\left(
1-\tilde{a}_{1}-\tilde{a}_{2}\right)  \varepsilon^{\alpha\beta\mu\nu}q_{1\mu
}q_{2\nu}\ .
\end{align}
or
\end{subequations}
\begin{subequations}
\label{AnoAAA}%
\begin{align}
D_{\alpha}\mathcal{\tilde{T}}_{AAA}^{\alpha\beta\gamma,abc}-2im\mathcal{\tilde
{T}}_{AAP}^{\beta\gamma}  &  =\frac{i\mathcal{I}_{3}d^{abc}}{8\pi^{2}}\left(
\tilde{a}_{1}+\frac{1}{3}\right)  \varepsilon^{\beta\gamma\mu\nu}q_{1\mu
}q_{2\nu}\ ,\\
D_{\beta}\mathcal{\tilde{T}}_{AAA}^{\alpha\beta\gamma,abc}-2im\mathcal{\tilde
{T}}_{AAP}^{\gamma\alpha}  &  =\frac{i\mathcal{I}_{3}d^{abc}}{8\pi^{2}}\left(
\tilde{a}_{2}+\frac{1}{3}\right)  \varepsilon^{\gamma\alpha\mu\nu}q_{1\mu
}q_{2\nu}\ ,\\
D_{\gamma}\mathcal{\tilde{T}}_{AAA}^{\alpha\beta\gamma,abc}-2im\mathcal{\tilde
{T}}_{AAP}^{\alpha\beta}  &  =\frac{i\mathcal{I}_{3}d^{abc}}{8\pi^{2}}\left(
\frac{1}{3}-\tilde{a}_{1}-\tilde{a}_{2}\right)  \varepsilon^{\alpha\beta\mu
\nu}q_{1\mu}q_{2\nu}\ .
\end{align}

The final step is to redefine the $\tilde{a}_{i},\tilde{b}_{i}\rightarrow
a_{i},b_{i}$ coefficients so that all the divergences are equal when
$a_{i}=1/3$, $b_{i}=1/4$. This has to be done individually for each $A,V$
configuration because, as said at the beginning of this Section, IR
singularities and pseudoscalar contributions occur at different levels, see
Eq.~(\ref{Interp1}) and~(\ref{Interp2}). This forces us to match the two at
the level of divergences, for each specific configuration. Yet, the $\tilde
{a}_{i},\tilde{b}_{i}\rightarrow a_{i},b_{i}$ substitutions do follow a
universal set of rules, with $\tilde{a}_{i}\rightarrow a_{i}-1$ whenever there
is an axial current on the $i^{th}$ leg, $\tilde{a}_{i}\rightarrow a_{i}-1/3$
if both $i^{th}$ and $j^{th}$ legs are axial, and $\tilde{a}_{i}\rightarrow
a_{i}$ if both are vector. Further, these same rules also relate the $b_{i}$
and $\tilde{b}_{i}$ coefficients. Actually, it is to make these rules uniform
that we adopted the somewhat strange-looking numbering convention in
Eqs.~(\ref{AmpBox}) and~(\ref{MassiveAno}): the subscript refers to a specific
leg of the box diagram. There are then only three box parameters since the
characters of three of its legs suffices to set that of the fourth leg,
exactly like there must be two parameters to fully determine a triangle configuration.

Altogether, these configuration-dependent shifts make the combinatorics of the
method a bit cumbersome, but this is a small price one should be willing to
pay in exchange for the simplicity of the massive loop calculations. Further,
in practice, there is no need to check the correspondence for all possible
$A,V$ configuration. We only have to do the exercise once to extract the
general form of the anomaly from the massive amplitudes. For that, we can
choose a configuration that makes the job easy. If there is only one axial
current, then $\tilde{a}_{i}=a_{i},\tilde{b}_{i}=b_{i}$ except if the $i^{th}$
leg is axial, in which case it has to be shifted by $-1$. For example,
substituting $\tilde{a}_{1,2}\rightarrow a_{1,2}$ in Eq.~(\ref{AnoVVA}) above
directly reproduces Eq.~(\ref{AnoLLL}) since the $1$ and $2$ legs are vector
(the overall factor $-1/2$ comes from $P_{L}=(1-\gamma_{5})/2$). At last, this
offers a simple recipe to derive the quartic terms of the covariant
derivatives. For example, to get that quoted in Eq.~(\ref{M0Pen}), we start by
plugging the massive box amplitude of Eq.~(\ref{MassiveAno}) into the
covariant derivative in Eq.~(\ref{CovPen}). For the configuration in which
only the $q_{1}$ leg is axial, we add $2im\times$ $\mathcal{\tilde{T}%
}(P_{q_{1}}^{e}V_{q_{2}}^{\beta,b}V_{q_{3}}^{\gamma,c}V_{q_{4}}^{\delta
,d}V_{q_{5}}^{\varepsilon,e})$ from Eq.~(\ref{PnV}). Since the $q_{1}$ leg is
always the fourth one for all the boxes in Eq.~(\ref{CovPen}), all that remain
to do is to set $\tilde{b}_{2,3}^{ijk}\rightarrow b_{2,3}^{ijk}$, $\tilde
{b}_{4}^{ijk}\rightarrow b_{4}^{ijk}-1$, giving Eq.~(\ref{M0Pen}). The other
five divergences quoted in Appendix~\ref{AppChiral} are derived in this way.

Wrapping up, the massive amplitudes do permit to reconstruct the anomalous
divergences for the most general chiral anomaly, including the quadratic,
cubic, and quartic terms. The loop calculations are very simple, but care is
needed to deal with the arbitrary coefficients because in the massive case,
the various configurations of axial and vector fields end up decorrelated.
Though not indispensable, working out their relationship to reconstruct the
massless case makes the subsequent imposition of physical constraints much
more transparent, with for example the non-abelian anomaly in Eq.~(\ref{AVC})
simply obtained by enforcing the Bose-symmetric $a_{i}=1/3$, $b_{i}=1/4$.
There is however an exception to the superiority of the massless expressions
over the massive ones, which is the situation in which $\tilde{a}_{i}%
,\tilde{b}_{i}=0$. This brings us to the topic of the next Section.

\section{The Bardeen anomaly}\label{SecBardeen}

Phenomenologically, the $SU(N)_{L}\otimes SU(N)_{R}$ model is mostly used in
the context of QCD, where it is called the chiral symmetry. As will be briefly
introduced later on, an important feature is then that QED has to be
identified with a combination of $SU(N)_{V}$ generators. For this model to be
viable, it is thus compulsory to move anomalies out of the vector currents. To
achieve this, the strategy of Ref.~\cite{Bardeen:1969md} is to change the
action by adding the so-called \textbf{Bardeen counterterms}:%
\end{subequations}
\begin{equation}
\mathcal{L}_{Bardeen}=-\frac{\varepsilon^{\mu\nu\rho\sigma}}{24\pi^{2}%
}\left\langle X_{1}\{\mathbf{V}_{\mu},\mathbf{A}_{\nu}\}\mathbf{F}_{\rho
\sigma}^{V}+3X_{2}i\mathbf{V}_{\mu}\mathbf{A}_{\nu}\mathbf{A}_{\rho}%
\mathbf{A}_{\sigma}+X_{3}i\mathbf{V}_{\mu}\mathbf{V}_{\nu}\mathbf{V}_{\rho
}\mathbf{A}_{\sigma}\right\rangle \ . \label{BCT}%
\end{equation}
Because these counterterms are not gauge invariant, the gauge variation of the
action supplemented with $\mathcal{L}_{Bardeen}$ defines new currents
$\tilde{J}_{V,A}^{\mu}$, see Eq.~(\ref{ConsCurr}). Explicitly, adding
$\delta\mathcal{L}_{Bardeen}/\delta\Lambda_{A,V}$ to Eq.~(\ref{AVC}), we get
(see Fig.~\ref{DiagsSUNxSUN})
\begin{subequations}
\label{AVC2}%
\begin{align}
(D_{\mu}\tilde{J}_{V}^{\mu})^{a}  &  =0\ ,\\
(D_{\mu}\tilde{J}_{A}^{\mu})^{a}  &  =-\frac{\varepsilon^{\mu\nu\rho\sigma}%
}{16\pi^{2}}\left\langle T^{a}\left(  \mathbf{F}_{\mu\nu}^{V}\mathbf{F}%
_{\rho\sigma}^{V}+\frac{1}{3}\mathbf{F}_{\mu\nu}^{A}\mathbf{F}_{\rho\sigma
}^{A}+i\frac{8}{3}\{\mathbf{F}_{\mu\nu}^{V},\mathbf{A}_{\rho},\mathbf{A}%
_{\sigma}\}-\frac{32}{3}\mathbf{A}_{\mu}\mathbf{A}_{\nu}\mathbf{A}_{\rho
}\mathbf{A}_{\sigma}\right)  \right\rangle \ ,
\end{align}
by setting $X_{1,2,3}=1$. Consistency is ensured because these currents still
correspond to the gauge variation of an action.

One may be uneasy about adding these counterterms, since they change the
physics. The idea is that once gauge invariance is known to be lost, there is
no reason for all the possible counterterms not to be initially present in the
Lagrangian of the $SU(N)_{L}\otimes SU(N)_{R}$ model. Exactly like any other
free parameter, the $X_{1,2,3}$ are to be fixed by imposing some
renormalization conditions, and this order by order in perturbation theory.
Though at tree-level, one could still preserve both axial and vector currents
by setting $X_{1,2,3}=0$, this is no longer possible at one loop, and the best
we can do is to preserve the vector one. The only peculiarity for the Bardeen
counterterms is first that they do not need to absorb infinities, and second
that they are not really adjusted to measured values but rather to ensure
specific physical conditions like the Bose symmetry in the non-abelian case of
Eq.~(\ref{AVC}), or the conservation of the vector current for Eq.~(\ref{AVC2}%
). In this latter case, the axial current is usually restricted to represent a
global symmetry in phenomenological applications since there is no hope of
preserving its gauge symmetry. The typical physical case one has in mind is
that where pseudoscalar mesons are coupled to the axial currents. In that
context, their anomalous interactions still derive from the Bardeen anomaly,
but take the form of the so called \textbf{Wess-Zumino-Witten term}, covered
later on in this section.

\subsection{The point-splitting method}\label{SecPointSplit}

The Bardeen calculation in Ref.~\cite{Bardeen:1969md} is done using techniques
that are no longer very common nowadays (it was reproduced using a generalized
Pauli-Villars regularization soon after in Ref.~\cite{Brown:1969kmw}). In
particular, it relies on the so-called \textbf{point-splitting method},
introduced by Schwinger in 1951~\cite{Schwinger:1951nm}. Since this is
comparatively less discussed than the other methods in the literature, let us
give some details in a slightly modernized language (see e.g.
Ref.~\cite{Banerjee:1986ri,Peskin:1995ev,Bertlmann:1996xk}). The starting
point is to notice that a current like $\bar{\psi}(x)\gamma^{\mu}\gamma
_{5}\psi(x)$ is a singular quantity involving fields at the same space-time
point. So, it should be defined by a limiting procedure, something like
$\bar{\psi}(x+\varepsilon/2)\gamma^{\mu}\gamma_{5}\psi(x-\varepsilon/2)$ in
the limit $\varepsilon\rightarrow0$. If the fermion is charged under some
gauge symmetry, this procedure is problematic because gauge fields transform
differently at each space-time point, so gauge invariance is lost for
$\varepsilon>0$. To cure for that, one way to proceed is to insert a Wilson
line running from $x-\varepsilon/2$ to $x+\varepsilon/2$, i.e., to write%
\end{subequations}
\begin{equation}
J_{A}^{\mu}=\bar{\psi}(x)\gamma^{\mu}\gamma_{5}\psi(x)\rightarrow
J_{A,\mathrm{reg}}^{\mu}=~\underset{\varepsilon\rightarrow0}{\lim}\bar{\psi
}(x+\varepsilon/2)\gamma^{\mu}\gamma_{5}\psi(x-\varepsilon/2)\exp\left[
ie\int_{x-\varepsilon/2}^{x+\varepsilon/2}d\sigma^{\nu}A_{\nu}(\zeta)\right]
\ .
\end{equation}
Under a gauge transformation, the exponential factor transports the gauge
variation of $\psi$ from $x-\varepsilon/2$ to $x+\varepsilon/2$, where it
cancels with that of $\bar{\psi}$.

\begin{figure}[t]
\centering\includegraphics[width=0.80\textwidth]{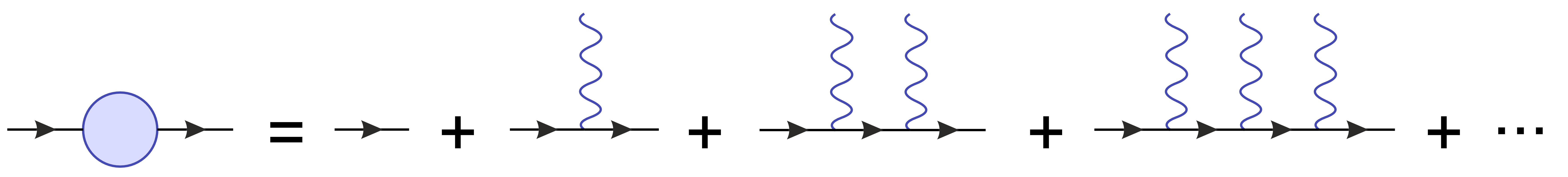}\caption{Graphical
representation of the electron propagator in an external electromagnetic
field, see Eqs.~(\ref{FullPropSer}) and~(\ref{FullProp}).}%
\label{Fig7}%
\end{figure}

If we take the divergence of $J_{A,\mathrm{reg}}^{\mu}$ and use the classical
equation of motion $(i \slashed D -m)\psi=0$ with $D^{\mu}=\partial^{\mu}-ieA^{\mu}$, the
$\slashed A$ of the covariant derivatives combine with the variation of the phase factor
to give the covariant expression
\begin{equation}
\partial_{\mu}J_{A,\mathrm{reg}}^{\mu}(x)=2imP_{\mathrm{reg}}(x)+ieF_{\mu
\alpha}(x)\underset{\varepsilon\rightarrow0}{\lim}\varepsilon^{\alpha
}J_{A,\mathrm{reg}}^{\mu}(x)+\mathcal{O}(\varepsilon^{2})\ , \label{RegWard}%
\end{equation}
where $P_{\mathrm{reg}}(x)$ is the regulated $\bar{\psi}\gamma_{5}\psi$
current. If the $\mathcal{O}(\varepsilon)$ and higher terms drop out when
$\varepsilon\rightarrow0$, we recover the classical Ward identity. But, as we
will now check, this does not happen because the matrix element of
$J_{A,\mathrm{reg}}^{\mu}$ develop a UV singularity in $1/\varepsilon$.
Adopting Schwinger's notations, one can represent all the matrix elements of
$J_{A,\mathrm{reg}}^{\mu}$ in a photon background by
\begin{equation}
\langle J_{A,\mathrm{reg}}^{\mu}\rangle_{\gamma}=\operatorname*{Tr}\left(
\gamma^{\mu}\gamma_{5}G(x+\varepsilon/2,x-\varepsilon/2)\right)  \exp\left[
ie\int_{x-\varepsilon/2}^{x+\varepsilon/2}d\sigma^{\nu}A_{\nu}(\zeta)\right]
\ ,
\end{equation}
with $G(x,y)$ is the full propagator, i.e., the Green function 
$(i\slashed D-m)G(x,y)=i\delta(x-y)$. Symbolically, its perturbative solution 
is the geometric series (see Fig.~\ref{Fig7}):
\begin{equation}
\frac{1}{i\slashed D-m}=\frac{1}{i\slashed \partial-m+e\slashed A}
=\frac{1}{i\slashed \partial-m}-\frac{1}{i\slashed \partial-m}e\slashed
A\frac{1}{i\slashed \partial-m}+...\ ,
\label{FullPropSer}
\end{equation}
which, in configuration space, takes the form
\begin{equation}
G(x+\varepsilon/2, x-\varepsilon/2)
=\int\frac{d^{4}p}{(2\pi)^{4}}\frac{i e^{ip\varepsilon}}
{\slashed p-m}+\int\frac{d^{4}p}{(2\pi)^{4}}\frac{d^{4}q}{(2\pi)^{4}}
\frac{e^{-ip(x-\varepsilon/2)}}{\slashed p-m}e\slashed A(p-q)
\frac{e^{iq(x+\varepsilon/2)}}{\slashed q-m}+...
\label{FullProp}
\end{equation}
where $\slashed A(p-q)$ is the Fourier transform of the $\slashed A(x)$ 
occurring in $i\slashed D$.

Plugging this expansion in $\langle J_{A,\mathrm{reg}}^{\mu}\rangle_{\gamma}$,
the Dirac trace kills the first term, while all the higher order terms are
regular and disappear as $\varepsilon\rightarrow0$. The only surviving
contribution comes from the second term. Switching integration variables to
$p\rightarrow p+q/2$ and $q\rightarrow p-q/2$, integrating by part, and
extracting the leading term in $q$,%
\begin{align}
\underset{\varepsilon\rightarrow0}{\lim}  &  ~\varepsilon^{\alpha
}\operatorname*{Tr}\left(  \gamma^{\mu}\gamma_{5}G(x+\varepsilon
/2,x-\varepsilon/2)\right) \nonumber\\
&  =\frac{e}{2}\int\frac{d^{4}q}{(2\pi)^{4}}e^{-iqx}A_{\nu}(q)\int\frac
{d^{4}p}{(2\pi)^{4}}\frac{\partial}{\partial p_{\alpha}}\operatorname*{Tr}%
\left[  \gamma^{\mu}\gamma_{5}\frac{1}{\slashed p+\tfrac{1}{2}
\slashed q-m}\gamma^{\nu}\frac{1}{\slashed p-\tfrac{1}{2}\slashed q-m}\right] \nonumber\\
&  =2ie\varepsilon^{\mu\nu\rho\sigma}\int\frac{d^{4}q}{(2\pi)^{4}}%
e^{-iqx}A_{\nu}(q)q_{\sigma}\int\frac{d^{4}p}{(2\pi)^{4}}\frac{\partial
}{\partial p_{\alpha}}\frac{p_{\rho}}{(p^{2}-m^{2})^{2}}\nonumber\\
&  =\frac{e\varepsilon^{\mu\nu\alpha\sigma}}{16\pi^{2}}\int\frac{d^{4}%
q}{(2\pi)^{4}}e^{-iqx}q_{\sigma}A_{\nu}(q)=\frac{ie}{8\pi^{2}}\tilde{F}%
^{\mu\alpha}(x)\ .
\end{align}
In the first line, notice the similarity with the surface term appearing in
Eq.~(\ref{SurfT}). To go from the second to the third line, we switch to
Euclidian momentum and use Stokes theorem. Plugging this in Eq.~(\ref{RegWard}%
) gives back the abelian Ward identity, Eq.~(\ref{AnoAbelian2}).

With a modern eye, Schwinger clearly stumbled on the anomaly nearly twenty
years before the seminal works of Adler, Bell and Jackiw. Yet, obviously, this
was less clear at the time. Schwinger's purpose was to propose a specific
regularization of the divergence of the axial current able to give back the
pseudoscalar current, so both could be used consistently to construct an
effective $\pi^{0}\gamma\gamma$ vertex able to account for the experimental
$\pi^{0}\rightarrow\gamma\gamma$ rate. In our usual notation, his calculation
did not include the Wilson line, and boils down to $(\partial_{\mu}A^{\mu
})_{\mathrm{reg}}=2imP_{\mathrm{reg}}$, with $(\partial_{\mu}A^{\mu
})_{\mathrm{reg}}=\partial_{\mu}A^{\mu}+e^{2}F_{\mu\nu}\tilde{F}^{\mu\nu
}/(8\pi^{2})$. Schwinger never gave any special meaning to these two terms of
$(\partial_{\mu}A^{\mu})_{\mathrm{reg}}$ since in his view, only
$(\partial_{\mu}A^{\mu})_{\mathrm{reg}}$ was physically well-defined. As
mentioned in Sec.~\ref{SVT}, the necessity to dig deeper only arose later,
once the axial current $A^{\mu}$ was understood as one of the Noether currents
of the hypothesized strong interaction chiral symmetry. With $A^{\mu}$ related
to the physical pion field, one is forced to give some physical meaning to
$(\partial_{\mu}A^{\mu})_{\mathrm{reg}}-\partial_{\mu}A^{\mu}$. Though
Schwinger's calculation of this difference is correct, we now understand it as
resulting from a fundamental physical effect: the breaking of the axial
symmetry by quantization.

\subsection{Sutherland-Veltman, again}

The point-splitting approach used by Bardeen in Ref.~\cite{Bardeen:1969md}
leads to quite complicated loop diagrams. Indeed, his purpose was to derive
the consistent anomaly, so all the vector and axial vector couplings had to be
split in a fully symmetric way. This provides a Bose-symmetric regularization
scheme, leading to Eq.~(\ref{AnoCons}), or equivalently to Eq.~(\ref{AVC}).

Instead of adding counterterms to get to Eq.~(\ref{AVC2}), we can recover the
Bardeen anomaly from our most general expression of the chiral anomaly in
Eqs.~(\ref{AnoLLL}),~(\ref{AnoLLLL}), and~(\ref{M0Pen}) by setting the $a_{i}%
$, $b_{i}$ coefficients to appropriate values. Specifically, we have to set
$a_{1}^{ij}=1$ if the $i$ current is axial, $a_{2}^{ij}=1$ if the $j$ current
is axial, $a_{k}^{ij}=1/3$ if both the $i$ and $j$ current are axial, and
$a_{k}^{ij}=0$ otherwise. Similarly, $b_{2}^{ijk}=1$ if the $j$ current is
axial, $b_{3}^{ijk}=1$ if the $k$ current is axial, and $b_{4}^{ijk}=1$ if the
$i,j,k$ are all vector since the fourth current is then axial. Finally, we
also have to set $b_{l}^{ijk}=1/3$ for those currents which are axial when
there are several, and $b_{l}^{ijk}=0$ otherwise.

The combinatorics may look a bit intricate, but everything works out very
straightforwardly once written down. In the present case, it is even not worth
the effort because switching to the massive case, we know that the
Sutherland-Veltman theorem must hold since the vector currents are conserved.
The Bardeen anomaly is very similar to the singlet anomaly: all the
divergences of the massive axial loops have to vanish because their mass
dependent parts cancel exactly their anomalous parts. This is indeed what
happens if we set all the $\tilde{a}$ and $\tilde{b}$ coefficients to zero in
Eq.~(\ref{MassiveAno}). Parametrically, the Bardeen anomaly is then equal to
the non-anomalous pseudoscalar contributions, up to a $-2im$ factor, a fact
first pointed out long ago in Ref.~\cite{Chanowitz:1979zu}. One can indeed
check that Eq.~(\ref{AVC2}) reproduce all the non-anomalous amplitudes in
Eq.~(\ref{PnV}), once multiplied by $-2im$.

All in all, since one does not need to explicit the $SU(N)$ traces in terms of
invariants to reconstruct the Bardeen anomaly from the pseudoscalar loops,
this makes the full calculation a rather trivial exercise. Knowing the Bardeen
counterterms, one could even work all the way back to the full consistent
anomaly. Yet, it is also satisfying to circumvent these counterterms
completely and understand the consistent and Bardeen anomalies as arising from
specific constraints set on the most general chiral anomaly.

\subsection{Bardeen counterterms and Bose symmetry\label{CTBose}}

Not all possible counterterms can impact the anomaly. For instance, with a
single chiral gauge field, one could try counterterms like $\varepsilon
^{\mu\nu\rho\sigma}\langle\boldsymbol{\partial}_{\mu}\mathbf{A}_{\nu}^{L}%
\partial_{\rho}\mathbf{A}_{\sigma}^{L}\rangle$, $\varepsilon^{\mu\nu\rho
\sigma}\langle\mathbf{A}_{\mu}^{L}\mathbf{A}_{\nu}^{L}\partial_{\rho
}\mathbf{A}_{\sigma}^{L}\rangle$, or $\varepsilon^{\mu\nu\rho\sigma}%
\langle\mathbf{A}_{\mu}^{L}\mathbf{A}_{\nu}^{L}\mathbf{A}_{\rho}^{L}%
\mathbf{A}_{\sigma}^{L}\rangle$, but their gauge variations do not leave any
trace, being components of $\varepsilon^{\mu\nu\rho\sigma}\mathbf{F}_{\mu\nu
}^{L}\mathbf{F}_{\rho\sigma}^{L}$, see Eq.~(\ref{AnoSinglet1}). To get a
non-trivial impact on the anomaly, more than one field is required.

This is where the connection between the Bardeen counterterms and Bose
symmetrization starts to make sense. The defining feature of the consistent
anomaly is to be Bose symmetric because it derives from the gauge variation of
a Lagrangian. Now, if there are more than one type of fields, symmetrization
is not automatic under the interchange of different types of fields. Covariant
derivatives of currents may still be symmetric under these generalized
interchanges, but only for some specific combinations of couplings in the
Lagrangian. That is the role of the counterterms: they tune the symmetry
relationships between currents of different types. With the two fields
$\mathbf{A}$ and $\mathbf{V}$, they encode a threefold ambiguity in the
anomaly of the two currents $\mathbf{J}_{V,A}^{\mu}$. This threefold ambiguity
is precisely the freedom one is left with if the Bose symmetry is imposed
separately under the interchange of either $\mathbf{A}$'s or $\mathbf{V}$'s,
but not under those interchanging $\mathbf{A}$'s and $\mathbf{V}$'s, in the
general form of the chiral anomaly. Indeed, that symmetry constraint is not
sufficient to fully fix all the arbitrary coefficients, and leaves three
combinations undetermined. In both cases, the threefold ambiguity is fixed by
requiring extra physical conditions, like an extended Bose symmetry under
$\mathbf{A}\leftrightarrow\mathbf{V}$ for the non-abelian anomaly in
Eq.~(\ref{AVC}), or the conservation of the vector current for the Bardeen
anomaly of Eq.~(\ref{AVC2}).

Let us be a bit more specific. If we leave the $X_{1,2,3}$ counterterms free,
the covariant divergences are found to be%
\begin{equation}
(D_{\mu}J_{V,A}^{\mu})^{a}=-\frac{\varepsilon^{\mu\nu\rho\sigma}}{48\pi^{2}%
}\langle T^{a}\mathcal{A}_{V,A}\rangle\ ,
\end{equation}
with%
\begin{align}
\mathcal{A}_{V}  &  =(1-X_{1})(\mathbf{F}_{\mu\nu}^{A}\mathbf{F}_{\rho\sigma
}^{V}+\mathbf{F}_{\mu\nu}^{V}\mathbf{F}_{\rho\sigma}^{A})+i(1-X_{3}%
)\{\mathbf{F}_{\mu\nu}^{A},\mathbf{V}_{\rho},\mathbf{V}_{\sigma}%
\}-6i(X_{1}-X_{2})\mathbf{A}_{\mu}\mathbf{F}_{\nu\rho}^{A}\mathbf{A}_{\sigma
}\nonumber\\
&  +i(2X_{1}-3X_{2}+1)\{\mathbf{F}_{\mu\nu}^{A},\mathbf{A}_{\rho}%
,\mathbf{A}_{\sigma}\}+i(1-2X_{1}+X_{3})(\{\mathbf{F}_{\mu\nu}^{V}%
,\mathbf{A}_{\rho},\mathbf{V}_{\sigma}\}+\{\mathbf{F}_{\mu\nu}^{V}%
,\mathbf{V}_{\rho},\mathbf{A}_{\sigma}\})\nonumber\\
&  +2i(X_{1}-X_{3})(\mathbf{A}_{\mu}\mathbf{V}_{\nu}\mathbf{F}_{\rho\sigma
}^{V}+\mathbf{F}_{\mu\nu}^{V}\mathbf{V}_{\rho}\mathbf{A}_{\sigma}%
)+2(X_{3}-1)\{\mathbf{V}_{\mu},\mathbf{A}_{\nu},\mathbf{A}_{\rho}%
,\mathbf{A}_{\sigma}\}\nonumber\\
&  +2(X_{3}-1)\{\mathbf{A}_{\mu},\mathbf{V}_{\nu},\mathbf{V}_{\rho}%
,\mathbf{V}_{\sigma}\}+6(X_{2}-X_{3})(\mathbf{A}_{\mu}\mathbf{A}_{\nu
}\mathbf{A}_{\rho}\mathbf{V}_{\sigma}+\mathbf{V}_{\mu}\mathbf{A}_{\nu
}\mathbf{A}_{\rho}\mathbf{A}_{\sigma})\ ,\\
\mathcal{A}_{A}  &  =\mathbf{F}_{\mu\nu}^{A}\mathbf{F}_{\rho\sigma}%
^{A}+\left(  2X_{1}+1\right)  \mathbf{F}_{\mu\nu}^{V}\mathbf{F}_{\rho\sigma
}^{V}+i(4X_{1}+3X_{2}+1)\{\mathbf{F}_{\mu\nu}^{V},\mathbf{A}_{\rho}%
,\mathbf{A}_{\sigma}\}+2i(X_{1}-X_{3})\mathbf{V}_{\mu}\mathbf{F}_{\nu\rho}%
^{V}\mathbf{V}_{\sigma}\nonumber\\
&  +i(2X_{1}-3X_{2}+1)(\{\mathbf{F}_{\mu\nu}^{A},\mathbf{A}_{\rho}%
,\mathbf{V}_{\sigma}\}+\{\mathbf{F}_{\mu\nu}^{A},\mathbf{V}_{\rho}%
,\mathbf{A}_{\sigma}\})-6i(X_{1}-X_{2})(\mathbf{F}_{\mu\nu}^{A}\mathbf{A}%
_{\rho}\mathbf{V}_{\sigma}+\mathbf{V}_{\mu}\mathbf{A}_{\nu}\mathbf{F}%
_{\rho\sigma}^{A})\nonumber\\
&  +i(1-2X_{1}+X_{3})\{\mathbf{F}_{\mu\nu}^{V},\mathbf{V}_{\rho}%
,\mathbf{V}_{\sigma}\}+2(X_{3}-1)\{\mathbf{V}_{\mu},\mathbf{V}_{\nu
},\mathbf{A}_{\rho},\mathbf{A}_{\sigma}\}+2(X_{3}-1)\mathbf{V}_{\mu}%
\mathbf{V}_{\nu}\mathbf{V}_{\rho}\mathbf{V}_{\sigma}\nonumber\\
&  +6(X_{2}-X_{3})\{\mathbf{V}_{\mu}\mathbf{V}_{\nu},\mathbf{A}_{\rho
},\mathbf{A}_{\sigma}\}-(2+30X_{2})\mathbf{A}_{\mu}\mathbf{A}_{\nu}%
\mathbf{A}_{\rho}\mathbf{A}_{\sigma}\ . \label{AcallA}%
\end{align}
This is the most general form of the consistent anomaly in the presence of $A$
and $V$ gauge fields. It clearly interpolates between the non-abelian and
Bardeen results, Eqs.~(\ref{AVC}) and~(\ref{AVC2}). Notice in particular how
the $X_{1}$ parameter permits to move the $AVV$ anomaly either on the axial
current, for $X_{1}=1$, or symmetrically on the $V$ currents for $X_{1}=-1/2$.
This is the well-known feature of the triangle anomaly described in most
introduction on the subject to emphasize the need to impose conditions on the
anomaly to get a definite result. Now, starting instead from the general
chiral anomaly and imposing the Bose symmetry separately on $A$ and $V$
interchanges, we end up with the relationships%
\begin{equation}
AVV:a_{1}=a_{2}=\frac{1-X_{1}}{3}\ ,\ \ AAA:a_{1}=a_{2}=\frac{1}{3}\ ,
\end{equation}
from the triangles with the $q_{1}$ current being axial, and%
\begin{equation}
AVVV:b_{2,3,4}=\frac{1}{4}-\frac{X_{1}}{3}+\frac{X_{3}}{12}%
\ ,\ \ VAAA:b_{2,3,4}=\frac{1}{4}+\frac{X_{1}}{3}-\frac{X_{2}}{4}\ ,
\end{equation}
from the boxes with the $q_{1}$ current being axial for $AVVV$, and vector for
$VAAA$. This shows that both procedures agree totally, hence that counterterms
do encode the same freedom as the arbitrary parameters. As an aside, notice
that the first term of $\mathcal{A}_{A}$ in Eq.~(\ref{AcallA}) is fixed. In
the $\mathbf{A}$, $\mathbf{V}$ model, it is impossible to turn off the anomaly
in the axial current because the $AAA$ triangle has to remain Bose symmetric.
To have a conserved axial gauge interaction would require at least two sets of
axial gauge fields, so that the anomaly may be entirely moved to one of them.
This kind of situation will be discussed in the next section.

Pushing the argument further, one may wonder what happens if absolutely no
Bose symmetry is imposed. This could be because the external gauge fields are
truly different (in which case the generators would be reducible), or because
we want to treat them separately. In both cases, the consistency condition
becomes a bit void since the fields completely cease to transform into each
other, so the final anomaly would not qualify as consistent. Yet, it is worth
to see what happens in this limiting case. The first step is to construct all
possible counterterms with three or four fields. For example, with only left
chiral interactions, the counterterms are made of $\varepsilon^{\mu\nu
\rho\sigma}\langle\mathbf{A}_{1\mu}^{L}\mathbf{A}_{2\nu}^{L}\partial_{\rho
}\mathbf{A}_{3\sigma}^{L}\rangle$, $\varepsilon^{\mu\nu\rho\sigma}%
\langle\mathbf{A}_{1\mu}^{L}\mathbf{A}_{2\nu}^{L}\mathbf{A}_{3\rho}%
^{L}\mathbf{A}_{4\sigma}^{L}\rangle$, and all their possible permutations.
Looking back at Eq.~(\ref{EffAg}), these counterterms would then contribute to
the currents. Importantly, being different (or treated as such), only one
field at a time would undergo a gauge transformation, so there would be as
many different currents as there are different fields. It is clear then that
these counterterms represent new local interactions between three or four
currents, that should be added to the loops. In practice, we can directly
construct these local interactions, there is no need to write down the
counterterms. With three currents, the local interaction $\mathcal{T}%
(J_{L1}^{a,\alpha}J_{L2}^{b,\beta}J_{L3}^{c,\gamma})$ must be proportional to
the $SU(N)$ tensor $d^{abc}$, and there are two possible Lorentz structures
$\varepsilon^{\alpha\beta\gamma\mu}q_{1\mu}$ and $\varepsilon^{\alpha
\beta\gamma\mu}q_{2\mu}$ since momentum conservation sets $q_{3}=-q_{1}-q_{2}%
$, so
\begin{equation}
\mathcal{T}(J_{L1}^{a,\alpha}J_{L2}^{b,\beta}J_{L3}^{c,\gamma})\sim
d^{abc}(X_{1}\varepsilon^{\alpha\beta\gamma\mu}q_{1\mu}+X_{2}\varepsilon
^{\alpha\beta\gamma\mu}q_{2\mu})\ .
\label{GenXi1}
\end{equation}
With four currents, the local interaction $\mathcal{T}(J_{L1}^{a,\alpha}%
J_{L2}^{b,\beta}J_{L3}^{c,\gamma}J_{L4}^{d,\delta})$ has the unique Lorentz
structure $\varepsilon^{\alpha\beta\gamma\delta}$ but three linearly
independent $SU(N)$ tensors of the $d\times f$ type:%
\begin{equation}
\mathcal{T}(J_{L1}^{a,\alpha}J_{L2}^{b,\beta}J_{L3}^{c,\gamma}J_{L4}%
^{d,\delta})\sim(X_{3}d^{abe}f^{cde}+X_{4}d^{ace}f^{bde}+X_{5}d^{ade}%
f^{bce})\varepsilon^{\alpha\beta\gamma\delta}\ .
\label{GenXi2}
\end{equation}
Finally, there can be no five-current local interaction at this order since
there is no way to construct an antisymmetric Lorentz structure with five
indices. Obviously, this matches precisely the free parameters in the chiral
anomaly, Eqs.~(\ref{AmpTri}) and~(\ref{AmpBox}), and even coincide with the
full result in the massive case, Eq.~(\ref{MassiveAno}). The overall
normalization is irrelevant since the size of the anomaly is ultimately set by
the IR obstructions, either in their singular forms or when expressed in terms
of pseudoscalar loops. This shows that counterterms are
just another route to keep track of the UV ambiguities, alongside the
$\gamma_{5}$ positions using dimensional regularization, the subtraction
points using dispersion relations, or the momentum routing when computing
surface terms.

Looking back, it is quite remarkable that the UV ambiguity inherent to 
the chiral anomaly matches Eqs.~(\ref{GenXi1}) and~(\ref{GenXi2}). In this 
sense, it can be said to be maximal since the whole five-dimensional
parameter-space spanned by $X_1$ to $X_5$ is attainable. Of course, this is
expected. Had we found that only a subspace of this space is reached, we
would then have had to identify some physical principle to explain that.
But, since no value of the $X_i$ would make the three, four, or five-point
amplitudes unphysical, this principle could then always be relaxed. Enforcing
it is thus a choice, and it must be interpreted as defining a specific form
of the chiral anomaly. So, fundamentally, the chiral anomaly must be as free
as it could be.

\subsection{Wess-Zumino-Witten action}\label{SecWZW}

The Bardeen anomaly describes anomalous interactions among gauge bosons. A
natural question to ask then is what happens if some of these symmetries are
spontaneously broken. Most interesting is the case in which the spontaneously
broken symmetry is actually global, because the Goldstone theorem couples its
currents directly to some Goldstone bosons. The anomaly must then translate
into new interactions among these Goldstone bosons, or between them and the
gauge fields of the remaining unbroken symmetries, and it is our purpose here
to derive them.

The main strategy can be illustrated with a simple example. Consider the usual
abelian Higgs model, with Lagrangian%
\begin{equation}
\mathcal{L}=-\frac{1}{4}F_{\mu\nu}F^{\mu\nu}+D_{\mu}\phi^{\dagger}D^{\mu}%
\phi-V(\phi^{\dagger}\phi)\;.
\end{equation}
Assuming $V(\phi^{\dagger}\phi)$ forces $\phi$ to acquire a vacuum expectation
value $v$, and adopting the polar representation
\begin{equation}
\phi=\frac{1}{\sqrt{2}}(v+\eta)e^{i\xi/v}\;, \label{ExpPara}%
\end{equation}
where $\eta$ and $\xi$ are real scalar fields, we get%
\begin{equation}
\mathcal{L}=-\frac{1}{4}F_{\mu\nu}F^{\mu\nu}+\frac{v^{2}}{2}\left(  A_{\mu
}-\frac{\partial^{\mu}\xi}{v}\right)  \left(  A^{\mu}-\frac{\partial^{\mu}\xi
}{v}\right)  \left(  1+\frac{\eta}{v}\right)  ^{2}+\frac{1}{2}\partial_{\mu
}\eta\partial^{\mu}\eta-V((v+\eta)^{2}/2)\ . \label{Stueckel}%
\end{equation}
Expanding the potential, $\eta$ must now have a mass term of the correct sign.
If this mass is sent to infinity, $\eta$ is removed from the dynamical degrees
of freedom and we get the so-called \textbf{Stueckelberg Lagrangian}~\cite{Ruegg:2003ps}.
Notice that only derivatives of $\xi$ can occur because $\mathcal{L}$ must be
shift-symmetric. Indeed, the vacua of $V(\phi^{\dagger}\phi)$ form a circle of
radius $\left\vert \phi^{\dagger}\phi\right\vert =v^{2}/2$, and any constant
shift $\xi\rightarrow\xi+\lambda$ corresponds to a different but equivalent
choice of vacuum along that circle. Non-constant shifts, on the other hand,
correspond to gauge transformations. For instance, invariance under
$\phi\rightarrow\exp(i\Omega)\phi$ with $\Omega$ a function of space-time is
guaranteed in the broken phase by%
\begin{equation}
A_{\mu}\rightarrow A_{\mu}+\partial_{\mu}\Omega\ ,\ \ \xi\rightarrow
\xi+v\Omega\ .
\end{equation}
The unitary gauge sets $\Omega=-\xi/v$, in which case the Goldstone mode is
fully absorbed into the gauge field as its longitudinal component and the
Lagrangian collapses to that for the Proca theory.

If we could do the same for a non-abelian symmetry, performing such a $\xi
$-dependent gauge transformation would naturally bring the Goldstone modes
into our previous derivation of the consistent anomaly. The main difficulty
though is that the consistent anomaly is already obtained from the gauge
variation of the action without Goldstone bosons, so it is too late to include
them. Historically, this problem was solved in two ways. Wess and Zumino in
Ref.~\cite{Wess:1971yu} were able to express the complete one-loop effective
action $\Gamma\lbrack A,\xi]$ in terms of $\Gamma\lbrack A]$, ensuring the
consistency condition holds. Its gauge variation then reproduces the
consistent anomaly and accounts for the Goldstone modes. The second approach
is due to Witten~\cite{Witten:1983tw}. Instead of working with the one-loop
effective action, which is a complicated non-local functional, the same
consistent anomaly can be obtained directly from the gauge variation of the
local Chern-Simons action in five dimensions. In that case, we can directly do
the $\xi$-dependent gauge transformation to get the consistent anomaly in the
presence of the Goldstone modes. This is the approach we shall follow.

For definiteness, let us specialize to the case of QCD at energies below the
hadronic scale (for an introductory review, see e.g.
Ref.~\cite{Leutwyler:2012}). At such low energies, the heavy $c,b,t$ quarks
are no longer dynamical while the $q^{1,2,3}=u,d,s$ quarks can be considered
massless in a first approximation. Because QCD does not distinguish flavors,
its Lagrangian limited to these $N_{f}=3$ active but massless flavors has the
large \textbf{chiral symmetry} $SU(N_{f})_{L}\otimes SU(N_{f})_{R}$. It is
however not realized in the usual sense since the spectrum of mesons and
baryons does not exhibit the $L\leftrightarrow R$ parity symmetry. Rather, it
is dynamically broken down to its $SU(N_{f})_{V=L+R}$ subgroup by the
condensate $\bar{q}_{L}^{i}q_{R}^{j}$ developing a vacuum expectation value.
There is then a set of Goldstone bosons associated to the axial currents
$A=R-L$, transforming in the adjoint representation of $SU(N_{f})_{V}$.
Adopting the same exponential parametrization as in Eq.~(\ref{ExpPara}), and
for $N_{f}=3$, those are identified with the angular components of the
condensate, $\langle0|\bar{q}_{L}^{i}q_{R}^{j}|0\rangle\sim\boldsymbol{U}%
^{ij}$ with%
\begin{equation}
\boldsymbol{U}=\exp\left(  i\frac{\sqrt{2}\boldsymbol{\xi}}{F}\right)
\ ,\ \boldsymbol{\xi}=\frac{1}{\sqrt{2}}\lambda^{a}\xi^{a}=\left(
\begin{array}
[c]{ccc}%
\frac{1}{\sqrt{2}}\pi^{0}+\frac{1}{\sqrt{6}}\eta_{8} & \pi^{+} & K^{+}\\
\pi^{-} & -\frac{1}{\sqrt{2}}\pi^{0}+\frac{1}{\sqrt{6}}\eta_{8} & K^{0}\\
K^{-} & \bar{K}^{0} & -\frac{2}{\sqrt{6}}\eta_{8}%
\end{array}
\right)  \ , \label{DefU}%
\end{equation}
where $F$ is a constant related to the spontaneous symmetry breaking scale,
and $\lambda^{a}$ are the Gell-Mann matrices. Under a generic chiral
transformation $(g_{L},g_{R})\in SU(3)_{L}\otimes SU(3)_{R}$, the
$\boldsymbol{U}$ matrix transforms as $\bar{q}_{L}^{i}q_{R}^{j}$:
\begin{equation}
\boldsymbol{U}\rightarrow g_{L}\boldsymbol{U}g_{R}^{\dagger}=g_{L}\left(
1+\frac{i\sqrt{2}\boldsymbol{\xi}}{F}+\frac{i\sqrt{2}\boldsymbol{\xi}}{F}%
\frac{i\sqrt{2}\boldsymbol{\xi}}{F}+...\right)  g_{R}^{\dagger}\;.
\label{NLU1}%
\end{equation}
This shows that the vector transformations $g_{V}\equiv g_{L}=g_{R}$ are
linearly represented on the fields%
\begin{equation}
\frac{i\sqrt{2}\boldsymbol{\xi}}{F}\rightarrow g_{V}\frac{i\sqrt
{2}\boldsymbol{\xi}}{F}g_{V}^{\dagger}\;, \label{NLU2}%
\end{equation}
as expected for the adjoint representation of $SU(3)_{V}$. On the other hand,
under the axial transformations $g_{A}\equiv g_{L}=g_{R}^{\dagger}$, the
Goldstone fields transform in a very complicated non-linear manner. In
particular, there must exist a Goldstone-dependent axial transformation that
renders $\boldsymbol{U}$ trivial, and indeed with%
\begin{equation}
g_{A}(\boldsymbol{\xi})\equiv g_{L}(\boldsymbol{\xi})=g_{R}^{\dagger
}(\boldsymbol{\xi})=\exp\left(  -i\frac{\boldsymbol{\xi}}{\sqrt{2}F}\right)
\ ,
\end{equation}
we get $g_{L}(\boldsymbol{\xi})\boldsymbol{U}g_{R}^{\dagger}(\boldsymbol{\xi
})=\mathbf{1}$, in complete analogy with the usual unitary gauge. Conversely,
the meson matrix can be understood as the pure gauge configuration
$\boldsymbol{U}=g_{L}(\boldsymbol{\xi})g_{R}^{\dagger}(\boldsymbol{\xi
})=(g_{A}(\boldsymbol{\xi}))^{2}$, which is reminiscent of the \textbf{coset
construction}~\cite{Coleman:1969sm, Callan:1969sn}.

\subsubsection{Topological couplings}

Let us first consider the situation in the absence of any external current. At
leading order in the number of derivatives~\cite{Weinberg:1978kz}, and because
Goldstone bosons can only couple derivatively, their dynamics has to be
encoded into the effective Lagrangian%
\begin{equation}
\mathcal{L}_{eff}=\dfrac{F^{2}}{4}\langle\partial^{\mu}\boldsymbol{U}%
\partial_{\mu}\boldsymbol{U}^{\dagger}\rangle\;,
\end{equation}
that is invariant under $SU(3)_{L}\otimes SU(3)_{R}$. The $F^{2}/4$ is just
there for dimensional reasons, and to ensure canonically-normalized kinetic
terms. This is however insufficient to fully specify the theory since the
$\boldsymbol{U}=g_{L}(\boldsymbol{\xi})g_{R}^{\dagger}(\boldsymbol{\xi})$
configuration may have a non-trivial winding number. Setting $\boldsymbol{\Omega
}=\boldsymbol{U}$ and $n=3$ in Eq.~(\ref{CSInteg2}), this can be accounted for
by adding the \textbf{Wess-Zumino-Witten action} (WZW in
short)~\cite{Wess:1971yu,Witten:1983tw}:%
\begin{equation}
\mathcal{S}_{eff}^{WZW}=(2\pi\nu)\frac{-i}{480\pi^{3}}\int_{D^{5}}%
d^{5}x\ \varepsilon^{\nu\rho\sigma\alpha\beta}\langle\boldsymbol{U}^{\dagger
}\partial_{\nu}\boldsymbol{UU}^{\dagger}\partial_{\rho}\boldsymbol{UU}%
^{\dagger}\partial_{\sigma}\boldsymbol{UU}^{\dagger}\partial_{\alpha
}\boldsymbol{UU}^{\dagger}\partial_{\beta}\boldsymbol{U}\rangle\ ,
\label{WZW1}%
\end{equation}
where $2\pi\nu$ with $\nu$ an integer is introduced to make $\exp
(i\mathcal{S}_{eff}^{WZW})$ invariant, given that the five-dimensional
integral has to give back an integer. Actually, that integral is necessarily
even since $\boldsymbol{U}=(g_{A}(\boldsymbol{\xi}))^{2}$ with $g_{A}%
(\boldsymbol{\xi})$ a $SU(3)$ transformation, see Fig.~\ref{FigSU2}. This is
rather irrelevant since the overall integral coefficient $\nu$ remains
undetermined. We will see below that matching to the real mesonic world sets
$\nu=N_{C}$ with $N_{C}$ the number of QCD colors. Notice that $\mathcal{S}%
_{eff}^{WZW}$ can be simplified somewhat using $\partial_{\mu}\boldsymbol{UU}%
^{\dagger}=-\boldsymbol{U}\partial_{\mu}\boldsymbol{U}^{\dagger}$ since
$\boldsymbol{UU}^{\dagger}=1$, after which the contribution of $\mathcal{S}%
_{eff}^{WZW}$ to the equation of motion is easily found to be a
five-dimensional total derivative:%
\begin{equation}
\partial_{\mu}\frac{\partial\mathcal{L}_{eff}^{WZW}}{\partial\partial_{\mu
}\boldsymbol{U}^{\dagger}}=3\varepsilon^{\mu\rho\sigma\alpha\beta}%
\partial_{\mu}(\boldsymbol{U}\partial_{\rho}\boldsymbol{U}^{\dagger}%
\partial_{\sigma}\boldsymbol{U}\partial_{\alpha}\boldsymbol{U}^{\dagger
}\partial_{\beta}\boldsymbol{U})\ . \label{EoM}%
\end{equation}
As such, it can only lead to a non-trivial dynamic on the boundary of $D^{5}$,
which we take as our four-dimensional space-time, $\partial D^{5}=V^{4}$.

When expanded, $\mathcal{L}_{eff}$ produces only couplings involving even
numbers of mesons, while $\mathcal{S}_{eff}^{WZW}$ gives those with odd
numbers of mesons. Let us illustrate that omitting the $K^{0}$, $\bar{K}^{0}$,
and $\eta_{8}$ fields. Then, all the couplings obtained by expanding
$\mathcal{L}_{eff}^{WZW}$ have the structure%
\begin{equation}
\Gamma_{\nu\rho\sigma\alpha\beta}^{nmp}=\frac{c_{nmp}}{4\pi^{2}F^{2n+2m+2p-1}%
}\partial_{\nu}(\pi^{0})^{2n-1}\partial_{\rho}(\pi^{+})^{m}\partial_{\sigma
}(\pi^{-})^{m}\partial_{\alpha}(K^{+})^{p}\partial_{\beta}(K^{-})^{p}\ ,
\end{equation}
where $n,m,p=1,2,...$ and $c_{nmp}$ are combinations of Clebsch-Gordan
coefficients. Thanks to the antisymmetric contraction, all these couplings can
be expressed as total derivatives, allowing to extract a proper
four-dimensional vertex involving mesons as
\begin{align}
\mathcal{S}_{eff}^{WZW}  &  =\frac{c_{nmp}}{4\pi^{2}F^{2n+2m+2p-1}}\int%
_{D^{5}}d^{5}x\ \varepsilon^{\nu\rho\sigma\alpha\beta}\Gamma_{\nu\rho
\sigma\alpha\beta}^{nmp}\nonumber\\
&  =\frac{c_{nmp}}{4\pi^{2}F^{2n+2m+2p-1}}\int_{\partial D^{5}}d^{4}%
x\ \varepsilon^{\rho\sigma\alpha\beta}(\pi^{0})^{2n-1}\partial_{\rho}(\pi
^{+})^{m}\partial_{\sigma}(\pi^{-})^{m}\partial_{\alpha}(K^{+})^{p}%
\partial_{\beta}(K^{-})^{p}\ ,
\end{align}
from which Feynman rules can be extracted, and cross sections computed like
e.g. for $K^{+}K^{-}\rightarrow\pi^{+}\pi^{-}\pi^{0}$. Note that it is much
less simple to write higher order terms including all eight meson fields as
total derivatives, though we know from Eq.~(\ref{EoM}) that is always possible.

\subsubsection{External field couplings}

A more powerful matching between QCD and its effective meson theory is
obtained by gauging the $SU(3)_{L}\otimes SU(3)_{R}$
symmetry~\cite{Gasser:1983yg,Gasser:1984gg}. For that, we introduce two octets
of gauge fields $\mathbf{A}^{L,R}$, out of which vector and axial gauge fields
are defined as usual, see Sec.~\ref{SecVAdefs}. This can be done at the level of QCD, through the
introduction of the couplings $\bar{q}^{i}(\mathbf{V}_{\mu}\gamma^{\mu
}+\mathbf{A}_{\mu}\gamma^{\mu}\gamma^{5})q^{j}$, as well as at the level of
the effective theory as%
\begin{equation}
\mathcal{L}_{eff}=\dfrac{F^{2}}{4}\langle D^{\mu}\boldsymbol{U}D_{\mu
}\boldsymbol{U}^{\dagger}\rangle\;,\ \ D_{\mu}\boldsymbol{U}=\partial_{\mu
}\boldsymbol{U}-i\left[  \mathbf{V}_{\mu},\boldsymbol{U}\right]  -i\left\{
\mathbf{A}_{\mu},\boldsymbol{U}\right\}  \ . \label{LeffChPT}%
\end{equation}
These flavored gauge interactions do not have physical realizations in the
real world, except for electromagnetic and weak interactions. In particular,
the former corresponds to $\mathbf{V}_{\mu}=-eA_{\mu}\boldsymbol{Q}$ with the
quark electric charges $\boldsymbol{Q}=\operatorname*{diag}(2/3,-1/3,-1/3)$.
To unclutter the notation, $e$ will mostly be set to one in the following.

At this stage, the matching is incomplete though. In QCD, the fermion loops
generate anomalous interactions among the vector and axial gauge fields. Since
we need to conserve the vector current $\mathbf{V}_{\mu}=-eA_{\mu
}\boldsymbol{Q}$, these anomalous interactions have to be in their Bardeen
form\footnote{For the weak interactions, the situation is more complicated
because it is also axial, and the Bardeen anomaly spreads the anomaly
symmetrically on all axial currents. To deal with that, extra counterterms are
required to break the axial Bose symmetry, but this will not be done
here~\cite{Harvey:2007ca}.}, Eq.~(\ref{AVC2}). Clearly, those anomalies are
not altered by the hadronization, so they must be present in the effective
theory also~\cite{tHooft:1979rat}. Yet, with only bosonic degrees of freedom,
$\mathcal{L}_{eff}$ cannot generate anomalous interactions at any order. So,
one needs to add new terms to $\mathcal{L}_{eff}$ to match the anomaly content
of QCD. Clearly, we first need to include $\mathcal{S}_{eff}^{WZW}$ of
Eq.~(\ref{WZW1}), since the presence of external field is irrelevant for that
term. Yet, that cannot be the end of the story because $\mathcal{S}%
_{eff}^{WZW}$ involves normal derivatives, and thus breaks all the
$SU(3)_{L}\otimes SU(3)_{R}$ gauge invariances, even that for QED,
$\boldsymbol{U}\rightarrow g_{V}\boldsymbol{U}g_{V}^{\dagger}$ with
$g_{V}=\exp(-i\lambda\boldsymbol{Q})$. Worse, we cannot simply promote normal
derivatives to covariant ones since $\boldsymbol{U}$ in Eq.~(\ref{WZW1}) lives
in five dimensions, and so do the partial derivatives, but the gauge fields do not.

In Ref.~\cite{Witten:1983tw}, gauge invariance is restored by hand, by adding 
appropriate couplings to the external fields%
\begin{equation}
\mathcal{S}_{eff}^{Ano}=\mathcal{S}_{eff}^{WZW}+\frac{\nu}{48\pi^{2}}%
\int_{\partial D^{5}}d^{4}x\varepsilon^{\mu\nu\rho\sigma}\left\langle
\mathbf{Z}_{\mu\nu\rho\sigma}\right\rangle \ , \label{WZWExt}%
\end{equation}
The explicit form of $\left\langle \mathbf{Z}_{\mu\nu\rho\sigma}\right\rangle $ is 
found via a Noether procedure. Specifically, performing $\boldsymbol{U}%
\rightarrow\boldsymbol{U}-i\lambda\lbrack\boldsymbol{Q},\boldsymbol{U}]$ in
$\mathcal{S}_{eff}^{WZW}$ with $\lambda=\lambda(x)$, the variation to
$\mathcal{O}(\lambda)$ is a total derivative%
\begin{align}
\delta_{\lambda}\left\langle \varepsilon^{\alpha\mu\nu\rho\sigma
}\boldsymbol{U}^{\dagger}\partial_{\nu}\boldsymbol{UU}^{\dagger}\partial
_{\rho}\boldsymbol{UU}^{\dagger}\partial_{\sigma}\boldsymbol{UU}^{\dagger
}\partial_{\alpha}\boldsymbol{UU}^{\dagger}\partial_{\beta}\boldsymbol{U}%
\right\rangle  &  =\nonumber\\
-5i\varepsilon^{\alpha\mu\nu\rho\sigma}\partial_{\mu}\lambda\partial_{\alpha}
&  \left\langle \partial_{\nu}\boldsymbol{UU}^{\dagger}\partial_{\rho
}\boldsymbol{UU}^{\dagger}\partial_{\sigma}\boldsymbol{U}\{\boldsymbol{U}%
^{\dagger},\boldsymbol{Q}\}\right\rangle \ .
\end{align}
Since $A_{\mu}\rightarrow A_{\mu}+\partial_{\mu}\lambda$, it can be
compensated by including in $\mathbf{Z}_{\mu\nu\rho\sigma}$ the term%
\begin{equation}
\mathbf{Z}_{\mu\nu\rho\sigma}^{1\gamma}=-A_{\mu}\left\langle \partial_{\nu
}\boldsymbol{UU}^{\dagger}\partial_{\rho}\boldsymbol{UU}^{\dagger}%
\partial_{\sigma}\boldsymbol{U}\{\boldsymbol{U}^{\dagger},\boldsymbol{Q}%
\}\right\rangle \ . \label{Z1gamma}%
\end{equation}
This same procedure has to be repeated because the gauge variation of the
$\mathbf{Z}_{\mu\nu\rho\sigma}^{1\gamma}$ includes not only that coming from
$A_{\mu}\rightarrow A_{\mu}+\partial_{\mu}\lambda$, but also that from the
variation of the trace:%
\begin{equation}
\delta_{\lambda}\left\langle \partial_{\nu}\boldsymbol{UU}^{\dagger}%
\partial_{\rho}\boldsymbol{UU}^{\dagger}\partial_{\sigma}\boldsymbol{U}%
\{\boldsymbol{U}^{\dagger},\boldsymbol{Q}\}\right\rangle =-i\varepsilon
^{\mu\nu\rho\sigma}\partial_{\nu}\lambda\partial_{\rho}\left\langle
2\boldsymbol{QQ}\{\boldsymbol{U}^{\dagger},\partial_{\sigma}\boldsymbol{U}%
\}+\boldsymbol{QU}^{\dagger}\boldsymbol{Q}\partial_{\sigma}\boldsymbol{U}%
-\boldsymbol{QUQ}\partial_{\sigma}\boldsymbol{U}^{\dagger}\right\rangle \ .
\end{equation}
Notice that there is an ambiguity when pulling out the derivative, because
\begin{equation}
\varepsilon^{\mu\nu\rho\sigma}\partial_{\rho}\left\langle \boldsymbol{QU}%
^{\dagger}\boldsymbol{Q}\partial_{\sigma}\boldsymbol{U}+\boldsymbol{Q}%
\partial_{\sigma}\boldsymbol{U}^{\dagger}\boldsymbol{QU}\right\rangle =0\ .
\end{equation}
The above choice maintains the overall antisymmetry of $\mathbf{Z}_{\mu\nu
\rho\sigma}$ under $\boldsymbol{U}\leftrightarrow\boldsymbol{U}^{\dagger}$
(which is related to charge conjugation invariance). After integrating by
part, this variation is cancelled by further adding%
\begin{equation}
\mathbf{Z}_{\mu\nu\rho\sigma}^{2\gamma}=i\varepsilon^{\mu\nu\rho\sigma}%
F_{\mu\nu}A_{\rho}\left\langle \boldsymbol{QQ}\{\boldsymbol{U}^{\dagger
},\partial_{\sigma}\boldsymbol{U}\}+\frac{1}{2}(\boldsymbol{QU}^{\dagger
}\boldsymbol{Q}\partial_{\sigma}\boldsymbol{U}-\boldsymbol{QUQ}\partial
_{\sigma}\boldsymbol{U}^{\dagger})\right\rangle \ , \label{Z2gamma}%
\end{equation}
where we have integrated by part and discarded the total derivative since
$\partial\partial D^{5}=\varnothing$. This procedure stops here as the trace
is now invariant, and anyway, there would be no way to construct a
three-photon vertex since $\varepsilon^{\mu\nu\rho\sigma}F_{\mu\nu}A_{\rho
}A_{\sigma}=0$.

At this stage, it is interesting to reflect on how the anomaly is reproduced
by the bosonic theory. The $\mathbf{Z}_{\mu\nu\rho\sigma}^{1\gamma}$ and
$\mathbf{Z}_{\mu\nu\rho\sigma}^{2\gamma}$ terms add local couplings involving
an odd number of mesons with one or two photons. As such, they are not by
themselves anomalous. Rather, one could see their presence as necessary to
generate interactions between the axial currents and photons that do reproduce
the anomalous Ward identities of QCD. The key for that are the IR poles, now
arising from the meson propagators. In that sense, it is the masslessness of
the Goldstone bosons that ensures the axial to photon amplitudes have the same
IR singularities as in the full QCD with massless quarks, see
Eqs.~(\ref{AmpTri}) and~(\ref{AmpBox}).

When quark masses are introduced, the mesons are no longer massless and become
so-called pseudo-Goldstone bosons. There are no IR poles anymore, exactly like
for the QCD amplitudes. Actually, a peculiarity of the Bardeen anomaly is that
the Sutherland-Veltman theorem holds. This means that at leading order, the
axial to photon amplitudes disappears in QCD (it is subleading in the $1/m$
expansion), and all that is left are the pseudoscalar loops of Eq.~(\ref{PnV}%
). It is now to reproduce these non-anomalous loops in the meson theory that
one has to add the $\mathbf{Z}_{\mu\nu\rho\sigma}^{1\gamma}$ and
$\mathbf{Z}_{\mu\nu\rho\sigma}^{2\gamma}$ couplings. The mechanisms at play
are thus quite different with or without massless quarks, and this is not that
surprising since, as explained before, the massless limit is all but smooth.
Yet, this does not change anything in practice, but simply translates in a
modern language why starting with a $\pi^{0}\bar{\psi}\gamma^{5}\psi$ coupling
and massive quarks gives the same result as starting with the $\partial_{\mu
}\pi^{0}\bar{\psi}\gamma^{\mu}\gamma^{5}\psi$ coupling and massless quarks.

\subsubsection{Goldstone-Wilczek current}

In the previous analysis, no constraint was put on $\boldsymbol{Q}$, so let us
now consider the case $\boldsymbol{Q}=\mathbf{1}/3$. This corresponds to
gauging the $U(1)$ vector symmetry corresponding to baryon number, with
$u,d,s$ all assigned a charge $1/3$ (we set the coupling constant to one).
Denoting the corresponding gauge field as $B_{\mu}$, the extra terms emerging
from the Noether procedure simplify into~\cite{Witten:1983tw,Manes:1984gk}%
\begin{subequations}
\begin{align}
\mathbf{Z}_{\mu\nu\rho\sigma}^{1B}  &  =-\frac{\nu/3}{24\pi^{2}}\int_{\partial
D^{5}}d^{4}x\varepsilon^{\mu\nu\rho\sigma}B_{\mu}\left\langle \partial_{\nu
}\boldsymbol{UU}^{\dagger}\partial_{\rho}\boldsymbol{UU}^{\dagger}%
\partial_{\sigma}\boldsymbol{UU}^{\dagger}\right\rangle \ ,\\
\mathbf{Z}_{\mu\nu\rho\sigma}^{2B}  &  =-\frac{\nu/3}{16\pi^{2}}\int_{\partial
D^{5}}d^{4}x\varepsilon^{\mu\nu\rho\sigma}B_{\mu\nu}B_{\rho}\left\langle
\partial_{\sigma}\boldsymbol{UU}^{\dagger}\right\rangle \ .
\end{align}
In the first expression, one identifies the so-called
\textbf{Goldstone-Wilczek current}~\cite{Goldstone:1981kk}%
\end{subequations}
\begin{equation}
J_{GW}^{\mu}=-\frac{\nu/3}{24\pi^{2}}\varepsilon^{\mu\nu\rho\sigma
}\left\langle \partial_{\nu}\boldsymbol{UU}^{\dagger}\partial_{\rho
}\boldsymbol{UU}^{\dagger}\partial_{\sigma}\boldsymbol{UU}^{\dagger
}\right\rangle \ .
\end{equation}
Being coupled to $B_{\mu}$, it is the mesonic representation of the vectorial
baryon number current $1/3\times(\bar{u}\gamma^{\mu}u+\bar{d}\gamma^{\mu
}d+\bar{s}\gamma^{\mu}s)$ which emerges if the same gauging is done in the
three-flavor QCD. Actually, this provides a nice interpretation of baryon
number, first proposed by Skyrme~\cite{Skyrme:1962vh}. The charge associated
to $J_{GW}^{\mu}$, that is, baryon number $\mathcal{B}$, is thus%
\begin{equation}
\mathcal{B}=\int d^{3}xJ_{GW}^{0}=-\frac{\nu/3}{24\pi^{2}}\int d^{3}%
x\varepsilon^{ijk}\left\langle \partial_{i}\boldsymbol{UU}^{\dagger}%
\partial_{j}\boldsymbol{UU}^{\dagger}\partial_{k}\boldsymbol{UU}^{\dagger
}\right\rangle \ ,
\end{equation}
in which we recognize the topological measure of the winding of $SU(2)$ gauge
configurations over $S^{3}$ that we computed in Sec.~\ref{SecChernSimons}, 
see Eq.~(\ref{CSInteg2}). At this stage, we can also see that $\nu=3$ properly
normalizes baryon number. Actually, denoting the number of QCD colors as $N_{C}$,
we should set $\nu=N_{C}$ since a baryon is then built from the antisymmetric
color contraction of $N_{C}$ quark fields, hence its generator is 
$\boldsymbol{Q}=\mathbf{1}/N_{C}$.

The current $J_{GW}^{\mu}$ is conserved in the absence of gauge interactions.
This is easily seen using $\partial_{\mu}\boldsymbol{UU}^{\dagger
}=-\boldsymbol{U}\partial_{\mu}\boldsymbol{U}^{\dagger}$ to write%
\begin{equation}
\varepsilon^{\mu\nu\rho\sigma}\left\langle \partial_{\mu}\boldsymbol{U}%
\partial_{\nu}\boldsymbol{U}^{\dagger}\partial_{\rho}\boldsymbol{U}%
\partial_{\sigma}\boldsymbol{U}^{\dagger}\right\rangle =-\varepsilon^{\mu
\nu\rho\sigma}\left\langle \partial_{\mu}\boldsymbol{U}\partial_{\nu
}\boldsymbol{U^{\dagger}}\partial_{\rho}\boldsymbol{U}\partial_{\sigma
}\boldsymbol{U}^{\dagger}\right\rangle \ .
\end{equation}
Now, since $J_{GW}^{\mu}$ is defined in four dimensions, there is no
obstruction to simply promote $\partial_{\mu}\rightarrow D_{\mu}$. The
surprise though is that this breaks current conservation:%
\begin{equation}
\partial_{\mu}J_{GW}^{\mu}=\frac{i\nu}{16\pi^{2}}\varepsilon^{\mu\nu\rho
\sigma}\left\langle \mathbf{F}_{\mu\nu}^{R}D_{\rho}\boldsymbol{U}^{\dagger
}D_{\sigma}\boldsymbol{U}\right\rangle -\frac{i\nu}{16\pi^{2}}\varepsilon
^{\mu\nu\rho\sigma}\left\langle \mathbf{F}_{\mu\nu}^{L}D_{\rho}\boldsymbol{U}%
D_{\sigma}\boldsymbol{U}^{\dagger}\right\rangle \ .
\end{equation}
To arrive at this expression, notice that $\partial_{\mu}$ becomes $D_{\mu}$
once acting inside the trace, and with $D_{\mu}\boldsymbol{U}=\partial_{\mu
}\boldsymbol{U}-i\mathbf{A}_{\mu}^{L}\boldsymbol{U}+i\boldsymbol{U}%
\mathbf{A}_{\mu}^{R}$, the field strengths appear from%
\begin{equation}
\lbrack D_{\mu},D_{\nu}]\boldsymbol{U}=-i\mathbf{F}_{L}^{\mu\nu}%
\boldsymbol{U}+i\boldsymbol{U}\mathbf{F}_{R}^{\mu\nu}\ ,\ \ [D_{\mu},D_{\nu
}]\boldsymbol{U}^{\dagger}=-i\mathbf{F}_{R}^{\mu\nu}\boldsymbol{U}^{\dagger
}+i\boldsymbol{U}^{\dagger}\mathbf{F}_{L}^{\mu\nu}\ .
\end{equation}
These field strengths could be expressed in terms of vector and axial gauge
interactions as in Sec.~\ref{SecVAdefs}, see Eq.~(\ref{BardeenCurv}).

The divergence $\partial_{\mu}J_{GW}^{\mu}$ is not very appealing and we can
do better~\cite{DHoker:1983ujb}. Imagine that instead of $J_{GW}^{\mu}$, we
consider a new current, also called Goldstone-Wilczek,%
\begin{equation}
\mathcal{J}_{GW}^{\mu}=J_{GW}^{\mu}+\frac{i\nu}{16\pi^{2}}\varepsilon^{\mu
\nu\rho\sigma}\left\langle \mathbf{F}_{\nu\rho}^{L}D_{\sigma}\boldsymbol{UU}%
^{\dagger}\right\rangle -\frac{i\nu}{16\pi^{2}}\varepsilon^{\mu\nu\rho\sigma
}\left\langle \mathbf{F}_{\nu\rho}^{R}D_{\sigma}\boldsymbol{U}^{\dagger
}\boldsymbol{U}\right\rangle \ . \label{ShiftGW}%
\end{equation}
Taking its divergence, with the field strengths satisfying the Bianchi
identity $\varepsilon^{\mu\nu\rho\sigma}\partial_{\rho}\mathbf{F}_{\mu\nu
}^{L,R}=0$, what remains becomes very elegant:%
\begin{equation}
\partial_{\mu}\mathcal{J}_{GW}^{\mu}=\frac{\nu}{16\pi^{2}}\varepsilon^{\mu
\nu\rho\sigma}\left\langle \mathbf{F}_{\mu\nu}^{L}\mathbf{F}_{\sigma\rho}%
^{L}\right\rangle -\frac{\nu}{16\pi^{2}}\varepsilon^{\mu\nu\rho\sigma
}\left\langle \mathbf{F}_{\rho\sigma}^{R}\mathbf{F}_{\mu\nu}^{R}\right\rangle
\ .
\end{equation}
Further, for purely vectorial interactions like electromagnetism,
$\mathbf{F}_{\mu\nu}^{L}=\mathbf{F}_{\mu\nu}^{R}$ and we do recover
$\partial_{\mu}\mathcal{J}_{GW}^{\mu}=0$. Quite naturally, it is only in the
presence of axial gauge interactions that the vectorial baryonic current
becomes anomalous, so it should actually be identified with $\mathcal{J}%
_{GW}^{\mu}$ and not with $J_{GW}^{\mu}$. We will see in the next section that
this precisely matches the situation in the SM provided $\nu=N_{C}$, with the
baryonic current coupled to electroweak gauge bosons because of the so-called
\textbf{covariant anomaly}. At that stage, it will also be clear why shifting
the current as in Eq.~(\ref{ShiftGW}) is acceptable, even unavoidable.

\subsubsection{Chern-Simons gauge variation}

Instead of the Noether procedure, it is also
possible~\cite{Chou:1983qy,Kawai:1984mx,Manes:1984gk} to get the effective
interactions directly from the general expression of the gauge variation of
the Chern-Simons action, see Eq.~(\ref{VarCS2}). The calculation is a bit
tricky though and proceeds in two steps.

The first thing to do is to identify the Chern-Simons action corresponding to
the Bardeen anomaly. Since it is consistent, it must be possible to construct
a five-dimensional action,%
\begin{equation}
S_{CS}^{5}[\mathbf{A},\mathbf{V}]=\int_{D^{5}}d\sigma_{\mu}\ G_{5}^{\mu
}[\mathbf{A},\mathbf{V}]\ ,
\end{equation}
such that its variation $\delta G_{5}^{\mu}[\boldsymbol{\Lambda}_{A}%
,\boldsymbol{\Lambda}_{V},\mathbf{A},\mathbf{V}]$ under infinitesimal gauge
transformations, Eq.~(\ref{GaugeAV}), gives back the Bardeen anomaly of
Eq.~(\ref{AVC2}). By trial and error, $G_{5}^{\mu}[\mathbf{A},\mathbf{V}]$ is
easily found to be~\cite{Manes:2018llx}
\begin{equation}
G_{5}^{\mu}[\mathbf{A},\mathbf{V}]=12\varepsilon^{\mu\nu\rho\sigma\alpha\beta
}\left\langle \mathbf{A}_{\nu}\mathbf{F}_{\rho\sigma}^{V}\mathbf{F}%
_{\alpha\beta}^{V}+\frac{1}{3}\mathbf{A}_{\nu}\mathbf{F}_{\rho\sigma}%
^{A}\mathbf{F}_{\alpha\beta}^{A}+\frac{8}{3}i\mathbf{A}_{\nu}\mathbf{A}_{\rho
}\mathbf{A}_{\sigma}\mathbf{F}_{\alpha\beta}^{V}-\frac{32}{15}\mathbf{A}_{\nu
}\mathbf{A}_{\rho}\mathbf{A}_{\sigma}\mathbf{A}_{\alpha}\mathbf{A}_{\beta
}\right\rangle \ . \label{BardeenCS}%
\end{equation}
Indeed, plugging in Eq.~(\ref{GaugeAV}), $\boldsymbol{\Lambda}_{V}$ immediately
drops out while the terms proportional to $\boldsymbol{\Lambda}_{A}$ only do so
after partial integration. For a non-trivial boundary, the surface term is
then indeed found proportional to the Bardeen anomaly of Eq.~(\ref{AVC2}).

Importantly, remark that $G_{5}^{\mu}[\mathbf{A},\mathbf{V}]$ is not simply
the difference between the chiral $G_{5}^{\mu}[\mathbf{A}^{R,L}]$ of
Eq.~(\ref{CSforms}). The mismatch however can be encoded into a total
derivative:%
\begin{equation}
G_{5}^{\mu}[\mathbf{A},\mathbf{V}]=G_{5}^{\mu}[\mathbf{A}^{L},\mathbf{A}%
^{R}]=G_{5}^{\mu}[\mathbf{A}^{R}]-G_{5}^{\mu}[\mathbf{A}^{L}]+\partial_{\nu
}H_{4}^{\mu\nu}[\mathbf{A}^{R},\mathbf{A}^{L}]\ ,
\end{equation}
where the difference is~\cite{Kawai:1984mx,Manes:1984gk}%
\begin{align}
H_{4}^{\mu\nu}[\mathbf{A}^{R},\mathbf{A}^{L}]  &  =8\varepsilon^{\mu\nu
\rho\sigma\alpha\beta}\left\langle \frac{1}{4}(\mathbf{A}_{\rho}^{L}%
\mathbf{A}_{\sigma}^{R}-\mathbf{A}_{\rho}^{R}\mathbf{A}_{\sigma}%
^{L})(\mathbf{F}_{\alpha\beta}^{L}+\mathbf{F}_{\alpha\beta}^{R})\right.
\nonumber\\
&  \ \ \ \ \ \ \ \ \ \ \ \ \ \ \ \left.  +\frac{i}{2}\mathbf{A}_{\rho}%
^{L}\mathbf{A}_{\sigma}^{L}\mathbf{A}_{\alpha}^{L}\mathbf{A}_{\beta}^{R}%
+\frac{i}{2}\mathbf{A}_{\rho}^{L}\mathbf{A}_{\sigma}^{R}\mathbf{A}_{\alpha
}^{R}\mathbf{A}_{\beta}^{R}+\frac{i}{4}\mathbf{A}_{\rho}^{R}\mathbf{A}%
_{\sigma}^{L}\mathbf{A}_{\alpha}^{R}\mathbf{A}_{\beta}^{L}\right\rangle \ ,
\end{align}
and we have used the fact that $H_{4}^{\mu\nu}$ itself is defined up to a
total derivative to make it antisymmetric under $L\leftrightarrow R$. The
meaning of this slightly complicated $H_{4}^{\mu\nu}$ becomes transparent once
written in terms of $\mathbf{A},\mathbf{V}$:%
\begin{equation}
H_{4}^{\mu\nu}[\mathbf{A},\mathbf{V}]=8\varepsilon^{\mu\nu\rho\sigma
\alpha\beta}\left\langle \{\mathbf{V}_{\mu},\mathbf{A}_{\nu}\}\mathbf{F}%
_{\rho\sigma}^{V}+3i\mathbf{V}_{\mu}\mathbf{A}_{\nu}\mathbf{A}_{\rho
}\mathbf{A}_{\sigma}+i\mathbf{V}_{\mu}\mathbf{V}_{\nu}\mathbf{V}_{\rho
}\mathbf{A}_{\sigma}\right\rangle \ ,
\end{equation}
which precisely matches the Bardeen counterterms, Eq.~(\ref{BCT}) for
$X_{1,2,3}=1$. Looking back, the equivalence of $H_{4}^{\mu\nu}$ with the
Bardeen counterterms should not be surprising. Remember that Chern-Simons
forms are defined by asking $\partial_{\mu}G_{2n-1}^{\mu}$ to give back a
degree $n$ antisymmetric polynomial in the field strength. But adding a total
derivative to $G_{2n-1}^{\mu}$ does not change that since $\varepsilon^{\mu
\nu\rho\sigma\alpha\beta}\partial_{\mu}\partial_{\nu}$ gives zero. There is
thus an ambiguity in the Chern-Simons form. Yet, it is not possible to exploit
it with only one gauge field because $\varepsilon^{\rho\sigma\alpha\beta
}\langle\mathbf{A}_{\rho}^{L}\mathbf{A}_{\sigma}^{L}\mathbf{A}_{\alpha}%
^{L}\mathbf{A}_{\beta}^{L}\rangle=0$ and $\varepsilon^{\rho\sigma\alpha\beta
}\langle\mathbf{A}_{\rho}^{L}\mathbf{A}_{\sigma}^{L}\mathbf{F}_{\alpha\beta
}^{L}\rangle\sim\varepsilon^{\rho\sigma\alpha\beta}\partial_{\alpha}%
\langle\mathbf{A}_{\rho}^{L}\mathbf{A}_{\sigma}^{L}\mathbf{A}_{\beta}%
^{L}\rangle$. With the two fields $\mathbf{A}$ and $\mathbf{V}$, this freedom
matches that encoded into the Bardeen counterterms, and so, more generally, it
is yet another way to keep track of the UV ambiguities in anomalies.

The second step is to perform a gauge variation of $G_{5}^{\mu}[\mathbf{A}%
,\mathbf{V}]$. Indeed, $G_{5}^{\mu}[\mathbf{A},\mathbf{V}]$ represents the
full Chern-Simons action both within QCD or in the meson world in the unitary
gauge, in which the Goldstone bosons have been absorbed into the axial gauge
fields. Thus, all we have to do is to express this unitary gauge $G_{5}^{\mu
}[\mathbf{A},\mathbf{V}]$ back into the gauge of Eq.~(\ref{LeffChPT}) using
the Goldstone-boson-dependent transformation $\boldsymbol{U}=g_{L}%
(\boldsymbol{\xi})g_{R}^{\dagger}(\boldsymbol{\xi})$. Indeed, in the
Lagrangian of Eq.~(\ref{LeffChPT}), the gauge boson-Goldstone boson mixing is
maximal, with
\begin{equation}
-i\frac{F^{2}}{2}\langle\mathbf{A}^{\mu}\{\partial_{\mu}\boldsymbol{U}%
,\boldsymbol{U}^{\dagger}\}\rangle=\frac{F}{\sqrt{2}}\langle\mathbf{A}^{\mu
}\partial_{\mu}\boldsymbol{\xi}\rangle+...\ , \label{GoldTheo}%
\end{equation}
in complete analogy with Eq.~(\ref{Stueckel}). By the way, this equation
translates the Goldstone theorem statement that matrix elements of the axial
currents with a single Goldstone field do not vanish, i.e. $\left\langle
0\left\vert A_{\mu}^{a}\right\vert \xi^{b}\left(  p\right)  \right\rangle \sim
iFC^{ab}p_{\mu}$ where $C^{ab}=C\delta^{ab}$ with $C$ a Glebsch-Gordan
coefficient. This is the so-called \textbf{partially conserved axial current}
hypothesis, or PCAC. The constant $F$ is thus identified on one hand with the
vacuum expectation value driving the chiral symmetry breaking, and on the
other, phenomenologically, with the meson decay constants, all equal at this
order $F=F_{\pi}=F_{K}=F_{\eta}\approx100$ MeV.

The plan is clear but technically cumbersome because $\boldsymbol{U}%
=g_{L}(\boldsymbol{\xi})g_{R}^{\dagger}(\boldsymbol{\xi})$ is a
non-infinitesimal $SU(3)_{L}\otimes SU(3)_{R}$ transformation, under which
$\mathbf{A}$ and $\mathbf{V}$ do not transform in a simple way. So, it is
necessary first to express $G_{5}^{\mu}[\mathbf{A},\mathbf{V}]$ back in terms
of $\mathbf{A}_{\mu}^{R,L}=\mathbf{V}_{\mu}\pm\mathbf{A}_{\mu}$, and then
perform the $(\boldsymbol{U},1)$ or the $(1,\boldsymbol{U}^{\dagger})\in
SU(3)_{L}\otimes SU(3)_{R}$ gauge transformation. Both are equivalent to
$(g_{L}(\boldsymbol{\xi}),g_{R}(\boldsymbol{\xi}))$ since $SU(3)_{V}$ is
conserved. Schematically, under $\mathbf{A}_{\mu}^{L}\rightarrow
\boldsymbol{U}^{\dagger}\mathbf{A}_{\mu}^{L}\boldsymbol{U}+i\boldsymbol{U}%
^{\dagger}\partial_{\mu}\boldsymbol{U}$ and $\mathbf{A}_{\mu}^{R}%
\rightarrow\mathbf{A}_{\mu}^{R}$, the Chern-Simons action varies according to
Eq.~(\ref{VarCS}):%
\begin{align}
\delta G_{5}^{\mu}[\mathbf{A}^{L},\mathbf{A}^{R}]  &  =-\partial_{\nu}%
G_{4}^{\mu\nu}[\boldsymbol{U},\mathbf{A}_{\mu}^{L}]+\partial_{\nu}H^{\mu\nu
}[\mathbf{A}^{R},\boldsymbol{U}^{\dagger}\mathbf{A}_{\mu}^{L}\boldsymbol{U}%
+i\boldsymbol{U}^{\dagger}\partial_{\mu}\boldsymbol{U}]\nonumber\\
&  \ \ \ -ic_{3}\varepsilon^{\nu\rho\sigma\alpha\beta}\langle\boldsymbol{U}%
^{\dagger}\partial_{\nu}\boldsymbol{UU}^{\dagger}\partial_{\rho}%
\boldsymbol{UU}^{\dagger}\partial_{\sigma}\boldsymbol{UU}^{\dagger}%
\partial_{\alpha}\boldsymbol{UU}^{\dagger}\partial_{\beta}\boldsymbol{U}%
\rangle\ .
\end{align}
The topological term reproduces Eq.~(\ref{WZW1}), while the total derivative
comes from the gauge variation of the left Chern-Simons form, $\partial_{\nu
}G_{4}^{\mu\nu}[\boldsymbol{U},\mathbf{A}_{\mu}^{L}]=G_{5}^{\mu}%
[\boldsymbol{U}^{\dagger}\mathbf{A}_{\mu}^{L}\boldsymbol{U}+i\boldsymbol{U}%
^{\dagger}\partial_{\mu}\boldsymbol{U}]-G_{5}^{\mu}[\mathbf{A}^{L}]$. Though
these expressions appear rather asymmetric in $\mathbf{A}^{L}\leftrightarrow
\mathbf{A}^{R}$, one can add a total derivative to get a more symmetric
expression. The non-topological terms integrate to give the effective
anomalous interactions in four dimensions we are after. In practice, the
algebra is so cumbersome that we did not try to reproduce the rather lengthy
full result, which one can find e.g. in
Refs.~\cite{Chou:1983qy,Kawai:1984mx,Manes:1984gk,Pak:1984bn,Bijnens:1993xi}.
Instead, let us derive the anomalous interactions in the presence of a vector
field only, so as to check the result of the Noether procedure. Plugging%
\begin{equation}
\mathbf{V}_{\mu},\mathbf{A}_{\mu}\rightarrow\frac{1}{2}\left(  \mathbf{V}%
_{\mu}\pm\left(  \boldsymbol{U}^{\dagger}\mathbf{V}_{\mu}\boldsymbol{U}%
+i\boldsymbol{U}^{\dagger}\partial_{\mu}\boldsymbol{U}\right)  \right)
\ ,\ \ \mathbf{F}_{\mu\nu}^{V,A}\rightarrow\frac{1}{2}\left(  \mathbf{F}%
_{\mu\nu}^{V}\pm\boldsymbol{U}^{\dagger}\mathbf{F}_{\mu\nu}^{V}\boldsymbol{U}%
\right)  \ ,
\end{equation}
into Eq.~(\ref{BardeenCS}), it is possible by inspection to express its gauge
variation as the WZW action of Eq.~(\ref{WZW1}) together with a total
derivative, that is, Eq.~(\ref{WZWExt}) with%
\begin{align}
\mathbf{Z}_{\mu\nu\rho\sigma}  &  =\{\mathbf{V}_{\mu},\boldsymbol{U}^{\dagger
}\}\partial_{\nu}\boldsymbol{UU}^{\dagger}\partial_{\rho}\boldsymbol{UU}^{\dagger}
\partial_{\sigma}\boldsymbol{U}\nonumber\\
&  \ \ \ \ +i\partial_{\mu}\mathbf{V}_{\nu}\boldsymbol{U}^{\dagger}
\mathbf{V}_{\rho}\partial_{\sigma}\boldsymbol{U}+2i\partial_{\mu}
\mathbf{V}_{\nu}\mathbf{V}_{\rho}\{\boldsymbol{U}^{\dagger},
\partial_{\sigma}\boldsymbol{U}\}+i\partial_{\mu}\mathbf{V}_{\nu}
\boldsymbol{U}\mathbf{V}_{\rho}\boldsymbol{U}^{\dagger}\partial_{\sigma}
\boldsymbol{UU}^{\dagger}\nonumber\\
&  \ \ \ \ +i\partial_{\mu}\mathbf{V}_{\nu}\boldsymbol{U}\{\mathbf{V}_{\rho},\boldsymbol{U}^{\dagger}\}
\partial_{\rho}\boldsymbol{UU}^{\dagger}+i\partial_{\mu}\mathbf{V}_{\nu}
\{\boldsymbol{U}^{\dagger},\mathbf{V}_{\rho}\}\partial_{\sigma}\boldsymbol{U}
+i\partial_{\mu}\mathbf{V}_{\nu}\{\boldsymbol{U}^{\dagger},\partial_{\sigma}
\boldsymbol{U}\}\mathbf{V}_{\rho}\nonumber\\
&  \ \ \ \ +i[\boldsymbol{U}^{\dagger},\mathbf{V}_{\mu}\mathbf{V}_{\nu}]
\partial_{\rho}\boldsymbol{UU}^{\dagger}\partial_{\sigma}\boldsymbol{U}
+\frac{i}{2}\mathbf{V}_{\mu}\boldsymbol{U}^{\dagger}\partial_{\nu}\boldsymbol{U}
\mathbf{V}_{\rho}\boldsymbol{U}^{\dagger}\partial_{\sigma}\boldsymbol{U}
-\frac{i}{2}\mathbf{V}_{\mu}\partial_{\nu}\boldsymbol{UU}^{\dagger}
\mathbf{V}_{\rho}\partial_{\sigma}\boldsymbol{UU}^{\dagger}\nonumber\\
&  \ \ \ \ +i\boldsymbol{U}^{\dagger}\mathbf{V}_{\mu}\boldsymbol{U}
\mathbf{V}_{\nu}\boldsymbol{U}^{\dagger}\partial_{\rho}\boldsymbol{UU}^{\dagger}
\partial_{\sigma}\boldsymbol{U}-\boldsymbol{U}^{\dagger}\mathbf{V}_{\mu}
\boldsymbol{U}\mathbf{V}_{\nu}\partial_{\rho}\mathbf{V}_{\sigma}
+\{\boldsymbol{U}^{\dagger},\mathbf{V}_{\mu}\}\mathbf{V}_{\nu}
\boldsymbol{U}\partial_{\rho}\mathbf{V}_{\sigma}\nonumber\\
&  \ \ \ \ +2\boldsymbol{U}\mathbf{V}_{\mu}\boldsymbol{U}^{\dagger}
\mathbf{V}_{\nu}\partial_{\rho}\mathbf{V}_{\sigma}
+\mathbf{V}_{\mu}\mathbf{V}_{\nu}\mathbf{V}_{\rho}\{\partial_{\sigma}
\boldsymbol{U},\boldsymbol{U}^{\dagger}\}-\mathbf{V}_{\mu}\boldsymbol{U}^{\dagger}
\mathbf{V}_{\nu}\mathbf{V}_{\rho}\partial_{\sigma}\boldsymbol{U}%
\nonumber\\
&  \ \ \ \ +\boldsymbol{U}^{\dagger}\mathbf{V}_{\mu}\mathbf{V}_{\nu}\boldsymbol{U}
\mathbf{V}_{\rho}\boldsymbol{U}^{\dagger}\partial_{\sigma}\boldsymbol{U}
+\boldsymbol{U}^{\dagger}\mathbf{V}_{\mu}\boldsymbol{U}\mathbf{V}_{\nu}
\boldsymbol{U}^{\dagger}\{\mathbf{V}_{\rho},\partial_{\sigma}\boldsymbol{U}\}\nonumber\\
&  \ \ \ \ +i\boldsymbol{U}^{\dagger}\mathbf{V}_{\mu}\boldsymbol{U}\mathbf{V}_{\nu}
\mathbf{V}_{\rho}\mathbf{V}_{\sigma}+i\boldsymbol{U}^{\dagger}\mathbf{V}_{\mu}
\mathbf{V}_{\nu}\mathbf{V}_{\rho}\boldsymbol{U}\mathbf{V}_{\sigma}
+\frac{i}{2}\boldsymbol{U}^{\dagger}\mathbf{V}_{\mu}\boldsymbol{U}\mathbf{V}_{\nu}
\boldsymbol{U}^{\dagger}\mathbf{V}_{\rho}\boldsymbol{U}\mathbf{V}_{\sigma}\ .
\end{align}
In the abelian case, there is only one vector current $\mathbf{V}_{\mu
}\rightarrow-eA_{\mu}\boldsymbol{Q}$, so only the terms in the first two lines
survive the antisymmetric contraction and we recover Eqs.~(\ref{Z1gamma})
and~(\ref{Z2gamma}).

\subsection{Anomalies in meson phenomenology}\label{SecMesons}

There are two different chiral anomalies showing up at low energy, below the
GeV scale. The QCD singlet anomaly described in Sec.~\ref{Singlet}, from which the
isospin singlet meson $\eta^{\prime}$ gets its large mass, and the Bardeen
anomaly responsible for example for $\pi^{0}\rightarrow\gamma\gamma$. Our goal
in the present section is to put them together, construct the simplest
effective meson theory, and identify the main mesonic signatures of these anomalies.

\subsubsection{Lagrangian including the singlet}

The first step is to introduce the singlet $\eta_{0}$ into the game. For that,
it suffices to include $\lambda^{0}=\sqrt{2/3}\times\mathbf{1}$ and extend the
summation $\sqrt{2}\boldsymbol{\xi}=\lambda^{a}\xi^{a}$ to $a=0,...,8$ in the
definition of $\boldsymbol{U}$, see Eq.~(\ref{DefU}). This generator is
normalized such that the net effect is%
\begin{equation}
\boldsymbol{\xi\rightarrow\xi}+\frac{\eta_{0}}{\sqrt{3}}\boldsymbol{1\ }.
\end{equation}
Once this is done, the leading order Lagrangian of Eq.~(\ref{LeffChPT}) is
extended to~\cite{DiVecchia:1980yfw}%
\begin{align}
\mathcal{L}_{eff}  &  =\dfrac{F^{2}}{4}\langle D^{\mu}\boldsymbol{U}D_{\mu
}\boldsymbol{U}^{\dagger}+\chi\boldsymbol{U}^{\dagger}+\boldsymbol{U}%
\chi^{\dagger}\rangle\nonumber\\
&  \ \ \ +\frac{i}{4}\langle\log\boldsymbol{U}-\log\boldsymbol{U}^{\dagger
}\rangle\partial^{\mu}G_{\mu}+\frac{1}{2}\theta\partial^{\mu}G_{\mu}+\frac
{1}{2\chi(0)}(\partial^{\mu}G_{\mu})^{2}\;. \label{ChPTCS}%
\end{align}
In the first line, $\chi\equiv2B\left(  s+ip\right)  $ introduces scalar and
pseudoscalar explicit chiral symmetry breaking terms, with $B$ a free hadronic
parameter. Here, we will always set $p=0$, $s=M$ with $M=\operatorname*{diag}%
(m_{u},m_{d},m_{s})$, so as to generate the Gell-Mann--Oakes--Renner mass
terms for the pseudoscalar mesons~\cite{Gell-Mann:1968hlm}:%
\begin{equation}
m_{\pi}^{2}=2Bm_{u}\ ,\ \ m_{K}^{2}=B(m_{u}+m_{s})\ , \label{GOM}%
\end{equation}
in the isospin limit $m_{u}=m_{d}$. In the second line, $\partial^{\mu}G_{\mu
}$ is the Chern-Simons form, and the final term accounts for the topological
susceptibility. Up to the normalization, it matches that in
Eq.~(\ref{LEffTopo}). The first and second coupling enforce a correct behavior
under $U(1)_{A}$ transformations since $\theta$ shifts by $2\alpha N_{f}$ when
$U\rightarrow\exp(-i(\alpha/2)\mathbf{1})\cdot U\cdot\exp(-i(\alpha
/2)\mathbf{1})$. This reproduces the consequence of the singlet anomaly in the
real QCD with $N_{f}$ flavored quarks. Specifically, if we compute divergence
of the axial current $\mathbf{A}_{\mu}\sim\{\partial_{\mu}\boldsymbol{U}%
,\boldsymbol{U}^{\dagger}\}$, see Eq.~(\ref{GoldTheo}), using the classical
equation of motion derived from Eq.~(\ref{ChPTCS}),%
\begin{equation}
D_{\mu}\left(  \boldsymbol{U}D^{\mu}\boldsymbol{U}^{\dagger}\right)
+\dfrac{1}{2}\left(  \chi\boldsymbol{U}^{\dagger}-\boldsymbol{U}\chi^{\dagger
}\right)  +\frac{1}{F^{2}}\partial_{\mu}K^{\mu}=0\ ,
\end{equation}
we find
\begin{equation}
\partial^{\mu}A_{\mu}^{a}=i\frac{F^{2}}{8}\left[  \langle\{\lambda_{a}%
,\chi\}\boldsymbol{U}^{\dagger}-\{\lambda_{a},\chi^{\dagger}\}\boldsymbol{U}%
\rangle\right]  +\langle\lambda_{a}/2\rangle\partial_{\mu}K^{\mu}\ .
\label{SingletAnoChPT}%
\end{equation}
This is indeed the mesonic version of $\partial^{\mu}A_{\mu}=2imP$, corrected
by the QCD anomaly in the singlet channel $A_{\mu}^{0}$. By the way, this also
shows that axial transformations change the phase of $\chi$. In practice,
$m_{u,d,s}$ are assumed real so $\theta$ actually stands for a combination of
the initial QCD $\theta$ term and of the quark mass phase\ (and is often
denoted $\theta_{eff}$, see Eq.~(\ref{ThetaEff})).

There is a hidden but important assumption in this Lagrangian. Strictly
speaking, $\langle\log\boldsymbol{U}-\log\boldsymbol{U}^{\dagger}\rangle
\sim\eta_{0}/F$ has no mass dimension, so we should include infinitely many
couplings involving factors of $(\langle\log\boldsymbol{U}-\log\boldsymbol{U}%
^{\dagger}\rangle-2i\theta)$, and there would be no hope to be predictive.
This is where so-called large $N_{C}$ arguments enter. Indeed, terms involving
a single trace are always dominant compared to multi-trace operators because a
flavor trace necessarily corresponds a closed fermion loop (why a fermion loop
costs a relative factor of $1/N_{C}$ would be interesting to demonstrate, but
this would be somewhat off-topic so we refer instead for example to
Ref.~\cite{Manohar:1998xv}). Keeping only the single-trace terms with the
minimal number of derivatives or quark mass terms gives Eq.~(\ref{ChPTCS}).

\subsubsection{Mixings in the isospin singlet channel}

Treated as a field, $\partial^{\mu}G_{\mu}$ has no true kinetic term so it can
be eliminated by solving its classical equation of motion, leading to%
\begin{equation}
\mathcal{L}_{eff}=\dfrac{F^{2}}{4}\langle D^{\mu}\boldsymbol{U}D_{\mu
}\boldsymbol{U}^{\dagger}+\chi\boldsymbol{U}^{\dagger}+\boldsymbol{U}%
\chi^{\dagger}\rangle-\frac{\chi(0)}{8}\left(  \theta+\frac{1}{2}i\langle
\log\boldsymbol{U}-\log\boldsymbol{U}^{\dagger}\rangle\right)  ^{2}\;.
\label{ChPTCS2}%
\end{equation}
Since $\langle\log\boldsymbol{U}-\log\boldsymbol{U}^{\dagger}\rangle\sim
\eta_{0}/F$, the last term relates the singlet $\eta_{0}$ mass to the
topological susceptibility,
\begin{equation}
\chi(0)=\frac{2}{3}m_{\eta0}^{2}F^{2}\ ,
\label{ChPTWV}
\end{equation}
which is the Veneziano-Witten relation of Eq.~(\ref{VeneWitt1}), but for
$F_{\eta}=\sqrt{2/3}F$ coming from the normalization of the $\lambda^{0}$
generator. Actually, at this stage, everything parallels the discussion in
Sec.~\ref{SecSuscTopo}, leading to Eq.~(\ref{DysonCS}). The only new feature is that
$\langle\chi\boldsymbol{U}^{\dagger}+\boldsymbol{U}\chi^{\dagger}\rangle$ does
not immediately produces the $6B/\omega_{m}$ quark-mass contribution to the
singlet mass, Eq.~(\ref{Eq6Bw}), because it is not diagonal in the meson basis.
Even in the isospin limit $m_{u}=m_{d}$, the singlet state $\eta_{0}$ mixes with 
the octet state $\eta_{8}$. Writing%
\begin{equation}
\left(
\begin{array}
[c]{c}%
\eta_{8}\\
\eta_{0}%
\end{array}
\right)  =\left(
\begin{array}
[c]{cc}%
\cos\theta_{P} & \sin\theta_{P}\\
-\sin\theta_{P} & \cos\theta_{P}%
\end{array}
\right)  \left(
\begin{array}
[c]{c}%
\eta\\
\eta^{\prime}%
\end{array}
\right)  \ , \label{hhmixing}%
\end{equation}
the only free parameter is $\chi(0)$, or equivalently $m_{\eta0}^{2}$ defined 
via Eq.~(\ref{ChPTWV}), and we can write~\cite{Gerard:2004gx}
\begin{subequations}
\label{EqEta3}%
\begin{equation}
\tan2\theta_{P}=\frac{2\sqrt{2}R}{R-9}\ ,\ m_{\eta^{\prime},\eta}^{2}=m_{\pi
}^{2}+m_{\eta0}^{2}\frac{R+3\pm\sqrt{9-2R+R^{2}}}{6}\ ,
\end{equation}
where we have introduced $R\equiv6\left(  m_{K}^{2}-m_{\pi}^{2}\right)
/m_{\eta0}^{2}$. When $\chi(0)\rightarrow0$, the mixing is said to be ideal,
$\theta_{P}\approx35{{}^\circ}$, and the mass eigenstates are then $\eta=(\bar{u}u+\bar{d}d)/\sqrt{2}$,
$\eta^{\prime}=\bar{s}s$. This is adequate for example for the $\omega$ and
$\phi$ vector mesons, but not for pseudoscalar mesons because instantons do
generate $\chi(0)\neq0$.

To confirm this, a nice and intuitive extraction of the mixing angle is
possible from quarkonium decays~\cite{Novikov:1979uy}, using%
\end{subequations}
\begin{equation}
R_{J/\psi}\equiv\frac{\mathcal{B}\left(  J/\psi\rightarrow\eta\gamma\right)
}{\mathcal{B}\left(  J/\psi\rightarrow\eta^{\prime}\gamma\right)  }=\left\vert
\frac{\left\langle 0\right\vert \partial^{\mu}G_{\mu}\left\vert \eta
\right\rangle }{\left\langle 0\right\vert \partial^{\mu}G_{\mu}\left\vert
\eta^{\prime}\right\rangle }\right\vert ^{2}\left(  \frac{p_{\eta}}%
{p_{\eta^{\prime}}}\right)  ^{3}\overset{\exp}{=}0.200\pm0.023\ ,
\end{equation}
where $p_{P}=M_{J/\psi}(1-m_{P}^{2}/M_{J/\psi}^{2})$ accounts for the
phase-space difference. Physically, the idea is that the $c$ and $\bar{c}$
quarks have to fully annihilate since $\eta$ and $\eta^{\prime}$ are made of
light quarks. So, the light mesons must arise purely from gluons, and the only
available gluon state of the right quantum number is the Chern-Simons term. At
the same time, the equation of motion derived from Eq.~(\ref{ChPTCS}) gives
$\partial^{\mu}G_{\mu}\sim\eta_{0}$, so%
\begin{equation}
\frac{\left\langle 0\right\vert \partial^{\mu}G_{\mu}\left\vert \eta
\right\rangle }{\left\langle 0\right\vert \partial^{\mu}G_{\mu}\left\vert
\eta^{\prime}\right\rangle }=\frac{\left\langle 0\left\vert \eta
_{0}\right\vert \eta\right\rangle }{\left\langle 0\left\vert \eta
_{0}\right\vert \eta^{\prime}\right\rangle }=-\tan\theta_{P}\overset{R_{J/\psi
}^{\exp}}{=}-22.0{{}^\circ}\pm 1.2{{}^\circ}\ ,
\end{equation}
to which corresponds $m_{\eta0}^{2}\approx800~$MeV.

\subsubsection{Two-photon decay modes from the WZW term}

To the Lagrangian in Eq.~(\ref{ChPTCS2}), we now add the odd-parity couplings
corresponding to the Bardeen anomaly. Our targets are the $\pi,\eta
,\eta^{\prime}\rightarrow\gamma\gamma$ processes, so taking $\mathbf{Z}%
_{\mu\nu\rho\sigma}^{2\gamma}$ of Eq.~(\ref{Z2gamma}) with $\boldsymbol{Q}%
=\operatorname*{diag}(2/3,-1/3,-1/3)$ and retaining only terms linear in the
mesons,%
\begin{equation}
\frac{\nu}{48\pi^{2}}\varepsilon^{\mu\nu\rho\sigma}\left\langle \mathbf{Z}%
_{\mu\nu\rho\sigma}^{2\gamma}\right\rangle \supset\frac{\nu\alpha}{12\pi
F}\left(  \pi^{0}+\sqrt{\frac{1}{3}}\eta_{8}+\sqrt{\frac{8}{3}}\eta
_{0}\right)  F_{\mu\nu}\tilde{F}^{\mu\nu}\ . \label{ChPTgg}%
\end{equation}
The decay rates are then%
\begin{equation}
\Gamma\left(  P\rightarrow\gamma\gamma\right)  =\frac{\nu^{2}C_{P}^{2}%
\alpha^{2}m_{P}^{3}}{576\pi^{3}F^{2}}\ ,
\end{equation}
with $C_{\pi}=1,C_{\eta8}=\sqrt{1/3},C_{\eta0}=\sqrt{8/3}$, and%
\begin{equation}
C_{\eta}=\cos\theta_{P}C_{\eta8}-\sin\theta_{P}C_{\eta0},\;\;C_{\eta^{\prime}%
}=\sin\theta_{P}C_{\eta8}+\cos\theta_{P}C_{\eta0}\ .
\end{equation}
The experimental measurement $\Gamma\left(  \pi^{0}\rightarrow\gamma
\gamma\right)  \approx7.8\;$eV fixes $\nu=3$, which is often interpreted as
the number of QCD colors $N_{C}$. Indeed, though we derived these terms in a
rather complicated way, from the gauge variation of the WZW term, the
two-photon couplings can also be obtained directly from the abelian anomaly.
From Eq.~(\ref{AnoAbelian2}), putting $u,d,s$ quarks of each color in the
triangle diagram,
\begin{equation}
\partial^{\mu}A_{\mu}^{a}=N_{C}\frac{e^{2}}{16\pi^{2}}F_{\mu\nu}\tilde{F}%
^{\mu\nu}\left\langle \lambda^{a}\{\boldsymbol{Q},\boldsymbol{Q}%
\}\right\rangle \ ,\; \label{QEDAnoChpt}%
\end{equation}
which for $\nu=N_{C}$ gives the coefficients in Eq.~(\ref{ChPTgg}) for
$a=0,3,8$.

It should be said though that this overall factor of $N_{C}$ does not suffice
to prove that $\pi^{0}\rightarrow\gamma\gamma$ measures $N_{C}$. One should
first figure out what $\boldsymbol{Q}$ should be if there are $N_{C}$ colors.
In particular, if the SM gauge anomalies are to cancel, we cannot keep
$\boldsymbol{Q}=\operatorname*{diag}(2/3,-1/3,-1/3)$. Rather, from
Eq.~(\ref{NcQuQd}), we should set%
\begin{equation}
\boldsymbol{Q}=\frac{1}{2}\lambda^{3}+\frac{1}{2\sqrt{3}}\lambda^{8}%
+\sqrt{\frac{3}{8}}\left(  \frac{1}{N_{C}}-\frac{1}{3}\right)  \lambda^{0}\ .
\label{QNc}%
\end{equation}
The piece proportional to $\lambda^{0}$ shows that the electromagnetic current
inherits a term aligned with baryon number. Plugging this expression in
Eq.~(\ref{Z2gamma}), we get instead of Eq.~(\ref{ChPTgg}) the couplings:%
\begin{equation}
\frac{\nu}{48\pi^{2}}\varepsilon^{\mu\nu\rho\sigma}\left\langle \mathbf{Z}%
_{\mu\nu\rho\sigma}^{2\gamma}\right\rangle \supset\frac{\nu}{N_{C}}%
\frac{\alpha}{4\pi F}\left(  \pi^{0}+\sqrt{\frac{1}{3}}\eta_{8}+\sqrt{\frac
{8}{3}}\frac{3N_{C}^{2}-2N_{C}+3}{8N_{C}}\eta_{0}\right)  F_{\mu\nu}\tilde
{F}^{\mu\nu}\ . \label{NcGG}%
\end{equation}
With $\nu=N_{C}$, $\pi^{0}\rightarrow\gamma\gamma$ ends up actually
fixed~\cite{Bar:2001qk}. It does not permit to access $N_{C}$ because the
$N_{C}$ dependence cancels between the contribution of the WZW action, with
$Q$ fixed to its $N_{C}=3$ value, and of the electromagnetic coupling to the
Goldstone-Wilczek current, tuned by $N_{C}$ times the coefficient of
$\lambda^{0}$ in Eq.~(\ref{QNc}). This current also explains why $\pi
^{0}\rightarrow\gamma\gamma$ still occurs if $N_{f}=2$, when the WZW action
vanishes along with the Bardeen anomaly since $SU(2)$ has no $d$ tensor. In
that case, $\boldsymbol{Q}=\operatorname*{diag}(2/3,-1/3)$ necessarily has a
singlet piece, so instead of the WZW action, one has to add the
Goldstone-Wilczek current coupled to photons to match the anomaly structure of
the two-flavor QCD. At the end of the day, the same result for $\pi
^{0}\rightarrow\gamma\gamma$ is obtained, with no explicit $N_{C}$
factor~\cite{Kaiser:2000ck,Bar:2001qk,Borasoy:2004ua}.

Concerning the singlet mesons, we can now extract the mixing angle from
$\Gamma^{\exp}\left(  \eta\rightarrow\gamma\gamma\right)  \approx0.51\;$keV
and $\Gamma^{\exp}\left(  \eta^{\prime}\rightarrow\gamma\gamma\right)
\approx4.3\;$keV, and get $\theta_{P}\approx-19{{}^\circ}$. 
The agreement is not perfect with the value extracted from quarkonium
decays, but already quite impressive for a leading order estimate.\ Remember
that not only is Eq.~(\ref{ChPTCS}) limited to the couplings involving the
least number of derivatives, but also to those that are dominant in $1/N_{C}$.
In this respect, one should take the $\eta_{0}$ coefficient in Eq.~(\ref{NcGG}%
) with a grain of salt, including the $N_{C}$ dependence, since many other
operators could contribute like e.g. $\varepsilon^{\mu\nu\rho\sigma}F_{\mu\nu
}A_{\rho}\left\langle \mathbf{Q}^{2}\right\rangle \left\langle \partial
_{\sigma}\boldsymbol{U}^{\dagger}\boldsymbol{U}\right\rangle $.

It should be said finally that besides the $\gamma\gamma$ modes, the Bardeen
anomaly also predicts many other radiative processes, for example $\eta
,\eta^{\prime}\rightarrow\pi^{+}\pi^{-}\gamma$, see e.g.
Ref.~\cite{Gan:2020aco}. Further, once effective operators representing the
quark flavor transitions induced by the weak interactions are included, many
anomalous $K$ meson decay processes can occur, e.g. $K_{L}\rightarrow
\gamma\gamma$ or $K_{S}\rightarrow\pi^{0}\gamma\gamma$, see e.g.
Ref.~\cite{Gerard:2005yk}.

\subsubsection{Two-pion decay modes from the $\theta$ term}

For all practical purposes, $\theta$ is set to zero since it is known
experimentally to be tiny. Yet, if it is non-zero, it allows for CP-violating
processes, chief among them are the $\eta,\eta^{\prime}\rightarrow\pi\pi$
decays. The Lagrangian in Eq.~(\ref{ChPTCS2}) is however not convenient to
compute these processes because of the presence of the tadpole $\chi
(0)\theta\eta_{0}$. To compute $\eta,\eta^{\prime}\rightarrow\pi\pi$, we would
need to sum all the $\eta,\eta^{\prime}\rightarrow\pi\pi+(2n+1)\times\eta_{0}$ processes
where $\eta_{0}\rightarrow\varnothing$. Such a situation is typical of
perturbation theory done around the wrong vacuum. Looking back at
Eq.~(\ref{DefU}), we did not select any particular vacuum when constructing
our effective theory. This made sense at the time since all are equivalent
under chiral transformation. But, the anomaly as well as quark masses do break
the chiral symmetry, tilting the vacuum space. Now, it appears that the
specific vacuum we chose within this space does not have minimum energy
because we have to radiate $\eta_{0}$ mesons to reach the correct one.

To solve this problem, we have to align the vacuum with the chiral symmetry
breaking terms, a procedure often called \textbf{Dashen theorem} in the
context of the chiral symmetry~\cite{Dashen:1970et} (see also
Ref.~\cite{Cheng:1987gp}). In the original article, a nice analogy is made
with the magnetization of a piece of iron as its temperature drops below the
critical value. Obviously, all final orientations of its magnetization are
equivalent. But if there is an external magnetic field, however small, the
rotational symmetry is explicitly broken, and it is best to immediately orient
the magnetization with the external field.

The goal is thus to correctly align $\boldsymbol{U}^{ij}\sim\langle0|\bar
{q}_{L}^{i}q_{R}^{j}|0\rangle$ with the chiral symmetry breaking terms. For
that, we start by introducing three free parameters $\boldsymbol{U}%
\rightarrow\boldsymbol{UU}_{\phi}^{\dagger}$ with $U_{\phi}\equiv\delta
^{ij}\exp(-i\phi_{i})$. There is no need to keep track of more parameters
since we know $SU(3)_{V}$ remains exact, so a generic $U_{\phi}$ can always be
brought to that form. Then, the non-derivative terms in Eq.~(\ref{ChPTCS2})
become%
\begin{align}
\mathcal{V}(\boldsymbol{U},\phi_{f}) &  =-\frac{F^{2}B}{2}\langle MC_{\phi
}(\boldsymbol{U}^{\dagger}+\boldsymbol{U})+iMS_{\phi}(\boldsymbol{U}%
-\boldsymbol{U}^{\dagger})\rangle\nonumber\\
&  \ \ \ +\frac{\chi(0)}{8}(\theta-\sum\phi_{f}+\frac{1}{2}i\langle
\log\boldsymbol{U}-\log\boldsymbol{U}^{\dagger}\rangle)^{2}\ ,\label{ChPTPot}%
\end{align}
where $C_{\phi}^{ij}=\delta^{ij}\cos\phi_{i}$ and $S_{\phi}^{ij}=\delta
^{ij}\sin\phi_{i}$. Enforcing Dashen theorem is equivalent to asking for
$E(\phi_{f})=\langle\Omega|\mathcal{V}(\boldsymbol{U},\phi_{f})|\Omega
\rangle=\mathcal{V}(\boldsymbol{1},\phi_{f})$ to be minimal, so the $\phi_{i}$
have to satisfy%
\begin{equation}
4F^{2}Bm_{i}\sin\phi_{i}=\chi(0)(\theta-\sum\nolimits_{f}\phi_{f}%
)\ ,\ \ i=u,d,s\ .\label{SolPhii}%
\end{equation}
Exactly the same condition is found by asking the potential not to be able to
produce any of the Goldstone bosons, including the $\eta_{0}$, i.e.
$\langle\Omega|\mathcal{V}(\boldsymbol{U},\phi_{f})|P\rangle=0$ for $P$ any of
the mesons. Imposing the constraint of Eq.~(\ref{SolPhii}) permits to
eliminate $S_{\phi}$, and the final Lagrangian takes the form first derived in
Ref.~\cite{DiVecchia:1980yfw}:%
\begin{align}
\mathcal{L}_{eff}(\phi_{f}) &  =\dfrac{F^{2}}{4}\langle D^{\mu}\boldsymbol{U}%
D_{\mu}\boldsymbol{U}^{\dagger}\rangle-\frac{F^{2}B}{2}\langle MC_{\phi
}(\boldsymbol{U}^{\dagger}+\boldsymbol{U})\rangle+\frac{\chi(0)}{32}%
\langle\log\boldsymbol{U}-\log\boldsymbol{U}^{\dagger}\rangle^{2}\nonumber\\
&  \ \ \ \ +i\frac{\chi(0)}{8}(\theta-\sum\nolimits_{f}\phi_{f})\langle
\log\boldsymbol{U}-\log\boldsymbol{U}^{\dagger}-(\boldsymbol{U}-\boldsymbol{U}%
^{\dagger})\rangle\ .\label{LagrO2}%
\end{align}
where $\omega_{m}=m_{u}^{-1}+m_{d}^{-1}+m_{s}^{-1}$. It is interesting to note
that $\phi_{u,d,s}$ actually parametrize the usual QCD freedom to move $\theta$
between the $\partial^{\mu}G_{\mu}$ term represented by $\langle
\log\boldsymbol{U}-\log\boldsymbol{U}^{\dagger}\rangle$, and the phase of the
quark mass term represented by $\langle\boldsymbol{U}-\boldsymbol{U}^{\dagger
}\rangle$. This latter quantity is the hadronic representation of the
pseudoscalar singlet current $\bar{u}\gamma_{5}u+\bar{d}\gamma_{5}d+\bar
{s}\gamma_{5}s$, and is also called the \textbf{Baluni term}%
~\cite{Baluni:1978rf,Cheng:1990pi}. Yet, minimizing the energy, a very
specific spread of the CP-violating effects is required at the meson level
between these two terms.

The only difficulty in this approach is that Eq.~(\ref{SolPhii}) can be solved
analytically only if $\phi_{f}\ll1$. Specifically, if we approximate $\sin
\phi_{i}\approx\phi_{i}$, move $m_{i}$ to the right, and sum on both sides
over $i$, we find%
\begin{equation}
\theta-\sum\nolimits_{f}\phi_{f}=\frac{\theta}{1+\kappa}\ \ \ \text{with}%
\ \kappa\equiv\frac{\chi(0)\omega_{m}}{4F^{2}B}=\frac{\chi(0)}{2F^{2}m_{\pi
}^{2}}\frac{4m_{K}^{2}-m_{\pi}^{2}}{2m_{K}^{2}-m_{\pi}^{2}}\ .
\end{equation}
While not manifest in Eq.~(\ref{ChPTCS2}), we now see that the CP-violating
interaction in $\mathcal{L}_{eff}(\phi_{f})$ does vanish if the topological
susceptibility or any of the quark mass does, as it should. Note that plugging
this expression in Eq.~(\ref{ChPTPot}) and calculating $\chi_{full}%
(0)=4\partial^{2}E(\phi_{f})/\partial\theta^{2}$ reproduces Eq.~(\ref{DysonCS}%
), showing that we also resummed the $\partial^{\mu}G_{\mu}\eta_{0}$ mixing
effects by correctly aligning the vacuum. It is then a simple exercise to
expand Eq.~(\ref{ChPTPot}) and find the cubic vertices%
\begin{equation}
\mathcal{L}_{eff}(\phi_{f})\supset\frac{\chi(0)}{4F^{3}}\frac{\theta}%
{1+\kappa}\left(  \sqrt{\frac{1}{3}}\eta_{8}+\sqrt{\frac{2}{3}}\eta
_{0}\right)  (2\pi^{+}\pi^{-}+\pi^{0}\pi^{0})\ ,
\end{equation}
from which, using Eq.~(\ref{hhmixing}), the $\pi^{+}\pi^{-}$ and $\pi^{0}%
\pi^{0}$ decay amplitudes are identical (but there is an extra $1/2$ factor
for the $\pi^{0}\pi^{0}$ decay rates),%
\begin{align}
\mathcal{M}(\eta &  \rightarrow\pi\pi)=\frac{\chi(0)}{2\sqrt{3}F^{3}}%
\frac{\theta}{1+\kappa}(\cos\theta_{P}-\sqrt{2}\sin\theta_{P})\ ,\\
\mathcal{M}(\eta^{\prime} &  \rightarrow\pi\pi)=\frac{\chi(0)}{2\sqrt{3}F^{3}%
}\frac{\theta}{1+\kappa}(\sin\theta_{P}+\sqrt{2}\cos\theta_{P})\ .
\end{align}
The same result could have been obtained starting from Eq.~(\ref{ChPTCS2}),
provided the tadpoles are appropriately included~\cite{Cheng:1990pi}. Notice
that these amplitudes manifestly vanish if $\chi(0)\rightarrow0$ or if any of
the quark is massless since then $\kappa\rightarrow\infty$.

These processes are unfortunately well beyond experimental reach given the
bound on $\theta$ from the neutron EDM. Actually, this latter observable is
superior to all others in constraining $\theta$. Yet, we will not describe how
to treat it here because it is a very delicate topic in Chiral Perturbation
Theory~\cite{Crewther:1979pi}. Other techniques can help, like sum rules (see
e.g. Ref.~\cite{Pospelov:1999ha}) or lattice calculations (see e.g.
Refs.~\cite{Yamanaka:2017mef,Dekens:2018bci} for a review), but this is beyond
our scope.

\section{The covariant anomaly}\label{SecCovAno}

The consistent anomaly necessarily imposes some Bose symmetry among the gauge
fields. This severely restricts the form of the covariant divergences, with
for example the impossibility to conserve the axial currents. The goal here is
to give up the Bose symmetry and ask instead that all but the axial and vector
current for one specific $SU(N)$ index are conserved. This is a bit peculiar
since a gauge transformation would mix up the currents, but we will see that
it nicely connects the general chiral anomaly of Sec.~\ref{SecChiralAno} with
the singlet anomaly discussed in Sec.~\ref{Singlet}.

Let us start with the consistent anomaly in the chiral gauge current of
Eq.~(\ref{AnoCons}). We can compensate the gauge dependence of $D_{\mu}%
J_{R,L}^{a,\mu}$ by adding to $J_{R,L}^{a,\mu}$ a new carefully chosen
gauge-dependent current, which is named the \textbf{Bardeen-Zumino
polynomial}~\cite{Bardeen:1984pm}%
\begin{equation}
X_{R,L}^{a,\mu}=\pm\frac{\varepsilon^{\mu\nu\rho\sigma}}{48\pi^{2}%
}\left\langle T^{a}\left(  \{\mathbf{A}_{\nu}^{R,L},\mathbf{F}_{\rho\sigma
}^{R,L}\}+i\mathbf{A}_{\nu}^{R,L}\mathbf{A}_{\rho}^{R,L}\mathbf{A}_{\sigma
}^{R,L}\frac{{}}{{}}\right)  \right\rangle \ . \label{BZcurr}%
\end{equation}
Then, we get a covariant result for the covariant divergence of $\mathcal{J}%
_{R,L}^{a,\mu}=J_{R,L}^{a,\mu}+X_{R,L}^{a,\mu}$,
\begin{equation}
(D_{\mu}^{L,R}\mathcal{J}_{L,R}^{\mu})^{a}=\pm\frac{\varepsilon^{\mu\nu
\rho\sigma}}{32\pi^{2}}\left\langle T^{a}\mathbf{F}_{\mu\nu}^{L,R}%
\mathbf{F}_{\rho\sigma}^{L,R}\right\rangle \ , \label{CovLLL}%
\end{equation}
by using $D_{\mu}^{R,L}X_{R,L}^{\mu}=\partial_{\mu}X_{R,L}^{\mu}%
-i[\mathbf{A}_{\mu}^{R,L},X_{R,L}^{\mu}]$. This is called the
\textbf{covariant anomaly}, while $\mathcal{J}_{R,L}^{\mu}$ are the covariant
currents. These currents cannot be obtained as the gauge variation of an
action since the $a$ index is singled out and treated differently.
Consequently, this anomaly does not satisfy the Wess-Zumino consistency
condition. Notice that for an abelian group, the only difference between the
consistent and covariant anomalies is the overall numerical factor, with the
former being suppressed by a factor $1/3$ coming from the Bose symmetrization.

From the chiral expressions in Eq.~(\ref{CovLLL}), covariant vector and axial
vector currents can be defined as $D_{\mu}\mathcal{J}_{V,A}^{\mu}=D_{\mu}%
^{R}\mathcal{J}_{R}^{\mu}\pm D_{\mu}^{L}\mathcal{J}_{L}^{\mu}$, such that by
direct substitution
\begin{subequations}
\label{AVcov}%
\begin{align}
(D_{\mu}\mathcal{J}_{V}^{\mu})^{a}  &  =-\frac{\varepsilon^{\mu\nu\rho\sigma}%
}{16\pi^{2}}\left\langle T^{a}\{\mathbf{F}_{\mu\nu}^{A},\mathbf{F}_{\rho
\sigma}^{V}\}\right\rangle \ ,\\
(D_{\mu}\mathcal{J}_{A}^{\mu})^{a}  &  =-\frac{\varepsilon^{\mu\nu\rho\sigma}%
}{16\pi^{2}}\left\langle T^{a}(\mathbf{F}_{\mu\nu}^{A}\mathbf{F}_{\rho\sigma
}^{A}+\mathbf{F}_{\mu\nu}^{V}\mathbf{F}_{\rho\sigma}^{V})\right\rangle \ .
\end{align}
Let us stress once more that these currents are defined out of conflicting
conditions: we ask for a gauge covariant expression while singling out one
gauge index, which manifestly breaks the gauge symmetry.

\subsection{Bardeen-Zumino current and geometry}

We have seen in previous sections that the consistent anomaly can be obtained
from the gauge variation of the Chern-Simons action. Still, this variation is
a surface term, so it requires a special geometric setting not to vanish
trivially, in which the Chern-Simons Lagrangian does not itself live already
on some boundary. Our goal here is to set that up, and in doing so, we will
see that it is possible to further identify two different sources for the
consistent anomaly, one of them being related to the covariant anomaly and the
other to the Bardeen-Zumino current. Why this is so has to do with the
peculiar equation of motion derived from the Chern-Simons action.\ In this
section, we follow essentially Ref.~\cite{Stone:2012ud}, translated back into
tensorial notation.

The first step is to derive the equation of motion. Instead of writing down
the Euler-Lagrange equations directly, let us proceed functionally and look at
the variation of the Chern-Simons action under $\mathbf{A}_{\mu}%
\rightarrow\mathbf{A}_{\mu}+\delta\mathbf{A}_{\mu}$, assuming it is defined on
some generic $2n-1$ dimensional space $V$. Computing $\delta G_{2n-1}^{\mu
}[\delta\mathbf{A}_{\mu},\mathbf{A}_{\mu}]=G_{2n-1}^{\mu}[\mathbf{A}_{\mu
}+\delta\mathbf{A}_{\mu}]-G_{2n-1}^{\mu}[\mathbf{A}_{\mu}]$ to first order in
$\delta\mathbf{A}_{\mu}$, and after integrating by part, we can split the
variation into a \textbf{bulk part and a boundary term}:%
\end{subequations}
\begin{align}
\delta S_{CS}[\delta\mathbf{A},\mathbf{A}]\overset{}{=}\int_{V}\delta
G_{2n-1}^{\mu}[\delta\mathbf{A},\mathbf{A}]d\sigma_{\mu}  &  =\int%
_{V}\left\langle \delta\mathbf{A}_{\nu}E_{2n-1}^{\mu\nu}\right\rangle
d\sigma_{\mu}+\int_{V}\partial_{\nu}\left\langle \delta\mathbf{A}_{\rho
}X_{2n-1}^{\mu\nu\rho}\right\rangle d\sigma_{\mu}\nonumber\\
&  =\int_{V}\left\langle \delta\mathbf{A}_{\nu}E_{2n-1}^{\mu\nu}\right\rangle
d\sigma_{\mu}+\int_{\partial V}\left\langle \delta\mathbf{A}_{\rho}%
X_{2n-1}^{\mu\nu\rho}\right\rangle d\sigma_{\mu\nu}\ . \label{deltaSCS}%
\end{align}
The equation of motion is found by requiring the action to be at its minimum,
which means that this variation must vanish for all $\delta\mathbf{A}$.
Assuming as usual that the fields all vanish on the boundary $\partial V$, the
bulk term then produces the covariant equations of motion:%
\begin{subequations}
\begin{align}
E_{1}^{\mu\nu}  &  =2\varepsilon^{\mu\nu}\ ,\\
E_{3}^{\mu\nu}  &  =4\varepsilon^{\mu\nu\rho\sigma}\mathbf{F}_{\rho\sigma
}\ ,\ \ \\
E_{5}^{\mu\nu}  &  =6\varepsilon^{\mu\nu\rho\sigma\alpha\beta}\mathbf{F}%
_{\rho\sigma}\mathbf{F}_{\alpha\beta}\ ,\ \ \\
E_{7}^{\mu\nu}  &  =8\varepsilon^{\mu\nu\rho\sigma\alpha\beta\gamma\delta
}\mathbf{F}_{\rho\sigma}\mathbf{F}_{\alpha\beta}\mathbf{F}_{\gamma\delta}\ .
\end{align}
The dynamics is trivial in two dimensions, and requires $\mathbf{A}_{\mu}$ to
be flat (pure gauge) in three dimensions since $\mathbf{F}_{\rho\sigma}=0$.
This makes the three-dimensional Chern-Simons theory a \textbf{topological
field theory}~\cite{Witten:1988ze}. Actually, Witten was awarded the Field
Medal in part for his discovery that this theory is able to tell us a lot
about the mathematics of knots~\cite{Witten:1988hf} (see also e.g.
Ref.~\cite{Baez:1995sj} for an introduction). For higher dimensions, the
equation of motion does have non-trivial solutions, and are for example
relevant to the study of higher-dimensional extensions to general relativity
(for a review, see e.g. Ref.~\cite{Zanelli:2012px}).

In the special case in which $\delta\mathbf{A}_{\mu}$ is induced by an
infinitesimal gauge variation, $\delta\mathbf{A}_{\mu}=\delta\mathbf{A}_{\mu
}^{gauge}=\partial_{\mu}\boldsymbol{\Lambda}-i[\mathbf{A}_{\mu},\boldsymbol{\Lambda}%
]$, the whole variation of the action becomes a surface term proportional to
the consistent anomaly, as demonstrated in Sec.~\ref{ChainofDescent}. But we now see from
Eq.~(\ref{deltaSCS}) that this anomaly has a bulk contribution and a boundary
contribution. Plugging in $\delta\mathbf{A}_{\mu}^{gauge}$, the bulk term
generates a surface term given by%
\end{subequations}
\begin{align}
\int_{V}\left\langle \delta\mathbf{A}_{\nu}^{gauge}E_{2n-1}^{\mu\nu
}\right\rangle d\sigma_{\mu}  &  =\int_{V}\left\langle (\partial_{\nu
}\boldsymbol{\Lambda}-i[\mathbf{A}_{\nu},\boldsymbol{\Lambda}])E_{2n-1}^{\mu\nu
}\right\rangle d\sigma_{\mu}\nonumber\\
&  =\int_{V}\partial_{\nu}\left\langle \boldsymbol{\Lambda}E_{2n-1}^{\mu\nu
}\right\rangle d\sigma_{\mu}+\left\langle \boldsymbol{\Lambda}(-\partial_{\nu
}E_{2n-1}^{\mu\nu})-i[\mathbf{A}_{\nu},\boldsymbol{\Lambda}]E_{2n-1}^{\mu\nu
})\right\rangle d\sigma_{\mu}\nonumber\\
&  =\int_{V}\partial_{\nu}\left\langle \boldsymbol{\Lambda}E_{2n-1}^{\mu\nu
}\right\rangle d\sigma_{\mu}-\left\langle \boldsymbol{\Lambda(}D_{\nu}%
E_{2n-1}^{\mu\nu})\right\rangle d\sigma_{\mu}\nonumber\\
&  =\int_{\partial V}\left\langle \boldsymbol{\Lambda}E_{2n-1}^{\mu\nu
}\right\rangle d\sigma_{\mu\nu}\ ,
\end{align}
where $D_{\nu}E_{2n-1}^{\mu\nu}=\partial_{\nu}E_{2n-1}^{\mu\nu}-i[\mathbf{A}%
_{\nu},E_{2n-1}^{\mu\nu}]$ vanishes by covariance (as can be checked
explicitly). Since the equation of motion is covariant, so is this surface
term. Taking the derivative with respect to $\Lambda^{a}$, $E_{5}^{\mu\nu}$
precisely reproduces the covariant anomaly of Eq.~(\ref{CovLLL}).

Turning to the boundary term of Eq.~(\ref{deltaSCS}), it is already a surface
term so we can freely integrate by part as $\partial\partial V=\varnothing$.
Its contribution to the total variation is%
\begin{align}
\int_{\partial V}\left\langle \delta\mathbf{A}_{\rho}^{gauge}X_{2n-1}^{\mu
\nu\rho}\right\rangle d\sigma_{\mu\nu}  &  =\int_{\partial V}\left\langle
(\partial_{\rho}\boldsymbol{\Lambda}-i[\mathbf{A}_{\nu},\boldsymbol{\Lambda}%
])X_{2n-1}^{\mu\nu\rho}\right\rangle d\sigma_{\mu\nu}\nonumber\\
&  =\int_{\partial V}\left\langle \boldsymbol{\Lambda(}D_{\rho}X_{2n-1}^{\mu
\nu\rho})\right\rangle d\sigma_{\mu\nu}\ .
\end{align}
Explicitly, from the Chern-Simons actions in Eq.~(\ref{CSactions}), we find:%
\begin{subequations}
\begin{align}
X_{1}^{\mu\nu\rho}  &  =0\ ,\\
X_{3}^{\mu\nu\rho}  &  =4\varepsilon^{\mu\nu\rho\sigma}\mathbf{A}_{\sigma
}\ ,\\
X_{5}^{\mu\nu\rho}  &  =4\varepsilon^{\mu\nu\rho\sigma\alpha\beta}\left(
\{\mathbf{A}_{\sigma},\mathbf{F}_{\alpha\beta}\}+i\mathbf{A}_{\sigma
}\mathbf{A}_{\alpha}\mathbf{A}_{\beta}\frac{{}}{{}}\right)  \ ,\\
X_{7}^{\mu\nu\rho}  &  =4\varepsilon^{\mu\nu\rho\sigma\alpha\beta\gamma\delta
}\left(  \{\mathbf{A}_{\sigma},\mathbf{F}_{\alpha\beta},\mathbf{F}%
_{\gamma\delta}\}+\frac{4}{5}\{\mathbf{A}_{\sigma},\mathbf{A}_{\alpha
},\{\mathbf{A}_{\beta},\mathbf{F}_{\gamma\delta}\}\}-\frac{4}{5}%
\mathbf{A}_{\sigma}\mathbf{A}_{\alpha}\mathbf{A}_{\beta}\mathbf{A}_{\gamma
}\mathbf{A}_{\delta}\right)  \ ,
\end{align}
where, as in Eq.~(\ref{AVC}), $\{...\}$ denotes cyclic permutations. The
contribution of the boundary to the total (infinitesimal) gauge variation is
thus encoded into the covariant divergence of the \textbf{boundary current}
$X_{2n-1}$. Since we know that once added to the bulk contribution, we must
get back the consistent anomaly, this proves that conversely, subtracting it
from the consistent anomaly leaves the covariant anomaly. Without surprise,
one can recognize in $X_{5}^{\mu\nu\rho}$ the Bardeen-Zumino current of
Eq.~(\ref{BZcurr}). Actually, translated into the language of differential
geometry, this is precisely how the covariant anomaly and the Bardeen-Zumino
current were first derived in Ref.~\cite{Bardeen:1984pm}. Interestingly, this
geometric relationship between the consistent and covariant anomalies is also
of phenomenological use in certain systems. For instance, when the fermion
gets its mass from a scalar field acquiring a vacuum expectation value, a
$1+1$-dimensional string-like boundary can appear where said fermion remains
massless. In such a setting, called an \textbf{axion string}, the massless
fermions on the string generate the two-dimensional boundary part of the
anomaly, compensated by an \textbf{anomaly inflow} from the
bulk~\cite{Callan:1984sa,Naculich:1987ci}.

\subsection{Bardeen-Zumino current and diagrams}

The covariant anomaly can be obtained from a direct calculation of the
triangle, box, and pentagon diagrams by imposing the classical Ward identity
on all but one leg. Specifically, starting with the general form of the chiral
anomaly, it suffices to set to $1$ the $a_{i}$, $b_{i}$ coefficient
corresponding to the $SU(N)$ index carrying the whole anomaly, and leave all
the others to zero since they are covariantly conserved. For example, let us
construct the covariant derivatives assuming the current with $SU(N)$ index
$a$ and momentum $q_{1}$ is anomalous. We start by setting $a_{1}=1$,
$a_{2}=0$ for the quadratic terms in Eq.~(\ref{AnoLLL}), giving
\end{subequations}
\begin{subequations}
\label{TriCov}%
\begin{align}
D_{\alpha}^{L}\mathcal{T}_{LLL}^{\alpha\beta\gamma,abc}  &  =-\frac
{\mathcal{I}_{3}d^{abc}}{16\pi^{2}}\varepsilon^{\beta\gamma\mu\nu}q_{1\mu
}q_{2\nu}\ ,\ \\
D_{\beta}^{L}\mathcal{T}_{LLL}^{\alpha\beta\gamma,abc}  &  =D_{\gamma}%
^{L}\mathcal{T}_{LLL}^{\alpha\beta\gamma,abc}=0\ \ .
\end{align}
From the box covariant divergences in Eq.~(\ref{AnoLLLL}), we set
$b_{2,3,4}=0$ since these legs are not anomalous. The only non-zero
coefficients are in the triangles with $a_{1}^{1i}=1$, giving
\end{subequations}
\begin{subequations}
\label{BoxCov}%
\begin{align}
D_{\alpha}^{L}\mathcal{T}_{LLLL}^{\alpha\beta\gamma\delta}  &  =-\frac
{i\mathcal{I}_{3}}{16\pi^{2}}\varepsilon^{\beta\gamma\delta\alpha}%
(d^{abe}f^{cde}q_{2\alpha}-d^{ace}f^{bde}q_{3\alpha}+d^{ade}f^{bce}q_{4\alpha
})\ ,\\
D_{\beta}^{L}\mathcal{T}_{LLLL}^{\alpha\beta\gamma\delta}  &  =D_{\gamma}%
^{L}\mathcal{T}_{LLLL}^{\alpha\beta\gamma\delta}=D_{\delta}^{L}\mathcal{T}%
_{LLLL}^{\alpha\beta\gamma\delta}=0\ .
\end{align}
Finally, for the pentagon diagrams, we set all the $b_{l}^{ijk}=0$ except for
$b_{4}^{ijk}=1$ for $i,j,k\neq1$ in Eq.~(\ref{M0Pen}) since the fourth leg is
then that with momentum $q_{1}$, and obtain
\end{subequations}
\begin{subequations}
\label{PenCov}%
\begin{align}
iD_{\alpha}^{L}\mathcal{T}_{LLLLL}^{\alpha\beta\gamma\delta\varepsilon,abcde}
&  =-\frac{i\mathcal{I}_{3}}{16\pi^{2}}\varepsilon^{\beta\gamma\delta
\varepsilon}d^{efg}(f^{adf}f^{bcg}-f^{acf}f^{bdg}+f^{abf}f^{cdg})\ ,\\
iD_{\beta}^{L}\mathcal{T}_{LLLLL}^{\alpha\beta\gamma\delta\varepsilon,abcde}
&  =iD_{\gamma}^{L}\mathcal{T}_{LLLLL}^{\alpha\beta\gamma\delta\varepsilon
,abcde}=iD_{\delta}^{L}\mathcal{T}_{LLLLL}^{\alpha\beta\gamma\delta
\varepsilon,abcde}=iD_{\varepsilon}^{L}\mathcal{T}_{LLLLL}^{\alpha\beta
\gamma\delta\varepsilon,abcde}=0\ .
\end{align}
These divergences reproduce precisely Eq.~(\ref{CovLLL}). The fact that all
the other covariant divergences give zero is implied in Eq.~(\ref{CovLLL}),
but now emerges explicitly.

This is an important point. It seems to show that when the Bardeen-Zumino
polynomial redefines the current with $SU(N)$ index $a$, all the others end up
being affected too since their divergences have to be brought to zero. In
reality, this reflects an intrinsically different perspective on what is a
current in the functional or diagrammatic approach, as was already alluded to
in Sec.~\ref{CTBose}. Specifically, when we define the consistent current as
$J_{\mu}^{a}\equiv\delta\Gamma\lbrack A]/\delta A^{a,\mu}$, $a$ is a dummy
index (as is $\mu$) and the current is in this sense unique. Its anomalous
divergence is then directly expressed in terms of gauge fields, and there is a
unique Bardeen-Zumino polynomial able to make this current covariant. Gauge
fields are not redefined in any way, only the current is shifted. Actually,
the effective action is not redefined either, so the consistent anomaly is
still there, just in a different current.

By contrast, in the diagrammatic approach, one computes the three, four, and
five-point effective amplitude without distinguishing between currents and
gauge fields, on the basis that $\bar{\psi}_{L}\gamma^{\mu}\mathbf{A}_{L}%
^{\mu}\psi_{L}=A_{L}^{a,\mu}J_{\mu}^{a}$. The consistent anomaly is then
obtained by asking for the same Ward identity for all the legs of these
diagrams, since it cannot matter which leg is the current and which others are
the gauge fields. For the covariant anomaly, we enforce the conservation of
all but one external current, with the idea that gauge fields are to be
attached to all these conserved legs.

In the functional approach, the various anomalies have clear meanings in terms
of symmetries and dynamics, from which they gather specific properties. It is
the opposite in the diagrammatic approach, with the various anomalies derived
by imposing first some specific properties on the single generic form of the
chiral anomaly, and later relating them with the more fundamental definitions
on the basis of their properties.

There is a way however to reconcile these different perspectives. We know from
Sec.~\ref{SecBardeen} that parameters are equivalent to counterterms. 
Eqs.~(\ref{TriCov}--\ref{PenCov}) proves that the covariant anomaly can be obtained
by a choice of parameters, so adding instead some counterterms to the action should also
lead to these equations. The requirement though is that one must distinguish
one of the gauge field from the others in the total action. In other words,
one should extract the covariant current by performing a gauge variation for a
specific "covariant" field (e.g. that with $SU(N)$ index $a$), keeping all the
others fixed, and the gauge current by varying all the others, keeping the
"covariant" one fixed. With appropriate counterterms, these latter currents
would be conserved and the former would match Eq.~(\ref{CovLLL}). Let us
stress that because this procedure is not legally a true gauge transformation,
the covariant anomaly does not satisfy the consistency condition.

There is an advantage to counterterms compared to shifting the anomalous
current via the Bardeen-Zumino polynomial. Only the former procedure permits
to get a complete and coherent picture of the anomaly in all the currents.
This is already evident in Eqs.~(\ref{TriCov}--\ref{PenCov}), but let us take
a more exotic case to illustrate this point. We start by adding generic free
parameters $Y_{1,2}$ in the Bardeen-Zumino polynomial:%
\end{subequations}
\begin{equation}
\tilde{X}_{L}^{a,\mu}=-\frac{\varepsilon^{\mu\nu\rho\sigma}}{48\pi^{2}%
}\left\langle T^{a}\left(  Y_{1}\{\mathbf{A}_{\nu}^{L},\mathbf{F}_{\rho\sigma
}^{L}\}+Y_{2}\frac{i}{2}\mathbf{A}_{\nu}^{L}[\mathbf{A}_{\rho}^{L}%
,\mathbf{A}_{\sigma}^{L}]\right)  \right\rangle \ , \label{CurrBZv}%
\end{equation}
We then find for $\mathcal{J}_{L}^{a,\mu}=J_{L}^{a,\mu}+\tilde{X}_{L}^{a,\mu}%
$,%
\begin{align}
(D_{\mu}^{L}\mathcal{J}_{L}^{\mu})^{a}  &  =\frac{\varepsilon^{\mu\nu
\rho\sigma}}{32\pi^{2}}\left\langle T^{a}\left(  \frac{1+2Y_{1}}{3}%
\mathbf{F}_{\mu\nu}^{L}\mathbf{F}_{\rho\sigma}^{L}+i\frac{1-Y_{2}}%
{3}\mathbf{A}_{\mu}^{L}\mathbf{F}_{\nu\rho}^{L}\mathbf{A}_{\sigma}^{L}\right.
\right. \nonumber\\
&  \ \ \ \ \ \ \ \ \ \ \left.  \left.  +i\frac{1-2Y_{1}+Y_{2}}{3}%
(\mathbf{F}_{\mu\nu}^{L}\mathbf{A}_{\rho}^{L}\mathbf{A}_{\sigma}%
^{L}+\mathbf{A}_{\mu}^{L}\mathbf{A}_{\nu}^{L}\mathbf{F}_{\rho\sigma}%
^{L})+2\frac{Y_{2}-1}{3}\mathbf{A}_{\mu}^{L}\mathbf{A}_{\nu}^{L}%
\mathbf{A}_{\rho}^{L}\mathbf{A}_{\sigma}^{L}\right)  \right\rangle \ ,
\label{CurrBZg}%
\end{align}
which interpolates between the consistent and covariant anomalies. An
interesting situation arises for $Y_{1}=-1/2$, when the quadratic term cancels
out. In the abelian case, this current would even be conserved. This is a
situation similar to that encountered for the singlet anomaly, where it was
noticed that the anomaly disappears if the Chern-Simon current is absorbed
into the axial current. Of course, we know that the anomaly has not
disappeared, but has been moved in the gauge currents. The question though is
to find an explicit expression for the gauge anomaly in this generic
situation, i.e., for arbitrary $Y_{1,2}$.

To compute this explicitly is actually easy: starting from the general form of
the chiral anomaly, it suffices to work out the value of the free parameters
able to reproduce Eq.~(\ref{CurrBZg}) for one current, and generating Bose
symmetric divergences for all the gauge currents. Let us assign the covariant
current to the $q_{1}$ leg with Lorentz and $SU(N)$ indices $\alpha$ and $a$.
We find from the $LLL$ triangle Eq.~(\ref{AnoLLL}),%
\begin{equation}
a_{1}=\frac{2Y_{1}+1}{3}\ ,\ a_{2}=\frac{1-Y_{1}}{3}\ ,
\end{equation}
This shows clearly that for $Y_{1}=-1/2$, the anomaly is indeed moved out of
the $q_{1}$ leg, but ends up symmetrically split in the $q_{2}$ and $q_{3}$
legs (notice that with these values, $1-a_{1}-a_{2}=a_{2}$). For the $LLLL$
box Eq.~(\ref{AnoLLLL}),%
\begin{equation}
b_{2,3,4}=\frac{3-4Y_{1}+Y_{2}}{12}\ , \label{biBZ}%
\end{equation}
together with $a_{1,2}^{ij}=(1-Y_{1})/3$ except for $a_{1}^{1j}=(2Y_{1}+1)/3$.
Interestingly, the covariant derivatives for the gauge currents are all linear
in their momentum, as they were for the consistent anomaly. For example, we
find%
\begin{equation}
D_{\beta}^{L}\mathcal{T}_{LLLL}^{\alpha\beta\gamma\delta}=-\frac
{\mathcal{I}_{3}}{192\pi^{2}}(-(1-Y_{2})d^{ade}f^{bce}+(1-Y_{2})d^{ace}%
f^{bde}+(3-4Y_{1}+Y_{2})d^{abe}f^{cde})\varepsilon^{\alpha\gamma\delta\beta
}q_{2\beta}\ ,
\end{equation}
which collapses to the consistent result Eq.~(\ref{AnoLLLL2}) for $Y_{1,2}=0$
and vanishes for $Y_{1,2}=1$. Finally, for the pentagon Eq.~(\ref{M0Pen}), we
set all $b_{l}^{ijk}$ to the value in Eq.~(\ref{biBZ}), except for
$b_{4}^{ijk}=(1+4Y_{1}-Y_{2})/4$ for $i,j,k\neq1$. This cancels out the
covariant derivatives for all but the $q_{1}$ current no matter $Y_{1,2}$. As
a check, notice that these parameter values interpolate between those for the
consistent anomaly, $a_{k}^{ij}=1/3$, $b_{l}^{ijk}=1/4$, and those giving the
covariant anomaly in Eqs.~(\ref{TriCov}--\ref{PenCov}).

Having found the explicit expressions for all the covariant derivatives, they
can be used to reconstruct the operator form of the anomaly in the gauge
currents:%
\begin{align}
(D_{\mu}^{L}J_{L}^{\mu})^{a}  &  =\frac{\varepsilon^{\mu\nu\rho\sigma}}%
{24\pi^{2}}\left\langle T^{a}\partial_{\mu}\left(  (1-Y_{1})(\mathbf{A}_{\nu
}^{L}\partial_{\rho}\mathbf{A}_{\sigma}^{C}+\mathbf{A}_{\nu}^{C}\partial
_{\rho}\mathbf{A}_{\sigma}^{L})\frac{{}}{{}}\right.  \right. \nonumber\\
&  \ \ \ \ \ \ \ \ \ \ \ \ \ \ \ \ \ \ \ \left.  \left.  +i\frac{1-Y_{2}}%
{2}\mathbf{A}_{\nu}^{L}\mathbf{A}_{\rho}^{C}\mathbf{A}_{\sigma}^{L}%
-i\frac{1-2Y_{1}+Y_{2}}{2}(\mathbf{A}_{\nu}^{C}\mathbf{A}_{\rho}^{L}%
\mathbf{A}_{\sigma}^{L}+\mathbf{A}_{\nu}^{L}\mathbf{A}_{\rho}^{L}%
\mathbf{A}_{\sigma}^{C})\right)  \right\rangle \ , \label{CurrBZgauge}%
\end{align}
where $\mathbf{A}^{C}$ stands for the field coupled to the covariant current,
and $\mathbf{A}^{L}$ for the true gauge fields. This expression vanishes for
$Y_{1,2}=1$, in agreement with Eqs.~(\ref{TriCov}--\ref{PenCov}), and
collapses to the consistent anomaly for $Y_{1,2}=0$ and $\mathbf{A}%
^{L}=\mathbf{A}^{C}$. Importantly, it does not vanish for $Y_{1}=-1/2$,
showing explicitly why in the abelian case, it is really not a good idea to
cancel the anomaly by a current redefinition. Doing that necessarily breaks
the gauge symmetry. Finally, it must be remarked that because we started by
shifting one of the currents by $\tilde{X}_{L}^{a,\mu}$, Eq.~(\ref{CurrBZv}),
the gauge anomaly can still be expressed as a total derivative.

Summing up, to fully specify the anomaly in a general case, it is necessary to
give not only the divergence of the shifted current, Eq.~(\ref{CurrBZg}), but
also that of the gauge currents, Eq.~(\ref{CurrBZgauge}). The only situation
in which this is somewhat redundant is that of the Bardeen-Zumino shift since
the gauge currents are then, by construction, covariantly conserved.

\subsection{Fujikawa's anomaly derivation} \label{SecFujikawa}

Having described the covariant anomaly, it is now a good time to discuss the
path integral method first proposed by
Fujikawa~\cite{Fujikawa:1979ay,Fujikawa:1980eg}.\ To set the stage, let us
recall a few definitions. For a given model, specified by some fields and
interactions, quantization can be performed by writing the path integral
instead of via canonical quantization rules. For instance, imagine a simple
theory involving a real scalar field $\phi$ and a Dirac fermion $\psi$, with
Lagrangian
\begin{equation}
\mathcal{L}[\phi,\psi,\bar{\psi}]=-\frac{1}{2}\phi(\square+M^{2})\phi
+\bar{\psi}\left(  i \slashed \partial-m\right)  \psi+\lambda\phi\bar{\psi}\psi\ .
\end{equation}
Its generating functional is constructed as the path integral
\begin{equation}
Z[j,\eta,\bar{\eta}]=\int D\phi D\psi D\bar{\psi}\exp i\int\mathcal{L}%
[\phi,\psi,\bar{\psi}]+j\phi+\bar{\eta}\psi+\bar{\psi}\eta\ .
\end{equation}
Specific Green functions (or correlators) are then obtained by functional
differentiation with respect to the sources $j$, $\eta$, and $\bar{\eta}$,
setting them to zero afterwards. Mathematically, the precise meaning of the
path integral measure, and the convergence properties of the integral, are
delicate issues. These will not be addressed here, and instead, we will be
happy with simple axioms sufficient to show that the usual perturbative series
in terms of Feynman diagrams can be reproduced~(in the spirit of
Ref.~\cite{Rivers:1987hi}). Specifically, we can pull the interaction
$\lambda\phi\bar{\psi}\psi$ out of $Z[j,\eta,\bar{\eta}]$ and replace it by
functional derivatives with respect to $j$, $\eta$, and $\bar{\eta}$ at given
points,%
\begin{equation}
Z[J,\eta,\bar{\eta}]=\left(  1+\lambda\int dy_{1}\mathcal{L}_{I}%
(y_{1})+\lambda^{2}\int dy_{1}\mathcal{L}_{I}(y_{1})\int dy_{2}\mathcal{L}%
_{I}(y_{2})+...\right)  Z[j,\eta,\bar{\eta}]_{\lambda=0}\ , \label{PertSer}%
\end{equation}
where $\mathcal{L}_{I}$ is given by $\phi\bar{\psi}\psi$ with $\phi
\rightarrow-i\delta/\delta j$, $\psi\rightarrow-i\delta/\delta\bar{\eta}$, ,
$\bar{\psi}\rightarrow-i\delta/\delta\eta$. Only the scalar and fermion
kinetic terms remain in $Z[J,\eta,\bar{\eta}]_{\lambda=0}$. For them, by
analogy with finite-dimensional Gaussian integrals, we prescribe the result%
\begin{equation}
\int D\phi\exp i\left\{  \int-\frac{1}{2}\phi\left(  \square+M^{2}\right)
\phi+j\phi\right\}  =\frac{1}{\sqrt{\det\left(  \square+M^{2}\right)  }}%
\exp\{-\frac{i}{2}j\frac{1}{\square+M^{2}}j\}\;, \label{ScalarProp}%
\end{equation}
and, owing to their grassmanian nature,%
\begin{equation}
\int D\psi D\bar{\psi}\exp i\left\{  \int\bar{\psi}\left(  i \slashed
\partial-m\right)  \psi+\bar{\eta}\psi+\bar{\psi}\eta\right\}  =\det\left(  i \slashed
\partial-m\right)  \exp\{-i\bar{\eta}\frac{1}{i \slashed \partial-m}\eta\}\;. 
\label{FermionProp}%
\end{equation}
The determinants are independent of the sources, and could be discarded by
properly normalizing $Z[J,\eta,\bar{\eta}]$ by $Z[0,0,0]$. Then, going back to
Eq.~(\ref{PertSer}), the usual perturbative series is indeed reproduced, with
the various interactions stitched together by the propagators in
Eqs.~(\ref{ScalarProp}) and~(\ref{FermionProp}). For gauge fields, this
procedure needs to be slightly amended to deal with the gauge freedom since
the kinetic term is not invertible. In practice, it is necessary to restrict
the measure $DA^{\mu}$ over gauge non-equivalent configuration, bringing in
the Fadeev-Popov ghosts. We refer e.g. to Ref.~\cite{Peskin:1995ev} for more details.

\subsubsection{Anomalies and Jacobian}

Within the path integral approach, the fields are not quantized but remain
classical functions of space-time. As a result, that part of the path integral
stays invariant under all the classical symmetries exhibited by the
Lagrangian. The only place where something could go wrong is in the
integration measure, and we will indeed see that anomalous symmetry
transformations do leave a non-trivial Jacobian generating an extra term in
the Ward identities. In this sense, the path integral formalism exhibits in
the most transparent way possible the statement that anomalies arise from an
incompatibility between a symmetry and quantization, since it is by writing
down a path integral with its functional measure that the theory becomes quantized.

It is only in very specific circumstances that the path integral measure is
not invariant under a symmetry transformation. Let us consider a massless
fermion, whose kinetic term is invariant under both the vector and axial
transformations. The latter then induce the Jacobian%
\begin{equation}
\int D\psi^{\prime}D\bar{\psi}^{\prime}=\int D\psi D\bar{\psi}\mathcal{J}%
\;,\;\;\ln\mathcal{J}=\ln\left(  \det e^{-i\beta\gamma_{5}}\det e^{-i\beta
\gamma_{5}}\right)  =-2i\operatorname*{Tr}(\beta\gamma_{5})+\mathcal{O}%
(\beta^{2})\;, \label{Jac1}%
\end{equation}
where the trace is to be carried over all degrees of freedom, both space-time
and internal. At first sight, one could be tempted to conclude that
$\ln\mathcal{J}\sim\operatorname*{Tr}(\gamma_{5})=0$, and thus the absence of
any anomaly. Indeed, for each Dirac state $\psi=\psi_{L}\oplus\psi_{R}$, there
is a state of chirality $-1$ and $+1$, hence their total contribution gives
zero. However, there are infinitely many states, hence infinitely many $+1$s
and $-1$s. It is only when each $-1$ is paired with a $+1$ that the sum
vanishes. In itself, this pairing prescription serves as a regularization
procedure, and it shows that for free fermions, the Jacobian is indeed
trivial. However, if fermions are coupled to a gauge field, we will see that
this naive regularization is incompatible with the gauge symmetry, and curing
for that ultimately prevents $\operatorname*{Tr}(\gamma_{5})$ from vanishing.

Let us proceed with the evaluation of $\operatorname*{Tr}(\beta\gamma_{5})$,
assuming the fermion is coupled to a non anomalous vector gauge interaction
that we wish to preserve. In that case, a good starting point is to write
$\mathcal{J}$ entirely in terms of gauge-invariant Gaussian integrals. The
generating functional with all the sources set to zero gives simply%
\begin{equation}
Z[0,0]=\int D\psi D\bar{\psi}\exp i\int\bar{\psi}i \slashed D\psi=
\det i \slashed D\;.
\end{equation}
This same result should be attainable after having performed the axial
rotation with $\beta=\beta(x)$. So, provided $\mathcal{J}$ does not depend on
$\psi$ (which will be seen to be the case),%
\begin{equation}
Z[0,0]=\int D\psi^{\prime}D\bar{\psi}^{\prime}\exp i\int\bar{\psi}^{\prime}i\slashed D\psi^{\prime}
=\int D\psi D\bar{\psi}\mathcal{J}\exp i\int\bar{\psi}e^{i\beta\gamma_{5}}i \slashed D e^{i\beta\gamma_{5}}\psi
=\mathcal{J}\det(e^{i\beta\gamma_{5}}i\slashed De^{i\beta\gamma_{5}})\;,
\end{equation}
and thus, equating the two expressions to first order in $\beta$,%
\begin{equation}
\ln\mathcal{J}=\ln\frac{\det(e^{i\beta\gamma_{5}}i\slashed D
e^{i\beta\gamma_{5}})}{\det(i\slashed D)}=
-\frac{1}{2}\operatorname*{Tr}\frac{\{\slashed D,i(\slashed D\beta)
\gamma_{5}\}}{\slashed D^{2}}\;, \label{DetTr}%
\end{equation}
by using $\ln\det X=\operatorname*{Tr}\ln X$. Still to first order in $\beta$,
after exponentiating $\slashed D^{-2}$ and writing the space-time part of the trace explicitly,%
\begin{align}
\ln\mathcal{J}  &  =-\frac{1}{2}\int_{0}^{\infty}d\tau\int dx\;\beta
(x)~\left\{  \frac{\delta}{\delta\beta(x)}\int dz~\operatorname*{Tr}\langle
z|e^{-\tau \slashed D^{2}}\{\slashed D,i(\slashed D\beta)\gamma_{5}\}|z\rangle\right\}_{\beta=0}\nonumber\\
&  =-2i\int_{0}^{\infty}d\tau\int dx\;\beta(x)\;\operatorname*{Tr}\langle
x|e^{-\tau \slashed D^{2}}\gamma_{5} \slashed D^{2}|x\rangle\nonumber\\
&  =2i\int_{0}^{\infty}d\tau\int dx\;\beta(x)\;\frac{d}{d\tau}%
\operatorname*{Tr}\langle x|e^{-\tau \slashed D^{2}}\gamma_{5}|x\rangle\nonumber\\
&  =-2i\underset{\tau\rightarrow0}{\lim}\int dx\;\beta(x)\;\operatorname*{Tr}%
\langle x|e^{-\tau \slashed D^{2}}\gamma_{5}|x\rangle=-2i\underset{\tau\rightarrow0}{\lim}%
\;\operatorname*{Tr}\left(  e^{-\tau \slashed D^{2}}\beta\gamma_{5}\right)  \ . \label{DetTr2}%
\end{align}
This coincides with Eq.~(\ref{Jac1}) but for the exponential factor, which
acts as a \textbf{gauge-invariant UV regulator}. It is its presence which
prevents $\operatorname*{Tr}(\gamma_{5})=0$. To proceed with the evaluation of
$\ln\mathcal{J}$ is now a standard matter (see e.g. Ref.~\cite{Peskin:1995ev}%
). The $\tau$ parametrization is particularly-well suited to the heat kernel
method, leading directly to the result, but for this simple situation we can
go through the calculation explicitly. After writing the operator $\slashed D^{2}$ as
\begin{equation}
\slashed D^{2}=\gamma^{\mu}\gamma^{\nu}D_{\mu}D_{\nu}=D_{\mu}D^{\mu}\mathbf{1}+\frac
{e}{2}\sigma^{\mu\nu}F_{\mu\nu}\;,
\end{equation}
and switching to momentum space by inserting $\langle k|x\rangle=\exp(ikx)$,
\begin{equation}
\ln\mathcal{J}=-2i\underset{\tau\rightarrow0}{\lim}\int dx\ \beta\ \langle
x|\left(  \int\frac{d^{4}k}{(2\pi)^{4}}\langle x|k\rangle\operatorname*{Tr}
\left(  e^{-\tau \slashed D^{2}}\gamma_{5}\right)  \langle k|x\rangle\right) 
 |x\rangle=\frac{ie^{2}}{8\pi^{2}}\int dx\;\beta\;F_{\mu\nu}\tilde{F}^{\mu\nu}\ . 
\label{FinalFuji}%
\end{equation}
Only the $\tau^{2}$ term with two field strength survives the Dirac trace and
the $\tau\rightarrow0$ limit, after which the momentum integral is done
shifting $D_{\mu}\rightarrow D_{\mu}+k_{\mu}$ and rescaling $k_{\mu
}\rightarrow k_{\mu}/\sqrt{\tau}$. This result agrees with the triangle
diagram calculation, Eq.~(\ref{AnoAbelian2}).

\subsubsection{Massless fermions and the index theorem}

The previous method generalizes to all the other anomalies for which gauge
invariance act as a constraint. Specifically, we can immediately recover the
singlet anomaly by inserting the appropriate group generators in the fermion
gauge couplings, and force the traces to extend to gauge indices. It is also
possible to reproduce the covariant anomaly, though this requires a bit more
work. Indeed, the Dirac operator $\slashed D^{2}$ is not hermitian when
both the vector and axial currents are present.
However, $\slashed D \slashed D^{\dagger}$ and $\slashed D^{\dagger}\slashed D$ 
are hermitian, so we have to use them to form gauge-invariant regulators.
The Jacobian for an axial and vector transformations are $\operatorname*{Tr}%
(\beta\gamma_{5})$ and $\operatorname*{Tr}(\beta)$, respectively, and those
can be regulated as (see e.g. Ref.~\cite{Bertlmann:1996xk})%
\begin{subequations}
\begin{align}
\gamma_{5}  &  \rightarrow\left(  e^{-\tau(\slashed D
\slashed D^{\dagger})}+e^{-\tau( \slashed D^{\dagger}\slashed D)}\right) \gamma_{5}\ ,\\
1  &  \rightarrow\left(  e^{-\tau(\slashed D \slashed D^{\dagger})}
-e^{-\tau(\slashed D^{\dagger}\slashed D)}\right)  1\ .
\end{align}
From here, projecting on fermion left and right components, the calculation
collapses to that of the abelian anomaly, leading to Eq.~(\ref{AVcov}). Notice
how the vector current is automatically conserved if $\slashed D= \slashed D^{\dagger}$, 
in agreement with the singlet anomaly. 

The Fujikawa method permits to show another feature of the anomaly, which is
its deep connection with the index theorem. In the previous derivation, we
used a trick to perform the proper time $\tau$ integral in Eq.~(\ref{DetTr2}).
One way to ground these manipulations on more solid mathematical grounds is to
go to some finite Euclidian space, in which one could imagine that $i \slashed D$ 
has a discrete set of eigenstates $|n\rangle$ of eigenvalue $\lambda_{n}$.
We can then write the Jacobian as%
\end{subequations}
\begin{equation}
\ln\mathcal{J}=-2i\underset{\tau\rightarrow0}{\lim}\;\operatorname*{Tr}\left(
e^{-\tau \slashed D^{2}}\beta\gamma_{5}\right)  =-2i\underset{\tau\rightarrow0}{\lim}\int
dx\;\operatorname*{Tr}\sum_{n}e^{-\tau\lambda_{n}^{2}}\langle x|n\rangle
\beta\gamma_{5}\langle n|x\rangle\ . \label{ZeroM}%
\end{equation}
Most states cancel in the sum because they are not simultaneous eigenstates of
$i \slashed D$ and $\gamma_{5}$. Indeed, $\{i \slashed D,\gamma_{5}\}=0$ implies that if 
$i \slashed D\langle n|x\rangle=\lambda_{n}\langle n|x\rangle$, then
$i \slashed D\gamma_{5}\langle n|x\rangle=-\lambda_{n}\gamma_{5}\langle n|x\rangle$.
Having different eigenvalues, $\langle n|x\rangle$ and $\gamma_{5}\langle
n|x\rangle$ are orthogonal. They are obviously not eigenfunctions of
$\gamma_{5}$; those would rather be $\langle n|x\rangle\pm\gamma_{5}\langle
n|x\rangle$. Their contributions thus drop out of Eq.~(\ref{ZeroM}).

What remains are those eigenstates for which $\lambda_{n}=0$, the so-called
\textbf{fermionic zero modes}. In that case, $i \slashed D$ and $\gamma_{5}$ can be 
diagonalized simultaneously. Since the eigenvalues
of $\gamma_{5}$ are $+1$ or $-1$, the sum in Eq.~(\ref{ZeroM}) gives the
difference in the numbers of $+1$ and $-1$ eigenvalues, $n_{+}-n_{-}$. This is
called the \textbf{index} of the $i \slashed D$ operator, and its connection to the anomaly is the famous
\textbf{Atiyah-Singer index theorem}~\cite{Atiyah:1968mp}%
\begin{equation}
\operatorname*{ind}(i \slashed D)=\frac{1}{8\pi^{2}}\int_{V}d^{4}x\ F_{\mu\nu}\tilde{F}^{\mu\nu}\ .
\label{ASTheorem}%
\end{equation}
This connection between the index and the anomaly was first caught using the
point-splitting method in Ref.~\cite{Nielsen:1977aw}. It indirectly proves the
statement that the anomaly does not receive higher order corrections since the
index is necessarily an integer. More profoundly, it also connects the index
with the topology of the gauge group. From the perspective of the right-hand
side of Eq.~(\ref{ASTheorem}), the integer nature of the index comes from
Chern's theorem, that is, the fact that the volume integral of $F_{\mu\nu
}\tilde{F}^{\mu\nu}$ measures the winding of the gauge configuration at infinity.

For the consistent anomaly, the situation is more complex for three reasons.
First, the purely chiral Dirac operator $\slashed D_{L}$ is not only not hermitian, *
it is even not well-defined: when acting on
left spinors, it produces a right spinor. Second, we know the regulator cannot
be gauge invariant. Instead, it needs to break the gauge symmetry in just the
right way to ensure the anomaly identified from $\ln\mathcal{J}$ satisfies the
consistency condition. Third, there is no index theorem since both sides of
Eq.~(\ref{ASTheorem}) are gauge invariant, so one cannot directly rely on
geometric arguments. There is, however, an elegant way to treat this problem,
at least conceptually because the algebra remains quite involved and will not
be described here. We know that the consistent anomaly in $d$ dimensions is
related to the singlet anomaly in $d+2$ dimensions, see Sec.~\ref{SecChernSimons}. 
So, starting with a Dirac operator in $d+2$ dimensions, by carefully identifying a
$d$-dimensional component and constraining the geometry of the two extra
dimensions, it is possible to get to the consistent
anomaly~\cite{Alvarez-Gaume:1983ict}. This is a profound result that does
extend the connection with topology to the consistent anomaly.

\subsubsection{Massive fermions and Sutherland-Veltman}

Introducing a fermion mass term does not alter much the derivation of the
abelian anomaly using the Fujikawa method. The interpretation in terms of the
index is lost though, since the massive Dirac operator has no zero modes. At
the same time, technically, having an extra $\exp(-\tau m^{2})$ helps with the
convergence of Eq.~(\ref{DetTr2}) at $\tau\rightarrow\infty$.

\begin{figure}[t]
\centering\includegraphics[width=0.80\textwidth]{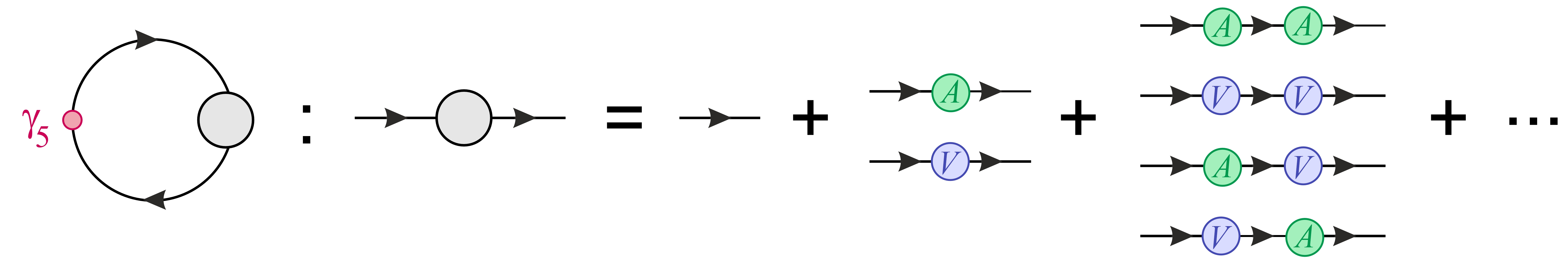}\caption{Graphical
representation of the pseudoscalar loops of Eq.~(\ref{BardeenSVF}).}%
\label{Fig8}%
\end{figure}

There is however a new class of anomalies directly accessible within the
functional formalism when $m\neq0$ whenever the Sutherland-Veltman theorem
holds. Remember that in that case, performing a mass expansion of
\begin{equation}
\partial_{\mu}A^{\mu}=2imP+\frac{e^{2}}{8\pi^{2}}F_{\mu\nu}\tilde{F}^{\mu\nu
}\ ,
\end{equation}
the axial current is $\mathcal{O}(m^{-2})$, but the anomaly stays
$\mathcal{O}(m^{0})$ because there are now pseudoscalar diagrams which are
$\mathcal{O}(m^{-1})$. In practice, this means there is no need to compute the
axial term at all. To remove it from the Jacobian in Eq.~(\ref{DetTr}), it
suffices to take $\beta$ constant, leading to%
\begin{equation}
\ln\mathcal{J}=\ln\frac{\det(e^{i\beta\gamma_{5}}(i \slashed D-m)
e^{i\beta\gamma_{5}})}{\det(i \slashed D-m)}\overset{\beta=c^{st}}{=}
\operatorname*{Tr}\frac{2im\beta\gamma_{5}}{i
\slashed D-m}+\mathcal{O}(\beta^{2})\;. \label{BardeenSVF}%
\end{equation}
By expanding the fermion propagator as a geometric series in the gauge fields,
one reproduces the whole series of pseudoscalar loop diagrams, see
Fig.~\ref{Fig8}. This permits to recover the singlet or Bardeen anomalies as
given by Eq.~(\ref{AnoSinglet1}) and Eq.~(\ref{AVC2}). Remember though that
none of these diagrams are anomalous, so they do not need a regulator to be
computed. Of course, despite of this, let us stress that one specific kind of
regulator is implicitly required to ensure the Sutherland-Veltman theorem
holds in the first place.

When the Sutherland-Veltman theorem does not hold, the functional formalism
does not bring much if one wishes to consistently keep track of all the
ambiguities. Certainly, the compact result given in Eq.~(\ref{MassiveAno}) can
be recovered if, after including both axial and vector currents in $\slashed D$, 
one expands $\ln\det(i\slashed D-m)$ as a geometric series and extract the $\mathcal{O}(m^{0})$ terms of the
three-, four-, and five-point diagrams. There is nothing new here as this is
precisely the usual perturbative calculation. Further,
this is still a long way from the most general form of the chiral anomaly in
Eqs.~(\ref{AnoLLL}),~(\ref{AnoLLLL}), and~(\ref{M0Pen}), because each
configuration of axial and vector currents ends up with its own coefficients,
and pseudoscalar loops still need to be added.

\subsection{Global anomalies in the Standard Model}

The expression for the covariant anomaly, or any other for that matter,
remains correct if the Lie algebra is not simple. Further, if it factorizes
into $G_{1}\otimes G_{2}$ say, there is no problem to cease coupling the
generators of $G_{1}$ to true gauge fields. With $G_{1}$ a global symmetry, it
makes most sense phenomenologically for its generators to carry the whole
anomaly, leaving the gauge symmetry $G_{2}$ intact, thus calling in the chiral
anomaly in its covariant form. In practice, for a set of the fermions
transforming in a representation of $G_{1}\otimes G_{2}$, there will be an
anomaly provided%
\begin{equation}
\left\langle T_{G_{1}}^{a}\{T_{G_{2}}^{b},T_{G_{2}}^{c}\}\right\rangle
\neq0\ .
\end{equation}
In this expression, it is understood that $T_{G_{1}}^{a}$ actually stands for
$T_{G_{1}}^{a}\otimes1_{G_{2}}$ and $T_{G_{2}}^{b,c}$ for $1_{G_{1}}\otimes
T_{G_{2}}^{b,c}$. Thus, $T_{G_{1}}^{a}$ has to be either a singlet or one of
the diagonal generators (those of the Cartan subalgebra), and for each of its
diagonal entries, one takes the trace over $\{T_{G_{2}}^{b},T_{G_{2}}^{c}\}$.
In this way, the $d$ symbol of the general chiral anomaly collapses to a sum
of quadratic Casimir invariants.

In the absence of axial gauge fields, i.e., with $F_{\mu\nu}^{A}=0$ in
Eq.~(\ref{AVcov}), the covariant anomaly then matches precisely onto the
singlet anomaly of Eq.~(\ref{AnoSinglet1}). For example, imagine that $G_{1}$
stands for the $U(2)_{A}$ part of the $N_{f}=2$ chiral symmetry, $G_{2}$ for
the $U(1)_{em}\otimes SU(3)_{C}$ gauge interactions, and consider the set of
fermions $(u^{1},u^{2},u^{3},d^{1},d^{2},d^{3})^{T}$ where the superscript
spells out the color indices. The relevant generators are then the six-by-six
matrices%
\begin{equation}
T_{C}^{a}=\frac{1}{2}\left(
\begin{array}
[c]{cc}%
\boldsymbol{\lambda}^{a} & 0\\
0 & \boldsymbol{\lambda}^{a}%
\end{array}
\right)  \;,\;\;T_{em}=\left(
\begin{array}
[c]{cc}%
2/3\times\mathbf{1} & 0\\
0 & -1/3\times\mathbf{1}%
\end{array}
\right)  \;,\;\;T_{3}=\frac{1}{2}\left(
\begin{array}
[c]{cc}%
\mathbf{1} & 0\\
0 & -\mathbf{1}%
\end{array}
\right)  \;,\ T_{0}=\left(
\begin{array}
[c]{cc}%
\mathbf{1} & 0\\
0 & \mathbf{1}%
\end{array}
\right)  \;.
\end{equation}
Note that $[T_{em},T_{C}^{a}]=0$, and $\langle T_{C}^{a}\rangle=0$. With this,
we can construct two non-vanishing traces:
\begin{align}
\partial^{\mu}A_{\mu}^{0}  &  =\frac{g_{s}^{2}}{16\pi^{2}}G_{\mu\nu}^{b}%
\tilde{G}^{c,\mu\nu}\operatorname*{Tr}(T^{0}\{T_{C}^{b},T_{C}^{c}%
\})=\frac{\alpha_{s}}{2\pi}G_{\mu\nu}^{a}\tilde{G}^{a,\mu\nu}\ ,\;\\
\partial^{\mu}A_{\mu}^{3}  &  =\frac{e^{2}}{16\pi^{2}}F_{\mu\nu}\tilde{F}%
^{\mu\nu}\operatorname*{Tr}(T^{3}\{T_{em},T_{em}\})=\frac{\alpha}{4\pi}%
F_{\mu\nu}\tilde{F}^{\mu\nu}\ ,\;
\end{align}
which are, up to the different normalization of the generators, the two-flavor
expressions of the axial anomaly of Eq.(\ref{SingletAnoChPT}) and of the
abelian anomaly of Eq.~(\ref{QEDAnoChpt}), respectively.

In the presence of an axial gauge interaction, a similar reduction permits to
describe two new situations: that of an anomalous global axial current coupled
to axial gauge interactions, and that of an anomalous global vector current in
the presence of both axial and vector gauge interactions. We will see below
that this is very relevant for the SM, for which the baryon and lepton number
currents are vectorial but anomalous because the weak interaction has an axial
component. We could not address these two situations in the first section
because they cannot be described using a Pauli-Villars regulator, which
automatically enforce the Sutherland-Veltman theorem. Indeed, whenever this
theorem holds, the anomaly is moved out of the vector currents, and spread
symmetrically over all the axial currents. Such configurations correspond to
the Bardeen anomaly, not to the covariant anomaly.

\subsubsection{The flavor symmetries and their anomalies}\label{SecSMFlavorSym}

The gauge interactions of the SM, as well as of many of its extensions like
the Two-Higgs-Doublet Model or the Minimal Supersymmetric Standard Model, do
not distinguish between the three families of fermions.\ Technically, that
sector of the theory is thus invariant under the large global \textbf{flavor
symmetry group~}\cite{Chivukula:1987py}%
\begin{equation}
G_{F}=U(3)^{5}=U(3)_{q}\otimes U(3)_{u}\otimes U(3)_{d}\otimes U(3)_{\ell
}\otimes U(3)_{e}\;. \label{GF}%
\end{equation}
These $U(3)$s act on flavor indices, i.e., they independently reorganize the
three generations for each fermion species, i.e., the left quark and lepton
doublets $q_{L}$ and $\ell_{L}$, and the right quark and lepton singlets
$u_{R}$, $d_{R}$, $e_{R}$. In the SM, this symmetry is explicitly broken in
the scalar sector by the Yukawa couplings, but this does not make it useless.
Whenever a term breaks a symmetry, it can be treated as spurions, i.e., a
non-dynamical field with the appropriate symmetry properties to restore the
symmetry. The interest of artificially restoring the $G_{F}$ symmetry in this
way is twofold. First, it permits to predict the flavor structure of any SM
process, and second, it can be used to set limits on the flavor structures of
new physics scenarios. In this latter case, even if such a model has many new
couplings breaking $G_{F}$, one can nevertheless restricts the available
spurions to just the SM Yukawa couplings. This forces all the new couplings to
be expressible in terms of the Yukawa couplings. They thus automatically
inherit their peculiar hierarchies, which may help evade some experimental
constraints on flavor transitions. This hypothesis is then called
\textbf{Minimal Flavor Violation}~\cite{DAmbrosio:2002vsn} (see e.g.
Ref.~\cite{Smith:2015nxt} for a review).

Because $G_{F}$ treats separately the left and right-handed fermion fields,
part of it must again be broken by chiral anomalies. Said differently, this
huge global symmetry is not compatible with the introduction of gauge
interactions (even though it was identified precisely looking at the gauge
part of the SM Lagrangian). In full generality, the anomalies of the global
flavor symmetry are those of the combined $G_{F}\otimes G_{SM}$ symmetry
group, but with the constraint of being covariant with respect to $G_{SM}$. In
other words, they are generically of the form%
\begin{equation}
\partial^{\mu}\bar{\psi}_{L}\gamma_{\mu}T_{G_{F}}^{a}\psi_{L}=-\frac{g^{2}%
}{32\pi^{2}}F_{\mu\nu}^{b}\tilde{F}^{c,\mu\nu}\left\langle T_{G_{F}}%
^{a}\{T_{G_{SM}}^{b},T_{G_{SM}}^{c}\}\right\rangle \;, \label{AnoMFV}%
\end{equation}
where $T_{G_{SM}}^{b}$ is the SM generator corresponding to the fermion
$\psi_{L}$, $F_{\mu\nu}^{a}$ and $g$ the corresponding SM field strength and
coupling constant, and all the SM fermions are defined as left-handed fields,
see Eq.~(\ref{SMfermions}).

Writing $U(3)=SU(3)\otimes U(1)$, the five $SU(3)$ of $G_{F}$ are not
anomalous since
\begin{equation}
\operatorname*{Tr}(T_{SU(3)_{\psi}}^{a}\{T_{G_{SM}}^{b},T_{G_{SM}}%
^{c}\})=\operatorname*{Tr}T_{SU(3)_{\psi}}^{a}\operatorname*{Tr}\{T_{G_{SM}%
}^{b},T_{G_{SM}}^{c}\}=0\ ,
\end{equation}
so the covariant anomalies Eq.~(\ref{AnoMFV}) collapse to singlet anomalies
(there can be also Bardeen anomalies internal to $G_{F}$, see Eq.~(\ref{AVC})
or~(\ref{AVC2}), but this will not concern us here). Actually, none of the
five $U(1)$ symmetries survives quantization, with the anomalies in their
associated currents taking the generic form%
\begin{equation}
\partial^{\mu}\bar{\psi}_{L}\gamma_{\mu}\psi_{L}=-\frac{N_{f}}{16\pi^{2}%
}\left[  d_{L}\mathcal{I}_{2}^{C}g_{s}^{2}G_{\mu\nu}^{a}\tilde{G}^{a,\mu\nu
}+d_{C}\mathcal{I}_{2}^{L}g^{2}W_{\mu\nu}^{i}\tilde{W}^{i,\mu\nu}+d_{C}%
d_{L}\mathcal{I}_{2}^{Y}g^{\prime2}B_{\mu\nu}\tilde{B}^{\mu\nu}\right]  \;.
\end{equation}
The Casimir invariants $\mathcal{I}_{2}^{C,L,Y}$ are associated to the
$SU(3)_{C}$, $SU(2)_{L}$, $U(1)_{Y}$ representations carried by $\psi_{L}$,
and $d_{C,L}$ are their corresponding dimensions. Explicitly, the five $U(1)$
currents obey:%
\begin{equation}
\left(
\begin{array}
[c]{c}%
\partial_{\mu}J_{q_{L}}^{\mu}\\
\partial_{\mu}J_{u_{R}^{\mathrm{C}}}^{\mu}\\
\partial_{\mu}J_{d_{R}^{\mathrm{C}}}^{\mu}\\
\partial_{\mu}J_{\ell_{L}}^{\mu}\\
\partial_{\mu}J_{e_{R}^{\mathrm{C}}}^{\mu}%
\end{array}
\right)  =-\frac{N_{f}}{16\pi^{2}}\left(
\begin{array}
[c]{ccc}%
1 & 3/2 & 1/6\\
1/2 & 0 & 4/3\\
1/2 & 0 & 1/3\\
0 & 1/2 & 1/2\\
0 & 0 & 1
\end{array}
\right)  \cdot\left(
\begin{array}
[c]{c}%
g_{s}^{2}G_{\mu\nu}^{a}\tilde{G}^{a,\mu\nu}\\
g^{2}W_{\mu\nu}^{i}\tilde{W}^{i,\mu\nu}\\
g^{\prime2}B_{\mu\nu}\tilde{B}^{\mu\nu}%
\end{array}
\right)  \;. \label{U1mat1}%
\end{equation}

Since there are only three gauge groups, two anomaly-free combinations must
exist. As a first step, it is useful to reorganize the $U(1)$s to single out
those corresponding to the global hypercharge $Y$, baryon number $\mathcal{B}$
and lepton number $\mathcal{L}$:%
\begin{equation}
\left(
\begin{array}
[c]{c}%
J_{Y}^{\mu}\\
J_{\mathcal{B}}^{\mu}\\
J_{\mathcal{L}}^{\mu}\\
J_{PQ}^{\mu}\\
J_{E}^{\mu}%
\end{array}
\right)  =\left(
\begin{array}
[c]{ccccc}%
1/3 & -4/3 & 2/3 & -1 & 2\\
1/3 & -1/3 & -1/3 & 0 & 0\\
0 & 0 & 0 & 1 & -1\\
0 & 1 & 1 & 0 & 1\\
0 & 0 & 0 & 0 & 1
\end{array}
\right)  \cdot\left(
\begin{array}
[c]{c}%
\partial_{\mu}J_{q_{L}}^{\mu}\\
\partial_{\mu}J_{u_{R}^{\mathrm{C}}}^{\mu}\\
\partial_{\mu}J_{d_{R}^{\mathrm{C}}}^{\mu}\\
\partial_{\mu}J_{\ell_{L}}^{\mu}\\
\partial_{\mu}J_{e_{R}^{\mathrm{C}}}^{\mu}%
\end{array}
\right)  \;. \label{U1mat2}%
\end{equation}
In this basis, the currents now have the anomalies:%
\begin{equation}
\left(
\begin{array}
[c]{c}%
\partial_{\mu}J_{Y}^{\mu}\\
\partial_{\mu}J_{\mathcal{B}}^{\mu}\\
\partial_{\mu}J_{\mathcal{L}}^{\mu}\\
\partial_{\mu}J_{PQ}^{\mu}\\
\partial_{\mu}J_{E}^{\mu}%
\end{array}
\right)  =-\frac{N_{f}}{16\pi^{2}}\left(
\begin{array}
[c]{ccc}%
0 & 0 & 0\\
0 & 1/2 & -1/2\\
0 & 1/2 & -1/2\\
1 & 0 & 8/3\\
0 & 0 & 1
\end{array}
\right)  \cdot\left(
\begin{array}
[c]{c}%
g_{s}^{2}G_{\mu\nu}^{a}\tilde{G}^{a,\mu\nu}\\
g^{2}W_{\mu\nu}^{i}\tilde{W}^{i,\mu\nu}\\
g^{\prime2}B_{\mu\nu}\tilde{B}^{\mu\nu}%
\end{array}
\right)  \;. \label{AnoCurr}%
\end{equation}
Without surprise, $\partial_{\mu}J_{Y}^{\mu}=0$ since the fermionic charges
under this global flavor $U(1)$ are aligned with those of the gauge $U(1)_{Y}%
$, which has to be anomaly-free. The other anomaly-free combination is
$\mathcal{B}-\mathcal{L}$, so $U(1)_{\mathcal{B}-\mathcal{L}}$ could be a
viable candidate for a new gauge interaction. The problem though is that if it
is gauged, then flavor currents can have a new anomaly into pairs of
$U(1)_{\mathcal{B}-\mathcal{L}}$ gauge bosons:
\begin{equation}
\partial^{\mu}\bar{\psi}_{L}\gamma_{\mu}\psi_{L}=-\frac{N_{f}}{16\pi^{2}}%
d_{C}d_{L}\mathcal{I}_{2}^{\mathcal{B}-\mathcal{L}}g_{\mathcal{B}-\mathcal{L}%
}^{2}X_{\mu\nu}\tilde{X}^{\mu\nu}\;\rightarrow\partial_{\mu}J_{\mathcal{B}%
-\mathcal{L}}^{\mu}=-\frac{N_{f}}{16\pi^{2}}g_{\mathcal{B}-\mathcal{L}}%
^{2}X_{\mu\nu}\tilde{X}^{\mu\nu}\;,
\end{equation}
where $g_{\mathcal{B}-\mathcal{L}}$ and $X^{\mu\nu}$ are the
$U(1)_{\mathcal{B}-\mathcal{L}}$ coupling constant and field strength, and
$\mathcal{I}_{2}^{\mathcal{B}-\mathcal{L}}=2/3$, $1/3$, $1/3$, $2$, $1$ for
$q_{L},u_{R},d_{R},\ell_{L},e_{R}$. To actually gauge $U(1)_{\mathcal{B}%
-\mathcal{L}}$, the trick is to also introduce right-handed neutrino $\nu_{R}%
$. Its current $\partial^{\mu}(\bar{\nu}_{R}\gamma_{\mu}\nu_{R})$ only has a
$X_{\mu\nu}\tilde{X}^{\mu\nu}$ term since $\nu_{R}$ is neutral under the SM
gauge groups, and this term precisely cancel that of the other fermions so
that $\partial_{\mu}J_{\mathcal{B}-\mathcal{L}}^{\mu}=0$. This is the
situation in many Grand Unified extensions of the SM. As discussed in Sec.~\ref{ConsistSM},
the $SO(10)$ theory is anomaly free, $\operatorname*{Tr}(T^{a}\{T^{b}%
,T^{c}\})=0$ for all representations, and embeds the $U(1)_{\mathcal{B}%
-\mathcal{L}}$ symmetry among its generators. This is possible because for the
fermions to fit in the fundamental $16$ representation, a field with precisely
the $\nu_{R}$ quantum numbers has to be present.

\subsubsection{On the baryon and lepton number currents}

\begin{figure}[t]
\centering\includegraphics[width=0.97\textwidth]{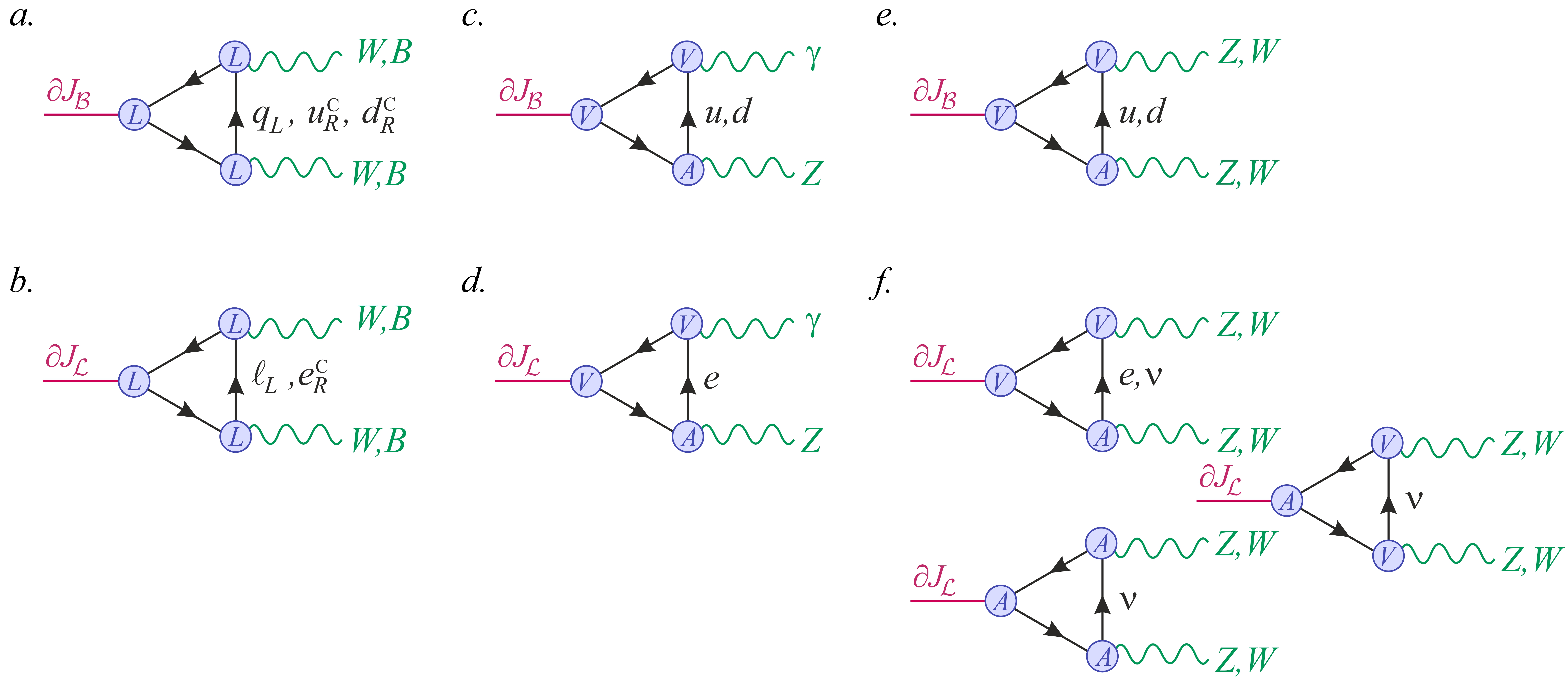}\caption{The
various triangle diagrams contributing to the baryon and lepton number current
anomalies. Diagrams $a.$ and $b.$ are in the chiral representation, while the
others are in the Dirac representation. For $f.$, the last two involving
neutrinos drop out if $\nu_{R}$ is introduced, while the $\nu$ contribution to
the first one gets doubled.}%
\label{FigBL}%
\end{figure}

As an application of the covariant anomaly, it is instructive to check
explicitly that $\partial_{\mu}J_{\mathcal{B}}^{\mu}$ and $\partial_{\mu
}J_{\mathcal{L}}^{\mu}$ take the same form before or after electroweak
symmetry breaking. Specifically, above the electroweak scale, the currents
take a chiral form%
\begin{equation}
J_{\mathcal{B}}^{\mu}=\frac{1}{3}\bar{q}_{L}\gamma^{\mu}q_{L}-\frac{1}{3}%
\bar{u}_{R}^{\mathrm{C}}\gamma^{\mu}u_{R}^{\mathrm{C}}-\frac{1}{3}\bar{d}%
_{R}^{\mathrm{C}}\gamma^{\mu}d_{R}^{\mathrm{C}}\ ,\ \ \ J_{\mathcal{L}}^{\mu
}=\bar{\ell}_{L}\gamma^{\mu}\ell_{L}-\bar{e}_{R}^{\mathrm{C}}\gamma^{\mu}%
e_{R}^{\mathrm{C}}\ ,
\end{equation}
and their anomalies are given in Eq.~(\ref{AnoCurr}), see Figs.~\ref{FigBL}$a$
and~\ref{FigBL}$b$. In the broken phase, it suffices to rotate to the
electroweak mass eigenstates $W_{\mu}^{3}=c_{W}Z_{\mu}+s_{W}A_{\mu}$, $B_{\mu
}=-s_{W}Z_{\mu}+c_{W}A_{\mu}$ to get the divergences:%
\begin{equation}
\partial_{\mu}J_{\mathcal{B}}^{\mu}=\partial_{\mu}J_{\mathcal{L}}^{\mu}%
=-\frac{N_{f}}{16\pi^{2}}\left(  \frac{e^{2}}{c_{W}s_{W}}Z_{\mu\nu}\tilde
{F}^{\mu\nu}+\frac{e^{2}}{2c_{W}^{2}s_{W}^{2}}\left(  1-2s_{W}^{2}\right)
Z_{\mu\nu}\tilde{Z}^{\mu\nu}+g^{2}W_{\mu\nu}^{+}\tilde{W}^{-,\mu\nu}\right)
\ , \label{dJBdJL2}%
\end{equation}
where $g^{\prime}c_{W}=e=gs_{W}$ with $c_{W}=\cos\theta_{W}$, $s_{W}%
=\sin\theta_{W}$. Now, the question is whether this same result can be
obtained directly in the broken phase, using the usual%
\begin{equation}
J_{\mathcal{B}}^{\mu}=\frac{1}{3}\bar{u}\gamma^{\mu}u+\frac{1}{3}\bar{d}%
\gamma^{\mu}d\ ,\ \ \ J_{\mathcal{L}}^{\mu}=\frac{1}{2}\bar{\nu}\gamma^{\mu
}\nu-\frac{1}{2}\bar{\nu}\gamma^{\mu}\gamma_{5}\nu+\bar{e}\gamma^{\mu}e\ ,
\label{BLDiracCurr}%
\end{equation}
together with the $\bar{\psi}_{f}(g_{X,V}^{f}\gamma^{\mu}+g_{X,A}^{f}%
\gamma^{\mu}\gamma_{5})\psi_{f}$ couplings to gauge bosons, where%
\begin{align}
g_{g,V}^{f}  &  =g_{s}T_{a}^{f}\ ,\ g_{\gamma,V}^{f}=eQ^{f}\ ,\ g_{g,A}%
^{f}=g_{\gamma,A}^{f}=0\ ,\\
g_{W,V}^{f}  &  =\dfrac{g}{\sqrt{2}}T_{3}^{f},\ \ g_{W,A}^{f}=-\dfrac{g}%
{\sqrt{2}}T_{3}^{f}\ ,\ \ g_{Z,V}^{f}=\dfrac{g}{2c_{W}}(T_{3}^{f}-2s_{W}%
^{2}Q^{f})\ ,\ g_{Z,A}^{f}=-\dfrac{g}{2c_{W}}T_{3}^{f}\ .
\end{align}
What makes this calculation non-trivial is the fact that gauge interactions
are both axial and vectorial, the $\mathcal{B}$ current is purely vectorial,
and the $\mathcal{L}$ current has again both components. Thus, we have to
consider all kinds of configurations for the triangle diagrams, each time
restricting the anomaly to the right leg.

Consider first the $Z_{\mu\nu}\tilde{F}^{\mu\nu}$ term. In the broken phase,
it necessarily comes from the covariant anomaly in the vector current since
the neutrino does not contribute, Figs.~\ref{FigBL}$c$ and~\ref{FigBL}$d$.
Adding the adequate couplings in Eq.~(\ref{AVcov}), one gets%
\begin{align}
\partial_{\mu}J_{\mathcal{B}}^{\mu}|_{V^{\mathcal{B}}A^{Z}V^{\gamma}}  &
=\frac{N_{f}}{8\pi^{2}}\sum_{f=u,d}\left[  2N_{C}\mathcal{B}_{f}g_{Z,A}%
^{f}g_{\gamma,V}^{f}\right]  Z_{\mu\nu}\tilde{F}^{\mu\nu}=-\frac{N_{f}}%
{16\pi^{2}}\dfrac{e^{2}}{c_{W}s_{W}}Z_{\mu\nu}\tilde{F}^{\mu\nu}\ ,\\
\partial_{\mu}J_{\mathcal{L}}^{\mu}|_{V^{\mathcal{L}}A^{Z}V^{\gamma}}  &
=\frac{N_{f}}{8\pi^{2}}\sum_{f=e,\nu}\left[  2\mathcal{L}_{f}g_{Z,A}%
^{f}g_{\gamma,V}^{f}\right]  Z_{\mu\nu}\tilde{F}^{\mu\nu}=-\frac{N_{f}}%
{16\pi^{2}}\dfrac{e^{2}}{c_{W}s_{W}}Z_{\mu\nu}\tilde{F}^{\mu\nu}\ .
\end{align}
Note that it is essential that $\mathcal{B}_{u,d}=1/N_{C}$. Thus, in both
cases, we recover exactly Eq.~(\ref{dJBdJL2}).

The calculation of the $Z_{\mu\nu}\tilde{Z}^{\mu\nu}$ and $W^{+,\mu\nu}%
\tilde{W}_{\mu\nu}^{-}$ terms is similar, though there are even more triangle
amplitudes to consider. Let us describe only $W^{+,\mu\nu}\tilde{W}_{\mu\nu
}^{-}$ as the $Z_{\mu\nu}\tilde{Z}^{\mu\nu}$ case is similar. For
$\partial_{\mu}J_{\mathcal{B}}^{\mu}$, accounting for the two ways to assign
$W$ bosons to the axial and vector couplings, Fig.~\ref{FigBL}$e$ corresponds
to%
\begin{equation}
\partial_{\mu}J_{\mathcal{B}}^{\mu}|_{V^{\mathcal{B}}A^{W}V^{W}}=\frac{N_{f}%
}{8\pi^{2}}\sum_{f=u,d}\left[  2N_{C}\mathcal{B}_{f}g_{W,A}^{f}g_{W,V}%
^{f}\right]  W^{+,\mu\nu}\tilde{W}_{\mu\nu}^{-}=-\frac{N_{f}}{16\pi^{2}}%
g^{2}W^{+,\mu\nu}\tilde{W}_{\mu\nu}^{-}\ ,
\end{equation}
which agrees with Eq.~(\ref{dJBdJL2}). The calculation for $\partial_{\mu
}J_{\mathcal{L}}^{\mu}$ is more involved because of its axial neutrino
component. We have to combine three different configurations: the
$V^{\mathcal{L}}A^{W}V^{W}$ triangle, the $A^{\mathcal{L}}V^{W}V^{W}$
triangle, and the $A^{\mathcal{L}}A^{Z}A^{Z}$ triangle, each time putting the
anomaly on the lepton number current, see Fig.~\ref{FigBL}$f$:%
\begin{align}
\partial_{\mu}J_{\mathcal{L}}^{\mu}|_{V^{\mathcal{L}}A^{W}V^{W}}  &
=\frac{N_{f}}{8\pi^{2}}\left[  2\mathcal{L}_{e}g_{W,A}^{e}g_{W,V}%
^{e}+\mathcal{L}_{\nu}g_{W,A}^{\nu}g_{W,V}^{\nu}\right]  W^{+,\mu\nu}\tilde
{W}_{\mu\nu}^{-}=-\frac{N_{f}}{16\pi^{2}}\dfrac{3g^{2}}{4}W^{+,\mu\nu}%
\tilde{W}_{\mu\nu}^{-}\ ,\\
\partial_{\mu}J_{\mathcal{L}}^{\mu}|_{A^{\mathcal{L}}V^{W}V^{W}}  &
=\frac{N_{f}}{8\pi^{2}}\left[  \mathcal{-L}_{\nu}g_{W,V}^{\nu}g_{W,V}^{\nu
}\right]  W^{+,\mu\nu}\tilde{W}_{\mu\nu}^{-}=\mathcal{-}\frac{N_{f}}{16\pi
^{2}}\dfrac{g^{2}}{8}W^{+,\mu\nu}\tilde{W}_{\mu\nu}^{-}\ ,\\
\partial_{\mu}J_{\mathcal{L}}^{\mu}|_{A^{\mathcal{L}}A^{W}A^{W}}  &
=\frac{N_{f}}{8\pi^{2}}\left[  -\mathcal{L}_{\nu}g_{W,A}^{\nu}g_{W,A}^{\nu
}\right]  W^{+,\mu\nu}\tilde{W}_{\mu\nu}^{-}=-\frac{N_{f}}{16\pi^{2}}%
\dfrac{g^{2}}{8}W^{+,\mu\nu}\tilde{W}_{\mu\nu}^{-}\ .
\end{align}
Summing these three pieces reproduces Eq.~(\ref{dJBdJL2}).

In the broken phase, fermions are massive but this does not alter the
divergence of the $\mathcal{B}$ and $\mathcal{L}$ currents. First, whenever
the anomalous current is vectorial, its divergence directly gives the anomaly.
For neutrinos, there is also an axial current and, by virtue of the classical
Ward identity $\partial^{\mu}A_{\mu}=2imP$, the divergences of the
$A^{\mathcal{L}}V^{W}V^{W}$ and $A^{\mathcal{L}}A^{Z}A^{Z}$ triangle
amplitudes are accompanied by pseudoscalar triangle diagrams. Yet, the
neutrino being massless, those pieces trivially vanish, and the final result
is the same as in the unbroken phase. Another way to see that these terms must
be absent is to actually make the neutrino massive by adding right-handed
neutrinos. If those have no Majorana mass terms, $\mathcal{L}$ stays exact but
becomes purely vectorial, so there is no pseudoscalar amplitude to consider.
Yet, the final result for the $Z_{\mu\nu}\tilde{Z}^{\mu\nu}$ and $W^{+,\mu\nu
}\tilde{W}_{\mu\nu}^{-}$ terms must be the same since right-handed neutrinos
are fully neutral under the SM gauge group.

\subsubsection{Strong but no weak CP puzzle}

We have seen in Sec.~\ref{ConsistSM} that the vacuum of non-abelian gauge theories is not
trivial. To keep track of that, one needs to add to the Yang-Mills Lagrangian
a new $\mathcal{CP}$-violating parameter, the $\theta$ term. Applied to the
SM, we should in principle add one such parameter for $SU(3)_{C}$ and one for
$SU(2)_{L}$. Even if it can be trivially rotated away, let us further add a
$\theta$ term for $U(1)_{Y}$, and consider%
\begin{equation}
\mathcal{L}_{\theta}=\theta_{C}\frac{\alpha_{s}}{8\pi}G_{\mu\nu}^{a}\tilde
{G}^{a,\mu\nu}+\theta_{L}\frac{g^{2}}{16\pi^{2}}W_{\mu\nu}^{i}\tilde{W}%
^{i,\mu\nu}+\theta_{Y}\frac{g^{\prime2}}{16\pi^{2}}B_{\mu\nu}\tilde{B}^{\mu
\nu}\;.
\end{equation}

If all the fermions are massless, these interactions can be rotated away
thanks to the covariant anomalies. Let us perform a $U(1)^{5}\in G_{F}$
rotation with parameters $N_{f}\alpha_{X}$, $X=q,u,d,\ell,e$, see
Eq.~(\ref{GF}). The Lagrangian of the SM is affected since the currents are
not conserved, and according to Noether's Theorem,%
\begin{equation}
\Delta\mathcal{L}=\sum_{\psi}\alpha_{\psi}\partial_{\mu}J_{\psi}^{\mu}%
=\frac{2\alpha_{q}+\alpha_{u}+\alpha_{d}}{2}\partial_{\mu}J_{PQ}^{\mu}%
+\frac{3\alpha_{q}+\alpha_{\ell}}{2}\partial_{\mu}J_{\mathcal{B}+\mathcal{L}%
}^{\mu}+(\alpha_{\ell}+\alpha_{e}-\alpha_{q}-\alpha_{d}\frac{{}}{{}}%
)\partial_{\mu}J_{E}^{\mu}\;,
\end{equation}
where we have used $\partial_{\mu}J_{\mathcal{B}-\mathcal{L}}^{\mu}%
=\partial_{\mu}J_{Y}^{\mu}=0$. Plugging Eq.~(\ref{AnoCurr}) for the
divergences of the three currents, the variation $\Delta\mathcal{L}$ can be
absorbed into shifts of the anomalous terms as%
\begin{equation}
\left\{
\begin{array}
[c]{l}%
\theta_{C}\rightarrow\theta_{C}-N_{f}\left(  2\alpha_{q}+\alpha_{u}+\alpha
_{d}\right)  \dfrac{{}}{{}}\ ,\\
\theta_{L}\rightarrow\theta_{L}-N_{f}\left(  3\alpha_{q}+\alpha_{\ell}\right)
\dfrac{{}}{{}}\ ,\\
\theta_{Y}\rightarrow\theta_{Y}-\dfrac{1}{3}N_{f}\left(  \alpha_{q}%
+8\alpha_{u}+2\alpha_{d}+3\alpha_{\ell}+6\alpha_{e}\right)  \ .
\end{array}
\right.  \label{thetashifts}%
\end{equation}
There is clearly enough freedom to set $\theta_{C}=\theta_{L}=0$, so the gauge
interactions become $\mathcal{CP}$-invariant.

The situation changes once $G_{F}$ is broken by the Yukawa couplings since all
phase rotations are no longer permitted. To see this, first note that without
loss of generality, the $G_{F}$ symmetry permits to express generic Yukawa
couplings in the gauge eigenstate basis as%
\begin{equation}
\mathcal{L}_{\text{mass}}=-v(\bar{u}_{R}\mathbf{Y}_{u}u_{L}+\bar{d}%
_{R}\mathbf{Y}_{d}d_{L}+\bar{e}_{R}\mathbf{Y}_{e}e_{L})+h.c.\ \overset{G_{F}%
}{\rightarrow}-\bar{u}_{R}\mathbf{m}_{u}u_{L}-\bar{d}_{R}\mathbf{m}_{d}%
V_{CKM}^{\dagger}d_{L}-\bar{e}_{R}^{I}\mathbf{m}_{e}e_{L}+h.c.\;,
\end{equation}
where $v$ is the electroweak vacuum expectation value, $\mathbf{m}_{u,d,e}$
are diagonal matrices, and $V_{CKM}$ is the Cabibbo-Kobayashi-Maskawa matrix.
The usual convention is to ask for real and positive fermion mass terms, i.e.,
$\arg\det\mathbf{m}_{u,d,e}=\arg\det V_{CKM}=0$\footnote{Notice that
conventional rephasing of the CKM matrix are irrelevant since they can be
achieved rotating right and left-handed fields by opposite phases.}. Assuming
$v$ is real, the $\alpha_{i}$ must thus satisfy%
\begin{equation}
\arg\det\mathbf{Y}_{u}=N_{f}(\alpha_{q}+\alpha_{u})\;,\;\arg\det\mathbf{Y}%
_{d}=N_{f}(\alpha_{q}+\alpha_{d})\;,\;\arg\det\mathbf{Y}_{e}=N_{f}%
(\alpha_{\ell}+\alpha_{e})\;. \label{SVDphase}%
\end{equation}
Eliminating $\alpha_{u,d,e}$, the shifts in the anomalous couplings generated
when enforcing real fermion masses are%
\begin{equation}
\left\{
\begin{array}
[c]{l}%
\theta_{C}\rightarrow\theta_{C}-\arg\det\mathbf{Y}_{u}-\arg\det\mathbf{Y}%
_{d}\dfrac{{}}{{}}\ ,\\
\theta_{L}\rightarrow\theta_{L}-N_{f}\left(  3\alpha_{q}+\alpha_{\ell}\right)
\dfrac{{}}{{}}\ ,\\
\theta_{Y}\rightarrow\theta_{Y}+N_{f}(3\alpha_{q}+\alpha_{\ell})-\dfrac{8}%
{3}\arg\det\mathbf{Y}_{u}-\dfrac{2}{3}\arg\det\mathbf{Y}_{d}-2\arg
\det\mathbf{Y}_{e}\ .
\end{array}
\right.
\end{equation}

No choice of $\alpha_{q}$ and $\alpha_{\ell}$ permits to remove both
$\theta_{L}$ and $\theta_{Y}$, but since the latter is harmless, we are free
to choose $\alpha_{\ell}+3\alpha_{q}=\theta_{L}/N_{f}$ and remove the
$\theta_{L}$ term. This freedom is clearly reminiscent of the invariance of
$\mathcal{L}_{\text{mass}}$ under the anomalous $U(1)_{\mathcal{B}%
+\mathcal{L}}$. Once this is done, there still remains a one-parameter freedom
in the choice of $3\alpha_{q}-\alpha_{\ell}$, this time reminiscent of the
invariance under the non-anomalous $U(1)_{\mathcal{B}-\mathcal{L}}$, which
thus cannot affect the $\theta$ terms. In contrast to $\theta_{L}$, the
requirement of real quark masses unambiguously freezes the $\theta_{C}$
anomalous interactions. This is the origin of the famous \textbf{strong
$\mathcal{CP}$ puzzle}: experimentally, the non-observation of an electric
dipole moment for the neutron sets the very strict bound~\cite{Abel:2020pzs} 
on the combination%
\begin{equation}
\theta_{eff}=\theta_{C}-\arg\det\mathbf{Y}_{u}-\arg\det\mathbf{Y}_{d}%
\lesssim10^{-10}\;. \label{ThetaEff}%
\end{equation}
With $\theta_{C}$ originating from the non-perturbative vacuum of the
$SU(3)_{C}$ gauge theory, and $\arg\det\mathbf{Y}_{u,d}$ from the Higgs
coupling to quarks, both a priori $\mathcal{O}(1)$, such a near-perfect
cancellation is unacceptable. Indeed, one could live with some parameters
being accidentally small. After all, the hierarchies in the Yukawa couplings
are quite strong. But, it is something else to accept such a strict
fine-tuning between two unrelated sectors of the Standard Model. There are
various approaches to this problem, most notably the axion mechanism which
would also provide a natural candidate for dark matter. For a recent brief
review, see e.g. Ref.~\cite{Smith:2024uer}; for a comprehensive review, see
e.g. Ref.~\cite{DiLuzio:2020wdo}.\pagebreak

\section{Conclusion}

The grand tour of the many forms of the chiral anomaly is now complete.
Central to our presentation is the derivation of a master equation for the
chiral anomaly, in the form of explicit expressions for the covariant
divergences of the triangle, box, and pentagon diagrams taken along any of
their external legs. Those expressions are generic in the sense that all the
inherent ambiguities are accounted for via a set of essentially five free
parameters. From them, any specific form of the chiral anomaly can be obtained
by imposing appropriate conditions, and this sums up to setting these
parameters to some specific values.

Now, as a matter of principle, we should stress that this core philosophy is
not in itself new. For example, it is stated by Weinberg in his book,
Ref.~\cite{Weinberg:1996kr}, that any form of the chiral anomaly can be
obtained by an appropriate choice of momentum routing, which is precisely what
our parameters embody. Nevertheless, this philosophy has never truly been set
up in practice, maybe because it was though it would be too complicated to be
truly useful. Our result show that the converse actually holds, and the final
picture is both simple and enlightening. Further, if one skips all the
phenomenological detours, the core of the reasoning is very straightforward
and compact, but permits to go well beyond the basic concepts usually covered
in introductory lectures. That was our initial goal.

Working out an explicit expression for the most generic chiral anomaly also
sheds light on several aspects. On the technical side, it is compulsory to
clarify the role of the fermion mass. This delicate issue is usually
circumvented because it is less relevant for the triangle diagram. Yet, the
nature of the $m\rightarrow0$ limit, especially in conjunction with
Sutherland-Veltman theorem, is a fundamental aspect of the chiral anomaly and
is truly critical to build a complete understanding. In this respect, let us
stress three points:

\begin{itemize}
\item The chiral anomaly represents the impact of quantization on the chiral
symmetry. As such, it cannot be confined to either the UV or the IR, but
necessarily involves both ends of the energy spectrum. From the UV, it gathers
its characteristic ambiguities which are to be fixed via physical conditions,
in the same spirit as in the renormalization program. From the IR, it gets
obstructions making it impossible to impose conditions in which all the
symmetries survive. Those take the form of IR singularities if the fermion is
massless, or pseudoscalar loop amplitudes if it is massive, in agreement with
Sutherland-Veltman theorem. As we have seen either directly or via the WZW
construct, this duality is crucial to fully understand $\pi^{0}\rightarrow
\gamma\gamma$.

\item A consistent treatment of the chiral anomaly requires to go beyond the
simple triangle diagram. Actually, most analyses do not even consider this
diagram but rather its divergence, that is, the triangle amplitude contracted
with an external momentum. This is insufficient for two reasons. First, the
box diagrams do play an important role, bringing additional UV ambiguities
with which one can tune the cubic and quartic terms of the anomaly. Second,
the original triangle amplitude is needed, not only its divergence, to
construct the covariant divergence of the box amplitudes. In practice,
fortunately, the final results for the triangle and box amplitudes are rather
simple. That for the pentagon diagram is more delicate, but becomes trivial if
the fermion is massive.

\item The abelian, singlet, and Bardeen anomalies are the only ones for which
the Sutherland-Veltman theorem can holds. This means that if the fermion is
massive, the various axial and vector triangle, box, and pentagon diagrams all
cancel exactly at the leading order in the inverse mass expansion. In this
quite accidental situation, the anomaly can then be obtained from the
non-anomalous pseudoscalar triangle, box, and pentagon diagrams. If the
fermion is massless, there is then no pseudoscalar loop diagrams but the
Sutherland-Veltman theorem no longer holds because the axial and vector loops
have IR singularities, leading to exactly the same expressions for the
corresponding anomalies.
\end{itemize}

On the more conceptual side, working at the diagrammatic level proves
rewarding because of the explicit nature of the various expressions and
derivations. For instance, the precise connection between the Bardeen
counterterms and Bose symmetry is usually not apparent, but emerges naturally
if one looks at the effective vertices involving three, four, or five vector
and axial currents. Similarly, the diagrammatic approach permits to work out
what happens to the other currents when one of them is shifted by some
arbitrary polynomial. This would be difficult to tackle in the differential
formalism. Yet, it sheds a complementary light on the covariant anomaly,
arising for the specific Bardeen-Zumino shift. Further, our systematic
avoidance of the differential formalisms certainly helps in making the
connection transparent between diagrams and formal properties. Altogether, it
is quite remarkable that the vast majority of concepts related to the chiral
anomaly turns out to be expressible both simply and straightforwardly, and
with a clear diagrammatic interpretation.

Before closing, let us suggest those using this work as an introduction to the
subject a few interesting directions to turn to:

\begin{itemize}
\item First and foremost, there are several aspects lying a bit out of our
main line of discussion that were very succinctly discussed. For instance, the
path integral approach and the index theorem would deserve further
exploration, as well as instantons and their applications.

\item At this stage, it is certainly worth to switch to the differential
language, and move on to more advanced topics like for example the whole chain
of descent equations, or higher form symmetries.

\item One best understands the peculiarities of the chiral anomaly by
comparing it to others, starting with the scale anomaly. In this context, it
is also worth to consider the chiral anomaly in curved space-time.

\item Finally, the most conspicuous absent in these notes is the axion,
introduced to solve the strong CP problem. The chiral anomaly is central to
its phenomenology, whether for its dynamics when added to the mesonic degrees
of freedom, or for its cosmological impacts for example via axion strings.
\end{itemize}

In conclusion, if there is one thing to gather from this work, it is that the
chiral anomaly, and all its forms and properties, can be understood rather
simply and compactly, without circumventing its many intricacies and subtleties.
Such a precise and solid foundation will undoubtedly prove useful in the future,
as even the oldest of all the anomalies certainly keeps many fascinating
aspects and manifestations yet to be discovered.

\subsubsection*{Acknowledgements:}

This work is supported by the IN2P3 Master project \textquotedblleft Axions
from Particle Physics to Cosmology\textquotedblright, and from the French
National Research Agency (ANR) in the framework of the \textquotedblleft
GrAHal\textquotedblright project (ANR-22-CE31-0025).

\appendix

\section{General expression for the anomaly\label{AppChiral}}

In this section, we give the most general forms of the chiral anomaly, for
both the massless and massive case. The quadratic terms have been already
given in the text, but for completeness
\begin{subequations}
\begin{align}
D_{\alpha}^{L}\mathcal{T}_{LLL}^{\alpha\beta\gamma,abc}  &  =-i\frac
{\mathcal{I}_{3}d^{abc}}{16\pi^{2}}a_{1}\varepsilon^{\beta\gamma\mu\nu}%
q_{1\mu}q_{2\nu}\ ,\\
D_{\beta}^{L}\mathcal{T}_{LLL}^{\alpha\beta\gamma,abc}  &  =-i\frac
{\mathcal{I}_{3}d^{abc}}{16\pi^{2}}a_{2}\varepsilon^{\gamma\alpha\mu\nu
}q_{1\mu}q_{2\nu}\ ,\\
D_{\gamma}^{L}\mathcal{T}_{LLL}^{\alpha\beta\gamma,abc}  &  =-i\frac
{\mathcal{I}_{3}d^{abc}}{16\pi^{2}}\left(  1-a_{1}-a_{2}\right)
\varepsilon^{\alpha\beta\mu\nu}q_{1\mu}q_{2\nu}\ .
\end{align}
\end{subequations}
To get the corresponding result for the $AVV$ and $AAA$ massive triangle case,
it suffices to drop the $1$ in $1-a_{1}-a_{2}$ and substitute $a_{i}%
\rightarrow\tilde{a}_{i}$. This $1$ originates from the IR singularities which
are automatically regulated once the fermion is massive. All three massive
divergences can then vanish if $\tilde{a}_{i}=0$, in agreement with
Sutherland-Veltman theorem, but the anomaly cannot be discarded because of the
pseudoscalar loop contributions. For the specific $AVV$ and $AAA$ triangle,
remember also to multiply the whole by $\pm2$ since the chiral result involves
$P_{L}=(1-\gamma_{5})/2$.

For the cubic terms, the covariant divergences are given by%
\begin{subequations}
\begin{align}
D_{\alpha}^{L}\mathcal{T}_{LLLL}^{\alpha\beta\gamma\delta,abcd}  &
=-iq_{1\alpha}\mathcal{T}_{LLLL}^{\alpha\beta\gamma\delta,abcd}-gf^{ade}%
\mathcal{T}(L_{q_{1}+q_{4}}^{\delta,e}L_{q_{2}}^{\beta,b}L_{q_{3}}^{\gamma
,c})_{a^{23}}\nonumber\\
&  \ \ \ -gf^{ace}\mathcal{T}(L_{q_{1}+q_{3}}^{\gamma,e}L_{q_{2}}^{\beta
,b}L_{q_{4}}^{\delta,d})_{a^{24}}-gf^{abe}\mathcal{T}(L_{q_{1}+q_{2}}%
^{\beta,e}L_{q_{3}}^{\gamma,c}L_{q_{4}}^{\delta,d})_{a^{34}}\ ,\\
D_{\beta}^{L}\mathcal{T}_{LLLL}^{\alpha\beta\gamma\delta,abcd}  &
=-iq_{2\beta}\mathcal{T}_{LLLL}^{\alpha\beta\gamma\delta,abcd}-gf^{bae}%
\mathcal{T}(L_{q_{2}+q_{1}}^{\alpha,e}L_{q_{3}}^{\gamma,c}L_{q_{4}}^{\delta
,d})_{a^{34}}\nonumber\\
&  \ \ \ \ -gf^{bce}\mathcal{T}(L_{q_{2}+q_{3}}^{\gamma,e}L_{q_{1}}^{\alpha
,a}L_{q_{4}}^{\delta,d})_{a^{14}}-gf^{bde}\mathcal{T}(L_{q_{2}+q_{4}}%
^{\delta,e}L_{q_{1}}^{\alpha,a}L_{q_{3}}^{\gamma,c})_{a^{13}}\ ,\\
D_{\gamma}^{L}\mathcal{T}_{LLLL}^{\alpha\beta\gamma\delta,abcd}  &
=-iq_{3\gamma}\mathcal{T}_{LLLL}^{\alpha\beta\gamma\delta,abcd}-gf^{cae}%
\mathcal{T}(L_{q_{3}+q_{1}}^{\alpha,e}L_{q_{2}}^{\beta,b}L_{q_{4}}^{\delta
,d})_{a^{24}}\nonumber\\
&  \ \ \ \ -gf^{cbe}\mathcal{T}(L_{q_{3}+q_{2}}^{\beta,e}L_{q_{1}}^{\alpha
,a}L_{q_{4}}^{\delta,d})_{a^{14}}-gf^{cde}\mathcal{T}(L_{q_{3}+q_{4}}%
^{\delta,e}L_{q_{1}}^{\alpha,a}L_{q_{2}}^{\beta,b})_{a^{12}}\ ,\\
D_{\delta}^{L}\mathcal{T}_{LLLL}^{\alpha\beta\gamma\delta,abcd}  &
=-iq_{4\delta}\mathcal{T}_{LLLL}^{\alpha\beta\gamma\delta,abcd}-gf^{dae}%
\mathcal{T}(L_{q_{4}+q_{1}}^{\alpha,e}L_{q_{2}}^{\beta,b}L_{q_{3}}^{\gamma
,c})_{a^{23}}\nonumber\\
&  \ \ \ \ -gf^{dce}\mathcal{T}(L_{q_{4}+q_{3}}^{\gamma,e}L_{q_{1}}^{\alpha
,a}L_{q_{2}}^{\beta,b})_{a^{12}}-gf^{dbe}\mathcal{T}(L_{q_{4}+q_{2}}^{\beta
,e}L_{q_{1}}^{\alpha,a}L_{q_{3}}^{\gamma,c})_{a^{13}}\ ,
\end{align}
\end{subequations}
where $q_{4}=-q_{1}-q_{2}-q_{3}$, $L^{\alpha,a}=A_{\alpha}^{L,a}T^{a}$, and in
subscript are indicated the corresponding triangle arbitrary parameters,
numbered according to their outgoing momenta. Plugging in the massless
triangle and box amplitudes, Eqs.~(\ref{AmpTri}) and~(\ref{AmpBox}), we find
\begin{subequations}
\label{AppBox}%
\begin{align}
D_{\alpha}^{L}\mathcal{T}_{Full,LLLL}^{\alpha\beta\gamma\delta}  &
=-\frac{\mathcal{I}_{3}\varepsilon^{\beta\gamma\delta\mu}}{16\pi^{2}}(%
\begin{array}
[c]{ccc}%
q_{1} & q_{2} & q_{3}%
\end{array}
)_{\mu}\cdot V_{1}\cdot V_{df},\\
D_{\beta}^{L}\mathcal{T}_{Full,LLLL}^{\alpha\beta\gamma\delta}  &
=-\frac{\mathcal{I}_{3}\varepsilon^{\alpha\gamma\delta\mu}}{16\pi^{2}}(%
\begin{array}
[c]{ccc}%
q_{1} & q_{2} & q_{3}%
\end{array}
)_{\mu}\cdot V_{2}\cdot V_{df}\ ,\\
D_{\gamma}^{L}\mathcal{T}_{Full,LLLL}^{\alpha\beta\gamma\delta}  &
=-\frac{\mathcal{I}_{3}\varepsilon^{\alpha\beta\delta\mu}}{16\pi^{2}}(%
\begin{array}
[c]{ccc}%
q_{1} & q_{2} & q_{3}%
\end{array}
)_{\mu}\cdot V_{3}\cdot V_{df}\ ,\\
D_{\delta}^{L}\mathcal{T}_{Full,LLLL}^{\alpha\beta\gamma\delta}  &
=-\frac{\mathcal{I}_{3}\varepsilon^{\alpha\beta\gamma\mu}}{16\pi^{2}}(%
\begin{array}
[c]{ccc}%
q_{1} & q_{2} & q_{3}%
\end{array}
)_{\mu}\cdot V_{4}\cdot V_{df}\ ,
\end{align}
\end{subequations}
where the $SU(N)$ basis is chosen as%
\begin{equation}
V_{df}=\left(  d^{ade}f^{bce}\ ,\ d^{ace}f^{bde}\ ,\ d^{abe}f^{cde}\right)
^{T}\ ,
\end{equation}
and the coefficients are
\begin{subequations}
\begin{align}
V_{1}  &  =\left(
\begin{array}
[c]{ccc}%
1-b_{4}-a_{1}^{24}-a_{1}^{34} & b_{3}-a_{1}^{34} & a_{1}^{24}-b_{2}\\
1-a_{1}^{24}-a_{2}^{24}-a_{1}^{34} & a_{2}^{23}-a_{1}^{34} & a_{2}^{23}%
+a_{1}^{24}+a_{2}^{24}-1\\
1-a_{1}^{24}-a_{1}^{34}-a_{2}^{34} & 1-a_{1}^{23}-a_{1}^{34}-a_{2}^{34} &
a_{1}^{24}-a_{1}^{23}%
\end{array}
\right)  \ ,\\
V_{2}  &  =\left(
\begin{array}
[c]{ccc}%
a_{1}^{14}+a_{2}^{14}+a_{1}^{34}-1 & a_{1}^{34}-a_{2}^{13} & 0\\
a_{1}^{14}+a_{1}^{34}+b_{4}-1 & a_{1}^{34}-b_{3} & b_{2}\\
a_{1}^{14}+a_{1}^{34}+a_{2}^{34}-1 & a_{1}^{13}+a_{1}^{34}+a_{2}^{34}-1 & 0
\end{array}
\right)  \ ,\\
V_{3}  &  =\left(
\begin{array}
[c]{ccc}%
1-a_{1}^{14}-a_{2}^{14}-a_{1}^{24} & 0 & a_{1}^{24}-a_{2}^{12}\\
1-a_{1}^{14}-a_{1}^{24}-a_{2}^{24} & 0 & a_{1}^{12}+a_{1}^{24}+2a_{2}^{24}-1\\
1-b_{4}-a_{1}^{14}-a_{1}^{24} & b_{3} & a_{1}^{24}-b_{2}%
\end{array}
\right)  \ ,\\
V_{4}  &  =\left(
\begin{array}
[c]{ccc}%
-b_{4} & b_{3}-a_{2}^{13} & a_{2}^{12}-b_{2}\\
-b_{4} & b_{3}-a_{2}^{23} & 1-b_{3}-a_{1}^{12}-a_{2}^{23}\\
-b_{4} & a_{1}^{13}+a_{1}^{23}+b_{3}-1 & a_{1}^{23}-b_{3}%
\end{array}
\right)  \ .
\end{align}
\end{subequations}
As explained in the text, the corresponding result for the massive case is
obtained plugging Eq.~(\ref{MassiveAno}) in the covariant divergences of
Eq.~(\ref{AppBox}). This gives the same expressions for $V_{\alpha
,\beta,\gamma,\delta}$ but with all the $1$ and $-1$ terms removed, and
$\tilde{a}_{k}^{ij},\tilde{b}_{i}$ instead of $a_{k}^{ij}$, $b_{i}$.

For the pentagon anomaly, the covariant divergences are
\begin{subequations}
\begin{align}
D_{\alpha}^{L}\mathcal{T}_{LLLLL}^{\alpha\beta\gamma\delta\varepsilon,abcde}
&  =-iq_{1\alpha}\mathcal{T}_{LLLLL}^{\alpha\beta\gamma\delta\varepsilon
,abcde}\nonumber\\
&  \ \ \ -gf^{aeg}\mathcal{T}(L_{q_{1}+q_{5}}^{\varepsilon,g}L_{q_{2}}%
^{\beta,b}L_{q_{3}}^{\gamma,c}L_{q_{4}}^{\delta,d})_{b^{234}}-gf^{adg}%
\mathcal{T}(L_{q_{1}+q_{4}}^{\delta,g}L_{q_{2}}^{\beta,b}L_{q_{3}}^{\gamma
,c}L_{q5}^{\varepsilon,e})_{b^{235}}\nonumber\\
&  \ \ \ -gf^{acg}\mathcal{T}(L_{q_{1}+q_{3}}^{\gamma,g}L_{q_{2}}^{\beta
,b}L_{q_{4}}^{\delta,d}L_{q5}^{\varepsilon,e})_{b^{245}}-gf^{abg}%
\mathcal{T}(L_{q_{1}+q_{2}}^{\beta,g}L_{q_{3}}^{\gamma,c}L_{q_{4}}^{\delta
,d}L_{q5}^{\varepsilon,e})_{b^{345}}\ ,\\
D_{\beta}^{L}\mathcal{T}_{LLLLL}^{\alpha\beta\gamma\delta\varepsilon,abcde}
&  =-iq_{2\beta}\mathcal{T}_{LLLLL}^{\alpha\beta\gamma\delta\varepsilon
,abcde}\nonumber\\
&  \ \ \ -gf^{beg}\mathcal{T}(L_{q_{2}+q_{5}}^{\varepsilon,g}L_{q_{1}}%
^{\alpha,a}L_{q_{3}}^{\gamma,c}L_{q_{4}}^{\delta,d})_{b^{134}}-gf^{bdg}%
\mathcal{T}(L_{q_{2}+q_{4}}^{\delta,g}L_{q_{1}}^{\alpha,a}L_{q_{3}}^{\gamma
,c}L_{q5}^{\varepsilon,e})_{b^{135}}\nonumber\\
&  \ \ \ -gf^{bcg}\mathcal{T}(L_{q_{2}+q_{3}}^{\gamma,g}L_{q_{1}}^{\alpha
,a}L_{q_{4}}^{\delta,d}L_{q5}^{\varepsilon,e})_{b^{145}}-gf^{bag}%
\mathcal{T}(L_{q_{2}+q_{1}}^{\alpha,g}L_{q_{3}}^{\gamma,c}L_{q_{4}}^{\delta
,d}L_{q5}^{\varepsilon,e})_{b^{345}}\ ,\\
D_{\gamma}^{L}\mathcal{T}_{LLLLL}^{\alpha\beta\gamma\delta\varepsilon,abcde}
&  =-iq_{3\gamma}\mathcal{T}_{LLLLL}^{\alpha\beta\gamma\delta\varepsilon
,abcde}\nonumber\\
&  \ \ \ -gf^{ceg}\mathcal{T}(L_{q_{3}+q_{5}}^{\varepsilon,g}L_{q_{1}}%
^{\alpha,a}L_{q_{2}}^{\beta,b}L_{q_{4}}^{\delta,d})_{b^{124}}-gf^{cdg}%
\mathcal{T}(L_{q_{3}+q_{4}}^{\delta,g}L_{q_{1}}^{\alpha,a}L_{q_{2}}^{\beta
,b}L_{q5}^{\varepsilon,e})_{b^{125}}\nonumber\\
&  \ \ \ -gf^{cbg}\mathcal{T}(L_{q_{3}+q_{2}}^{\beta,g}L_{q_{1}}^{\alpha
,a}L_{q_{4}}^{\delta,d}L_{q5}^{\varepsilon,e})_{b^{145}}-gf^{cag}%
\mathcal{T}(L_{q_{3}+q_{1}}^{\alpha,g}L_{q_{2}}^{\beta,b}L_{q_{4}}^{\delta
,d}L_{q5}^{\varepsilon,e})_{b^{245}}\ ,\\
D_{\delta}^{L}\mathcal{T}_{LLLLL}^{\alpha\beta\gamma\delta\varepsilon,abcde}
&  =-iq_{4\delta}\mathcal{T}_{LLLLL}^{\alpha\beta\gamma\delta\varepsilon
,abcde}\nonumber\\
&  \ \ \ -gf^{deg}\mathcal{T}(L_{q_{4}+q_{5}}^{\varepsilon,g}L_{q_{1}}%
^{\alpha,a}L_{q_{2}}^{\beta,b}L_{q_{3}}^{\gamma,c})_{b^{123}}-gf^{dcg}%
\mathcal{T}(L_{q_{4}+q_{3}}^{\gamma,g}L_{q_{1}}^{\alpha,a}L_{q_{2}}^{\beta
,b}L_{q5}^{\varepsilon,e})_{b^{125}}\nonumber\\
&  \ \ \ -gf^{dbg}\mathcal{T}(L_{q_{4}+q_{2}}^{\beta,g}L_{q_{1}}^{\alpha
,a}L_{q_{3}}^{\gamma,c}L_{q5}^{\varepsilon,e})_{b^{135}}-gf^{dag}%
\mathcal{T}(L_{q_{4}+q_{1}}^{\alpha,g}L_{q_{2}}^{\beta,b}L_{q_{3}}^{\gamma
,c}L_{q5}^{\varepsilon,e})_{b^{235}}\ ,\\
D_{\varepsilon}^{L}\mathcal{T}_{LLLLL}^{\alpha\beta\gamma\delta\varepsilon
,abcde}  &  =-iq_{5\varepsilon}\mathcal{T}_{LLLLL}^{\alpha\beta\gamma
\delta\varepsilon,abcde}\nonumber\\
&  \ \ \ -gf^{edg}\mathcal{T}(L_{q_{5}+q_{4}}^{\delta,g}L_{q_{1}}^{\alpha
,a}L_{q_{2}}^{\beta,b}L_{q_{3}}^{\gamma,c})_{b^{123}}-gf^{ecg}\mathcal{T}%
(L_{q_{5}+q_{3}}^{\gamma,g}L_{q_{1}}^{\alpha,a}L_{q_{2}}^{\beta,b}L_{q_{4}%
}^{\delta,d})_{b^{124}}\nonumber\\
&  \ \ \ -gf^{ebg}\mathcal{T}(L_{q_{5}+q_{2}}^{\beta,g}L_{q_{1}}^{\alpha
,a}L_{q_{3}}^{\gamma,c}L_{q_{4}}^{\delta,d})_{b^{134}}-gf^{dag}\mathcal{T}%
(L_{q_{5}+q_{1}}^{\alpha,g}L_{q_{2}}^{\beta,b}L_{q_{3}}^{\gamma,c}L_{q_{4}%
}^{\delta,d})_{b^{234}}\ ,
\end{align}
\end{subequations}
where $q_{5}=-q_{1}-q_{2}-q_{3}-q_{4}$. As explained in the text, we did not
compute the $\mathcal{T}_{LLLLL}^{\alpha\beta\gamma\delta\varepsilon,abcde}$
amplitude for $m=0$, but rely instead on the massive result for 
$\mathcal{\tilde{T}}_{LLLLL}^{\alpha\beta\gamma\delta\varepsilon,abcde}$.
Then, correcting for the pseudoscalar loops and shifting the $\tilde{b}_{i}$
coefficients, we could reconstruct the above divergences. 
Let us express the results as%
\begin{equation}
D_{\phi}^{L}\mathcal{T}_{LLLLL}^{\alpha\beta\gamma\delta\varepsilon
,abcde}=\frac{i\mathcal{I}_{3}\varepsilon^{\alpha\beta\gamma\delta
\varepsilon\backslash\phi}}{16\pi^{2}}V_{dff}\cdot V_{\phi}\ ,
\end{equation}
where $\phi=\alpha,\beta,\gamma,\delta,\varepsilon$ and the notation
$\varepsilon^{\alpha\beta\gamma\delta\varepsilon\backslash\phi}$ means the
index equal to $\phi$ is to be removed. A basis of five-index $SU(N)$
invariants sufficient to describe all divergences can be chosen as:%
\begin{align}
V_{dff}  &  =\left(  f^{aef}d^{bfg}f^{cdg}\ ,\ f^{adf}d^{bfg}f^{ceg}%
\ ,\ d^{afg}f^{bef}f^{cdg}\ ,\ d^{afg}f^{bdf}f^{ceg}\ ,\right. \nonumber\\
&  \ \ \ \ \ \ \ \ \ d^{cfg}f^{aef}f^{bdg},\ d^{cfg}f^{adf}f^{beg}%
\ ,\ d^{dfg}f^{aef}f^{bcg}\ ,\ d^{dfg}f^{acf}f^{beg}\ ,\ \nonumber\\
&  \ \ \ \ \ \ \ \ \ \ \ \left.  d^{efg}f^{adf}f^{bcg}\ ,\ d^{efg}%
f^{acf}f^{bdg}\ ,\ d^{efg}f^{abf}f^{cdg}\right)  \ .
\end{align}
Finally, the coefficients for the five divergences are%
\[
V_{\alpha}=\left(
\begin{array}
[c]{c}%
b_{2,3,4}^{234}-b_{2,3,4}^{245}\\
b_{2,3,4}^{245}-b_{2,3,4}^{235}\\
0\\
0\\
b_{2}^{234}+b_{2,3,4}^{345}-1\\
1-b_{2}^{235}-b_{2,3,4}^{345}\\
b_{2}^{345}-b_{3}^{234}\\
b_{2}^{245}-b_{2}^{345}\\
b_{3}^{235}-b_{2,3,4}^{245}-b_{3}^{345}-b_{4}^{345}+1\\
b_{2}^{245}+b_{4}^{245}+b_{3}^{345}+b_{4}^{345}-1\\
1-b_{2,3,4}^{245}-b_{4}^{345}%
\end{array}
\right)  ,\ V_{\beta}=\left(
\begin{array}
[c]{c}%
0\\
0\\
b_{2,3,4}^{134}-b_{2,3,4}^{145}\\
b_{2,3,4}^{145}-b_{2,3,4}^{135}\\
1-b_{2}^{135}-b_{2,3,4}^{345}\\
b_{2}^{134}+b_{2,3,4}^{345}-1\\
b_{2}^{145}-b_{2}^{345}\\
b_{2}^{345}-b_{3}^{134}\\
b_{2}^{145}+b_{4}^{145}+b_{3}^{345}+b_{4}^{345}-1\\
b_{3}^{135}-b_{2,3,4}^{145}-b_{3}^{345}-b_{4}^{345}+1\\
b_{2,3,4}^{145}+b_{4}^{345}-1
\end{array}
\right)  ,
\]

\begin{equation}
V_{\gamma}=\left(
\begin{array}
[c]{c}%
1-b_{2}^{125}-b_{2,3,4}^{245}\\
b_{2}^{124}+b_{2,3,4}^{245}-1\\
b_{2,3,4}^{145}-b_{2,3,4}^{125}\\
b_{2,3,4}^{124}-b_{2,3,4}^{145}\\
0\\
0\\
b_{3}^{124}-b_{2}^{145}\\
b_{2}^{245}-b_{3}^{124}\\
b_{3}^{124}-b_{2}^{145}-b_{4}^{145}-b_{2,3,4}^{245}+1\\
b_{2,3,4}^{145}-b_{3}^{124}+b_{2}^{245}+b_{4}^{245}-1\\
b_{3}^{124}+b_{3}^{125}-b_{2,3,4}^{145}-b_{2,3,4}^{245}+1
\end{array}
\right)  ,V_{\delta}=\left(
\begin{array}
[c]{c}%
b_{2}^{125}-b_{2}^{123}\\
b_{2}^{123}+b_{2,3,4}^{235}-1\\
b_{2,3,4}^{125}-b_{2,3,4}^{123}\\
b_{2,3,4}^{123}-b_{2,3,4}^{135}\\
b_{3}^{123}-b_{2}^{135}\\
b_{2}^{235}-b_{3}^{123}\\
0\\
0\\
b_{4}^{123}-b_{3}^{235}\\
b_{3}^{135}-b_{4}^{123}\\
b_{4}^{123}-b_{3}^{125}%
\end{array}
\right)  ,V_{\varepsilon}=\left(
\begin{array}
[c]{c}%
b_{2}^{123}+b_{2,3,4}^{234}-1\\
b_{2}^{124}-b_{2}^{123}\\
b_{2,3,4}^{123}-b_{2,3,4}^{134}\\
b_{2,3,4}^{124}-b_{2,3,4}^{123}\\
b_{2}^{234}-b_{3}^{123}\\
b_{3}^{123}-b_{2}^{134}\\
b_{3}^{124}-b_{3}^{234}\\
b_{3}^{134}-b_{3}^{124}\\
b_{3}^{124}-b_{4}^{123}\\
b_{4}^{123}-b_{3}^{124}\\
b_{3}^{124}-b_{4}^{123}%
\end{array}
\right)  \ ,
\end{equation}
where $b_{2,3,4}^{ijk}\equiv b_{2}^{ijk}+b_{3}^{ijk}+b_{4}^{ijk}$. Again, the
corresponding result for the massive case is obtained by removing all the $1$
and $-1$ terms and substituting $b_{l}^{ijk}\rightarrow\tilde{b}_{l}^{ijk}$.

\end{document}